# DIFFUSION−CONTROLLED GROWTH OF PHASES IN METAL−TIN SYSTEMS RELATED TO MICROELECTRONICS PACKAGING

A Thesis

Submitted for the degree of

## DOCTOR OF PHILOSOPHY

### IN THE FACULTY OF ENGINEERING

*by*

**Varun A Baheti**

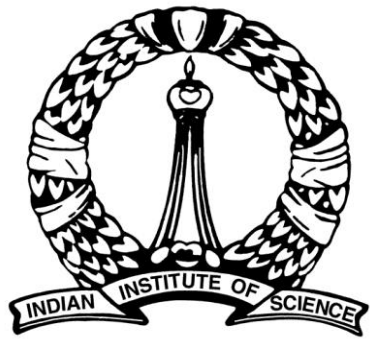

**Materials Engineering**
**Indian Institute of Science**
**BANGALORE 560012**

**JULY 2017**

*Dedicated to*

*My Parents*

*Madhuri Baheti*
*&*
*Arvind Baheti*

**Declaration of Originality**

I, Varun Arvind Baheti, hereby declare that I am the sole author of this thesis. I authorize Indian Institute of Science (IISc) Bangalore to lend this thesis to other institutions or individuals for the purpose of scholarly research. All the resources used or quoted by me are indicated and acknowledged with proper referencing in this thesis.

[Varun A. Baheti]

## Acknowledgements

In the title page, although only my name is present, but in fact, this thesis dissertation would not have been possible without the expert guidance of my esteemed supervisors, Dr. Aloke Paul (AP) and Dr. Praveen Kumar (PK). Although it is difficult to express my gratitude in words, I specially thank them for their concern. They had been so caring and patient and always there for me to give their valuable suggestions in professional life and beyond. I would also like to take this opportunity to express my gratitude and feeling of admiration to Dr. Ravi Raju, one of my M.E. thesis supervisors. Along with my guides, I am very grateful to Dr. Sanjay Kashyap for helping me with TEM, which has resulted in an immense contribution to my thesis, and Dr. Kamanio Chattopadhyay (popularly known as KC), who guided me on the TEM analysis as if my third guide.

AP is the strongest person I know. In spite of ups & down in his health, he has never let me know or feel about it, as he is always there for his inputs and useful discussions in personal meetings or even over internet. I find his writing to be in such a lucid manner that it feels like a face to face discussion while reading, and one such example is the recent handbook chapters written by him. I find myself very lucky and fortunate for my presence in AP group during preparation of his textbook and handbook on diffusion, which has benefited and enabled me to learn a lot of new diffusion concepts beyond classroom teaching. He has given me full freedom to think and act independently on various experimental and theoretical aspects related to my thesis. He has made it possible for me to finish work and write this thesis within 5 years, and it is very difficult for me to express my sincere gratitude in words towards him for his constant support throughout various crucial stages during my PhD. He is at first a wonderful human being with a pure soul and a very cool person, even before his role as my guide.

In spite of being very busy, PK has been able to give time. I have learnt more in personal discussions with him when he explained about basic concepts and cleared my doubts which aroused in the classroom. I got the opportunity to learn a lot from him. He taught me to make meticulous plans for future research work, but I couldn't learn it perfectly. I feel that his expectations are very high from his students, which encouraged to me put my best efforts for this thesis. His suggestions in periodic joint meetings have been very useful to enhance the analysis and quality of the work further. I thank him from the bottom of my heart, especially for revising this thesis with several important minute corrections and his valuable suggestions, which have enhanced its readability.

My sincere thanks to all the Professors of the department of materials engineering, IISc. Their class lectures and personal discussions were greatly helpful in comprehending the concepts of materials engineering. I must also thank all those who helped me in this work. I am thankful to Dr. Ashok M. Raichur and his group members for DI water which marked the start of every electroplating experiment. I especially thank my lab mates Anshu and Raja (Esakkiraja Neelamegan) for saving 1000 hrs experiment. I would like to express my sincere gratitude to Mrs. Vandana, Mrs. Sneha and Mr. Chandan for helping me in obtaining EPMA slots, and most importantly Raja as he was always there to help me out, with his kind words, "First you finish, since once I start using, later I don't want anyone to come in between!"; as he always had to put bigger EPMA line scans of longer durations. I would like to acknowledge the help of Dr. John M. Sosa, Ohio State University, Columbus, USA and his team for their kind help and support for the use of MIPAR. I also acknowledge the support of staff members in MNCF (Micro Nano Characterization Facility) and AFMM (Advanced Facility for Microscopy and Microanalysis) facilities at IISc for providing the usage of FIB and TEM, respectively. Special acknowledgement to Mrs. Suma for teaching me FIB and always being there for all the help I needed. I am grateful to my colleague Sandeep and my lab mate Raja for their valuable contribution to the Ni/Sn RT (room temperature) ageing experiment of 2.5 years. I thank my friend Rajib Kalsar for attempting several EBSD scans on my samples; however, the scans could not give successful results. I thank my all lab mates: Hari, Soma, Sai, Somasurendra, Soumitra, Sangeeta, Kiruthika, Tanaji, Urbi, Raja, Mangesh, Anshu, Ujjval, Bibhu and Sarfaraj for the freedom that I enjoyed in this lab and for all the help they have done throughout my research work. I greatly acknowledge the motivation, help and support I got from my parents and younger sisters throughout my career and during this PhD project. Thanks to my B.Tech. Professors – A. Sharma, A.K. Bhargava, K.L. Narang, M. Mandira, P.R. Soni, R.K. Duchaniya, Y.V.S.S. Prasad, and all others for having faith in me and motivating me for higher studies. Seven years of life at IISc would not have been so comfortable and fun–filled without my IISc hostel and department friends: Nirmal, Deshant, Sharun, Sourabh, Shekhar, Shailendra, Manu, Sambit, Girish, and others. I always cherish the friendship of my friends outside IISc: A.K. Choudhary, D. Kohli, H. Singh, P. Tayade, P.K. Barnwal and R. Kaushik. I would like to thank all who have been helpful directly or indirectly in my research work and peaceful stay here at IISc, which finally helped me in giving this shape to the thesis. Finally, I thank Almighty God for giving me various challenges in life and the courage to face them all, and come up with fruitful results at the end. This Thesis is one of them!

**Abstract**

The electro–mechanical connection between under bump metallization (UBM) and solder in flip–chip bonding is achieved by the formation of brittle intermetallic compounds (IMCs) during the soldering process. These IMCs continue to grow in the solid–state during storage at room temperature and service at an elevated temperature leading to degradation of the contacts. In this thesis, the diffusion–controlled growth mechanism of the phases and the formation of the Kirkendall voids at the interface of UBM (Cu, Ni, Au, Pd, Pt) and Sn (bulk/electroplated) are studied extensively.

Based on the microstructural analysis in SEM and TEM, the presence of bifurcation of the Kirkendall marker plane, a very special phenomenon discovered recently, is found in the Cu–Sn system. The estimated diffusion coefficients at these marker planes indicate one of the reasons for the growth of the Kirkendall voids, which is one of the major reliability concerns in a microelectronic component. Systematic experiments using different purity of Cu are conducted to understand the effect of impurities on the growth of the Kirkendall voids. It is conclusively shown that increase in impurity enhances the growth of voids.

The growth rates of the interdiffusion zone are found to be comparable in the Cu–Sn and the Ni–Sn systems. EPMA and TEM analyses indicate the growth of a metastable phase in the Ni–Sn system in the low temperature range. Following, the role of Ni addition in Cu on the growth of IMCs in the Cu–Sn system is studied based on the quantitative diffusion analysis. The analysis of thermodynamic driving forces, microstructure and crystal structure of $Cu_6Sn_5$ shed light on the atomic mechanism of diffusion. It does not change the crystal structure of phases; however, the microstructural evolution, the diffusion rates of components and the growth of the Kirkendall voids are strongly influenced in the presence of Ni. Considering microstructure of the product phases in various Cu/Sn and Cu(Ni)/Sn diffusion couples, it has been observed that (i) phases have smaller grains and nucleate repeatedly, when they grow from Cu or Cu(Ni) alloy, and (ii) the same phases have elongated grains, when they grow from another phase.

A difference in growth rate of the phases is found in bulk and electroplated diffusion couples in the Au–Sn system. The is explained in $AuSn_4$ based on the estimated tracer diffusion coefficients, homologous temperature of the experiments, grain size distribution and crystal structure of the phase. The growth rates of the phases in the Au–Sn system are compared with the Pd–Sn and the Pt–Sn systems. Similar to the Au–Sn system, the growth rate of the interdiffusion zone is found to be parabolic in the Pd–Sn system; however, it is linear in the Pt–Sn system. Following, the effect of addition of Au, Pd and Pt in Cu is studied on growth rate of the phases. An analysis on the formation of the Kirkendall voids indicates that the addition of Pd or Pt is deleterious to the structure compared to the addition of Au. This study indicates that formation of voids is equally influenced by the presence of inorganic as well as organic impurities.

# List of Contents

# List of Figures

# List of Tables

# Chapter 1

# Introduction

In the 1960's, IBM (International Business Machines) Corporation first introduced the flip–chip technology which is widely used for IC (integrated circuit) packaging. In this packaging technology, the chip is flipped 'upside–down' on the substrate and a direct connection is achieved between the chip and the substrate or a chip to another chip. A schematic of flip–chip joint is shown in Figure 1.1. During soldering, in these joints, it is very common to have a formation of intermetallic phase layer (which is often brittle) at the interface of the solder and the components being joined. As explained latter, quite often these intermetallic layer(s) become the bottleneck in improving the reliability of these solder joints.

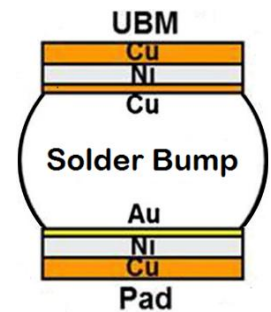

Figure 1.1: Schematic illustration of flip–chip joint used in microelectronics packaging [1].

It is well–known that lead (Pb) and its compounds being toxic substances are harmful for the environmental and human health; and therefore since past few decades, the use of lead–free solder (LFS) bumps is very common in the modern electronics industry. The LFS are mostly Sn–based, such as Sn–Ag–Cu (SAC) alloys. The use of SAC has an added advantage over Sn–Pb solders, especially in terms of their improved mechanical properties, which makes them suitable under harsh environments prevalent in automotive and aerospace applications, along with their widespread use in electronics industry.

## 1.1    Motivation for the choice of various material systems

The Cu–Sn system, mainly considered for this research project, finds application in the microelectronics industry, especially with respect to flip–chip or DCA (direct–chip attach) assembly. Here, the region of interest is under bump metallization (UBM) and solder bump (Sn–based) interface due to the formation of brittle intermetallic compounds (IMCs) there. Understanding the growth of IMCs between UBM and solder is important, as in many cases, it controls the electro–mechanical properties of the product. Cu, Ni and Au are the commonly used UBM materials, as shown in Figure 1.1 [1]. Cu is the most common base conductor UBM, which is used for good bonding because of its fast reaction with solder and it also possesses good solderability characteristics. Ni is often used as a reactive–diffusion barrier layer between Cu and solder due to its inherently slower reaction kinetics with Sn–based solders. Au protects the base conductor from corrosion and oxidation and thereby promote solderability; and being a noble metal, it further provides better shelf–life during storage. Pd and Pt, being noble metals like Au, can also serve the same purpose and could be potential replacement for Au. These 5 UBM namely Cu, Ni, Au, Pd and Pt are considered for this research work.

During soldering, storage and the active use of devices, IMCs are formed and grown; and in many cases, they contribute greatly to the failures occurring during operation of the devices. Note that due to the local heating of power components, high temperatures of around 150 °C can be easily reached locally in the electronic devices used for novel applications [2]. At present, because of miniaturization, the volume of the solder and the thicknesses of the metallization layers have become very small. For example, a decade ago, solder bump size was around 90 µm [1], while the same is 20 µm for microbumps as reported recently in 2015 [3]. Therefore, it is possible that during service, the entire solder joint could be converted into IMCs. Further, it is also a well–known fact that the presence of a third element might change the growth kinetics of the phases in a given binary system drastically [4-8] and hence will have a large impact on the reliability of the product. Therefore, the aim of this research work is to examine the growth of IMCs in the binary M–Sn [M = Cu, Ni, Au, Pd, Pt], and ternary Cu(M)–Sn [M = Ni, Au, Pd, Pt] systems to understand to the effect of M content on the growth of phases and the Kirkendall voids in the Cu–Sn system, where M denotes a metal in binary system or an alloying element in Cu for ternary system.

A typically example of Sn–based solder alloy is SAC305 (Sn–3Ag–0.5 Cu, in wt. %). Due to alloying addition in the solder, complexity of quantitative analysis increases in the multicomponent metal–solder system. Just for simplicity, Sn being a major solder constituent and the element of SAC, participating in the formation of intermetallic phases, is chosen instead of the actual solder alloy, to do systematic quantitative analysis and basic understanding of the process. Experiments in the solid–state are ideal for the various M–Sn and Cu(M)–Sn systems considered in this study, as homogeneous phase layers are easily grown. It is well–known that when Sn or solder is used in the liquid–state, some part of the product phase gets separated into the liquid phase, making it impossible to estimate the actual thickness of the product phase(s) and thereby impossible to do the systematic quantitative analysis.

Undoubtedly, the bulk M–Sn systems considered for this work are widely studied in the literature at temperatures $\geq 125\ °C$ [2] to understand the diffusion–controlled growth of the phases, which subsequently control the electro–mechanical performance of an electronic component. Irrespective of a very high number of publications every year, doubts still exist on the growth behaviour of phases mainly because of the different temperatures used in different studies and the temperature dependent growth mechanism particularly in these systems. Moreover, there are difficulties in the joining or proper bonding of bulk diffusion couples at lower temperatures, usually ≤ 100 °C. Therefore, in this study Sn is electroplated (EP) on the UMB substrate due to its application in microelectronics packaging, *i.e.*, first solder bump is EP on UBM and then soldering is achieved by the reflow step [3, 9, 10]. Mainly there are two types of solutions for Sn plating [10]. Doing it from acidic (divalent form) solutions consumes less electricity than alkali (tetravalent form) solutions. Also, acidic solutions can be used for electroplating on circuit boards with patterned photoresist. However, alkali solutions can cause the photoresist to delaminate, which limits their application in microelectronics packaging. Note that in the EP samples bonding happens at the time of electroplating itself by the formation of thin IMC layer in the systems considered in this work. Thus, the electroplating of Sn is done at room temperature (RT) in this work, following the acid plating method [11] due to its industrial relevance. Now, this allows to the study phase evolution at RT to 100 °C also, which has not been possible in earlier studies conducted using bulk condition of samples. For the first time, we have covered the whole temperature range in the solid–state (*i.e.*, RT to 215 °C) for the binary M–Sn systems with the aim of examining the growth of various product phases in these systems and highlighting the growth mechanisms based on the estimated diffusion coefficients. Additionally, the difference in growth behaviour for bulk and electroplated Sn is compared.

## 1.2   Thesis structure

This thesis is a broad study on the growth kinetics of phases in various metal–tin systems related to microelectronics packaging, consisting of 10 chapters. It is structured in the following manner:

Chapter 1 gives a general overview and introduction to this PhD thesis, and the motivation for choosing the research problem. The statement of problem and available studies in literature specific to the various systems are discussed later in their respective chapters.

Chapter 2 describes the experimental conditions and techniques used for this research work. Additionally, an important concept of *incremental diffusion couple* is discussed, which will be often referred in the subsequent chapters.

Chapter 3 demonstrates the difference, if any, between the relations used for the estimation of important diffusion parameters, which are derived following the concentration normalized variable and the composition normalized variable.

Chapter 4 describes the growth behaviour of the product phases in the Cu–Sn and the Ni–Sn systems over a wide temperature range, from room temperature to 215 °C, in the solid–state. Furthermore, the comparison of the growth of the product phases in these two systems is also reported.

Chapter 5 confirms the bifurcation, *i.e.*, presence of the Kirkendall marker plane in both the phases $Cu_3Sn$ and $Cu_6Sn_5$, in the Cu–Sn system. Also, the growth of the Kirkendall voids in the $Cu_3Sn$ phase, which is another important consequence of the Kirkendall effect, is studied systematically by considering different known concentration of impurities in Cu. The growth of voids (along with the brittle phases) is a major source of electro–mechanical failure in electronic components.

Chapter 6 systematically investigates the effect of Ni on the growth of the interfacial product phases between Cu(Ni)/Sn diffusion couples. A detailed analysis is presented on the growth kinetics, crystal structure, formation of the Kirkendall voids, microstructural changes, calculated thermodynamic driving forces and important diffusion parameters in the Cu(Ni)–Sn system.

Chapter 7 compares the temperature dependent evolution of the interdiffusion zone in the Au/Sn bulk and the Au/Sn electroplated diffusion couples.

Chapter 8 describes the growth behaviour of the product phases in the Pd–Sn and the Pt–Sn systems. Furthermore, it is compared with the Au–Sn system.

Chapter 9 investigates the role of Au, Pd and Pt in Cu on the formation of the Kirkendall voids and the growth of the product phases in the Cu–Sn system, and compares the same effect with that of Ni addition in Cu.

Chapter 10 summarizes the major findings and key observations found in this work. They are also concluded for various systems in their respective chapters.

# Chapter 2

# Experimental Techniques

Diffusion couple technique [12] was used to investigate the growth kinetics of phases in the binary M–Sn [M = Cu, Ni, Au, Pd, Pt] and ternary Cu(M)–Sn [M = Ni, Au, Pd, Pt] systems. M stands for a metal in a binary M/Sn couple or an alloying element in a binary Cu(M) alloy, which was used to make a ternary Cu(M)/Sn couple. Two methods were mainly used to fabricate the diffusion couples for this research work. First was by the electrodeposition method only for the binary systems and second was by the conventional method for both the binary and ternary systems, which are discussed later in Section 2.3. Using this diffusion couple technique, two dissimilar metals were brought into an intimate contact for allowing the species to interdiffuse by annealing the couple under high vacuum ($\sim 10^{-4}$ Pa) at a desired temperature for an appropriate time. Post annealing, a diffusion couple was cut along the desired cross–section, covering a maximum length along bonded interface. After cross–sectioning, the interdiffusion zone (IDZ) of a diffusion couple was analysed mainly using: (i) a field emission gun equipped scanning electron microscope (FE–SEM) for imaging and (ii) an electron probe micro–analyser (FE–EPMA) for composition measurements of the various product phases grown across an IDZ.

## 2.1  Electroplating Conditions for Tin (Sn) and Copper (Cu) Deposition

Electroplated (EP) Sn and EP Cu were produced by the electroplating chemicals supplied by Grauer & Weil (Growel), India. Following the suggestions provided in their technical data sheet [11, 13], all the chemicals were mixed in deionized (DI) water as per the composition reported in Table 2.1 and Table 2.2. With the purpose of studying the growth of phases during storage, all the electroplating experiments were conducted in an air–conditioned (AC) room which was maintained at $20 \pm 5$ °C and at a current density of 20 mA/cm$^2$. The thickness of the EP layers was kept in the range of 0.5–1 mm. This was done to ensure that ends of the diffusion couple (*i.e.*, end–members) are not affected by the diffusing components after annealing for the desired time and temperature; otherwise, it will not fulfill the mathematical boundary conditions (*i.e.*, un–affected compositions of end–members) in the relations (as derived in Chapter 3) used for the estimation of diffusion parameters. Sn electroplating solution contains $SnSO_4$, $H_2SO_4$, and stannolume additive and brightener [11]. Sn electroplating bath composition used for one litre volume is listed in Table 2.1. Similarly, commercial Cu plating chemicals used in industries (supplied by Growel) were used with the purpose of studying the effect of Cu electroplating bath composition on the Kirkendall voids formation in the $Cu_3Sn$ phase. Cu electroplating solution contains $CuSO_4$, $H_2SO_4$, HCl, and different cuprobrite additive and brightener [13]. Cu electroplating bath composition used for one litre volume is listed in Table 2.2.

| Chemical | Concentration |
|----------|---------------|
| Stannous Sulphate ($SnSO_4$) | 30 g/L |
| Sulphuric Acid ($H_2SO_4$) | 100 ml/L |
| Stannolume Carrier Additive | 30 ml/L |
| Stannolume Brightener | 3 ml/L |

Table 2.1: Bath composition for pure tin (Sn) electroplating.

| Chemical | Concentration |
|----------|---------------|
| Copper Sulphate ($CuSO_4$) | 225 g/L |
| Sulphuric Acid ($H_2SO_4$) | 35 ml/L |
| Hydrochloric acid (HCl) | 0.256 ml/L |
| Cuprobrite 3006 Make–up | 10 ml/L |
| Cuprobrite 3006 Part A | 0.5 ml/L |
| Cuprobrite 3006 Part B | 0.5 ml/L |

Table 2.2: Bath composition for pure copper (Cu) electroplating.

Two Cu substrates were electroplated with Cu using bath composition listed in Table 2.2. With the purpose of understanding the role of impurities in EP Cu on the growth of the Kirkendall voids, one of them (*i.e.*, Cu/EP–Cu) was heat treated (HT) in a calibrated ($\pm$ 2 °C) vacuum ($\sim 10^{-4}$ Pa) oven at 200 °C for 100 hrs. These two electroplated Cu layers were electroplated with Sn to prepare EP–Cu/EP–Sn diffusion couples and further annealed at 200 °C for 100 hrs. The results of this particular experiment are discussed later in Section 5.3.

## 2.2    Preparation of Starting Materials

In this study, pure elements and alloys prepared from them, both were used. Pure tin (Sn), copper (Cu), commercially pure (CP) Cu, nickel (Ni), gold (Au), palladium (Pd) and platinum (Pt) were used as starting materials. Detailed specifications of the same are listed in Table 2.3.

| Material | Supplier | Purity (wt. %) |
|----------|----------|----------------|
| Sn | Alfa Aesar | 99.99 |
| Cu | Sigma Aldrich | 99.98 |
| Ni | Sigma Aldrich | 99.98 |
| Au | Arora Matthey | 99.95 |
| Pd | Arora Matthey | 99.95 |
| Pt | Arora Matthey | 99.95 |
| Cu | Alfa Aesar | 99.999 |
| Ni | Alfa Aesar | 99.95 |
| CP Cu | Sadar Patrappa Road, Local Market | 98–99 [14, 15] |
| Cu | Alfa Aesar | 99.9 |
| Cu | Alfa Aesar | 99.99 |
| Cu | Alfa Aesar | 99.9999 |

Table 2.3: Specifications of the elements used in making alloys and diffusion couples.

Cu(Ni), Cu(Au), Cu(Pd) and Cu(Pt) alloys were prepared by adding 99.95 wt.% Ni, Au, Pd and Pt, respectively, into 99.999 wt.% Cu. These alloys were produced by melting in an arc melting unit under an argon atmosphere. To ensure homogeneity, the alloys were re–melted 4–5 times by flipping them each time. Following, the Cu(Ni), Cu(Au), Cu(Pd) and Cu(Pt) alloys were homogenized in the

solid–state at 1050 °C, 900 °C, 1050 °C and 1050 °C, respectively, for 50 hrs in a calibrated ($\pm$ 5 °C) vacuum ($\sim 10^{-4}$ Pa) tube furnace. To check the homogeneity of these Cu alloys, the compositions were measured randomly at many places using EPMA. The deviations from the target compositions were found to be within the acceptable limits and the same data are listed in Table 2.4.

| Alloying Element (M) in Cu(M) alloy | At.% of Element M along with deviation ($\pm$) | | | | | | | |
|---|---|---|---|---|---|---|---|---|
| | 0.5 | 1 | 2.5 | 3 | 5 | 7.5 | 8 | 15 |
| Ni | $\pm$ 0.02 | | $\pm$ 0.1 | $\pm$ 0.1 | $\pm$ 0.3 | $\pm$ 0.3 | $\pm$ 0.1 | |
| Au | | $\pm$ 0.06 | | | $\pm$ 0.2 | | $\pm$ 0.2 | |
| Pd | | $\pm$ 0.05 | | | $\pm$ 0.3 | | $\pm$ 0.3 | $\pm$ 0.7 |
| Pt | | $\pm$ 0.04 | | | | | | |

Table 2.4: Average composition of alloying element (M) in various Cu(M) alloys.

Cu(Pd) alloy with 15 at.% Pd in Cu was prepared to evaluate the effect of higher Pd content in Cu on the growth of phases, as discussed later in Chapter 9. Cu(Pt)/Sn diffusion couple could be produced successfully only for 1 at.% Pt in Cu.

## 2.3    Preparation of Diffusion Couples

After the standard metallographic preparation, either a pure bulk metal or the prepared Cu alloy was diffusion coupled with 99.99 wt.% pure Sn for making a bulk diffusion couple. The steps for the preparation of bulk couples can be found in *Section 3.17, Chapter 3, Volume 1* of 'Handbook of Solid State Diffusion' [12]. The inert particles, which acts as the Kirkendall marker, of $TiO_2$ or $Y_2O_3$ were also used in a few couples.

EP diffusion couples were prepared by electroplating a layer of Sn on a metal substrate. EP layers were directly used without any metallographic preparation, since they possess smooth surface.

The experiments conducted in this research work can be broadly divided into 4 different categories, to clarify the utility of both the methods employed to make a diffusion couple and the use of materials discussed above, as follows:

(1)     *Electroplated M/Sn couples*: Metal substrate (M = Cu, Ni, Au, Pd and Pt) was electroplated with Sn. All these M/EP–Sn diffusion couples were either annealed at 50–215 °C or stored up to the maximum time of 912 days, *i.e.*, 2.5 years.

(2)     *Bulk M/Sn couples*: Pure metal (M = Cu, Ni, Au, Pd and Pt) and bulk Sn were coupled. They were annealed at 50–215 °C for the similar time as that of M/EP–Sn couples, for the comparison purpose. Condition of the metal was same in both the types of couples, *i.e.*, bulk and electroplated.

(3)     *Bulk Cu/Sn couples*: CP Cu and 99.9–99.9999 wt.% Cu, a total of 5 purities, were coupled with bulk Sn and annealed at 200 °C, to study the effect of Cu purity (impurity in Cu) on growth of the Kirkendall voids in the $Cu_3Sn$ phase.

(4)     *Bulk Cu(M)/Sn couples*: Binary Cu(M) alloy and bulk Sn were coupled. Cu(Ni)/Sn couples were annealed at 150–200 °C, while Cu(Au)/Sn, Cu(Pd)/Sn and Cu(Pt)/Sn couples were annealed at 200 °C; for the purpose of comparison of the formation of the Kirkendall voids in $(Cu,M)_3Sn$ and the growth kinetics of the product phases with the Cu/Sn binary and Cu(Ni)/Sn ternary diffusion couples.

Room temperature (RT) experiments were conducted at around 25 °C by keeping the samples in a vacuum desiccator. A calibrated ($\pm$ 2 °C) high vacuum ($\sim 10^{-4}$ Pa) oven was used for conducting other experiments at higher temperatures, *i.e.*, 50–215 °C. Time dependent experiments were also conducted in various systems at different temperatures for different ranges of annealing time to examine the nature of growth of various phases.

With the aim of conducting experiments in the solid–state condition, the maximum suitable annealing temperature (*i.e.*, 215 or 200 °C) was selected, since a slight temperature overshoot for a very small time during the heating cycle before stability could otherwise lead to the melting of Sn.

## 2.4   Various Analysis of Diffusion Couples

After diffusion annealing, the diffusion couples were mounted in epoxy resin. For further analysis, they were cross–sectioned and prepared metallographically.

The microstructures of the IDZ in all the diffusion couples were examined using FE–SEM in both BSE (back–scattered electron) and SE (secondary electron) imaging modes, which is very important in particular for studying the growth of the Kirkendall voids, for example, as reported in Chapter 5 of this thesis. Sometimes while imaging in FE–SEM, the grain morphology present in an IDZ was revealed after tuning the contrast–brightness to extreme ranges of black and/or white using BSE mode, for example, *see* Figure 6.8.

The Kirkendall void statistics were estimated using an image analysis software named MIPAR (Materials Image Processing and Automated Reconstruction) [16].

Dual column FIB (focused ion beam), starting from 30 kV with final thinning at 5 kV, was used for the TEM (transmission electron microscope) sample preparation. TEM operating at 300 kV beam energy was employed for acquiring selected–area electron diffraction pattern (DP) along with the corresponding TEM micrographs, viz. dark–field (DF) and bright–field (BF) images. Recorded DPs were indexed using JEMS software, which can be considered as a Java version of Electron Microscopy Simulation. Indexing of DPs acquired from TEM and/or composition measurements done in FE–EPMA confirm the presence of various product phases grown in the IDZ.

The location of inert $TiO_2$ or $Y_2O_3$ particles in various product phases was identified using EDS (energy–dispersive X–ray spectroscopy) with the help of X–ray peak originating from Ti or Y and O.

## 2.5    Concept of Incremental Diffusion Couple

Note that although any experiments using incremental diffusion couples are not conducted in this thesis work; however, it is important to understand the preparation technique of such couples and the underlying concept since this term will be referred quite a few times in the subsequent chapters while discussing the results.

In a few systems, more than one phases are expected to grow (*i.e.*, multiphase growth) in the IDZ as per the equilibrium phase diagram. However, sometimes a few phase(s) might not be able to grow with detectable thickness because of consumption by the other neighbouring phase(s) with a much higher growth rate. In such a case of multiphase growth, to check the growth of other phase(s) usually one of the end–members is removed either by polishing or etching to make an *incremental diffusion couple*, so that other phases are allowed to grow in this couple as per the phase diagram. Sometimes, incremental couples are even prepared by coupling appropriate compositions for the growth of desired phase; for example, Cu and $Cu_6Sn_5$ were coupled for the growth of $Cu_3Sn$, while $Cu_3Sn$ and Sn were coupled for the growth of $Cu_6Sn_5$ by Paul et al. [17]. It should be noted here that the thickness of a single product phase in an incremental diffusion couple is higher than the thickness of the same phase in a multiphase IDZ when annealed for same time at the same temperature [18]. Hence, it might be found in an incremental couple, when the phase is not consumed by the neighbouring phases.

*To summarize, in this thesis: the temperature (RT–215 °C) dependent growth of phases in 5 binary M–Sn systems and the composition (Ni, Au, Pd, Pt) dependent growth of phases in 4 ternary Cu(M)–Sn systems are studied using SEM, EPMA and TEM; along with systematic experiments and analysis to understand the effect of (inorganic and organic) impurities in Cu on the formation of the Kirkendall voids.*

# Chapter 3

# Different approaches for the estimation of diffusion parameters

The interdiffusion coefficients are estimated either following the Wagner's method expressed with respect to the composition (mole or atomic fraction) normalized variable after considering the molar volume variation or the den Broeder's method expressed with respect to the concentration (composition divided by the molar volume) normalized variable. On the other hand, the relations for estimation of the intrinsic diffusion coefficients of components as established by van Loo and integrated diffusion coefficients in a phase with narrow homogeneity range as established by Wagner are currently available with respect to the composition normalized variable only. In this study, we have first derived the relation proposed by den Broeder following the line of treatment proposed by Wagner. Further, the relations for estimation of the intrinsic diffusion coefficients of the components and integrated interdiffusion coefficient are established with respect to the concentration normalized variable, which were not available earlier. The veracity of these methods is examined based on the estimation of data in Ni–Pd, Ni–Al and Cu–Sn systems. Our analysis indicates that both the approaches are logically correct and there is small difference in the estimated data in these systems although a higher difference could be found in other systems. The integrated interdiffusion coefficients with respect to the concentration (or concentration normalized variable) can only be estimated considering the ideal molar volume variation. This might be drawback in certain practical systems.

---

This chapter is written based on the article:
[1] V.A. Baheti and A. Paul: Development of different methods and their efficiencies for the estimation of diffusion coefficients following the diffusion couple technique, *accepted for publication*, Acta Materialia (2018).

### 3.1    Introduction and Statement of the Problem

Diffusion couple technique is a tool to study diffusion in inhomogeneous materials by coupling dissimilar materials at the temperature of interest [18]. As an added advantage, one can mimic the heterogeneous material systems in application for understanding the phase transformations and the growth of product phases by diffusion–controlled process, which control the various physico–mechanical properties and reliability of the structure [18]. This is even emerged as a research tool to screen a very wide range of compositions optimizing physical and mechanical properties for the development of a new material from only very few samples, which otherwise would need a large volume of samples and unusually high man–time [19].

The two major developments to establish this method as an efficient research tool for diffusion studies can be stated as:

(i)    The relation developed by Matano [20] for the estimation of composition dependent interdiffusion coefficients. It was developed by simplifying the partial differential equation of Fick's second law [21] to ordinary differential equation utilizing the Boltzmann parameter [22]. This is known as the Matano–Boltzmann analysis.

(ii)    The Darken–Manning relation [23, 24] developed based on the Kirkendall effect [25] for the estimation of the intrinsic diffusion coefficients (influenced by thermodynamic driving force) and tracer diffusion coefficients (indicating the self–diffusion coefficients) of components [18].

However, the use of Matano–Boltzmann method for the estimation of the interdiffusion coefficients $\widetilde{D}(C_B^*)$ as a function of concentration $(C_B)$ introduces error in calculations in most of the practical systems. This relation is expressed as

$$\tilde{D}(C_B^*) = -\frac{1}{2t\left(\frac{dC_B}{dx}\right)}\left[x^*(C_B^* - C_B^-) - \int_{x^-\infty}^{x^*}(C_B - C_B^-)\,dx\right] \tag{3.1}$$

where $t$ is the annealing time and $x^* = (x^* - x_o)$ since the location parameter is measured with respect to $x_o$, *i.e.*, the location of Matano (or initial contact) plane. The asterisk $(*)$ represents the location of interest.

Therefore, one of the very important pre–requisites for the use of Matano–Boltzmann analysis is the need to locate the Matano plane. This can be followed only when the molar volume varies ideally with composition or if we consider it as constant. However, it does not fulfill in most of the practical systems and hence, it is almost impossible to locate $x_o$ exactly. As explained mathematically in Reference [6], it gives different values of $x_o$ when estimated using different components and the difference between them is exactly the same as expansion (for the positive deviation of molar volume) or shrinkage (for the negative deviation of molar volume) of the diffusion couple in a binary system.

To circumvent this problem, mainly two relations are established independently:

(i)     The relation developed by Wagner [26] following an analytical approach based on simple algebraic equations, which is expressed as

$$\tilde{D}(Y_{N_B}^*) = \frac{V_m^*}{2t\left(\frac{dY_{N_B}}{dx}\right)}\left[\left(1 - Y_{N_B}^*\right)\int_{x^-\infty}^{x^*}\frac{Y_{N_B}}{V_m}\,dx + Y_{N_B}^*\int_{x^*}^{x^+\infty}\frac{(1 - Y_{N_B})}{V_m}\,dx\right] \tag{3.2}$$

where $Y_{N_B} = \frac{N_B - N_B^-}{N_B^+ - N_B^-}$ is the composition normalized variable. $N_B$ is the composition in mole (or atomic) fraction of component B. $V_m$ is the molar volume. "–" and "+" represents the un–reacted left– and right–hand side of the diffusion couple.

(ii)    The relation developed by den Broeder [27] by extending the Matano–Boltzmann analysis following a graphical approach, which is expressed as

$$\widetilde{D}\left(Y_{C_B}^*\right) = \frac{1}{2t\left(\frac{dY_{C_B}^*}{dx}\right)}\left[\left(1 - Y_{C_B}^*\right)\int_{x^-\infty}^{x^*} Y_{C_B}\, dx + Y_{C_B}^*\int_{x^*}^{x^{+\infty}}\left(1 - Y_{C_B}\right) dx\right] \qquad (3.3)$$

where $Y_{C_B} = \frac{C_B - C_B^-}{C_B^+ - C_B^-}$ is the concentration normalized variable. $C_B\left(= \frac{N_B}{V_m}\right)$ is the concentration of component B.

The main advantage of using any of the above two relations can be understood immediately that there is no need to locate the Matano plane, and hence it can also consider the actual variation of molar volume with composition.

Out of all the methods, the Wagner's method [26] draws a special attention, since in the same manuscript, the author established the concept of the integrated interdiffusion coefficient ($\widetilde{D}_{int}$) for the estimation of the diffusion coefficients in line compounds or the phases with narrow homogeneity range in which concentration gradient cannot be measured. Immediately after that, van Loo [28, 29] proposed the relations for intrinsic $D_i$ (or tracer $D_i^*$) diffusion coefficients of components, in which the Matano plane is not necessary to locate. Much later, Paul [6] derived these relations by extending the Wagner's approach. Both of these relations are derived with the composition normalized variable $Y_{N_B}$.

To summarize, the relations for the estimation of interdiffusion and integrated diffusion coefficients (derived by Wagner [26]), and intrinsic diffusion coefficients (derived by van Loo [29] and Paul [6]) are expressed with respect to composition (mole or atomic fraction) normalized variable $Y_{N_B}$ although the molar volume term to consider the change in total volume of the sample is included correctly during the derivation of these relations (for example, *see* Equation 3.2). On the other hand, den Broeder's relation [27] for the interdiffusion coefficient is derived based on concentration (composition divided by molar volume) normalized variable $Y_{C_B}$ in which the molar volume term is automatically included, *see* Equation 3.3. The relations for the estimation of other diffusion parameters (integrated and intrinsic diffusion coefficients) with respect to the variable $Y_{C_B}$ are not available. For a constant molar volume, it is easy to visualize from Equations 3.2 and 3.3 that both the relations of the interdiffusion coefficients lead to the same equation and therefore will give the same value.

These two methods (den Broeder and Wagner) are compared based on the estimated data only since these are derived completely differently (den Broeder: graphical and Wagner: algebraic formulations). Therefore, with the aim of examining the veracity of these two approaches, we do the following:

(i)     For the sake of efficient comparison, we follow the line of treatment proposed by Wagner to check if we can arrive at the den Broeder's relation following Wagner's line of treatment.

(ii)    This will then help to extend it to derive the relations for the estimation of the intrinsic diffusion coefficients of components and the integrated interdiffusion coefficient (for the phases with narrow homogeneity range) with respect to $Y_{C_B}$ which are not available at present.

(iii)   Following, we consider the experimental results in Ni–Pd (a system with solid solution), Ni–Al (in β–NiAl, a phase with the wide homogeneity range of composition) and Cu–Sn (a system with the narrow homogeneity range phases) to discuss efficiencies/limitations of the approaches.

## 3.2   Interdiffusion and intrinsic diffusion coefficients with respect to the concentration normalized variable

The derivation of relations (*see* Section 3.6) for the interdiffusion coefficients by Wagner [26] and the intrinsic diffusion coefficients by Paul [6] after extending the same line of treatment with respect to composition normalized variable $Y_{N_B}$ can be found in the respective references as mentioned or in the text book as mentioned in Reference [18]. In this section, we follow the Wagner's line of treatment to find if we can arrive at the den Broeder's relation with respect to $Y_{C_B}$. Then we extend it further to derive the relations for the intrinsic and tracer diffusion coefficients. These will then allow us to compare the data of a particular diffusion parameter when estimated following different relations utilizing $Y_{N_B}$ and $Y_{C_B}$. It should be noted here that the estimation of the tracer diffusion coefficients following the diffusion couple technique is considered indirect but reliable [17],[31],[32],[33],[34]. These are important to correlate the diffusion data with defects assisting the diffusion process in the absence of thermodynamic driving forces.

### 3.2.1 Derivation of the Interdiffusion Coefficient with respect to the concentration normalized variable

Interdiffusion coefficients are related to the interdiffusion fluxes following the Fick's first law with respect to component B as [21]

$$\tilde{J}_B = -\tilde{D}\frac{\partial C_B}{\partial x} \tag{3.4}$$

From the standard thermodynamic relation $C_A\bar{V}_A + C_B\bar{V}_B = 1$ [18], we can write

$$\tilde{D}\frac{\partial C_B}{\partial x} = -\tilde{J}_B = -(C_A\bar{V}_A + C_B\bar{V}_B)\tilde{J}_B \tag{3.5}$$

where $\bar{V}_i$ are the partial molar volumes of components A and B.

Using another standard thermodynamic equation $\bar{V}_A dC_A + \bar{V}_B dC_B = 0$ [18], we can relate the interdiffusion fluxes with respect to components A and B as

$$\tilde{J}_B = -\tilde{D}\frac{\partial C_B}{\partial x} = \frac{\bar{V}_A}{\bar{V}_B}\tilde{D}\frac{\partial C_A}{\partial x} = -\frac{\bar{V}_A}{\bar{V}_B}\tilde{J}_A$$

$$\bar{V}_B\tilde{J}_B = -\bar{V}_A\tilde{J}_A \tag{3.6}$$

Note here that the interdiffusion fluxes and the concentration gradients are different at one particular composition (with respect to a particular location in a diffusion couple) in a system with non–ideal molar volume variation. For a constant molar volume $\bar{V}_A = \bar{V}_B = V_m$, these are equal but with opposite sign [35]. On the other hand, the interdiffusion coefficient is the material constant and one will find the same value irrespective of any component considered for the estimation of the data.

Combining Equations (3.5) and (3.6), we can write

$$\widetilde{D} = \frac{-\tilde{J}_B}{\left(\frac{\partial C_B}{\partial x}\right)} = \frac{-(C_A \bar{V}_A + C_B \bar{V}_B)\tilde{J}_B}{\left(\frac{\partial C_B}{\partial x}\right)}$$

$$\widetilde{D} = \frac{\bar{V}_A (C_B \tilde{J}_A - C_A \tilde{J}_B)}{\left(\frac{\partial C_B}{\partial x}\right)}$$

$$\tilde{J}_B = -\bar{V}_A (C_B \tilde{J}_A - C_A \tilde{J}_B) \tag{3.7}$$

Following Boltzmann [22], compositions in an interdiffusion zone can be related to its position and annealing time by an auxiliary variable as

$$\lambda = \lambda(C_B) = \frac{x - x_o}{\sqrt{t}} = \frac{x}{\sqrt{t}} \tag{3.8}$$

where $x_o = 0$ is the location of the initial contact plane (Matano plane).

After differentiating Boltzmann parameter in Equation (3.8) with respect to $t$ and then utilizing the same relation again, we get

$$\frac{d\lambda}{dt} = -\frac{1}{2}\frac{x}{t^{3/2}} = -\frac{\lambda}{2t}$$

$$\frac{-1}{dt} = \frac{\lambda}{2t d\lambda} \tag{3.9}$$

The concentration normalized variable introduced by den Broeder [27] is expressed as

$$Y_{C_B} = \frac{C_B - C_B^-}{C_B^+ - C_B^-} \tag{3.10}$$

where $C_B^-$ and $C_B^+$ are the concentration of B at the un–affected left– and right–hand side of the diffusion couple. Note that $Y_{C_B}$ and $(1 - Y_{C_B})$ are equal to zero at these un–affected parts of the diffusion couple.

It can be rearranged to

$$C_B = C_B^+ Y_{C_B} + C_B^- (1 - Y_{C_B}) \tag{3.11a}$$

Using standard thermodynamic relation $C_A \bar{V}_A + C_B \bar{V}_B = 1$, Equation (3.11a) can be written as

$$\frac{1 - C_A \bar{V}_A}{\bar{V}_B} = C_B^+ Y_{C_B} + C_B^- \left(1 - Y_{C_B}\right)$$

$$1 - C_A \bar{V}_A = \bar{V}_B C_B^+ Y_{C_B} + \bar{V}_B C_B^- (1 - Y_{C_B})$$

$$C_A \bar{V}_A = 1 - \bar{V}_B C_B^+ Y_{C_B} - \bar{V}_B C_B^- (1 - Y_{C_B})$$

$$C_A \bar{V}_A = \left(1 - Y_{C_B}\right) + Y_{C_B} - \bar{V}_B C_B^+ Y_{C_B} - \bar{V}_B C_B^- (1 - Y_{C_B})$$

$$C_A = \frac{\left(1 - \bar{V}_B C_B^+\right) Y_{C_B} + \left(1 - \bar{V}_B C_B^-\right)\left(1 - Y_{C_B}\right)}{\bar{V}_A} \tag{3.11b}$$

From Fick's second law [21], we know that $\frac{\partial C_i}{\partial t} = \frac{\partial}{\partial x}\left(\widetilde{D} \frac{\partial C_i}{\partial x}\right) = -\frac{\partial \widetilde{J}_i}{\partial x}$. Therefore, with respect to components A and B and with the help of Equation (3.9), we can write

$$\frac{\partial \widetilde{J}_B}{\partial x} = -\frac{\partial C_B}{\partial t} = \frac{\lambda}{2t} \frac{d(C_B)}{d\lambda} \tag{3.12a}$$

$$\frac{\partial \widetilde{J}_A}{\partial x} = -\frac{\partial C_A}{\partial t} = \frac{\lambda}{2t} \frac{d(C_A)}{d\lambda} \tag{3.12b}$$

Note here that in Equations (3.11a) and (3.11b), the concentrations of component B and A, *i.e.*, $C_B$ and $C_A$ are expressed in terms of the concentration normalized variable ($Y_{C_B}$). So, next we aim to rewrite Fick's second law, *i.e.*, Equations (3.12a) and (3.12b) with respect to $Y_{C_B}$.

Replacing Equation (3.11a) in (3.12a) and Equation (3.11b) in (3.12b), we get

$$\frac{\partial \widetilde{J}_B}{\partial x} = \frac{\lambda}{2t}\left[C_B^+ \frac{dY_{C_B}}{d\lambda} + C_B^- \frac{d(1 - Y_{C_B})}{d\lambda}\right] \tag{3.13a}$$

$$\frac{\partial \widetilde{J}_A}{\partial x} = \frac{\lambda}{2t}\left[\left(\frac{1 - \bar{V}_B C_B^+}{\bar{V}_A}\right)\frac{dY_{C_B}}{d\lambda} + \left(\frac{1 - \bar{V}_B C_B^-}{\bar{V}_A}\right)\frac{d(1 - Y_{C_B})}{d\lambda}\right] \tag{3.13b}$$

Now, we aim to write the above equations with respect to $Y_{C_B}$ and $\left(1 - Y_{C_B}\right)$ separately.

Operating $\left[C_B^- \times \text{Eq. (3.13b)}\right] - \left[\left(\frac{1 - \bar{V}_B C_B^-}{\bar{V}_A}\right) \times \text{Eq. (3.13a)}\right]$ leads to

$$C_B^- \frac{\partial \tilde{J}_A}{\partial x} - \left(\frac{1-\overline{V}_B C_B^-}{\overline{V}_A}\right)\frac{\partial \tilde{J}_B}{\partial x} = \frac{\lambda}{2t}\left(\frac{C_B^- - C_B^+}{\overline{V}_A}\right)\frac{dY_{C_B}}{d\lambda} \tag{3.14a}$$

Operating $\left[C_B^+ \times \text{Eq. (3.13b)}\right] - \left[\left(\frac{1-\overline{V}_B C_B^+}{\overline{V}_A}\right) \times \text{Eq. (3.13a)}\right]$ leads to

$$C_B^+ \frac{\partial \tilde{J}_A}{\partial x} - \left(\frac{1-\overline{V}_B C_B^+}{\overline{V}_A}\right)\frac{\partial \tilde{J}_B}{\partial x} = \frac{\lambda}{2t}\left(\frac{C_B^+ - C_B^-}{\overline{V}_A}\right)\frac{d(1-Y_{C_B})}{d\lambda} \tag{3.14b}$$

After differentiating Boltzmann parameter in Equation (3.8) with respect to *x*, we get

$$d\lambda = \frac{dx}{\sqrt{t}} \tag{3.15}$$

Multiplying left–hand side by $\frac{dx}{\sqrt{t}}$ and right–hand side by $d\lambda$ of the Equation (3.14a) and (3.14b), respectively, we get

$$\frac{\overline{V}_A C_B^- \, dJ_A - (1-\overline{V}_B C_B^-) \, dJ_B}{\sqrt{t}} = \left(\frac{C_B^- - C_B^+}{2t}\right)\lambda \, d(Y_{C_B}) \tag{3.16a}$$

$$\frac{\overline{V}_A C_B^+ \, dJ_A - (1-\overline{V}_B C_B^+) \, dJ_B}{\sqrt{t}} = \left(\frac{C_B^+ - C_B^-}{2t}\right)\lambda \, d(1-Y_{C_B}) \tag{3.16b}$$

Equation (3.16a) is integrated for a fixed annealing time *t* from un–affected left–hand side of the diffusion couple, *i.e.*, $\lambda = \lambda^{-\infty}$ (corresponds to $x = x^{-\infty}$) to the location of interest $\lambda = \lambda^*$ (corresponds to $x = x^*$) for estimation of the diffusion coefficient. Following, we rearrange, with respect to integration by parts $\left[\int u dv = uv - \int (v du)\right]$.

$$\frac{1}{\sqrt{t}}\left[\overline{V}_A C_B^- \int_0^{\tilde{J}_A} d\tilde{J}_A - (1-\overline{V}_B C_B^-) \int_0^{\tilde{J}_B} d\tilde{J}_B\right] = \left(\frac{C_B^- - C_B^+}{2t}\right)\int_{\lambda-\infty}^{\lambda^*} \lambda \, d(Y_{C_B})$$

$$\frac{(\overline{V}_A^* C_B^-)\tilde{J}_A^* - (1-\overline{V}_B^* C_B^-)\tilde{J}_B^*}{\sqrt{t}} = \left(\frac{C_B^- - C_B^+}{2t}\right)\left[\lambda^* Y_{C_B}^* - \int_{\lambda-\infty}^{\lambda^*} Y_{C_B} \, d\lambda\right] \tag{3.17a}$$

Similarly, Equation (3.16b) is integrated from the location of interest $\lambda = \lambda^*$ to the un–affected right–hand side of the diffusion couple, *i.e.*, $\lambda = \lambda^{+\infty}$ (corresponds to $x = x^{+\infty}$).

$$\frac{1}{\sqrt{t}}\left[\overline{V}_A C_B^+ \int_{\tilde{J}_A}^0 d\tilde{J}_A - (1-\overline{V}_B C_B^+) \int_{\tilde{J}_B}^0 d\tilde{J}_B\right] = \left(\frac{C_B^+ - C_B^-}{2t}\right)\int_{\lambda^*}^{\lambda+\infty} \lambda \, d(1-Y_{C_B})$$

$$\frac{-(\overline{V}_A^* C_B^+)\tilde{J}_A^* + (1-\overline{V}_B^* C_B^+)\tilde{J}_B^*}{\sqrt{t}} = \left(\frac{C_B^+ - C_B^-}{2t}\right)\left[-\lambda^*\left(1-Y_{C_B}^*\right) - \int_{\lambda^*}^{\lambda+\infty}\left(1-Y_{C_B}\right) d\lambda\right] \tag{3.17b}$$

Note here that the interdiffusion fluxes $\tilde{J}_i$ is equal to zero at the un–affected parts of the diffusion couple, $x = x^{-\infty}$ and $x = x^{+\infty}$, while $\tilde{J}_i^*$ is the fixed value (for certain annealing time $t$) at the location of interest $x = x^*$ in the above Equations (3.17). Next, we aim to rewrite the above equations with respect to interdiffusion fluxes $\tilde{J}_i$ of both components to get an expression for the interdiffusion coefficient $\tilde{D}$.

Operating $\left[ Y_{C_B}^* \times \text{Eq.} \, (3.17\text{b}) \right] - \left[ \left(1 - Y_{C_B}^*\right) \times \text{Eq.} \, (3.17\text{a}) \right]$ leads to

$$\frac{\bar{V}_A^*\left(C_B^*\tilde{J}_A^* - C_A^*\tilde{J}_B^*\right)}{\sqrt{t}} = \left(\frac{C_B^+ - C_B^-}{2t}\right)\left[\left(1 - Y_{C_B}^*\right) \int_{\lambda^{-\infty}}^{\lambda^*} Y_{C_B} \, d\lambda + Y_{C_B}^* \int_{\lambda^*}^{\lambda^{+\infty}} \left(1 - Y_{C_B}\right) d\lambda\right] \qquad (3.18)$$

Numerator on the left–hand side can be derived, by using $C_B^* = C_B^+ Y_{C_B}^* + C_B^-\left(1 - Y_{C_B}^*\right)$ following Equation (3.11a) and standard thermodynamic relation $\bar{V}_A^* C_A^* = 1 - \bar{V}_B^* C_B^*$, following the steps:

$$Y_{C_B}^*\left\{-\left(\bar{V}_A^* C_B^+\right)\tilde{J}_A^* + \left(1 - \bar{V}_B^* C_B^+\right)\tilde{J}_B^*\right\} - \left(1 - Y_{C_B}^*\right)\left\{\left(\bar{V}_A^* C_B^-\right)\tilde{J}_A^* - \left(1 - \bar{V}_B^* C_B^-\right)\tilde{J}_B^*\right\}$$

$$= \left\{-\bar{V}_A^* C_B^+ Y_{C_B}^* - \bar{V}_A^* C_B^-\left(1 - Y_{C_B}^*\right)\right\}\tilde{J}_A^* + \left\{\left(1 - \bar{V}_B^* C_B^+\right)Y_{C_B}^* + \left(1 - Y_{C_B}^*\right)\left(1 - \bar{V}_B^* C_B^-\right)\right\}\tilde{J}_B^*$$

$$= -\bar{V}_A^*\left\{C_B^+ Y_{C_B}^* + C_B^-\left(1 - Y_{C_B}^*\right)\right\}\tilde{J}_A^* + \left\{Y_{C_B}^* - \bar{V}_B^* C_B^+ Y_{C_B}^* + 1 - \bar{V}_B^* C_B^- - Y_{C_B}^* + \bar{V}_B^* C_B^- Y_{C_B}^*\right\}\tilde{J}_B^*$$

$$= -\bar{V}_A^*\left\{C_B^+ Y_{C_B}^* + C_B^-\left(1 - Y_{C_B}^*\right)\right\}\tilde{J}_A^* + \left[1 - \bar{V}_B^*\left\{C_B^+ Y_{C_B}^* + C_B^-\left(1 - Y_{C_B}^*\right)\right\}\right]\tilde{J}_B^*$$

$$= -\bar{V}_A^*\left(C_B^*\tilde{J}_A^* - C_A^*\tilde{J}_B^*\right).$$

Utilizing $d\lambda = \frac{dx}{\sqrt{t}}$ from Equation (3.15), we get

$$\bar{V}_A^*\left(C_B^*\tilde{J}_A^* - C_A^*\tilde{J}_B^*\right) = \left(\frac{C_B^+ - C_B^-}{2t}\right)\left[\left(1 - Y_{C_B}^*\right) \int_{x^{-\infty}}^{x^*} Y_{C_B} \, dx + Y_{C_B}^* \int_{x^*}^{x^{+\infty}} \left(1 - Y_{C_B}\right) dx\right]$$

$$(3.19)$$

For $C_B = C_B^*$, from Equation (3.7) we know that $\tilde{J}_B^* = -\bar{V}_A^*\left(C_B^*\tilde{J}_A^* - C_A^*\tilde{J}_B^*\right)$ and hence the interdiffusion flux with respect to component B can be expressed as

$$\tilde{J}_B^* = \tilde{J}_B(C_B^*) = -\left(\frac{C_B^+ - C_B^-}{2t}\right)\left[\left(1 - Y_{C_B}^*\right) \int_{x^{-\infty}}^{x^*} Y_{C_B} \, dx + Y_{C_B}^* \int_{x^*}^{x^{+\infty}} \left(1 - Y_{C_B}\right) dx\right]$$

$$(3.20\text{a})$$

Similarly, one can derive the interdiffusion flux with respect to component A as

$$\tilde{J}_A^* = \tilde{J}_A(C_A^*) = \left(\frac{C_A^- - C_A^+}{2t}\right)\left[Y_{C_A}^* \int_{x^{-\infty}}^{x^*} \left(1 - Y_{C_A}\right)dx + \left(1 - Y_{C_A}^*\right)\int_{x^*}^{x^{+\infty}} Y_{C_A}\,dx\right]$$

(3.20b)

where $Y_{C_A} = \frac{C_A - C_A^+}{C_A^- - C_A^+}$. Note opposite sign of interdiffusion fluxes when estimated with respect to component A and B because of opposite direction of diffusion of these components.

From Equation (3.11a) we know that $C_B = C_B^+ Y_{C_B} + C_B^-(1 - Y_{C_B})$. By differentiating it with respect to *x*, we can write

$$\left(\frac{dC_B}{dx}\right)_{x=x^*} = C_B^+ \frac{dY_{C_B}}{dx} - C_B^- \frac{dY_{C_B}}{dx} = (C_B^+ - C_B^-)\left(\frac{dY_{C_B}}{dx}\right)_{x=x^*}$$

(3.21)

From Equation (3.7) for $C_B = C_B^*$, we know

$$\tilde{D}(C_B^*) = \frac{-\tilde{J}_B^*}{\left(\frac{dC_B}{dx}\right)_{x=x^*}}$$

(3.22)

Substituting for flux [Eq. (3.20a)] and gradient [Eq. (3.21)] in Fick's first law [Eq. (3.22)], we get the expression for the estimation of interdiffusion coefficient as

$$\tilde{D}(Y_{C_B}^*) = \frac{1}{2t\left(\frac{dY_{C_B}^*}{dx}\right)}\left[\left(1 - Y_{C_B}^*\right)\int_{x^-\infty}^{x^*} Y_{C_B}\,dx + Y_{C_B}^* \int_{x^*}^{x^{+\infty}} \left(1 - Y_{C_B}\right)dx\right]$$

(3.23)

den Broeder [27] derived this relation with respect to $Y_{C_B}$ following the graphical approach. It should be noted here that the interdiffusion coefficients ($\tilde{D}\left(Y_{C_A}^*\right)$ and $\tilde{D}\left(Y_{C_B}^*\right)$) estimated with respect to component A and B are the same [35]. In this study, we arrive at the same relation (*see* Equation 3.3) following the Wagner's [26] line of treatment, although Wagner derived the relation as expressed in Equation 3.2 with respect to $Y_{N_B}$. Therefore, in a sense, both the relations are logically correct.

Additionally, for a constant molar volume ($V_m = V_m^- = V_m^+$), both the den Broeder and the Wagner relations lead to

$$\widetilde{D}\left(Y_{N_B}^*\right) = \frac{1}{2t\left(\frac{dY_{N_B}}{dx}\right)}\left[\left(1 - Y_{N_B}^*\right)\int_{x^{-\infty}}^{x^*} Y_{N_B}\, dx + Y_{N_B}^*\int_{x^*}^{x^{+\infty}}\left(1 - Y_{N_B}\right)dx\right] \qquad (3.24)$$

Since, $Y_{C_B} = \dfrac{C_B - C_B^-}{C_B^+ - C_B^-} = \dfrac{\frac{N_B}{V_m} - \frac{N_B^-}{V_m}}{\frac{N_B^+}{V_m} - \frac{N_B^-}{V_m}} = \dfrac{N_B - N_B^-}{N_B^+ - N_B^-} = Y_{N_B}$

### 3.2.2 Derivation of the intrinsic and tracer diffusion coefficients with respect to the concentration normalized variable

As already mentioned, the relations for the intrinsic diffusion coefficients are available only with respect to $Y_{N_B}$. Therefore, these relations should be derived with respect $Y_{C_B}$ to examine the differences in the data when estimated following these different approaches, *i.e.*, with respect to $Y_{C_B}$ and $Y_{N_B}$. Previously, Paul [6] derived these relations with respect to $Y_{N_B}$ by extending the Wagner's line of treatment to derive the same relations as developed earlier by van Loo [29] differently. We now, extend the analysis to develop the relations for the intrinsic diffusion coefficients with respect to $Y_{C_B}$.

When the location of interest is the position of the Kirkendall marker plane (K), *i.e.*, $x^* = x^K$, we can write Equations (3.17a) and (3.17b), respectively as

$$\frac{\left(\bar{V}_A^K C_B^-\right)\tilde{J}_A^K - \left(1 - \bar{V}_B^K C_B^-\right)\tilde{J}_B^K}{\sqrt{t}} = \left(\frac{C_B^+ - C_B^-}{2t}\right)\left[-\lambda^K Y_{C_B}^K + \int_{\lambda^{-\infty}}^{\lambda^K} Y_{C_B}\, d\lambda\right] \qquad (3.25a)$$

$$\frac{-\left(\bar{V}_A^K C_B^+\right)\tilde{J}_A^K + \left(1 - \bar{V}_B^K C_B^+\right)\tilde{J}_B^K}{\sqrt{t}} = \left(\frac{C_B^+ - C_B^-}{2t}\right)\left[-\lambda^K\left(1 - Y_{C_B}^K\right) - \int_{\lambda^K}^{\lambda^{+\infty}}\left(1 - Y_{C_B}\right)d\lambda\right] \qquad (3.25b)$$

Now, we aim to rewrite the above equations with respect to $\tilde{J}_B^K$ and $\tilde{J}_A^K$ such that we can get an expression for intrinsic diffusion coefficient of component B and A, *i.e.*, $D_B$ and $D_A$, respectively, at the Kirkendall maker plane utilizing the Darken's equation [23] relating the interdiffusion flux $\left(\tilde{J}_i\right)$ with the intrinsic flux $\left(J_i\right)$ of component.

Operating $[\bar{V}_A^K C_B^+ \times \text{Eq. (3.25a)}] + [\bar{V}_A^K C_B^- \times \text{Eq. (3.25b)}]$ leads to

$$\frac{-(C_B^+ - C_B^-)\tilde{J}_B^K}{\sqrt{t}}$$

$$= \left(\frac{C_B^+ - C_B^-}{2t}\right)\left[-\lambda^K\{C_B^+ Y_{C_B}^K + C_B^-(1 - Y_{C_B}^K)\} + C_B^+ \int_{\lambda-\infty}^{\lambda^K} Y_{C_B}\, d\lambda - C_B^- \int_{\lambda^K}^{\lambda+\infty}(1 - Y_{C_B})\, d\lambda\right]$$

Note that $\bar{V}_A^K$ has been cancelled on both sides, since numerator on the left–hand side is

$$\{-(\bar{V}_A^K C_B^+)(1 - \bar{V}_B^K C_B^-) + \bar{V}_A^K C_B^-(1 - \bar{V}_B^K C_B^+)\}\tilde{J}_B^K = -\bar{V}_A^K(C_B^+ - C_B^-)\tilde{J}_B^K$$

Utilizing $C_B^K = C_B^+ Y_{C_B}^K + C_B^-(1 - Y_{C_B}^K)$ from Equation (3.11a) and after rearranging, we get

$$\tilde{J}_B^K = \frac{\sqrt{t}}{2t}\left[\lambda^K C_B^K - C_B^+ \int_{\lambda-\infty}^{\lambda^K} Y_{C_B}\, d\lambda + C_B^- \int_{\lambda^K}^{\lambda+\infty}(1 - Y_{C_B})\, d\lambda\right] \qquad (3.26a)$$

Similarly, operating $[(1 - \bar{V}_B^K C_B^+) \times \text{Eq. (3.25a)}] + [(1 - \bar{V}_B^K C_B^-) \times \text{Eq. (3.25b)}]$ and utilizing $\bar{V}_A^K C_A^K = (1 - \bar{V}_B^K C_B^+)Y_{C_B}^K + (1 - \bar{V}_B^K C_B^-)(1 - Y_{C_B}^K)$ from Equation (3.11b), we get

$$\frac{-\bar{V}_A^K(C_B^+ - C_B^-)\tilde{J}_A^K}{\sqrt{t}}$$

$$= \left(\frac{C_B^+ - C_B^-}{2t}\right)\left[-\lambda^K \bar{V}_A^K C_A^K + (1 - \bar{V}_B^K C_B^+)\int_{\lambda-\infty}^{\lambda^K} Y_{C_B}\, d\lambda - (1 - \bar{V}_B^K C_B^-)\int_{\lambda^K}^{\lambda+\infty}(1 - Y_{C_B})\, d\lambda\right]$$

since numerator on the left–hand side is

$$\{\bar{V}_A^K C_B^-(1 - \bar{V}_B^K C_B^+) - \bar{V}_A^K C_B^+(1 - \bar{V}_B^K C_B^-)\}\tilde{J}_A^K = -\bar{V}_A^K(C_B^+ - C_B^-)\tilde{J}_A^K$$

Dividing both sides of equation by a factor of $\bar{V}_A^K$ and after rearranging, we get

$$\tilde{J}_A^K = \frac{\sqrt{t}}{2t}\left[\lambda^K C_A^K - \left(\frac{1 - \bar{V}_B^K C_B^+}{\bar{V}_A^K}\right)\int_{\lambda-\infty}^{\lambda^K} Y_{C_B}\, d\lambda + \left(\frac{1 - \bar{V}_B^K C_B^-}{\bar{V}_A^K}\right)\int_{\lambda^K}^{\lambda+\infty}(1 - Y_{C_B})\, d\lambda\right] \quad (3.26b)$$

From Boltzmann parameter in Equation (3.8), we know that $\lambda^K = \frac{x^K}{\sqrt{t}}$ or $x^K = \lambda^K\sqrt{t}$.

Therefore, the velocity of the Kirkendall marker plane can be expressed as

$$v^K = \frac{dx^K}{dt} = \frac{d(\lambda^K\sqrt{t})}{dt} = \lambda^K \frac{d(\sqrt{t})}{dt} = \frac{\lambda^K}{2\sqrt{t}} = \frac{\lambda^K\sqrt{t}}{2t}$$

Also, differentiating Boltzmann parameter with respect to *x*, from Equation (3.15) we know that $\sqrt{t}\, d\lambda = dx$.

Putting $\frac{\lambda^K \sqrt{t}}{2t} = v^K$ and $\sqrt{t}\, d\lambda = dx$ in Equations (3.26), we get

$$\tilde{J}_B^K = v^K C_B^K - \frac{1}{2t}\left[ C_B^+ \int_{x^{-\infty}}^{x^K} Y_{C_B}\, dx - C_B^- \int_{x^K}^{x^{+\infty}} \left(1 - Y_{C_B}\right) dx \right] \tag{3.27a}$$

$$\tilde{J}_A^K = v^K C_A^K - \frac{1}{2t}\left[ \left(\frac{1 - \bar{V}_B^K C_B^+}{\bar{V}_A^K}\right) \int_{x^{-\infty}}^{x^K} Y_{C_B}\, dx - \left(\frac{1 - \bar{V}_B^K C_B^-}{\bar{V}_A^K}\right) \int_{x^K}^{x^{+\infty}} \left(1 - Y_{C_B}\right) dx \right] \tag{3.27b}$$

Following Darken's Analysis [23], we know that $\tilde{J}_B^K = J_B + v^K C_B^K$ and $\tilde{J}_A^K = J_A + v^K C_A^K$. Therefore, we can get an expression for intrinsic flux of component B and A, *i.e.*, $J_B$ and $J_A$, respectively, as follows:

$$J_B = \tilde{J}_B^K - v^K C_B^K$$

$$J_B = -\frac{1}{2t}\left[ C_B^+ \int_{x^{-\infty}}^{x^K} Y_{C_B}\, dx - C_B^- \int_{x^K}^{x^{+\infty}} \left(1 - Y_{C_B}\right) dx \right] \tag{3.28a}$$

$$J_A = \tilde{J}_A^K - v^K C_A^K$$

$$J_A = -\frac{1}{2t}\left[ \left(\frac{1 - \bar{V}_B^K C_B^+}{\bar{V}_A^K}\right) \int_{x^{-\infty}}^{x^K} Y_{C_B}\, dx - \left(\frac{1 - \bar{V}_B^K C_B^-}{\bar{V}_A^K}\right) \int_{x^K}^{x^{+\infty}} \left(1 - Y_{C_B}\right) dx \right] \tag{3.28b}$$

Using Fick's first law [21], we can write $D_B = \frac{-J_B}{\left(\frac{\partial C_B}{\partial x}\right)_{x^K}}$ and $D_A = \frac{-J_A}{\left(\frac{\partial C_A}{\partial x}\right)_{x^K}}$. Therefore, we can write an expression for intrinsic diffusion coefficient of component B and A, *i.e.*, $D_B$ and $D_A$, respectively, as follows:

$$D_B = \frac{1}{2t}\left(\frac{\partial x}{\partial C_B}\right)_K \left[ C_B^+ \int_{x^{-\infty}}^{x^K} Y_{C_B}\, dx - C_B^- \int_{x^K}^{x^{+\infty}} \left(1 - Y_{C_B}\right) dx \right] \tag{3.29a}$$

$$D_A = \frac{1}{2t}\left(\frac{\partial x}{\partial C_A}\right)_K \left[ \left(\frac{1 - \bar{V}_B^K C_B^+}{\bar{V}_A^K}\right) \int_{x^{-\infty}}^{x^K} Y_{C_B}\, dx - \left(\frac{1 - \bar{V}_B^K C_B^-}{\bar{V}_A^K}\right) \int_{x^K}^{x^{+\infty}} \left(1 - Y_{C_B}\right) dx \right] \tag{3.29b}$$

The same relation of $D_A$ with respect to $Y_{C_A}$ can be derived as

$$D_A = \frac{1}{2t}\left(\frac{\partial x}{\partial C_A}\right)_K \left[ C_A^- \int_{x^{+\infty}}^{x^K} Y_{C_A}\, dx - C_A^+ \int_{x^K}^{x^{-\infty}} \left(1 - Y_{C_A}\right) dx \right] \tag{3.29c}$$

Compared to Equation 3.29b, Equation 3.29c avoids the need for partial molar volumes and hence the error associated with the estimation of these values, as shown later in Section 3.2.3.

Using $\bar{V}_A dC_A + \bar{V}_B dC_B = 0$, we get $\frac{\partial C_B}{\partial x} = -\frac{\bar{V}_A}{\bar{V}_B}\frac{\partial C_A}{\partial x} \Longrightarrow \frac{\partial x}{\partial C_B} = -\frac{\bar{V}_B}{\bar{V}_A}\frac{\partial x}{\partial C_A}$.

Utilizing $\left(\frac{\partial x}{\partial c_B}\right)_K = -\frac{\bar{V}_B^K}{\bar{V}_A^K}\left(\frac{\partial x}{\partial c_A}\right)_K$ in Equations (3.29), the ratio of intrinsic diffusivities

can be written as

$$\frac{D_B}{D_A} = \frac{\bar{V}_B^K}{\bar{V}_A^K}\left[\frac{C_B^+ \int_{x-\infty}^{x^K} Y_{C_B} dx - C_B^- \int_{x^K}^{x+\infty}(1-Y_{C_B})dx}{-C_A^- \int_{x+\infty}^{x^K} Y_{C_A} dx + C_A^+ \int_{x^K}^{x-\infty}(1-Y_{C_A})dx}\right] \tag{3.29d}$$

This is derived, extending the den Broeder approach, in this thesis for the first time

using $Y_{C_B}$. The similar equations with respect to $Y_{N_B}$ as derived by van Loo [29] and

Paul [6] are expressed as

$$D_B = \frac{1}{2t}\left(\frac{\partial x}{\partial c_B}\right)_K \left[N_B^+ \int_{x-\infty}^{x^K} \frac{Y_{N_B}}{V_m} dx - N_B^- \int_{x^K}^{x+\infty} \frac{(1-Y_{N_B})}{V_m} dx\right] \tag{3.30a}$$

$$D_A = \frac{1}{2t}\left(\frac{\partial x}{\partial c_A}\right)_K \left[N_A^+ \int_{x-\infty}^{x^K} \frac{Y_{N_B}}{V_m} dx - N_A^- \int_{x^K}^{x+\infty} \frac{(1-Y_{N_B})}{V_m} dx\right] \tag{3.30b}$$

$$\frac{D_B}{D_A} = \frac{\bar{V}_B^K}{\bar{V}_A^K}\left[\frac{N_B^+ \int_{x-\infty}^{x^K} \frac{Y_{N_B}}{V_m} dx - N_B^- \int_{x^K}^{x+\infty} \frac{(1-Y_{N_B})}{V_m} dx}{-N_A^+ \int_{x-\infty}^{x^K} \frac{Y_{N_B}}{V_m} dx + N_A^- \int_{x^K}^{x+\infty} \frac{(1-Y_{N_B})}{V_m} dx}\right] \approx \frac{\bar{V}_B^K}{\bar{V}_A^K}\frac{D_B^*}{D_A^*} \tag{3.30c}$$

If a constant molar volume is considered (such that the molar volume and the partial

molar volumes at every composition are equal, *i.e.*, $V_m = \bar{V}_A = \bar{V}_B$, both the Equations

(3.29) and (3.30) will be reduced to the same equation

$$D_B = \frac{1}{2t}\left(\frac{\partial x}{\partial N_B}\right)_K \left[N_B^+ \int_{x-\infty}^{x^K} Y_{N_B} dx - N_B^- \int_{x^K}^{x+\infty}\left(1-Y_{N_B}\right) dx\right] \tag{3.31a}$$

$$D_A = \frac{1}{2t}\left(\frac{\partial x}{\partial N_A}\right)_K \left[N_A^+ \int_{x-\infty}^{x^K} Y_{N_B} dx - N_A^- \int_{x^K}^{x+\infty}\left(1-Y_{N_B}\right) dx\right] \tag{3.31b}$$

Following Darken–Manning Analysis [23, 24], the intrinsic $(D_i)$ and tracer $(D_i^*)$

diffusion coefficients are related as

$$D_A = \frac{V_m}{\bar{V}_B} D_A^* \Phi(1 + W_A) \tag{3.32a}$$

$$D_B = \frac{V_m}{\bar{V}_A} D_B^* \Phi(1 - W_B) \tag{3.32b}$$

where the terms $W_i = \frac{2N_i(D_A^* - D_B^*)}{M_0(N_A D_A^* + N_B D_B^*)}$ arise from the vacancy–wind effect, a constant $M_0$ depends on the crystal structure. $\Phi = \frac{\mathrm{d}\ln a_A}{\mathrm{d}\ln N_A} = \frac{\mathrm{d}\ln a_B}{\mathrm{d}\ln N_B}$ is the thermodynamic factor which (according to the Gibbs–Duhem relation) is same for both the components A and B in a binary system. $a_i$ is the activity of component $i$. Therefore, the tracer diffusion coefficients can be estimated from the known thermodynamic parameters following Equations 3.29 or 3.30 and 3.32.

### 3.2.3 Comparison of the interdiffusion and intrinsic diffusion coefficients estimated following the relations established with respect to concentration and composition normalized variables

We compare the estimated values based on the estimation of diffusion coefficients in the Ni–Pd system [36]. The interdiffusion zone developed after annealing Ni and Pd at 1100 °C for 196 hrs is shown in Figure 3.1a. The location of the Kirkendall marker plane is identified by the ThO$_2$ particles, at 40.3 at% Ni. The composition profile developed in the interdiffusion zone is shown in Figure 3.1b. This is measured in a direction perpendicular to the Kirkendall marker plane following the diffusion direction of the components. The variation of molar volume used for the estimation of diffusion coefficients is shown in Figure 3.1c. The partial molar volumes of the components at the composition of the Kirkendall marker plane ($N_{Ni}^K = 0.403$) are shown. Since this diffusion couple is prepared with pure components as the end–members, the composition normalized variables are the same as composition of the respective components, as shown in Figure 3.1b. The concentration normalized variables for component A and B are shown in Figure 3.1d. The estimated data are shown in Figure 3.2. To compare the data, the different parts of the den Broeder's and the Wagner's relations are plotted. Gradients of concentration normalized variable $\left(\frac{dY_{C_B}^*}{dx}\right)$ and composition normalized variable

$\left(\frac{dY_{N_B}^*}{dx}\right)$ are shown in Figure 3.2a. The bracketed terms $\left[\left(1 - Y_{C_B}^*\right) \int_{x-\infty}^{x^*} Y_{C_B}\, dx + \right.$

$\left. Y_{C_B}^* \int_{x^*}^{x+\infty} \left(1 - Y_{C_B}\right) dx\right] = 2t \times \widetilde{D}\left(Y_{C_B}^*\right) \times \left(\frac{dY_{C_B}^*}{dx}\right)$ and $V_m^* \left[\left(1 - Y_{N_B}^*\right) \int_{x-\infty}^{x^*} \frac{Y_{N_B}}{V_m}\, dx + \right.$

$\left. Y_{N_B}^* \int_{x^*}^{x+\infty} \frac{\left(1 - Y_{N_B}\right)}{V_m}\, dx\right] = 2t \times \widetilde{D}\left(Y_{N_B}^*\right) \times \left(\frac{dY_{N_B}^*}{dx}\right)$ are shown in Figure 3.2b. Following,

$\widetilde{D}\left(Y_{C_B}^*\right)$ and $\widetilde{D}\left(Y_{N_B}^*\right)$ are shown in Figure 3.2c. As expected based on the definition of terms $Y_{C_B}$ and $Y_{N_B}$, although there is difference in the slope and the bracket terms;

however, a very minor difference in the estimated diffusion coefficients with respect to $Y_{C_B}$ and $Y_{N_B}$ is evident.

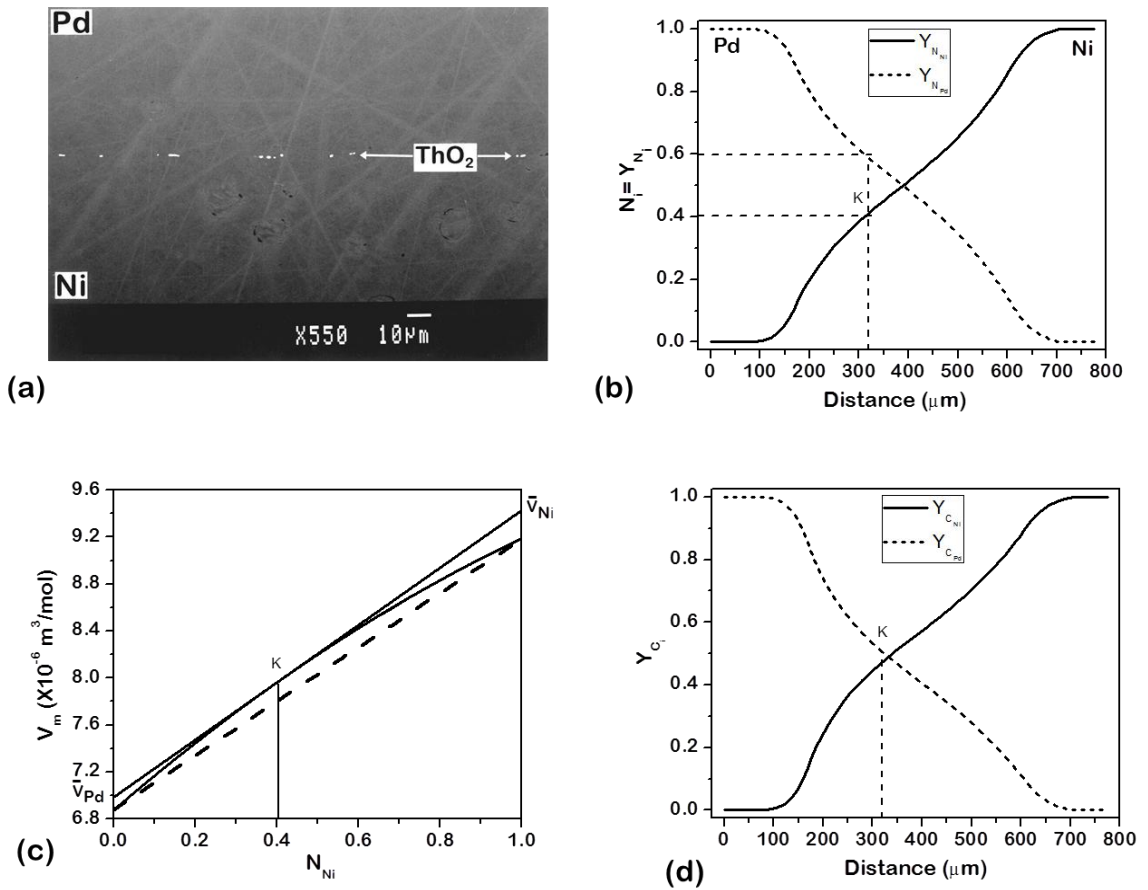

Figure 3.1: (a) Micrograph of the Ni/Pd diffusion couple annealed at 1100 °C for 196 hrs [36]. ThO₂ particles identifies the location of the Kirkendall marker plane, (b) the corresponding composition profile [36] (equal to the composition normalize variable) developed in the interdiffusion zone, (c) Molar volume variation in the Ni–Pd solid solution [36], (d) the concentration normalized variables.

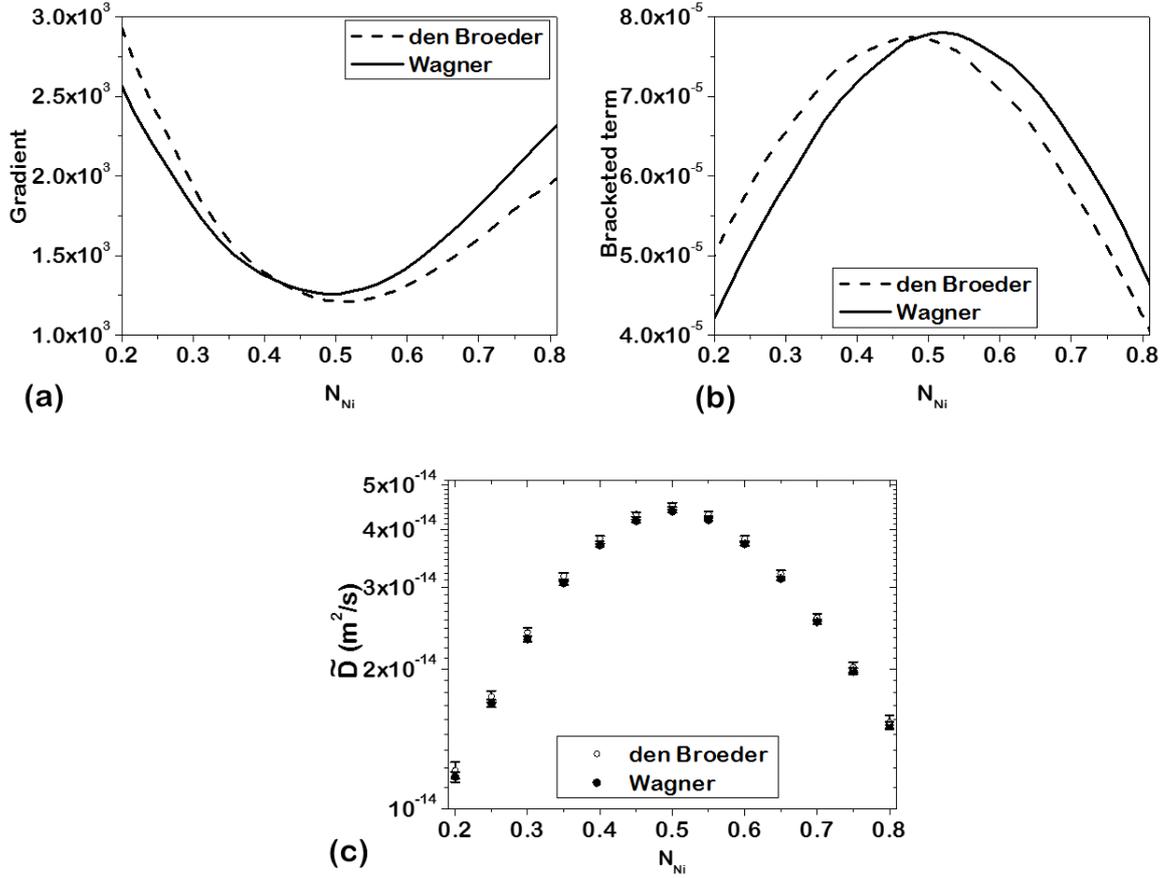

Figure 3.2: Estimated data in the Ni–Pd system using the Wagner and the den Broeder methods: (a) Gradients for Wagner and den Broeder relations (b) bracketed terms of Wagner and den Broeder relations and (c) the estimated interdiffusion coefficients.

Following, the intrinsic diffusion coefficients of components following den Broeder and Wagner methods are estimated. Since pure end–members are used, considering the composition profile in Figure 3.1b, we can write $N_B^-(= N_{Ni}^-) = 0$, $N_A^+(= N_{Pd}^+) = 0$, $N_B^+(= N_{Ni}^+) = 1$, and $N_A^-(= N_{Pd}^-) = 1$. Therefore, we have $C_A^-(= C_{Pd}^-) = \frac{N_{Pd}^-}{V_m^-} = \frac{1}{V_{Pd}}$, $C_B^-(= C_{Ni}^-) = \frac{N_{Ni}^-}{V_m^-} = \frac{0}{V_{Pd}} = 0$, $C_A^+(= C_{Pd}^+) = \frac{N_{Pd}^+}{V_m^+} = \frac{0}{V_{Ni}} = 0$ and $C_B^+(= C_{Ni}^+) = \frac{N_{Ni}^+}{V_m^+} = \frac{1}{V_{Ni}}$. Therefore, we can simplify the Equation 3.29a, b and c in the case of Ni–Pd diffusion couple as

$D_B(= D_{Ni}) = \frac{1}{2t}\left(\frac{\partial x}{\partial C_B}\right)_K \left[C_B^+ \int_{x-\infty}^{x^K} Y_{C_B} \, dx\right] = \frac{1}{2t}\left(\frac{\partial x}{\partial C_{Ni}}\right)_K \left[\frac{1}{V_{Ni}} \int_{x-\infty}^{x^K} Y_{C_{Ni}} \, dx\right] =$ $2.6{\times}10^{-14}$ m$^2$/s,

$$D_A(=D_{Pd}) = \frac{1}{2t}\left(\frac{\partial x}{\partial c_A}\right)_K \left[\left(\frac{1-\bar{V}_B^K c_B^+}{\bar{V}_A^K}\right)\int_{x-\infty}^{x^K} Y_{C_B}\, dx - \left(\frac{1}{\bar{V}_A^K}\right)\int_{x^K}^{x+\infty}\left(1-Y_{C_B}\right) dx\right] =$$

$$\frac{1}{2t}\left(\frac{\partial x}{\partial c_{Pd}}\right)_K \left[\left(\frac{1-\frac{\bar{V}_{Ni}^K}{V_{Ni}}}{\bar{V}_{Pd}^K}\right)\int_{x-\infty}^{x^K} Y_{C_{Ni}}\, dx - \left(\frac{1}{\bar{V}_{Pd}^K}\right)\int_{x^K}^{x+\infty}\left(1-Y_{C_{Ni}}\right) dx\right] = 5.2{\times}10^{-14}\ \text{m}^2/\text{s},$$

$$D_A(=D_{Pd}) = \frac{1}{2t}\left(\frac{\partial x}{\partial c_A}\right)_K \left[C_A^-\int_{x+\infty}^{x^K} Y_{C_A}\, dx\right] = \frac{1}{2t}\left(\frac{\partial x}{\partial c_{Pd}}\right)_K \left[\frac{1}{V_{Pd}}\int_{x+\infty}^{x^K} Y_{C_{Pd}}\, dx\right] =$$

$$4.9{\times}10^{-14}\ \text{m}^2/\text{s}.$$

The same can be estimated following the Wagner's method modifying the Equations 3.30a and b for the Ni–Pd diffusion couple at the Kirkendall marker plane as

$$D_B(=D_{Ni}) = \frac{1}{2t}\left(\frac{\partial x}{\partial c_B}\right)_K \left[\int_{x-\infty}^{x^K}\frac{Y_{N_B}}{V_m}\, dx\right] = \frac{1}{2t}\left(\frac{\partial x}{\partial c_{Ni}}\right)_K \left[\int_{x-\infty}^{x^K}\frac{Y_{N_{Ni}}}{V_m}\, dx\right] = 2.6{\times}10^{-14}\ \text{m}^2/\text{s},$$

$$D_A(=D_{Pd}) = \frac{1}{2t}\left(\frac{\partial x}{\partial c_A}\right)_K \left[-\int_{x^K}^{x+\infty}\frac{\left(1-Y_{N_B}\right)}{V_m}\, dx\right] = \frac{1}{2t}\left(\frac{\partial x}{\partial c_{Pd}}\right)_K \left[-\int_{x^K}^{x+\infty}\frac{\left(1-Y_{N_{Ni}}\right)}{V_m}\, dx\right] =$$

$$4.9{\times}10^{-14}\ \text{m}^2/\text{s}.$$

Therefore, there is no difference in the estimated intrinsic diffusion coefficients following den Broeder and Wagner methods. A small difference in values of intrinsic diffusion coefficient of Pd is found following the den Broeder method when estimated with respect to $Y_{C_{Ni}}$ and $Y_{C_{Pd}}$. This must be because of error associated with the calculation of partial molar volumes while estimating the data utilizing $Y_{C_{Ni}}$.

Following, we estimate the interdiffusion coefficients in the β–NiAl phase. The composition profile of a diffusion couple Ni$_{0.46}$Al$_{0.54}$ / Ni$_{0.575}$Al$_{0.425}$ after annealing at 1200 °C for 24 hrs is shown in Figure 3.3a. The molar volume variation in this intermetallic compound is shown in Figure 3.3b [6]. The estimated interdiffusion coefficients by two methods are shown in Figure 3.3c. The difference between the data estimated using both the methods in this system is higher compared to the Ni–Pd system.

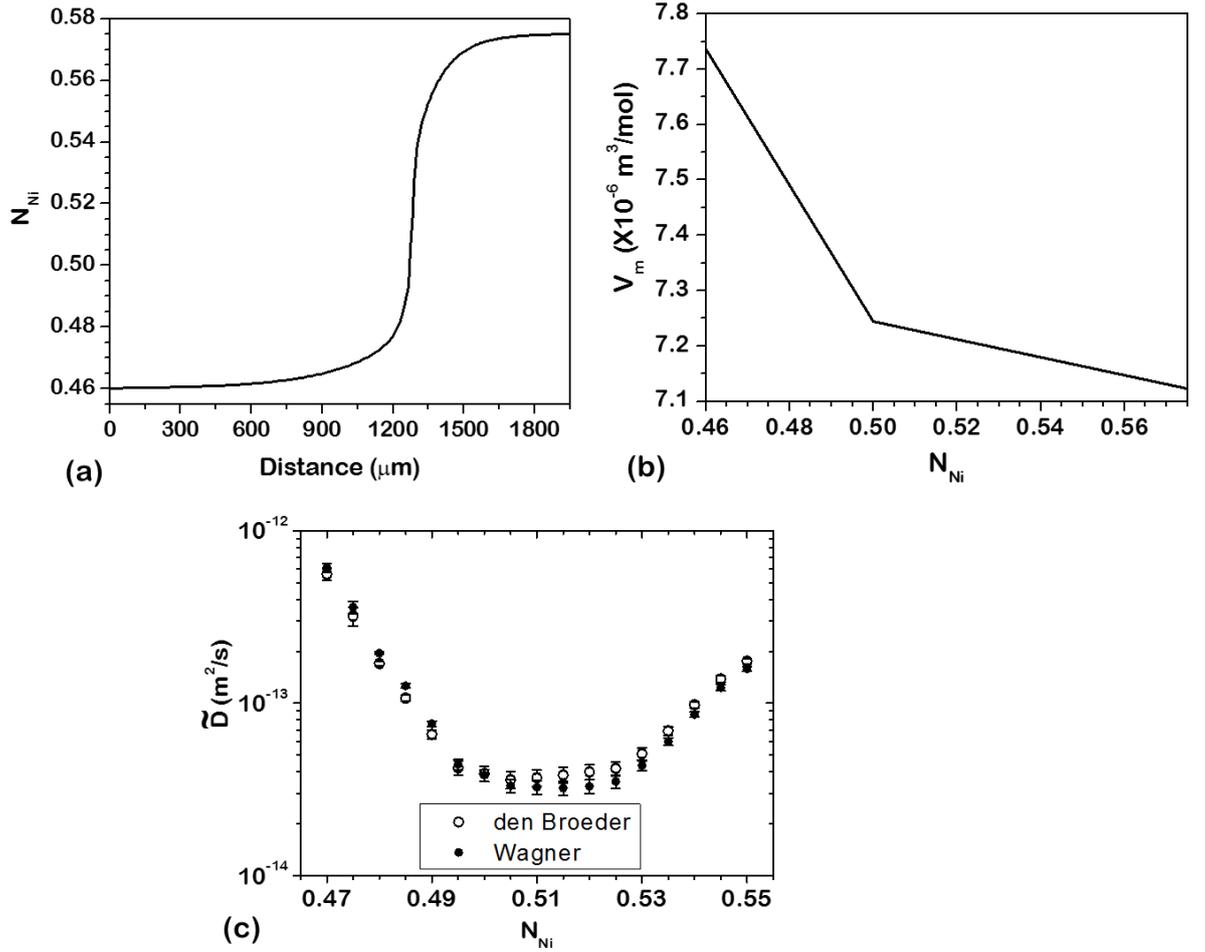

Figure 3.3: (a) Composition profile of β–NiAl phase grown in Ni$_{0.46}$Al$_{0.54}$ / Ni$_{0.575}$Al$_{0.425}$ diffusion couple after annealing at 1200 °C for 24 hrs, (b) the molar volume variation in the β–NiAl phase [6] and (c) the estimated interdiffusion coefficients following the Wagner and the den Broeder approaches.

## 3.3 Integrated Interdiffusion Coefficient

Wagner, in his seminal contribution [10], introduced the concept of the integrated interdiffusion coefficient in a phase with narrow homogeneity range since the concentration/composition gradient in such a phase cannot be determined. This is expressed with respect to $Y_{N_B}$ or $N_B$ as

$$\widetilde{D}_{int}^{\beta} = V_m^{\beta} \Delta x^{\beta} \left( \frac{N_B^{+} - N_B^{-}}{2t} \right) \left[ \frac{Y_{N_B}^{\beta} \left( 1 - Y_{N_B}^{\beta} \right)}{V_m^{\beta}} \Delta x^{\beta} + \left( 1 - Y_{N_B}^{\beta} \right) \int_{x^{-\infty}}^{x^{\beta_1}} \frac{Y_{N_B}}{V_m} dx + \right.$$

$$\left. Y_{N_B}^{\beta} \int_{x^{\beta_2}}^{x^{+\infty}} \frac{\left( 1 - Y_{N_B} \right)}{V_m} dx \right] \tag{3.33a}$$

$$\widetilde{D}_{int}^{\beta} = \frac{\left(N_B^{\beta}-N_B^{-}\right)\left(N_B^{+}-N_B^{\beta}\right)}{\left(N_B^{+}-N_B^{-}\right)}\frac{\left(\Delta x^{\beta}\right)^2}{2t} + \frac{V_m^{\beta}\Delta x^{\beta}}{2t}\left[\frac{\left(N_B^{\beta}-N_B^{-}\right)}{\left(N_B^{+}-N_B^{-}\right)}\int_{x^{-\infty}}^{x^{\beta_1}}\frac{\left(N_B-N_B^{-}\right)}{V_m}dx + \right.$$

$$\left.\frac{\left(N_B^{\beta}-N_B^{-}\right)}{\left(N_B^{+}-N_B^{-}\right)}\int_{x^{\beta_2}}^{x^{+\infty}}\frac{\left(N_B^{+}-N_B\right)}{V_m}dx\right] \tag{3.33b}$$

where $\Delta x^{\beta}$ is the thickness and $N_B^{\beta}$ is the average or stoichiometric composition of the phase of our interest.

At present, the relation to estimate the same diffusion parameter with respect to $Y_{C_B}$ or $C_B$ is not available. Therefore, as given in the supplementary section 3.5.1, we derived this relation by extending the den Broeder's relation for the interdiffusion coefficient. This is expressed with respect to $Y_{C_B}$ or $C_B$ as

$$\widetilde{D}_{int}^{\beta} = \frac{\left(V_m^{\beta}\right)^2}{\bar{V}_A^{\beta}}\Delta x^{\beta}\left(\frac{C_B^{+}-C_B^{-}}{2t}\right)\left[Y_{C_B}^{\beta}\left(1-Y_{C_B}^{\beta}\right)\Delta x^{\beta} + \left(1-Y_{C_B}^{\beta}\right)\int_{x^{-\infty}}^{x^{\beta_1}}Y_{C_B}\,dx + \right.$$

$$\left. Y_{C_B}^{\beta}\int_{x^{\beta_2}}^{x^{+\infty}}\left(1-Y_{C_B}\right)dx\right] \tag{3.34a}$$

$$\widetilde{D}_{int}^{\beta} = \frac{\left(V_m^{\beta}\right)^2}{\bar{V}_A^{\beta}}\frac{\left(C_B^{\beta}-C_B^{-}\right)\left(C_B^{+}-C_B^{\beta}\right)}{\left(C_B^{+}-C_B^{-}\right)}\frac{\left(\Delta x^{\beta}\right)^2}{2t} + \frac{\left(V_m^{\beta}\right)^2}{\bar{V}_A^{\beta}}\frac{\Delta x^{\beta}}{2t}\left[\frac{C_B^{+}-C_B^{\beta}}{C_B^{+}-C_B^{-}}\int_{x^{-\infty}}^{x^{\beta_1}}\left(C_B-C_B^{-}\right)dx + \right.$$

$$\left.\frac{C_B^{\beta}-C_B^{-}}{C_B^{+}-C_B^{-}}\int_{x^{\beta_2}}^{x^{+\infty}}\left(C_B^{+}-C_B\right)dx\right] \tag{3.34b}$$

It can be seen that an additional term of partial molar volume is present in the relation expressed with respect to $Y_{C_B}$ or $C_B$ as compared to the relation expressed with respect to $Y_{N_B}$ or $N_B$.

### 3.3.1 Comparison of the estimated diffusion coefficients in the phases with narrow homogeneity range

Now we compare the efficiencies and difficulties of estimation of the data utilizing the growth of the product phases, as shown in Figure 3.4a, in the Cu–Sn system. The Cu/Sn diffusion couple was annealed at 200 °C for 81 hrs (*i.e.*, $2t = 2\times81\times3600$ s) in which two phases Cu₃Sn and Cu₆Sn₅ grows in the interdiffusion zone [33]. The average thicknesses of the phases are estimated as 3.5 μm (= $\Delta x^{Cu_3Sn}$) for Cu₃Sn and

13 μm (= $\Delta x^{Cu_6Sn_5}$) for Cu₆Sn₅. The marker plane, detected by the presence of

duplex morphology, in the Cu₆Sn₅ phase is found at a distance of 7 μm from the

Cu₃Sn/Cu₆Sn₅ interface. The actual molar volumes of these phases are estimated as

$V_m^{Cu_3Sn} = 8.59 \times 10^{-6}$ and $V_m^{Cu_6Sn_5} = 10.59 \times 10^{-6}$ m³/mol. From the knowledge of

the molar volumes of the end–member components $V_m^- = V_m^{Cu} = 7.12 \times 10^{-6}$ and

$V_m^+ = V_m^{Sn} = 16.24 \times 10^{-6}$ m³/mol, we can estimate the ideal molar volume of the

product phase of interest ($\beta$) following the Vegard's law $V_m^\beta = N_{Cu}^\beta V_m^{Cu} + N_{Sn}^\beta V_m^{Sn}$.

This is estimated as $V_m^{Cu_3Sn}(ideal) = 9.4 \times 10^{-6}$ m³/mol for the Cu₃Sn phase

$\left( N_{Cu}^{Cu_3Sn} = \frac{3}{4}, N_{Sn}^{Cu_3Sn} = \frac{1}{4} \right)$ and $V_m^{Cu_6Sn_5}(ideal) = 11.3 \times 10^{-6}$ m³/mol for the Cu₆Sn₅

phase $\left( N_{Cu}^{Cu_6Sn_5} = \frac{6}{11}, N_{Sn}^{Cu_6Sn_5} = \frac{5}{11} \right)$. Therefore, as shown in Figure 3.4b, the negative deviations of the molar volumes are 8.6% for the Cu₃Sn phase and 6.3% for the Cu₆Sn₅ phase.

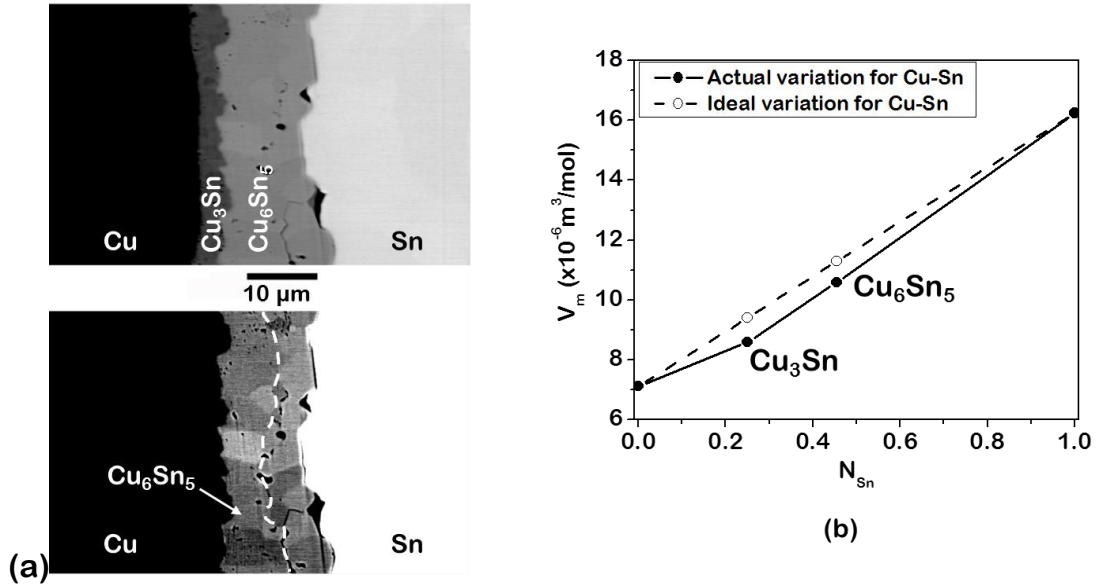

Figure 3.4: (a) BSE micrographs showing (top) the growth of Cu₃Sn and Cu₆Sn₅ phases in the interdiffusion zone of Cu/Sn diffusion couple annealed at 200 °C for 81 hrs [33], where (bottom) the location of the Kirkendall marker plane as denoted by a dashed line is indicated by duplex morphology inside Cu₆Sn₅, and (b) actual as well as ideal variations of molar volume in the Cu–Sn system.

| | Following composition normalize variable | Following concentration normalize variable | | |
|---|---|---|---|---|
| | Using $Y_{N_{Sn}}$ or $Y_{N_{Cu}}$ Actual $V_m$ | Using $Y_{C_{Sn}}$ Actual $V_m$ | Using $Y_{C_{Cu}}$ Actual $V_m$ | Using $Y_{C_{Sn}}$ Ideal $V_m$ |
| $\widetilde{D}_{int}^{Cu_3Sn}$ ($\times 10^{-17}$ m$^2$/s) | 1.26 $\pm 0.05$ | $\frac{V_m}{\overline{V}_{Cu}}8.69\times10^{-18}$ or $(8.69 \pm 0.05)\times10^{-18}$ Considering $\overline{V}_i^\beta = V_m^\beta$ | $\frac{V_m}{\overline{V}_{Sn}}1.90\times10^{-17}$ or $(1.90 \pm 0.05)\times10^{-17}$ Considering $\overline{V}_i^\beta = V_m^\beta$ | 1.28 $\pm 0.05$ |
| $\widetilde{D}_{int}^{Cu_6Sn_5}$ ($\times 10^{-17}$ m$^2$/s) | 8.49 $\pm 1$ | $\frac{V_m}{\overline{V}_{Cu}}4.72\times10^{-17}$ or $(4.72 \pm 1)\times10^{-17}$ Considering $\overline{V}_i^\beta = V_m^\beta$ | $\frac{V_m}{\overline{V}_{Sn}}1.16\times10^{-16}$ or $(1.16 \pm 1)\times10^{-16}$ Considering $\overline{V}_i^\beta = V_m^\beta$ | 8.53 $\pm 1$ |

Table 3.1: Diffusion parameters estimated in the Cu$_3$Sn and Cu$_6$Sn$_5$ phases using Cu and Sn profiles following both the Wagner and the den Broeder methods using the actual as well as the ideal variation of molar volumes in the Cu/Sn diffusion couple.

The detailed estimation procedure following the Wagner method can be found in books as mentioned in Refs. [12, 18]. As explained in detail in the supplementary section 3.5.2, the integrated diffusion coefficients of the phases following this method are estimated as $\widetilde{D}_{int}^{Cu_3Sn} = 1.26 \times 10^{-17} \ m^2/s$ and $\widetilde{D}_{int}^{Cu_6Sn_5} = 8.49 \times 10^{-17} \ m^2/s$. As it should be, the same values are estimated considering the components A and B following Equations S10 or S11 in the supplementary section 3.5.1. The ratio of diffusivities in the Cu$_6$Sn$_5$ phase is estimated as $\frac{D_{Sn}^*}{D_{Cu}^*} = 1.30 \pm 0.05$. It should be noted here that a different value of this ratio was reported in Reference [33], which was an average of data estimated at different locations in different diffusion couples, compared to the data reported in this study estimated based on the micrograph, as shown in Figure 3.4a.

Compared to the Wagner method (Equations S10 or S11 in the supplementary section 3.5.1), den Broeder method (Equations S7 or S8 in the supplementary section 3.5.1) has an additional complication because of the presence of partial molar volume terms in them. In a compound with narrow homogeneity range, the variation of the lattice parameter with respect to the composition is not known. The variation in such a small composition range might be small; however, the difference between the partial molar volumes could still be very high. To circumvent this problem, there could be two options: (i) consider $V_m^\beta = \bar{V}_A^\beta = \bar{V}_B^\beta$, *i.e.*, a constant molar volume in the phase of interest β or (ii) an ideal variation of the molar volume in the whole A–B system. To discuss the pros and cons of these two assumptions, we extend our analysis based on the estimated data in the Cu–Sn system. Following the first assumption, as listed in column number 2 and 3 of Table 3.1, we estimate two different values of the data when estimated following the composition/concentration profile of component A and B. The estimation steps can be found in the supplementary section 3.5.3. This comes from the fact that the assumption leads to different (absolute) values of the interdiffusion fluxes, *i.e.*, $\left|\tilde{J}_A^\beta\right| \neq \left|\tilde{J}_B^\beta\right|$, when estimated following the Equations S6a and S6b (*see* supplementary section 3.5.3) because of this assumption. Moreover, when a constant molar volume is considered, following Equation 3.6, we should have $\bar{V}_B \tilde{J}_B + \bar{V}_A \tilde{J}_A = \tilde{J}_A^\beta + \tilde{J}_B^\beta = 0$. In fact, the assumptions should be taken such that this relation is fulfilled. Therefore, this is not a valid assumption for the estimation of the integrated diffusion coefficients following the den Broeder method, *i.e.*, relations with respect to concentration normalized variable.

Therefore, the den Broeder method for estimation of the integrated diffusion coefficients can be used considering an ideal variation of the molar volume, as shown by dotted line in Figure 3.4b. This fulfills the condition $\bar{V}_B \tilde{J}_B + \bar{V}_A \tilde{J}_A = 0$, where the partial molar volumes are equal to the molar volumes of the end–member components. Following, we get a same value of the integrated diffusion coefficient in a particular phase as $\tilde{D}_{int}^{Cu_3Sn} = 1.28 \times 10^{-17}\ m^2/s$ and $\tilde{D}_{int}^{Cu_6Sn_5} = 8.53 \times 10^{-17}\ m^2/s$. The ratio of diffusivities $\frac{D_{Sn}^*}{D_{Cu}^*}$ is found to be $1.29 \pm 0.05$. It can be seen in Table 3.1 that there is very small difference in the estimated values following Wagner and den Broeder method. Therefore, one can practically follow any of the methods. However, it is advisable to follow the Wagner method since there is no need of considering the ideal molar volume variation instead of considering the actual molar volume variation, which might play a significant effect in certain systems.

## 3.4    Conclusion

The relation for the composition dependent interdiffusion coefficient was first proposed by Matano [20] in 1933, which was difficult to follow in most of the practical systems. As a result, there were many efforts to develop a better relation. Balluffi [37], Sauer–Freise [38], Wagner [26] and den Broeder [27] proposed relations, which played influential role in the field of solid–state diffusion. Currently, two approaches are followed with equal importance by different groups. One was proposed by Wagner with respect to composition normalized variable after considering the molar volume variation and another one was proposed by den Broeder with respect to the concentration normalized variable. Although, it is known to produce different values of the interdiffusion coefficient depending on the molar volume variation [30], the choice of a method by a particular research group is rather random. Incidentally both the methods were published in the same year 1969. The manuscript published by Wagner draws special attention since he put forward the concept of the integrated diffusion coefficient for the phases with narrow homogeneity range in which the interdiffusion coefficients cannot be determined because of unknown composition (or concentration) gradient. This relation is therefore naturally derived with respect to the composition normalized variable. Even the relations for the estimation of the intrinsic diffusion coefficients were also derived by van Loo [29] with respect to the composition normalized variable, which was later derived again by Paul [6] extending the Wagner's analysis.

To examine the veracity of the methods with respect composition and concentration normalized variables, the relation proposed by den Broeder is first derived following the line of treatment followed by Wagner. Following, this is extended to derive the relations for the intrinsic diffusion coefficients and the integrated diffusion coefficients to develop the relations with respect to the concentration normalized variable, which were not available earlier. We have shown further that an additional assumption of the ideal molar volume variation is required for the estimation of the integrated diffusion coefficient with respect to the concentration normalized variable when compared to the relation developed by Wagner with respect to the composition normalized variable, which can be used with actual molar volume variation.

## 3.5 Supplementary Section: Derivation of relations for the estimation of diffusion parameters and Estimation of diffusion data in the Cu–Sn system

### 3.5.1 Derivation of the relation for the Integrated Interdiffusion Coefficient with respect to concentration as well as composition normalized variables

The integrated diffusion coefficient $\left(\widetilde{D}_{int}^{\beta}\right)$ in a phase (β) with narrow homogeneity range is defined as the interdiffusion coefficient $\left(\widetilde{D}\right)$ integrated over the unknown composition range of the phase of interest such that

$$\widetilde{D}_{int}^{\beta} = \int_{N_B^{\beta_1}}^{N_B^{\beta_2}} \widetilde{D}\, dN_B = \widetilde{D} \int_{N_B^{\beta_1}}^{N_B^{\beta_2}} dN_B = \widetilde{D}\left(N_B^{\beta_2} - N_B^{\beta_1}\right)$$

$$\widetilde{D}_{int}^{\beta} = \widetilde{D}\,\Delta N_B^{\beta} \tag{S1}$$

where we can assume that the interdiffusion coefficient $\left(\widetilde{D}\right)$ does not vary significantly over the small composition range of the phase of interest.

Using standard thermodynamic relation $dC_B = \left(\frac{\overline{V}_A}{V_m^2}\right) dN_B$ [18] in Fick's first law [21], $\widetilde{D}_{int}^{\beta}$ can be related to the interdiffusion flux of component B as

$$\tilde{J}_B = -\widetilde{D}\frac{\partial C_B}{\partial x} = -\widetilde{D}\frac{\overline{V}_A}{V_m^2}\frac{\partial N_B}{\partial x} \tag{S2a}$$

$$\tilde{J}_B = -\widetilde{D}\frac{\overline{V}_A^{\beta}}{\left(V_m^{\beta}\right)^2}\frac{\Delta N_B^{\beta}}{\Delta x^{\beta}} \tag{S2b}$$

where $\Delta N_B^{\beta} = N_B^{\beta_2} - N_B^{\beta_1}$ is the narrow homogeneity range of the β phase and $\Delta x^{\beta} = x^{\beta_2} - x^{\beta_1}$ is the thickness of the β phase. Note here that for the phase with narrow homogeneity range, the unknown variation of the slope with the composition (or hence the location parameter *x*) is considered as linear, *i.e.*, $(\Delta N/\Delta x)$. Further, using $\overline{V}_B^{\beta}\tilde{J}_B^{\beta} = -\overline{V}_A^{\beta}\tilde{J}_A^{\beta}$ from Equation 3.6, and utilizing the definition of the integrated interdiffusion coefficient (Equation S1), we can relate it with the interdiffusion fluxes of component B and A as

$$\widetilde{D}_{int}^{\beta} = -\frac{\left(V_m^{\beta}\right)^2}{\overline{V}_A^{\beta}} \Delta x^{\beta} \tilde{J}_B^{\beta} = \frac{\left(V_m^{\beta}\right)^2}{\overline{V}_B^{\beta}} \Delta x^{\beta} \tilde{J}_A^{\beta} \tag{S3}$$

Note that in the case of normal downhill diffusion, the composition profile is plotted such that the component B diffuse from right to left and component A diffuse from left to right. In such a situation, the interdiffusion flux at one particular composition located at a particular location of the diffusion couple after annealing for time $t$, $\tilde{J}_B$ will have a negative sign and $\tilde{J}_A$ will have a positive sign leading equal and positive value of $\widetilde{D}_{int}^{\beta}$ irrespective of the composition profile considered for the estimation. With respect the concentration profiles of components B and A, following the Equations 3.20, we have

$$\tilde{J}_B^{\beta} = \tilde{J}_B\left(C_B^{\beta}\right) = -\left(\frac{C_B^+ - C_B^-}{2t}\right)\left[\left(1 - Y_{C_B}^{\beta}\right)\int_{x-\infty}^{x^*} Y_{C_B}\, dx + Y_{C_B}^{\beta}\int_{x^*}^{x+\infty}\left(1 - Y_{C_B}\right) dx\right] \tag{S4a}$$

$$\tilde{J}_A^{\beta} = \tilde{J}_A\left(C_A^{\beta}\right) = -\left(\frac{C_A^- - C_A^+}{2t}\right)\left[\left(1 - Y_{C_A}^{\beta}\right)\int_{x+\infty}^{x^*} Y_{C_A}\, dx + Y_{C_A}^{\beta}\int_{x^*}^{x-\infty}\left(1 - Y_{C_A}\right) dx\right] \tag{S4b}$$

where $Y_{C_B}^{\beta} = \frac{C_B^{\beta} - C_B^-}{C_B^+ - C_B^-}$ and $Y_{C_A}^{\beta} = \frac{C_A^{\beta} - C_A^+}{C_A^- - C_A^+}$.

The term inside square bracket is separated into 3 parts in the interdiffusion zone as the thickness related to the phase of interest and the other two parts for the interdiffusion zone before and after that:

$$\tilde{J}_B^{\beta} = -\left(\frac{C_B^+ - C_B^-}{2t}\right)\left[\left(1 - Y_{C_B}^{\beta}\right)\int_{x-\infty}^{x^{\beta_1}} Y_{C_B}\, dx + Y_{C_B}^{\beta}\int_{x^{\beta_1}}^{x^{\beta_2}}\left(1 - Y_{C_B}\right) dx + Y_{C_B}^{\beta}\int_{x^{\beta_2}}^{x+\infty}\left(1 - Y_{C_B}\right) dx\right]$$

$$\tag{S5a}$$

$$\tilde{J}_A^{\beta} = -\left(\frac{C_A^- - C_A^+}{2t}\right)\left[\left(1 - Y_{C_A}^{\beta}\right)\int_{x+\infty}^{x^{\beta_2}} Y_{C_A}\, dx + Y_{C_A}^{\beta}\int_{x^{\beta_2}}^{x^{\beta_1}}\left(1 - Y_{C_A}\right) dx + Y_{C_A}^{\beta}\int_{x^{\beta_1}}^{x-\infty}\left(1 - Y_{C_A}\right) dx\right]$$

$$\tag{S5b}$$

In the phase of interest $Y_{C_B}^\beta$ (or $Y_{C_A}^\beta$) is constant because of the growth of the phase with very narrow homogeneity range, *i.e.*, with almost a fixed composition $C_B^\beta$ (or $C_A^\beta$). Therefore, after rearranging, we can write

$$\tilde{J}_B^\beta = -\left(\frac{C_B^+ - C_B^-}{2t}\right)\left[Y_{C_B}^\beta\left(1 - Y_{C_B}^\beta\right)\Delta x^\beta + \left(1 - Y_{C_B}^\beta\right)\int_{x^{-\infty}}^{x^{\beta_1}} Y_{C_B}\, dx + Y_{C_B}^\beta \int_{x^{\beta_2}}^{x^{+\infty}}\left(1 - Y_{C_B}\right) dx\right]$$

(S6a)

$$\tilde{J}_A^\beta = \left(\frac{C_A^- - C_A^+}{2t}\right)\left[Y_{C_A}^\beta(1 - Y_{C_A}^\beta)\Delta x^\beta + \left(1 - Y_{C_A}^\beta\right)\int_{x^{\beta_2}}^{x^{+\infty}} Y_{C_A}\, dx + Y_{C_A}^\beta \int_{x^{-\infty}}^{x^{\beta_1}}\left(1 - Y_{C_A}\right) dx\right]$$

(S6b)

where $\Delta x^\beta = x^{\beta_2} - x^{\beta_1}$ is the thickness of the β phase. It should be noted there that the minus sign in $\tilde{J}_A^\beta$ is omitted because of changing the limits of integration. Therefore, from Equation S3, the integrated diffusion coefficient with respect to $C_B$ and $C_A$ can be expressed as

$$\tilde{D}_{int}^\beta = -\frac{\left(V_m^\beta\right)^2}{\bar{V}_A^\beta}\Delta x^\beta\, \tilde{J}_B^\beta = \frac{\left(V_m^\beta\right)^2}{\bar{V}_A^\beta}\Delta x^\beta\left(\frac{C_B^+ - C_B^-}{2t}\right)\left[Y_{C_B}^\beta\left(1 - Y_{C_B}^\beta\right)\Delta x^\beta + \left(1 - Y_{C_B}^\beta\right)\int_{x^{-\infty}}^{x^{\beta_1}} Y_{C_B}\, dx + Y_{C_B}^\beta \int_{x^{\beta_2}}^{x^{+\infty}}\left(1 - Y_{C_B}\right) dx\right]$$

(S7a)

$$\tilde{D}_{int}^\beta = \frac{\left(V_m^\beta\right)^2}{\bar{V}_B^\beta}\Delta x^\beta\, \tilde{J}_A^\beta = \frac{\left(V_m^\beta\right)^2}{\bar{V}_B^\beta}\Delta x^\beta\left(\frac{C_A^- - C_A^+}{2t}\right)\left[Y_{C_A}^\beta(1 - Y_{C_A}^\beta)\Delta x^\beta + \left(1 - Y_{C_A}^\beta\right)\int_{x^{\beta_2}}^{x^{+\infty}} Y_{C_A}\, dx + Y_{C_A}^\beta \int_{x^{-\infty}}^{x^{\beta_1}}\left(1 - Y_{C_A}\right) dx\right]$$

(S7b)

Further, expanding $Y_{C_B}$ and $Y_{C_A}$ (from Equation S4) with respect to $C_B$ and $C_A$, we get

$$\tilde{D}_{int}^\beta = \frac{\left(V_m^\beta\right)^2}{\bar{V}_A^\beta}\frac{\left(C_B^\beta - C_B^-\right)\left(C_B^+ - C_B^\beta\right)}{\left(C_B^+ - C_B^-\right)}\frac{\left(\Delta x^\beta\right)^2}{2t} + \frac{\left(V_m^\beta\right)^2}{\bar{V}_A^\beta}\frac{\Delta x^\beta}{2t}\left[\frac{C_B^+ - C_B^\beta}{C_B^+ - C_B^-}\int_{x^{-\infty}}^{x^{\beta_1}}(C_B - C_B^-)\, dx + \frac{C_B^\beta - C_B^-}{C_B^+ - C_B^-}\int_{x^{\beta_2}}^{x^{+\infty}}(C_B^+ - C_B)\, dx\right]$$

(S8a)

$$\widetilde{D}_{int}^{\beta} = \frac{\left(V_m^{\beta}\right)^2}{V_B^{\beta}} \frac{\left(C_A^{\beta}-C_A^{+}\right)\left(C_A^{-}-C_A^{\beta}\right)}{\left(C_A^{-}-C_A^{+}\right)} \frac{\left(\Delta x^{\beta}\right)^2}{2t} + \frac{\left(V_m^{\beta}\right)^2}{V_B^{\beta}} \frac{\Delta x^{\beta}}{2t} \left[\frac{C_A^{-}-C_A^{\beta}}{C_A^{-}-C_A^{+}} \int_{x^{\beta_2}}^{x^{+\infty}} \left(C_A - C_A^{+}\right) dx + \right.$$

$$\left. \frac{C_A^{\beta}-C_A^{+}}{C_A^{-}-C_A^{+}} \int_{x^{-\infty}}^{x^{\beta_1}} \left(C_A^{-} - C_A\right) dx \right] \tag{S8b}$$

Note that $C_A + C_B = \frac{1}{V_m}$ and $C_B^{+} > C_B^{-}$, $C_A^{-} > C_A^{+}$.

Therefore, Equations S7 or S8 are relations for the estimation of $\widetilde{D}_{int}^{\beta}$ with respect to $Y_{C_B}$ (and $Y_{C_A}$) or $C_B$ (and $C_A$), which was not available earlier. Previously, Wagner [26] derived the relation with respect to $Y_{N_B}$ or $N_B$ considering non–ideal variation of the molar volume, which can be expressed as [18]

The interdiffusion fluxes from the composition profiles of components B and A are

$$\tilde{J}_B^{\beta} = -\frac{\bar{V}_A^{\beta}}{V_m^{\beta}} \left(\frac{N_B^{+}-N_B^{-}}{2t}\right) \left[\frac{Y_{N_B}^{\beta}\left(1-Y_{N_B}^{\beta}\right)}{V_m^{\beta}}\Delta x^{\beta} + \left(1-Y_{N_B}^{\beta}\right) \int_{x^{-\infty}}^{x^{\beta_1}} \frac{Y_{N_B}}{V_m} dx + Y_{N_B}^{\beta} \int_{x^{\beta_2}}^{x^{+\infty}} \frac{\left(1-Y_{N_B}\right)}{V_m} dx \right]$$
$$\tag{S9a}$$

$$\tilde{J}_A^{\beta} = \frac{\bar{V}_B^{\beta}}{V_m^{\beta}} \left(\frac{N_A^{-}-N_A^{+}}{2t}\right) \left[\frac{Y_{N_A}^{\beta}\left(1-Y_{N_A}^{\beta}\right)}{V_m^{\beta}}\Delta x^{\beta} + \left(1-Y_{N_A}^{\beta}\right) \int_{x^{\beta_2}}^{x^{+\infty}} \frac{Y_{N_A}}{V_m} dx + Y_{N_A}^{\beta} \int_{x^{-\infty}}^{x^{\beta_1}} \frac{\left(1-Y_{N_A}\right)}{V_m} dx \right]$$
$$\tag{S9b}$$

where $Y_{N_B}^{\beta} = \frac{N_B^{\beta}-N_B^{-}}{N_B^{+}-N_B^{-}}$ and $Y_{N_A}^{\beta} = \frac{N_A^{\beta}-N_A^{+}}{N_A^{-}-N_A^{+}}$.

Therefore, from Equation S3, the $\widetilde{D}_{int}^{\beta}$ with respect to $N_B$ and $N_A$ can be expressed as

$$\widetilde{D}_{int}^{\beta} = -\frac{\left(V_m^{\beta}\right)^2}{V_A^{\beta}}\Delta x^{\beta} \tilde{J}_B^{\beta} = V_m^{\beta}\Delta x^{\beta} \left(\frac{N_B^{+}-N_B^{-}}{2t}\right) \left[\frac{Y_{N_B}^{\beta}\left(1-Y_{N_B}^{\beta}\right)}{V_m^{\beta}}\Delta x^{\beta} + \left(1-Y_{N_B}^{\beta}\right) \int_{x^{-\infty}}^{x^{\beta_1}} \frac{Y_{N_B}}{V_m} dx + \right.$$

$$\left. Y_{N_B}^{\beta} \int_{x^{\beta_2}}^{x^{+\infty}} \frac{\left(1-Y_{N_B}\right)}{V_m} dx \right] \tag{S10a}$$

$$\widetilde{D}_{int}^{\beta} = \frac{\left(V_m^{\beta}\right)^2}{V_B^{\beta}}\Delta x^{\beta} \tilde{J}_A^{\beta} = V_m^{\beta}\Delta x^{\beta} \left(\frac{N_A^{-}-N_A^{+}}{2t}\right) \left[\frac{Y_{N_A}^{\beta}\left(1-Y_{N_A}^{\beta}\right)}{V_m^{\beta}}\Delta x^{\beta} + \left(1-Y_{N_A}^{\beta}\right) \int_{x^{\beta_2}}^{x^{+\infty}} \frac{Y_{N_A}}{V_m} dx + \right.$$

$$\left. Y_{N_A}^{\beta} \int_{x^{-\infty}}^{x^{\beta_1}} \frac{\left(1-Y_{N_A}\right)}{V_m} dx \right] \tag{S10b}$$

Further, expanding $Y_{N_B}$ and $Y_{N_A}$ (from Equation S9) with respect to $N_B$ and $N_A$, we get

$$\widetilde{D}_{int}^{\beta} = \frac{\left(N_B^{\beta}-N_B^{-}\right)\left(N_B^{+}-N_B^{\beta}\right)}{\left(N_B^{+}-N_B^{-}\right)}\frac{(\Delta x^{\beta})^2}{2t} + \frac{V_m^{\beta}\Delta x^{\beta}}{2t}\left[\frac{\left(N_B^{\beta}-N_B^{-}\right)}{\left(N_B^{+}-N_B^{-}\right)}\int_{x-\infty}^{x^{\beta_1}}\frac{(N_B-N_B^{-})}{V_m}dx + \right.$$

$$\left.\frac{\left(N_B^{\beta}-N_B^{-}\right)}{\left(N_B^{+}-N_B^{-}\right)}\int_{x^{\beta_2}}^{x^{+\infty}}\frac{(N_B^{+}-N_B)}{V_m}dx\right] \tag{S11a}$$

$$\widetilde{D}_{int}^{\beta} = \frac{\left(N_A^{\beta}-N_A^{+}\right)\left(N_A^{-}-N_A^{\beta}\right)}{\left(N_A^{-}-N_A^{+}\right)}\frac{(\Delta x^{\beta})^2}{2t} + \frac{V_m^{\beta}\Delta x^{\beta}}{2t}\left[\frac{\left(N_A^{-}-N_A^{\beta}\right)}{\left(N_A^{-}-N_A^{+}\right)}\int_{x^{\beta_2}}^{x^{+\infty}}\frac{(N_A-N_A^{+})}{V_m}dx + \right.$$

$$\left.\frac{\left(N_A^{\beta}-N_A^{+}\right)}{\left(N_A^{-}-N_A^{+}\right)}\int_{x-\infty}^{x^{\beta_1}}\frac{(N_A^{-}-N_A)}{V_m}dx\right] \tag{S11b}$$

Note that $N_A + N_B = 1$ and $N_B^{+} > N_B^{-}$, $N_A^{-} > N_A^{+}$. Equations S10 or S11 was derived by Wagner [26], which is expressed with respect to $Y_{N_B}$ (and $Y_{N_A}$) or $N_B$ (and $N_A$).

In an intermetallic compound with narrow homogeneity range, we cannot estimate the composition or concentration gradients. Even we do not know the partial molar volumes of the components in a phase. Therefore, instead of following the Equation 3.29b, we can estimate the ratio of the tracer diffusion coefficients by neglecting the vacancy wind effect following Equations 3.29a, c and 3.32 as

$$\frac{D_B^{*}}{D_A^{*}} = \left[\frac{C_B^{+}\int_{x-\infty}^{x^K}Y_{C_B}dx - C_B^{-}\int_{x^K}^{x+\infty}\left(1-Y_{C_B}\right)dx}{-C_A^{-}\int_{x+\infty}^{x^K}Y_{C_A}dx + C_A^{+}\int_{x^K}^{x-\infty}\left(1-Y_{C_A}\right)dx}\right] \tag{S12}$$

Note here that the contribution of the vacancy wind effect does not contribute very significantly in most of the systems and the difference in estimated data could fall within the limit of experimental error [18]. The same relation with respect to $Y_{N_B}$ and $N_B$ (following Equations 3.30 and 3.32) can be expressed as

$$\frac{D_B^{*}}{D_A^{*}} = \left[\frac{N_B^{+}\int_{x-\infty}^{x^K}\frac{Y_{N_B}}{V_m}dx - N_B^{-}\int_{x^K}^{x+\infty}\frac{\left(1-Y_{N_B}\right)}{V_m}dx}{-N_A^{-}\int_{x-\infty}^{x^K}\frac{Y_{N_B}}{V_m}dx + N_A^{+}\int_{x^K}^{x+\infty}\frac{\left(1-Y_{N_B}\right)}{V_m}dx}\right] \tag{S13}$$

Since these relations are free from partial molar volume terms, there data can be estimated straightforwardly.

***Estimation of the integrated diffusion coefficients***

***following different methods in the Cu–Sn system***

The interdiffusion zone is shown in Figure 3.4a. The average thicknesses of the phases are $\Delta x^{Cu_3Sn} = 3.5$ μm and $\Delta x^{Cu_6Sn_5} = 13$ μm. The couple was annealed for 81 hrs and therefore $2t = 2 \times 81 \times 3600$ s. Marker plane in $Cu_6Sn_5$ phase is found at a distance of 7 μm from $Cu_3Sn/Cu_6Sn_5$ interface. The molar volumes of the phases are $V_m^{Cu_3Sn} = 8.59 \times 10^{-6}$ and $V_m^{Cu_6Sn_5} = 10.59 \times 10^{-6}$ m³/mol.

The composition profile with respect to component Sn and Cu is shown in Figure 3.5.

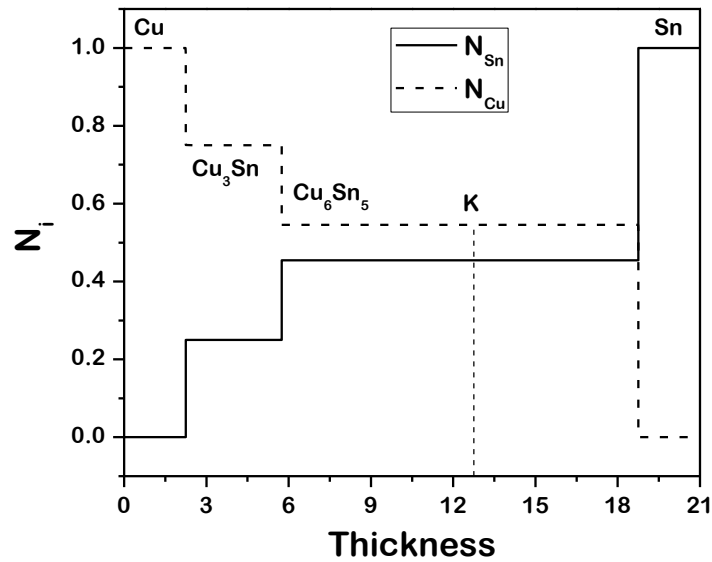

Figure 3.5: The average composition profiles of Cu and Sn in the Cu–Sn diffusion couple annealed at 200 °C for 81 hrs. The micrograph is shown in Figure 3.4a.

### 3.5.2. Estimation with respect to the composition normalized variable following the Wagner method:

Composition normalized variables with respect to component Sn and Cu are

$$Y_{N_{Sn}}^{Cu_3Sn} = \frac{N_{Sn}^{Cu_3Sn} - N_{Sn}^-}{N_{Sn}^+ - N_{Sn}^-} = \frac{\frac{1}{4} - 0}{1 - 0} = \frac{1}{4}, \ Y_{N_{Sn}}^{Cu_6Sn_5} = \frac{N_{Sn}^{Cu_6Sn_5} - N_{Sn}^-}{N_{Sn}^+ - N_{Sn}^-} = \frac{\frac{5}{11} - 0}{1 - 0} = \frac{5}{11}$$

$$Y_{N_{Cu}}^{Cu_3Sn} = \frac{N_{Cu}^{Cu_3Sn} - N_{Cu}^+}{N_{Cu}^- - N_{Cu}^+} = \frac{\frac{3}{4} - 0}{1 - 0} = \frac{3}{4}, \ Y_{N_{Cu}}^{Cu_6Sn_5} = \frac{N_{Cu}^{Cu_6Sn_5} - N_{Cu}^+}{N_{Cu}^- - N_{Cu}^+} = \frac{\frac{6}{11} - 0}{1 - 0} = \frac{6}{11}$$

### 3.5.2.1 Estimation in the Cu$_3$Sn phase

For the Cu$_3$Sn phase, since there is no phase in the interdiffusion zone between Cu and the phase of interest, the second term inside the bracket in Equation S9a becomes zero and we can write the interdiffusion flux from the composition profile of component Sn as

$$\tilde{J}_{Sn}^{Cu_3Sn} = -\frac{\bar{V}_{Cu}^{Cu_3Sn}}{V_m^{Cu_3Sn}}\left(\frac{N_{Sn}^+ - N_{Sn}^-}{2t}\right)\left[\frac{Y_{N_{Sn}}^{Cu_3Sn}\left(1-Y_{N_{Sn}}^{Cu_3Sn}\right)}{V_m^{Cu_3Sn}}\Delta x^{Cu_3Sn} + 0 + Y_{N_{Sn}}^{Cu_3Sn}\frac{\left(1-Y_{N_{Sn}}^{Cu_6Sn_5}\right)}{V_m^{Cu_6Sn_5}}\Delta x^{Cu_6Sn_5}\right]$$

$$\tilde{J}_{Sn}^{Cu_3Sn} = -\frac{\bar{V}_{Cu}^{Cu_3Sn}}{V_m^{Cu_3Sn}}\left(\frac{1-0}{2\times 81\times 3600}\right)\left[\frac{\frac{1}{4}\left(1-\frac{1}{4}\right)}{8.59\times 10^{-6}}3.5\times 10^{-6} + 0 + \frac{1}{4}\frac{\left(1-\frac{5}{11}\right)}{10.59\times 10^{-6}}13\times 10^{-6}\right]$$

$$\tilde{J}_{Sn}^{Cu_3Sn} = -\frac{\bar{V}_{Cu}^{Cu_3Sn}}{V_m^{Cu_3Sn}}\times 4.18\times 10^{-7}\ mol/m^2.s$$

Following Equation S3,

$$\tilde{D}_{int}^{Cu_3Sn} = \frac{\left(V_m^{Cu_3Sn}\right)^2}{\bar{V}_{Cu}^{Cu_3Sn}}\Delta x^{Cu_3Sn}\left(-\tilde{J}_{Sn}^{Cu_3Sn}\right)$$

$$\tilde{D}_{int}^{Cu_3Sn} = \frac{\left(V_m^{Cu_3Sn}\right)^2}{\bar{V}_{Sn}^{Cu_3Sn}}\Delta x^{Cu_3Sn}\left(\frac{\bar{V}_{Sn}^{Cu_3Sn}}{V_m^{Cu_3Sn}}\times 4.18\times 10^{-7}\right)$$

$$\tilde{D}_{int}^{Cu_3Sn} = V_m^{Cu_3Sn}\Delta x^{Cu_3Sn}(4.18\times 10^{-7}) = 8.59\times 10^{-6}\times 3.5\times 10^{-6}\times(4.18\times 10^{-7})$$

$$\tilde{D}_{int}^{Cu_3Sn} = 1.26\times 10^{-17}\ m^2/s$$

Following Equation S9b, since there is Cu$_6$Sn$_5$ phase in the interdiffusion zone between Sn and Cu$_3$Sn, and no phase is between Cu$_3$Sn and Cu, we can write the interdiffusion flux with respect to component Cu as

$$\tilde{J}_{Cu}^{Cu_3Sn} = \frac{\bar{V}_{Sn}^{Cu_3Sn}}{V_m^{Cu_3Sn}}\left(\frac{N_{Cu}^- - N_{Cu}^+}{2t}\right)\left[\frac{Y_{N_{Cu}}^{Cu_3Sn}\left(1-Y_{N_{Cu}}^{Cu_3Sn}\right)}{V_m^{Cu_3Sn}}\Delta x^{Cu_3Sn} + \left(1-Y_{N_{Cu}}^{Cu_3Sn}\right)\frac{Y_{N_{Cu}}^{Cu_6Sn_5}}{V_m^{Cu_6Sn_5}}\Delta x^{Cu_6Sn_5} + 0\right]$$

$$\tilde{J}_{Cu}^{Cu_3Sn} = \frac{\bar{V}_{Sn}^{Cu_3Sn}}{V_m^{Cu_3Sn}}\left(\frac{1-0}{2\times 81\times 3600}\right)\left[\frac{\frac{3}{4}\left(1-\frac{3}{4}\right)}{8.59\times 10^{-6}}3.5\times 10^{-6} + \left(1-\frac{3}{4}\right)\frac{\frac{6}{11}}{10.59\times 10^{-6}}13\times 10^{-6} + 0\right]$$

$$\tilde{J}_{Cu}^{Cu_3Sn} = \frac{\bar{V}_{Sn}^{Cu_3Sn}}{V_m^{Cu_3Sn}}\times 4.18\times 10^{-7}\ mol/m^2.s$$

Following Equation S3,

$$\widetilde{D}_{int}^{Cu_3Sn} = \frac{\left(V_m^{Cu_3Sn}\right)^2}{\bar{V}_{Cu}^{Cu_3Sn}} \Delta x^{Cu_3Sn} \left(\tilde{J}_{Cu}^{Cu_3Sn}\right)$$

$$\widetilde{D}_{int}^{Cu_3Sn} = \frac{\left(V_m^{Cu_3Sn}\right)^2}{\bar{V}_{Sn}^{Cu_3Sn}} \Delta x^{Cu_3Sn} \left(\frac{\bar{V}_{Sn}^{Cu_3Sn}}{V_m^{Cu_3Sn}} \times 4.18 \times 10^{-7}\right)$$

$$\widetilde{D}_{int}^{Cu_3Sn} = V_m^{Cu_3Sn} \Delta x^{Cu_3Sn} (4.18 \times 10^{-7}) = 8.59 \times 10^{-6} \times 3.5 \times 10^{-6} \times (4.18 \times 10^{-7})$$

$$\widetilde{D}_{int}^{Cu_3Sn} = 1.26 \times 10^{-17} \ m^2/s$$

Therefore, as it should be, the exactly same value is estimated following the composition profiles of Sn and Cu. Most importantly, one has to make sure that Equation 3.6 is fulfilled. This is indeed fulfilled since

$$\bar{V}_{Cu}^{Cu_3Sn} \tilde{J}_{Cu}^{Cu_3Sn} + \bar{V}_{Sn}^{Cu_3Sn} \tilde{J}_{Sn}^{Cu_3Sn} = \bar{V}_{Cu}^{Cu_3Sn} \frac{\bar{V}_{Sn}^{Cu_3Sn}}{V_m^{Cu_3Sn}} \times 4.18 \times 10^{-7} - \bar{V}_{Sn}^{Cu_3Sn} \frac{\bar{V}_{Cu}^{Cu_3Sn}}{V_m^{Cu_3Sn}} \times 4.18 \times 10^{-7} = 0$$

### 3.5.2.2 Estimation in the Cu$_6$Sn$_5$ phase

Following Equation S9a, since there is no phase in the interdiffusion zone between Cu$_6$Sn$_5$ and Sn, we can write the interdiffusion flux with respect to component Sn

$$\tilde{J}_{Sn}^{Cu_6Sn_5} = -\frac{\bar{V}_{Cu}^{Cu_6Sn_5}}{V_m^{Cu_6Sn_5}} \left(\frac{N_{Sn}^+ - N_{Sn}^-}{2t}\right) \left[\frac{Y_{N_{Sn}}^{Cu_6Sn_5}\left(1 - Y_{N_{Sn}}^{Cu_6Sn_5}\right)}{V_m^{Cu_6Sn_5}} \Delta x^{Cu_6Sn_5} + \left(1 - Y_{N_{Sn}}^{Cu_6Sn_5}\right) \frac{Y_{N_{Sn}}^{Cu_3Sn}}{V_m^{Cu_3Sn}} \Delta x^{Cu_3Sn} + 0\right]$$

$$\tilde{J}_{Sn}^{Cu_6Sn_5} = -\frac{\bar{V}_{Cu}^{Cu_6Sn_5}}{V_m^{Cu_6Sn_5}} \left(\frac{1-0}{2 \times 81 \times 3600}\right) \left[\frac{\frac{5}{11}\left(1 - \frac{5}{11}\right)}{10.59 \times 10^{-6}} 13 \times 10^{-6} + \left(1 - \frac{5}{11}\right) \frac{\frac{1}{4}}{8.59 \times 10^{-6}} 3.5 \times 10^{-6} + 0\right]$$

$$\tilde{J}_{Sn}^{Cu_6Sn_5} = -\frac{\bar{V}_{Cu}^{Cu_6Sn_5}}{V_m^{Cu_6Sn_5}} 6.17 \times 10^{-7} \ mol/m^2.s$$

Following Equation S3,

$$\widetilde{D}_{int}^{Cu_6Sn_5} = \frac{\left(V_m^{Cu_6Sn_5}\right)^2}{\bar{V}_{Cu}^{Cu_6Sn_5}} \Delta x^{Cu_6Sn_5} \left(-\tilde{J}_{Sn}^{Cu_6Sn_5}\right)$$

$$\widetilde{D}_{int}^{Cu_6Sn_5} = \frac{\left(V_m^{Cu_6Sn_5}\right)^2}{\bar{V}_{Cu}^{Cu_6Sn_5}} \Delta x^{Cu_6Sn_5} \left(\frac{\bar{V}_{Cu}^{Cu_6Sn_5}}{V_m^{Cu_6Sn_5}} 6.17 \times 10^{-7}\right)$$

$$\widetilde{D}_{int}^{Cu_6Sn_5} = V_m^{Cu_6Sn_5} \Delta x^{Cu_6Sn_5} (6.17 \times 10^{-7}) = 10.59 \times 10^{-6} \times 13 \times 10^{-6} \times (6.17 \times 10^{-7})$$

$$\widetilde{D}_{int}^{Cu_6Sn_5} = 8.49 \times 10^{-17} \ m^2/s$$

Following Equation S9b, since there is no phase in the interdiffusion zone between Sn and Cu$_6$Sn$_5$, and Cu$_3$Sn phase is between Cu$_6$Sn$_5$ and Cu, we can write the interdiffusion flux with respect to component Cu as

$$\tilde{J}_{Cu}^{Cu_6Sn_5} = \frac{\bar{V}_{Sn}^{Cu_6Sn_5}}{V_m^{Cu_6Sn_5}}\left(\frac{N_{Cu}^- - N_{Cu}^+}{2t}\right)\left[\frac{Y_{N_{Cu}}^{Cu_6Sn_5}\left(1-Y_{N_{Cu}}^{Cu_6Sn_5}\right)}{V_m^{Cu_6Sn_5}}\Delta x^{Cu_6Sn_5} + 0 + Y_{N_{Cu}}^{Cu_6Sn_5}\frac{\left(1-Y_{N_{Cu}}^{Cu_3Sn}\right)}{V_m^{Cu_3Sn}}\Delta x^{Cu_3Sn}\right]$$

$$\tilde{J}_{Cu}^{Cu_6Sn_5} = \frac{\bar{V}_{Sn}^{Cu_6Sn_5}}{V_m^{Cu_6Sn_5}}\left(\frac{1-0}{2\times 81\times 3600}\right)\left[\frac{\frac{6}{11}\left(1-\frac{6}{11}\right)}{10.59\times10^{-6}}13\times10^{-6} + 0 + \frac{6}{11}\frac{\left(1-\frac{3}{4}\right)}{8.59\times10^{-6}}3.5\times10^{-6}\right]$$

$$\tilde{J}_{Cu}^{Cu_6Sn_5} = \frac{\bar{V}_{Sn}^{Cu_6Sn_5}}{V_m^{Cu_6Sn_5}}6.17\times10^{-7}\ mol/m^2.s$$

Following Equation S3,

$$\tilde{D}_{int}^{Cu_6Sn_5} = \frac{\left(V_m^{Cu_6Sn_5}\right)^2}{\bar{V}_{Sn}^{Cu_6Sn_5}}\Delta x^{Cu_6Sn_5}\left(\tilde{J}_{Cu}^{Cu_6Sn_5}\right)$$

$$\tilde{D}_{int}^{Cu_6Sn_5} = \frac{\left(V_m^{Cu_6Sn_5}\right)^2}{\bar{V}_{Sn}^{Cu_6Sn_5}}\Delta x^{Cu_6Sn_5}\left(\frac{\bar{V}_{Sn}^{Cu_6Sn_5}}{V_m^{Cu_6Sn_5}}6.17\times10^{-7}\right)$$

$$\tilde{D}_{int}^{Cu_6Sn_5} = V_m^{Cu_6Sn_5}\Delta x^{Cu_6Sn_5}(6.17\times10^{-7}) = 10.59\times10^{-6}\times13\times10^{-6}\times(6.17\times10^{-7})$$

$$\tilde{D}_{int}^{Cu_6Sn_5} = 8.49\times10^{-17}\ m^2/s$$

Therefore, again we have the same values when estimated with respect to component Sn and Cu. We can also verify the Equation 3.6 following

$$\bar{V}_{Cu}^{Cu_6Sn_5}\tilde{J}_{Cu}^{Cu_6Sn_5} + \bar{V}_{Sn}^{Cu_6Sn_5}\tilde{J}_{Sn}^{Cu_6Sn_5} = \bar{V}_{Cu}^{Cu_6Sn_5}\frac{\bar{V}_{Sn}^{Cu_6Sn_5}}{V_m^{Cu_6Sn_5}}6.17\times10^{-7} - \bar{V}_{Sn}^{Cu_6Sn_5}\frac{\bar{V}_{Cu}^{Cu_6Sn_5}}{V_m^{Cu_6Sn_5}}6.17\times10^{-7} = 0$$

It is to be noted there that although the partial molar volume terms are unknown, we could still verify the condition in Equation 3.6 is indeed fulfill. However, the same is not true with respect to the concentration normalized variable, as shown in the next section.

### 3.5.3. Estimation with respect to the concentration normalized variable following the relations derived in the present work:

Since the partial molar volumes of components in the $\beta$ phase are unknown, it is evident from Equations S7 or S8 that we cannot estimate $\widetilde{D}_{int}^{\beta}$ directly with respect to $Y_{C_B}$ (and $Y_{C_A}$) or $C_B$ (and $C_A$). To facilitate the discussion on one of the important points as discussed in the manuscript, we estimate data for both the actual and the ideal molar volume of the phase of interest β.

**3.5.3.1 Estimation of the data considering the actual molar volume variation**

For the actual $V_m$ of phases, we can write

$$Y_{C_{Sn}}^{Cu_3Sn} = \frac{C_{Sn}^{Cu_3Sn} - C_{Sn}^-}{C_{Sn}^+ - C_{Sn}^-} = \frac{\left(\frac{1}{4}\right)/(8.59\times10^{-6}) - 0}{1/(16.24\times10^{-6}) - 0} = \frac{1\times16.24}{4\times8.59}$$

$$Y_{C_{Sn}}^{Cu_6Sn_5} = \frac{C_{Sn}^{Cu_6Sn_5} - C_{Sn}^-}{C_{Sn}^+ - C_{Sn}^-} = \frac{\left(\frac{5}{11}\right)/(10.59\times10^{-6}) - 0}{1/(16.24\times10^{-6}) - 0} = \frac{5\times16.24}{11\times10.59}$$

$$Y_{C_{Cu}}^{Cu_3Sn} = \frac{C_{Cu}^{Cu_3Sn} - C_{Cu}^+}{C_{Cu}^- - C_{Cu}^+} = \frac{\left(\frac{3}{4}\right)/(8.59\times10^{-6}) - 0}{1/(7.12\times10^{-6}) - 0} = \frac{3\times7.12}{4\times8.59}$$

$$Y_{C_{Cu}}^{Cu_6Sn_5} = \frac{C_{Cu}^{Cu_6Sn_5} - C_{Cu}^+}{C_{Cu}^- - C_{Cu}^+} = \frac{\left(\frac{6}{11}\right)/(10.59\times10^{-6}) - 0}{1/(7.12\times10^{-6}) - 0} = \frac{6\times7.12}{11\times10.59}$$

**3.5.3.1.1  Estimation in the Cu₃Sn phase**

Following Equation S6a, we can write the interdiffusion flux with respect to component Sn as

$$\tilde{J}_{Sn}^{Cu_3Sn} = -\left(\frac{C_{Sn}^+ - C_{Sn}^-}{2t}\right)\left[Y_{C_{Sn}}^{Cu_3Sn}\left(1 - Y_{C_{Sn}}^{Cu_3Sn}\right)\Delta x^{Cu_3Sn} + 0 + Y_{C_{Sn}}^{Cu_3Sn}\left(1 - Y_{C_{Sn}}^{Cu_6Sn_5}\right)\Delta x^{Cu_6Sn_5}\right]$$

$$\tilde{J}_{Sn}^{Cu_3Sn} = -\left(\frac{\frac{1}{16.24\times10^{-6}}}{2\times81\times3600}\right)\left[\frac{16.24}{4\times8.59}\left(1 - \frac{16.24}{4\times8.59}\right)3.5\times10^{-6} + \frac{16.24}{4\times8.59}\left(1 - \frac{5\times16.24}{11\times10.59}\right)13\times10^{-6}\right]$$

$$\tilde{J}_{Sn}^{Cu_3Sn} = -2.89\times10^{-7} \; mol/m^2.s$$

Following Equation S3,

$$\widetilde{D}_{int}^{Cu_3Sn} = \frac{\left(V_m^{Cu_3Sn}\right)^2}{\bar{V}_{Cu}^{Cu_3Sn}}\Delta x^{Cu_3Sn}\left(-\tilde{J}_{Sn}^{Cu_3Sn}\right)$$

$$\widetilde{D}_{int}^{Cu_3Sn} = \frac{\left(V_m^{Cu_3Sn}\right)^2}{\overline{V}_{Cu}^{Cu_3Sn}} \Delta x^{Cu_3Sn}(2.89{\times}10^{-7})$$

$$\widetilde{D}_{int}^{Cu_3Sn} = \frac{V_m^{Cu_3Sn}}{\overline{V}_{Cu}^{Cu_3Sn}} 8.59{\times}10^{-6}{\times}3.5{\times}10^{-6}{\times}(2.89{\times}10^{-7})$$

$$\widetilde{D}_{int}^{Cu_3Sn} = \frac{V_m^{Cu_3Sn}}{\overline{V}_{Cu}^{Cu_3Sn}}(8.69{\times}10^{-18})\ m^2/s.$$

Since partial molar volumes are not known it is a common practice to consider the partial molar volumes as equal to the molar volume in a phase with narrow homogeneity range. Therefore, the estimated value would be

$$\widetilde{D}_{int}^{Cu_3Sn} = 8.69{\times}10^{-18}\ m^2/s$$

However, a major problem is faced with this assumption (*i.e.*, considering $\overline{V}_i^{\beta} = V_m^{\beta}$), when the same data are estimated with respect to the composition profile of another component. Following Equation S6b, we can write the interdiffusion flux with respect to component Cu as

$$\tilde{J}_{Cu}^{Cu_3Sn} = \left(\frac{C_{Cu}^{-} - C_{Cu}^{+}}{2t}\right)\left[Y_{C_{Cu}}^{Cu_3Sn}(1 - Y_{C_{Cu}}^{Cu_3Sn})\Delta x^{Cu_3Sn} + \left(1 - Y_{C_{Cu}}^{Cu_3Sn}\right)Y_{C_{Cu}}^{Cu_6Sn_5}\Delta x^{Cu_6Sn_5} + 0\right]$$

$$\tilde{J}_{Cu}^{Cu_3Sn} = \left(\frac{\frac{1}{7.12{\times}10^{-6}}}{2{\times}81{\times}3600}\right)\left[\left(\frac{3{\times}7.12}{4{\times}8.59}\right)\left(1 - \frac{3{\times}7.12}{4{\times}8.59}\right)3.5{\times}10^{-6} + \left(1 - \frac{3{\times}7.12}{4{\times}8.59}\right)\left(\frac{6{\times}7.12}{11{\times}10.59}\right)13{\times}10^{-6}\right]$$

$$\tilde{J}_{Cu}^{Cu_3Sn} = 6.33{\times}10^{-7}\ mol/m^2.s$$

Following Equation S3,

$$\widetilde{D}_{int}^{Cu_3Sn} = \frac{\left(V_m^{Cu_3Sn}\right)^2}{\overline{V}_{Sn}^{Cu_3Sn}} \Delta x^{Cu_3Sn}\left(\tilde{J}_{Cu}^{Cu_3Sn}\right)$$

$$\widetilde{D}_{int}^{Cu_3Sn} = \frac{\left(V_m^{Cu_3Sn}\right)^2}{\overline{V}_{Sn}^{Cu_3Sn}} \Delta x^{Cu_3Sn}(6.33{\times}10^{-7})$$

$$\widetilde{D}_{int}^{Cu_3Sn} = \frac{V_m^{Cu_3Sn}}{\overline{V}_{Sn}^{Cu_3Sn}} 8.59{\times}10^{-6}{\times}3.5{\times}10^{-6}{\times}(6.33{\times}10^{-7})$$

$$\widetilde{D}_{int}^{Cu_3Sn} = \frac{V_m^{Cu_3Sn}}{\overline{V}_{Sn}^{Cu_3Sn}}(1.90{\times}10^{-17})\ m^2/s$$

Therefore, if we consider again that partial molar volumes are equal to the molar volume of the phase then we have $\widetilde{D}_{int}^{Cu_3Sn} = 1.90 \times 10^{-17}\ m^2/s$. This lead to different values when the diffusion coefficients are estimated with respect to the component Sn and Cu, which is not acceptable. This is resulted from the fact that the estimated interdiffusion fluxes do not fulfill the condition $\overline{V}_{Cu}^{Cu_3Sn} \widetilde{J}_{Cu}^{Cu_3Sn} + \overline{V}_{Sn}^{Cu_3Sn} \widetilde{J}_{Sn}^{Cu_3Sn} = 0$, which can be understood (considering the partial molar volumes as the same) from the estimated values of the interdiffusion fluxes.

### 3.5.3.1.2 Estimation in the Cu₆Sn₅ phase

Following Equation S6a, we can write the interdiffusion flux with respect to component Sn as

$$\widetilde{J}_{Sn}^{Cu_6Sn_5} = -\left(\frac{C_{Sn}^+ - C_{Sn}^-}{2t}\right)\left[Y_{C_{Sn}}^{Cu_6Sn_5}\left(1 - Y_{C_{Sn}}^{Cu_6Sn_5}\right)\Delta x^{Cu_6Sn_5} + \left(1 - Y_{C_{Sn}}^{Cu_6Sn_5}\right)Y_{C_{Sn}}^{Cu_3Sn}\Delta x^{Cu_3Sn} + 0\right]$$

$$\widetilde{J}_{Sn}^{Cu_6Sn_5} = -\left(\frac{1}{2 \times 81 \times 3600}\right)\left[\frac{5 \times 16.24}{11 \times 10.59}\left(1 - \frac{5 \times 16.24}{11 \times 10.59}\right)13 \times 10^{-6} + \left(1 - \frac{5 \times 16.24}{11 \times 10.59}\right)\frac{1 \times 16.24}{4 \times 8.59}3.5 \times 10^{-6}\right]$$

$$\widetilde{J}_{Sn}^{Cu_6Sn_5} = -3.43 \times 10^{-7}\ mol/m^2.s$$

Following Equation S3,

$$\widetilde{D}_{int}^{Cu_6Sn_5} = \frac{\left(V_m^{Cu_6Sn_5}\right)^2}{\overline{V}_{Cu}^{Cu_6Sn_5}}\Delta x^{Cu_6Sn_5}\left(-\widetilde{J}_{Sn}^{Cu_6Sn_5}\right)$$

$$\widetilde{D}_{int}^{Cu_6Sn_5} = \frac{\left(V_m^{Cu_6Sn_5}\right)^2}{\overline{V}_{Cu}^{Cu_6Sn_5}}\Delta x^{Cu_6Sn_5}(3.43 \times 10^{-7})$$

$$\widetilde{D}_{int}^{Cu_6Sn_5} = \frac{V_m^{Cu_6Sn_5}}{\overline{V}_{Cu}^{Cu_6Sn_5}}10.59 \times 10^{-6} \times 13 \times 10^{-6} \times (3.43 \times 10^{-7})$$

$$\widetilde{D}_{int}^{Cu_6Sn_5} = \frac{V_m^{Cu_6Sn_5}}{\overline{V}_{Cu}^{Cu_6Sn_5}}(4.72 \times 10^{-17})\ m^2/s$$

Since the partial molar volumes are not known, if we consider that partial molar volumes are equal to the actual molar volume, we have

$$\widetilde{D}_{int}^{Cu_6Sn_5} = 4.72 \times 10^{-17}\ m^2/s$$

Following Equation S6b, we can write the interdiffusion flux with respect to component Cu as

$$\tilde{J}_{Cu}^{Cu_6Sn_5} = \left(\frac{C_{Cu}^- - C_{Cu}^+}{2t}\right)\left[Y_{C_{Cu}}^{Cu_6Sn_5}\left(1 - Y_{C_{Cu}}^{Cu_6Sn_5}\right)\Delta x^{Cu_6Sn_5} + 0 + Y_{C_{Cu}}^{Cu_6Sn_5}\left(1 - Y_{C_{Cu}}^{Cu_3Sn}\right)\Delta x^{Cu_3Sn}\right]$$

$$\tilde{J}_{Cu}^{Cu_6Sn_5} = \left(\frac{\frac{1}{7.12\times10^{-6}}}{2\times81\times3600}\right)\left[\left(\frac{6\times7.12}{11\times10.59}\right)\left(1 - \frac{6\times7.12}{11\times10.59}\right)13\times10^{-6} + \left(\frac{6\times7.12}{11\times10.59}\right)\left(1 - \frac{3\times7.12}{4\times8.59}\right)3.5\times10^{-6}\right]$$

$$\tilde{J}_{Cu}^{Cu_6Sn_5} = 8.44\times10^{-7} \ mol/m^2.s$$

Following Equation S3,

$$\widetilde{D}_{int}^{Cu_6Sn_5} = \frac{\left(V_m^{Cu_6Sn_5}\right)^2}{\bar{V}_{Sn}^{Cu_6Sn_5}}\Delta x^{Cu_6Sn_5}\left(\tilde{J}_{Cu}^{Cu_6Sn_5}\right)$$

$$\widetilde{D}_{int}^{Cu_6Sn_5} = \frac{\left(V_m^{Cu_6Sn_5}\right)^2}{\bar{V}_{Sn}^{Cu_6Sn_5}}\Delta x^{Cu_6Sn_5}(8.44\times10^{-7})$$

$$\widetilde{D}_{int}^{Cu_6Sn_5} = \frac{V_m^{Cu_6Sn_5}}{\bar{V}_{Sn}^{Cu_6Sn_5}}10.59\times10^{-6}\times13\times10^{-6}\times(8.44\times10^{-7})$$

$$\widetilde{D}_{int}^{Cu_6Sn_5} = \frac{V_m^{Cu_6Sn_5}}{\bar{V}_{Sn}^{Cu_6Sn_5}}(1.16\times10^{-16}) \ m^2/s$$

Again, if we consider the partial molar volumes as equal to the molar volume, we have

$$\widetilde{D}_{int}^{Cu_6Sn_5} = 1.16\times10^{-16} \ m^2/s.$$

This situation arises, since the relation expressed in Equation 3.6, *i.e.*, $\bar{V}_{Cu}^{Cu_3Sn}\tilde{J}_{Cu}^{Cu_3Sn} + \bar{V}_{Sn}^{Cu_3Sn}\tilde{J}_{Sn}^{Cu_3Sn} = 0$ does not fulfill with this assumption.

Therefore, we cannot consider the actual molar volume variation for the estimation of the integrated diffusion coefficient following the relations derived with respect to concentration normalized variable. Now let us examine the situations considering the ideal molar volume variations.

### 3.5.3.2 Estimation of the data considering the ideal molar volume variation

Molar volumes of end–members: $V_m^{Cu} = 7.12\times10^{-6}$ and $V_m^{Sn} = 16.24\times10^{-6}$ m³/mol

Ideal molar volume of $\beta$ phase following the Vegard's law is $V_m^\beta = N_{Cu}^\beta V_m^{Cu} + N_{Sn}^\beta V_m^{Sn}$

Therefore, $V_m^{Cu_3Sn}(ideal) = 9.4{\times}10^{-6}$ m$^3$/mol and $V_m^{Cu_6Sn_5}(ideal) = 11.3{\times}10^{-6}$ m$^3$/mol

Considering the ideal variation of molar volume, we can write

$$Y_{C_{Sn}}^{Cu_3Sn} = \frac{c_{Sn}^{Cu_3Sn} - c_{Sn}^-}{c_{Sn}^+ - c_{Sn}^-} = \frac{\left(\frac{1}{4}\right)/(9.4 \times 10^{-6}) - 0}{1/(16.24 \times 10^{-6}) - 0} = \frac{1 \times 16.24}{4 \times 9.4}$$

$$Y_{C_{Sn}}^{Cu_6Sn_5} = \frac{c_{Sn}^{Cu_6Sn_5} - c_{Sn}^-}{c_{Sn}^+ - c_{Sn}^-} = \frac{\left(\frac{5}{11}\right)/(11.3 \times 10^{-6}) - 0}{1/(16.24 \times 10^{-6}) - 0} = \frac{5 \times 16.24}{11 \times 11.3}$$

$$Y_{C_{Cu}}^{Cu_3Sn} = \frac{c_{Cu}^{Cu_3Sn} - c_{Cu}^+}{c_{Cu}^- - c_{Cu}^+} = \frac{\left(\frac{3}{4}\right)/(9.4 \times 10^{-6}) - 0}{1/(7.12 \times 10^{-6}) - 0} = \frac{3 \times 7.12}{4 \times 9.4}$$

$$Y_{C_{Cu}}^{Cu_6Sn_5} = \frac{c_{Cu}^{Cu_6Sn_5} - c_{Cu}^+}{c_{Cu}^- - c_{Cu}^+} = \frac{\left(\frac{6}{11}\right)/(11.3 \times 10^{-6}) - 0}{1/(7.12 \times 10^{-6}) - 0} = \frac{6 \times 7.12}{11 \times 11.3}$$

### 3.5.3.2.1    Estimation in the Cu$_3$Sn phase

We first estimate $\tilde{J}_i^\beta$ (flux of component inside $\beta$ phase) and then we see how it affects $\tilde{D}_{int}^\beta$.

Following Equation S6a, we can write the interdiffusion flux with respect to component Sn as

$$\tilde{J}_{Sn}^{Cu_3Sn} = -\left(\frac{c_{Sn}^+ - c_{Sn}^-}{2t}\right)\left[Y_{C_{Sn}}^{Cu_3Sn}\left(1 - Y_{C_{Sn}}^{Cu_3Sn}\right)\Delta x^{Cu_3Sn} + 0 + Y_{C_{Sn}}^{Cu_3Sn}\left(1 - Y_{C_{Sn}}^{Cu_6Sn_5}\right)\Delta x^{Cu_6Sn_5}\right]$$

$$\tilde{J}_{Sn}^{Cu_3Sn} = -\left(\frac{\frac{1}{16.24 \times 10^{-6}}}{2 \times 81 \times 3600}\right)\left[\frac{16.24}{4 \times 9.4}\left(1 - \frac{16.24}{4 \times 9.4}\right)3.5 \times 10^{-6} + \frac{16.24}{4 \times 9.4}\left(1 - \frac{5 \times 16.24}{11 \times 11.3}\right)13 \times 10^{-6}\right]$$

$$\tilde{J}_{Sn}^{Cu_3Sn} = -2.96 \times 10^{-7} \ mol/m^2.s$$

Following Equation S3,

$$\tilde{D}_{int}^{Cu_3Sn} == \frac{\left(V_m^{Cu_3Sn}\right)^2}{\overline{V}_{Cu}^{Cu_3Sn}}\Delta x^{Cu_3Sn}\left(-\tilde{J}_{Sn}^{Cu_3Sn}\right)$$

$$\tilde{D}_{int}^{Cu_3Sn} = \frac{\left(V_m^{Cu_3Sn}\right)^2}{\overline{V}_{Cu}^{Cu_3Sn}}\Delta x^{Cu_3Sn}(2.96 \times 10^{-7})$$

$$\tilde{D}_{int}^{Cu_3Sn} = \frac{\left(9.4 \times 10^{-6}\right)^2}{7.12 \times 10^{-6}} \times 3.5 \times 10^{-6} \times (2.96 \times 10^{-7})$$

$$\tilde{D}_{int}^{Cu_3Sn} \approx 1.28 \times 10^{-17} \ m^2/s.$$

Following Equation S6b, we can write the interdiffusion flux with respect to component Cu as

$$\tilde{J}_{Cu}^{Cu_3Sn} = \left(\frac{C_{Cu}^{-} - C_{Cu}^{+}}{2t}\right)\left[Y_{C_{Cu}}^{Cu_3Sn}(1 - Y_{C_{Cu}}^{Cu_3Sn})\Delta x^{Cu_3Sn} + \left(1 - Y_{C_{Cu}}^{Cu_3Sn}\right)Y_{C_{Cu}}^{Cu_6Sn_5}\Delta x^{Cu_6Sn_5} + 0\right]$$

$$\tilde{J}_{Cu}^{Cu_3Sn} = \left(\frac{1}{\frac{7.12\times10^{-6}}{2\times81\times3600}}\right)\left[\left(\frac{3\times7.12}{4\times9.4}\right)\left(1 - \frac{3\times7.12}{4\times9.4}\right)3.5\times10^{-6} + \left(1 - \frac{3\times7.12}{4\times9.4}\right)\left(\frac{6\times7.12}{11\times11.3}\right)13\times10^{-6}\right]$$

$$\tilde{J}_{Cu}^{Cu_3Sn} = 6.71\times10^{-7} \; mol/m^2.s$$

Following Equation S3,

$$\widetilde{D}_{int}^{Cu_3Sn} = \frac{\left(V_m^{Cu_3Sn}\right)^2}{\overline{V}_{Sn}^{Cu_3Sn}}\Delta x^{Cu_3Sn}\left(\tilde{J}_{Cu}^{Cu_3Sn}\right)$$

$$\widetilde{D}_{int}^{Cu_3Sn} = \frac{\left(V_m^{Cu_3Sn}\right)^2}{\overline{V}_{Sn}^{Cu_3Sn}}\Delta x^{Cu_3Sn}(6.71\times10^{-7})$$

$$\widetilde{D}_{int}^{Cu_3Sn} = \frac{\left(9.4\times10^{-6}\right)^2}{16.24\times10^{-6}}\times3.5\times10^{-6}\times(6.71\times10^{-7})$$

$$\widetilde{D}_{int}^{Cu_3Sn} \approx 1.28\times10^{-17} \; m^2/s.$$

Therefore, we get the same value when estimated with respect to the component Sn and Cu, since the relation expressed in Equation 3.6 is also fulfilled

$$\overline{V}_{Cu}^{Cu_3Sn}\tilde{J}_{Cu}^{Cu_3Sn} + \overline{V}_{Sn}^{Cu_3Sn}\tilde{J}_{Sn}^{Cu_3Sn} = 7.12\times10^{-6}\times6.71\times10^{-7} - 16.24\times10^{-6}\times2.96\times10^{-7} \approx 0$$

### 3.5.3.2.2    Estimation in the Cu₆Sn₅ phase

Following Equation S6a, we can write the interdiffusion flux with respect to component Sn as

$$\tilde{J}_{Sn}^{Cu_6Sn_5} = -\left(\frac{C_{Sn}^{+} - C_{Sn}^{-}}{2t}\right)\left[Y_{C_{Sn}}^{Cu_6Sn_5}\left(1 - Y_{C_{Sn}}^{Cu_6Sn_5}\right)\Delta x^{Cu_6Sn_5} + \left(1 - Y_{C_{Sn}}^{Cu_6Sn_5}\right)Y_{C_{Sn}}^{Cu_3Sn}\Delta x^{Cu_3Sn} + 0\right]$$

$$\tilde{J}_{Sn}^{Cu_6Sn_5} = -\left(\frac{1}{\frac{16.24\times10^{-6}}{2\times81\times3600}}\right)\left[\frac{5\times16.24}{11\times11.3}\left(1 - \frac{5\times16.24}{11\times11.3}\right)13\times10^{-6} + \left(1 - \frac{5\times16.24}{11\times11.3}\right)\frac{16.24}{4\times9.4}3.5\times10^{-6}\right]$$

$$\tilde{J}_{Sn}^{Cu_6Sn_5} = -3.66\times10^{-7} \; mol/m^2.s$$

Following Equation S3,

$$\widetilde{D}_{int}^{Cu_6Sn_5} = \frac{\left(V_m^{Cu_6Sn_5}\right)^2}{\overline{V}_{Cu}^{Cu_6Sn_5}}\Delta x^{Cu_6Sn_5}\left(-\tilde{J}_{Sn}^{Cu_6Sn_5}\right)$$

$$\widetilde{D}_{int}^{Cu_6Sn_5} = \frac{\left(V_m^{Cu_6Sn_5}\right)^2}{\bar{V}_{Cu}^{Cu_6Sn_5}} \Delta x^{Cu_6Sn_5} (3.66\times10^{-7})$$

$$\widetilde{D}_{int}^{Cu_6Sn_5} = \frac{(11.3\times10^{-6})^2}{7.12\times10^{-6}} \times 13\times10^{-6} \times (3.66\times10^{-7})$$

$$\widetilde{D}_{int}^{Cu_6Sn_5} \approx 8.53\times10^{-17} \ m^2/s$$

Following Equation S6b, we can write the interdiffusion flux with respect to component Cu as

$$\tilde{J}_{Cu}^{Cu_6Sn_5} = \left(\frac{C_{Cu}^- - C_{Cu}^+}{2t}\right) \left[Y_{C_{Cu}}^{Cu_6Sn_5}\left(1 - Y_{C_{Cu}}^{Cu_6Sn_5}\right)\Delta x^{Cu_6Sn_5} + 0 + Y_{C_{Cu}}^{Cu_6Sn_5}\left(1 - Y_{C_{Cu}}^{Cu_3Sn}\right)\Delta x^{Cu_3Sn}\right]$$

$$\tilde{J}_{Cu}^{Cu_6Sn_5} = \left(\frac{\frac{1}{7.12\times10^{-6}}}{2\times81\times3600}\right)\left[\left(\frac{6\times7.12}{11\times11.3}\right)\left(1 - \frac{6\times7.12}{11\times11.3}\right)13\times10^{-6} + \left(\frac{6\times7.12}{11\times11.3}\right)\left(1 - \frac{3\times7.12}{4\times9.4}\right)3.5\times10^{-6}\right]$$

$$\tilde{J}_{Cu}^{Cu_6Sn_5} = 8.31\times10^{-7} \ mol/m^2.s$$

Following Equation S3,

$$\widetilde{D}_{int}^{Cu_6Sn_5} = \frac{\left(V_m^{Cu_6Sn_5}\right)^2}{\bar{V}_{Sn}^{Cu_6Sn_5}} \Delta x^{Cu_6Sn_5}\left(\tilde{J}_{Cu}^{Cu_6Sn_5}\right)$$

$$\widetilde{D}_{int}^{Cu_6Sn_5} = \frac{\left(V_m^{Cu_6Sn_5}\right)^2}{\bar{V}_{Sn}^{Cu_6Sn_5}} \Delta x^{Cu_6Sn_5}(8.31\times10^{-7})$$

$$\widetilde{D}_{int}^{Cu_6Sn_5} = \frac{(11.3\times10^{-6})^2}{16.24\times10^{-6}} \times 13\times10^{-6} \times (8.31\times10^{-7})$$

$$\widetilde{D}_{int}^{Cu_6Sn_5} \approx 8.53\times10^{-17} \ m^2/s$$

Therefore, we have the same value when estimated with respect to the component Sn and Cu, since Equation 3.6 fulfills following

$$\bar{V}_{Cu}^{Cu_6Sn_5}\tilde{J}_{Cu}^{Cu_6Sn_5} + \bar{V}_{Sn}^{Cu_6Sn_5}\tilde{J}_{Sn}^{Cu_6Sn_5} = 7.12\times10^{-6} \times 8.31\times10^{-7} - 16.24\times10^{-6}\times3.66\times10^{-7} \approx 0$$

Therefore, we can conclude that if the relation with respect to the concentration normalized variable is used for the estimation of the interdiffusion coefficient, we need to consider the ideal molar volume. We cannot consider the actual molar volumes of the phases.

*One important fact should be noted here that*

*For pure end–members, $N_B^- = 0$, $N_A^+ = 0$, $N_B^+ = 1$ and $N_A^- = 1$.*

*Therefore, we have $C_B^- = 0$, $C_A^+ = 0$, $C_B^+ = \frac{1}{V_m^+}$ and $C_A^- = \frac{1}{V_m^-}$ such that*

$$Y_{C_B} = \frac{C_B - C_B^-}{C_B^+ - C_B^-} = \frac{C_B}{C_B^+} = \frac{N_B}{V_m} \times V_m^+ = \frac{Y_{N_B}}{V_m} \times V_m^+$$

$$Y_{C_A} = \frac{C_A - C_A^+}{C_A^- - C_A^+} = \frac{C_A}{C_A^-} = \frac{N_A}{V_m} \times V_m^- = \frac{Y_{N_A}}{V_m} \times V_m^-$$

*While using the relations with respect to the composition normalized variables, only the molar volumes of phases is considered irrespective of the choice of diffusion profile of component A or B; however, the same is not true while using the relations with respect to the concentration normalized variables, since depending on the choice of diffusion profile, the molar volume ($V_m^+$ or $V_m^-$) of one of the end–members along with that of the phases is always being considered for the estimation of a particular diffusion parameter, leading to very minor difference in the estimated values.*

### 3.6    Derivation of relations with respect to the composition normalized variable

The relations developed by Wagner [26] and Paul [6] are given in this section, with equations labelled as (S–…). Minor modifications of a few steps have been made by the present author, so that it is possible to correlate the equations at each step starting from (3.5) and from (S–1), having a similar line of treatment (*i.e.*, algebraic operations) for the derivation of all relations.

### 3.6.1    Interdiffusion Coefficient

Using the standard thermodynamic relation $dC_B = \left(\frac{\bar{V}_A}{V_m^2}\right) dN_B$, we can write Fick's first law as [21] $\tilde{J}_B = -\widetilde{D} \frac{\partial C_B}{\partial x} = -\widetilde{D} \frac{\bar{V}_A}{V_m^2} \frac{\partial N_B}{\partial x}$.

From the standard thermodynamic relation $N_A \bar{V}_A + N_B \bar{V}_B = V_m$, we can write

$$\widetilde{D}\frac{\overline{V}_A}{V_m^2}\frac{\partial N_B}{\partial x} = -\tilde{J}_B = -\frac{N_A\overline{V}_A+N_B\overline{V}_B}{V_m}\tilde{J}_B \tag{S–1}$$

where $\overline{V}_i$ are the partial molar volumes of components A and B.

Using another standard thermodynamic equation $\overline{V}_A dC_A + \overline{V}_B dC_B = 0$ [18], we can relate the interdiffusion fluxes with respect to components A and B as

$$\tilde{J}_B = -\widetilde{D}\frac{\partial C_B}{\partial x} = \frac{\overline{V}_A}{\overline{V}_B}\widetilde{D}\frac{\partial C_A}{\partial x} = -\frac{\overline{V}_A}{\overline{V}_B}\tilde{J}_A$$

$$\Rightarrow \overline{V}_B\tilde{J}_B = -\overline{V}_A\tilde{J}_A \tag{S–2}$$

Combining Equations (S–1) and (S–2), we can write

$$\widetilde{D} = \frac{-\tilde{J}_B}{\left(\frac{\partial C_B}{\partial x}\right)} = \frac{\frac{-(N_A\overline{V}_A+N_B\overline{V}_B)\tilde{J}_B}{V_m}}{\frac{\overline{V}_A}{V_m^2}\left(\frac{\partial N_B}{\partial x}\right)} = \frac{\frac{(-N_B\overline{V}_B\tilde{J}_B-N_A\overline{V}_A\tilde{J}_B)}{V_m}}{\frac{\overline{V}_A}{V_m^2}\left(\frac{\partial N_B}{\partial x}\right)}$$

$$\Rightarrow \widetilde{D} = \frac{\frac{\overline{V}_A(N_B\tilde{J}_A-N_A\tilde{J}_B)}{V_m}}{\frac{\overline{V}_A}{V_m^2}\left(\frac{\partial N_B}{\partial x}\right)} = \frac{V_m(N_B\tilde{J}_A-N_A\tilde{J}_B)}{\left(\frac{\partial N_B}{\partial x}\right)}$$

or $\quad \tilde{J}_B = -\frac{\overline{V}_A}{V_m}(N_B\tilde{J}_A - N_A\tilde{J}_B) \tag{S–3}$

Following Boltzmann [22], compositions in an interdiffusion zone can be related to its position and annealing time by an auxiliary variable as

$$\lambda = \lambda(C_B) = \frac{x-x_o}{\sqrt{t}} = \frac{x}{\sqrt{t}} \tag{S–4}$$

where $x_o = 0$ is the location of the initial contact plane (Matano plane), *i.e.*, bonding interface of the diffusion couple at annealing time $t = 0$.

After differentiating Boltzmann parameter in Equation (S–4) with respect to *t* and then utilizing the same relation again, we get

$$d\lambda = -\frac{1}{2}\frac{x}{t^{3/2}}dt = -\frac{\lambda}{2t}dt \quad \Rightarrow \quad \frac{-1}{dt} = \frac{\lambda}{2td\lambda} \tag{S–5}$$

Wagner [26] introduced the composition normalized variable, which is expressed as

$$Y_N = \frac{N_B-N_B^-}{N_B^+-N_B^-} \tag{S–6}$$

where $N_B^-$ and $N_B^+$ are the un–affected compositions of B on left– and right–hand side of the diffusion couple. Note that $Y_N$ and $(1 - Y_N)$ are equal to zero at these un–affected parts of the diffusion couple.

It can be rearranged to

$$N_B = N_B^+ Y_N + N_B^-(1 - Y_N) \tag{S–7a}$$

Using standard thermodynamic relation $N_A + N_B = 1$, Equation (S–7a) can be written as

$$1 - N_A = N_B^+ Y_N + N_B^-(1 - Y_N)$$

$$\Rightarrow 1 - N_A = N_B^+ Y_N + N_B^-(1 - Y_N) + Y_N - Y_N$$

$$\Rightarrow N_A = 1 - N_B^+ Y_N - N_B^-(1 - Y_N) - Y_N + Y_N$$

$$\Rightarrow N_A = (Y_N - N_B^+ Y_N) + [1 - Y_N - N_B^-(1 - Y_N)]$$

$$\Rightarrow N_A = (1 - N_B^+)Y_N + (1 - N_B^-)(1 - Y_N) \tag{S–7b}$$

From Fick's second law [21], we know that $\frac{\partial C_i}{\partial t} = \frac{\partial}{\partial x}\left(\widetilde{D}\frac{\partial C_i}{\partial x}\right) = -\frac{\partial \tilde{J}_i}{\partial x}$. Therefore, with respect to components A and B, after utilizing $\left[\frac{-1}{dt} = \frac{\lambda}{2t\,d\lambda}\right]$ from Equation (S–5) we can write

$$\frac{\partial \tilde{J}_B}{\partial x} = -\frac{\partial C_B}{\partial t} = -\frac{\partial}{\partial t}\left(\frac{N_B}{V_m}\right) = \frac{\lambda}{2t}\frac{d}{d\lambda}\left(\frac{N_B}{V_m}\right) \tag{S–8a}$$

$$\frac{\partial \tilde{J}_A}{\partial x} = -\frac{\partial C_A}{\partial t} = -\frac{\partial}{\partial t}\left(\frac{N_A}{V_m}\right) = \frac{\lambda}{2t}\frac{\partial}{d\lambda}\left(\frac{N_A}{V_m}\right) \tag{S–8b}$$

Note here that in Equations (S–7a) and (S–7b), the compositions of component B and A, *i.e.*, $N_B$ and $N_A$ are expressed in terms of the composition normalized variable $(Y_N)$. So, next we aim to rewrite Fick's second law, *i.e.*, Equations (S–8a) and (S–8b) with respect to $Y_N$.

Replacing Equation (S–7a) in (S–8a) and Equation (S–7b) in (S–8b), we get

$$\frac{\partial \tilde{J}_B}{\partial x} = \frac{\lambda}{2t} \left[ N_B^+ \frac{d}{d\lambda} \left( \frac{Y_N}{V_m} \right) + N_B^- \frac{d}{d\lambda} \left( \frac{1 - Y_N}{V_m} \right) \right] \tag{S-9a}$$

$$\frac{\partial \tilde{J}_A}{\partial x} = \frac{\lambda}{2t} \left[ (1 - N_B^+) \frac{d}{d\lambda} \left( \frac{Y_N}{V_m} \right) + (1 - N_B^-) \frac{d}{d\lambda} \left( \frac{1 - Y_N}{V_m} \right) \right] \tag{S-9b}$$

Now, we aim to write the above equations with respect to $\left( \frac{Y_N}{V_m} \right)$ and $\left( \frac{1 - Y_N}{V_m} \right)$ separately.

Operating $[N_B^- \times \text{Eq.} \, (\text{S-9b})] - [(1 - N_B^-) \times \text{Eq.} \, (\text{S-9a})]$ leads to

$$N_B^- \frac{\partial \tilde{J}_A}{\partial x} - (1 - N_B^-) \frac{\partial \tilde{J}_B}{\partial x} = \frac{\lambda}{2t} (N_B^- - N_B^+) \frac{d}{d\lambda} \left( \frac{Y_N}{V_m} \right) \tag{S-10a}$$

Operating $[N_B^+ \times \text{Eq.} \, (\text{S-9b})] - [(1 - N_B^+) \times \text{Eq.} \, (\text{S-9a})]$ leads to

$$N_B^+ \frac{\partial \tilde{J}_A}{\partial x} - (1 - N_B^+) \frac{\partial \tilde{J}_B}{\partial x} = \frac{\lambda}{2t} (N_B^+ - N_B^-) \frac{d}{d\lambda} \left( \frac{1 - Y_N}{V_m} \right) \tag{S-10b}$$

After differentiating Boltzmann parameter in Equation (S-4) with respect to *x*, we get

$$d\lambda = \frac{dx}{\sqrt{t}} \tag{S-11}$$

Thus, multiplying left–hand side by $\left( \frac{dx}{\sqrt{t}} \right)$ and right–hand side by $(d\lambda)$ of any Equation does not affect the Equation of our interest.

Multiplying left–hand side by $\frac{dx}{\sqrt{t}}$ and right–hand side by $d\lambda$ of the Equation (S-10a) and (S-10b), respectively, we get

$$\frac{N_B^- \, d\tilde{J}_A - (1 - N_B^-) \, d\tilde{J}_B}{\sqrt{t}} = \left( \frac{N_B^- - N_B^+}{2t} \right) \lambda \, d \left( \frac{Y_N}{V_m} \right) \tag{S-12a}$$

$$\frac{N_B^+ \, d\tilde{J}_A - (1 - N_B^+) \, d\tilde{J}_B}{\sqrt{t}} = \left( \frac{N_B^+ - N_B^-}{2t} \right) \lambda \, d \left( \frac{1 - Y_N}{V_m} \right) \tag{S-12b}$$

Integration by parts,

$$\int u \, dv = u \int dv - \int \left( du \int dv \right) = uv - \int (v \, du) \tag{S-13}$$

This will be applied on the right–hand side of Equation (S-12a) such that $u = \lambda$ and $v = \left( \frac{Y_N}{V_m} \right)$, while to Equation (S-12b) such that $u = \lambda$ and $v = \left( \frac{1 - Y_N}{V_m} \right)$.

Using integration by parts [Eq. (S–13)], Equation (S–12a) is integrated for a fixed annealing time $t$ from un–affected left–hand side of the diffusion couple, *i.e.*, $\lambda = \lambda^{-\infty}$ (corresponds to $x = x^{-\infty}$) to the location of interest $\lambda = \lambda^*$ (corresponds to $x = x^*$) for estimation of the diffusion coefficient.

$$\frac{1}{\sqrt{t}}\left[N_B^- \int_0^{\tilde{J}_A^*} d\tilde{J}_A - (1 - N_B^-) \int_0^{\tilde{J}_B^*} d\tilde{J}_B\right] = \left(\frac{N_B^- - N_B^+}{2t}\right) \int_{\lambda^{-\infty}}^{\lambda^*} \lambda\, d\left(\frac{Y_N}{V_m}\right)$$

$$\Rightarrow \frac{(N_B^-)\tilde{J}_A^* - (1 - N_B^-)\tilde{J}_B^*}{\sqrt{t}} = \left(\frac{N_B^- - N_B^+}{2t}\right)\left[\frac{\lambda^* Y_N^*}{V_m^*} - \int_{\lambda^{-\infty}}^{\lambda^*} \frac{Y_N}{V_m} d\lambda\right] \qquad \text{(S–14a)}$$

Similarly, Equation (S–12b) is integrated from the location of interest $\lambda = \lambda^*$ to the un–affected right–hand side of the diffusion couple, *i.e.*, $\lambda = \lambda^{+\infty}$ (corresponds to $x = x^{+\infty}$).

$$\frac{1}{\sqrt{t}}\left[N_B^+ \int_{\tilde{J}_A^*}^0 d\tilde{J}_A - (1 - N_B^+) \int_{\tilde{J}_B^*}^0 d\tilde{J}_B\right] = \left(\frac{N_B^+ - N_B^-}{2t}\right) \int_{\lambda^*}^{\lambda^{+\infty}} \lambda\, d\left(\frac{1 - Y_N}{V_m}\right)$$

$$\Rightarrow \frac{-(N_B^+)\tilde{J}_A^* + (1 - N_B^+)\tilde{J}_B^*}{\sqrt{t}} = \left(\frac{N_B^+ - N_B^-}{2t}\right)\left[-\frac{\lambda^*(1 - Y_N^*)}{V_m^*} - \int_{\lambda^*}^{\lambda^{+\infty}} \frac{(1 - Y_N)}{V_m} d\lambda\right] \qquad \text{(S–14b)}$$

Note here that the interdiffusion fluxes $\tilde{J}_i$ is equal to zero at the un–affected parts of the diffusion couple, $x = x^{-\infty}$ and $x = x^{+\infty}$, while $\tilde{J}_i^*$ is the fixed value (for certain annealing time $t$) at the location of interest $x = x^*$ in the above Equations (S–14). Also, note that values as zero of $Y_N^{-\infty}$ and $(1 - Y_N^{+\infty})$ from Equation (S–6). Next, we aim to rewrite the above equations with respect to interdiffusion fluxes $\tilde{J}_i$ of both components and finally get an expression for the interdiffusion coefficient $\tilde{D}$.

Operating $[Y_N^* \times \text{Eq. (S–14b)}] - [(1 - Y_N^*) \times \text{Eq. (S–14a)}]$ leads to

$$\frac{N_B^* \tilde{J}_A^* - N_A^* \tilde{J}_B^*}{\sqrt{t}} = \left(\frac{N_B^+ - N_B^-}{2t}\right)\left[(1 - Y_N^*) \int_{\lambda^{-\infty}}^{\lambda^*} \frac{Y_N}{V_m} d\lambda + Y_N^* \int_{\lambda^*}^{\lambda^{+\infty}} \frac{(1 - Y_N)}{V_m} d\lambda\right] \qquad \text{(S–15)}$$

Note here the absence of minus sign on right–hand side of above Equation and numerator on the left–hand side can be found by using $N_B^* = N_B^+ Y_N^* + N_B^-(1 - Y_N^*)$ from Equation (S–7a) and standard relation $N_A^* = 1 - N_B^*$, as follows:

$$Y_N^*\{-(N_B^+)\tilde{J}_A^* + (1 - N_B^+)\tilde{J}_B^*\} - (1 - Y_N^*)\{(N_B^-)\tilde{J}_A^* - (1 - N_B^-)\tilde{J}_B^*\}$$

$$= \{-N_B^+ Y_N^* - N_B^-(1 - Y_N^*)\}\tilde{J}_A^* + \{(1 - N_B^+)Y_N^* + (1 - Y_N^*)(1 - N_B^-)\}\tilde{J}_B^*$$

$$= -\{N_B^+ Y_N^* + N_B^-(1 - Y_N^*)\}\tilde{J}_A^* + \{Y_N^* - N_B^+ Y_N^* + 1 - N_B^- - Y_N^* + N_B^- Y_N^*\}\tilde{J}_B^*$$

$$= -\{N_B^+ Y_N^* + N_B^-(1 - Y_N^*)\}\tilde{J}_A^* + [1 - \{N_B^+ Y_N^* + N_B^-(1 - Y_N^*)\}]\tilde{J}_B^*$$

$$= -(N_B^* \tilde{J}_A^* - N_A^* \tilde{J}_B^*).$$

Utilizing $d\lambda = \frac{dx}{\sqrt{t}}$ from Equation (S–11), we get

$$N_B^* \tilde{J}_A^* - N_A^* \tilde{J}_B^* = \left(\frac{N_B^+ - N_B^-}{2t}\right)\left[(1 - Y_N^*)\int_{x^{-\infty}}^{x^*}\frac{Y_N}{V_m}dx + Y_N^*\int_{x^*}^{x^{+\infty}}\frac{(1 - Y_N)}{V_m}dx\right] \qquad \text{(S–16)}$$

For $N_B = N_B^*$, from Equation (S–3) we know that $\tilde{J}_B^* = -\frac{\bar{V}_A^*}{V_m^*}(N_B^* \tilde{J}_A^* - N_A^* \tilde{J}_B^*)$ and hence

the interdiffusion flux with respect to component B can be expressed as

$$\tilde{J}_B^* = \tilde{J}_B(N_B^*) = -\frac{\bar{V}_A^*}{V_m^*}\left(\frac{N_B^+ - N_B^-}{2t}\right)\left[(1 - Y_N^*)\int_{x^{-\infty}}^{x^*}\frac{Y_N}{V_m}dx + Y_N^*\int_{x^*}^{x^{+\infty}}\frac{(1 - Y_N)}{V_m}dx\right] \qquad \text{(S–17a)}$$

Also, using $\tilde{J}_A^* = -\frac{\bar{V}_B^*}{\bar{V}_A^*}\tilde{J}_B^*$ from Equation (S–2), we can also write

$$\tilde{J}_A^* = \tilde{J}_A(N_B^*) = \frac{\bar{V}_B^*}{V_m^*}\left(\frac{N_B^+ - N_B^-}{2t}\right)\left[(1 - Y_N^*)\int_{x^{-\infty}}^{x^*}\frac{Y_N}{V_m}dx + Y_N^*\int_{x^*}^{x^{+\infty}}\frac{(1 - Y_N)}{V_m}dx\right] \qquad \text{(S–17b)}$$

From Equation (S–7a) we know that $N_B = N_B^+ Y_N + N_B^-(1 - Y_N)$.

By differentiating it with respect to $x$, we can write

$$\left(\frac{dN_B}{dx}\right)_{x=x^*} = N_B^+ \frac{dY_N}{dx} - N_B^- \frac{dY_N}{dx} = (N_B^+ - N_B^-)\left(\frac{dY_N}{dx}\right)_{x=x^*} \qquad \text{(S–18)}$$

From Equation (S–3) for $N_B = N_B^*$, we know

$$\widetilde{D}(N_B^*) = \frac{-\tilde{J}_B^*}{\left(\frac{\bar{V}_A}{V_m^2} \times \frac{dN_B}{dx}\right)_{x=x^*}} \qquad \text{(S–19)}$$

Substituting for flux [Eq. (S–17a)] and gradient [Eq. (S–18)] in Fick's first law [Eq.

(S–19)], we get the expression (which is same for both the components A and B in a

binary system) for the estimation of interdiffusion coefficient as

$$\widetilde{D}(Y_N^*) = \frac{V_m^*}{2t\left(\frac{dY_N^*}{dx}\right)}\left[(1-Y_N^*)\int_{x^{-\infty}}^{x^*}\frac{Y_N}{V_m}dx + Y_N^*\int_{x^*}^{x^{+\infty}}\frac{(1-Y_N)}{V_m}dx\right] \qquad \text{(S–20)}$$

where $Y_N = \frac{N_i - N_i^-}{N_i^+ - N_i^-}$ is the composition normalizing variable, $N_i$ is the mole fraction of the component $i$, and $N_i^-$ and $N_i^+$ are the un–affected end–member compositions on the left– and right–hand side of a diffusion couple, respectively. $t$ (s) is the annealing time, $x$ (m) is the location parameter. $x^{-\infty}$ and $x^{+\infty}$ correspond to the un–affected parts of the diffusion couple. The asterisk $(*)$ represents the location of interest.

### 3.6.2    Intrinsic (and Tracer) Diffusion Coefficients

When the location of interest is the position of the Kirkendall marker plane (K), *i.e.*,

$x^* = x^K$, then we can write Equations (S–14a) and (S–14b), respectively as

$$\frac{(N_B^-)\tilde{J}_A^K - (1-N_B^-)\tilde{J}_B^K}{\sqrt{t}} = \left(\frac{N_B^+ - N_B^-}{2t}\right)\left[-\frac{\lambda^K Y_N^K}{V_m^K} + \int_{\lambda^{-\infty}}^{\lambda^K}\frac{Y_N}{V_m}d\lambda\right] \qquad \text{(S–21a)}$$

$$\frac{-(N_B^+)\tilde{J}_A^K + (1-N_B^+)\tilde{J}_B^K}{\sqrt{t}} = \left(\frac{N_B^+ - N_B^-}{2t}\right)\left[-\frac{\lambda^K(1-Y_N^K)}{V_m^K} - \int_{\lambda^K}^{\lambda^{+\infty}}\frac{(1-Y_N)}{V_m}d\lambda\right] \qquad \text{(S–21b)}$$

Now, we aim to rewrite the above equations with respect to $\tilde{J}_B^K$ and $\tilde{J}_A^K$ such that we can get an expression for intrinsic diffusion coefficient of component B and A, *i.e.*,

$D_B$ and $D_A$, respectively, at the Kirkendall maker plane utilizing the Darken's equation

[23] relating the interdiffusion flux $(\tilde{J}_i)$ with the intrinsic flux $(J_i)$ of component.

Operating $[N_B^+ \times \text{Eq. (S– 21a)}] + [N_B^- \times \text{Eq. (S– 21b)}]$ leads to

$$\frac{-(N_B^+ - N_B^-)\tilde{J}_B^K}{\sqrt{t}}$$

$$= \left(\frac{N_B^+ - N_B^-}{2t}\right)\left[-\lambda^K\frac{\{N_B^+ Y_N^K + N_B^-(1-Y_N^K)\}}{V_m^K} + N_B^+\int_{\lambda^{-\infty}}^{\lambda^K}\frac{Y_N}{V_m}d\lambda - N_B^-\int_{\lambda^K}^{\lambda^{+\infty}}\frac{(1-Y_N)}{V_m}d\lambda\right]$$

since numerator on the left–hand side is

$\{-N_B^+(1-N_B^-) + N_B^-(1-N_B^+)\}\tilde{J}_B^K = -(N_B^+ - N_B^-)\tilde{J}_B^K$

Utilizing $C_B^K = \frac{N_B^K}{V_m^K} = \frac{N_B^+Y_N^K + N_B^-(1-Y_N^K)}{V_m^K}$ from Equation (S–7a) and after rearranging, we get

$$\tilde{J}_B^K = \frac{\sqrt{t}}{2t}\left[\lambda^K C_B^K - N_B^+ \int\limits_{\lambda^{-\infty}}^{\lambda^K} \frac{Y_N}{V_m} d\lambda + N_B^- \int\limits_{\lambda^K}^{\lambda^{+\infty}} \frac{(1-Y_N)}{V_m} d\lambda\right] \tag{S–22a}$$

Similarly, operating $[(1-N_B^+)\times\text{Eq.}\,(\text{S–21a})] + [(1-N_B^-)\times\text{Eq.}\,(\text{S–21b})]$ and

utilizing $C_A^K = \frac{N_A^K}{V_m^K} = \frac{(1-N_B^+)Y_N^K + (1-N_B^-)(1-Y_N^K)}{V_m^K}$ from Equation (S–7b), we get

$\frac{-(N_B^+ - N_B^-)J_A^K}{\sqrt{t}} = \left(\frac{N_B^+ - N_B^-}{2t}\right)\left[-\lambda^K C_A^K + (1-N_B^+)\int_{\lambda^{-\infty}}^{\lambda^K}\frac{Y_N}{V_m}d\lambda - (1-N_B^-)\int_{\lambda^K}^{\lambda^{+\infty}}\frac{(1-Y_N)}{V_m}d\lambda\right]$

since numerator on the left–hand side is

$\{N_B^-(1-N_B^+) - N_B^+(1-N_B^-)\}\tilde{J}_A^K = -(N_B^+ - N_B^-)\tilde{J}_A^K$

Using $N_A^+ = 1 - N_B^+$, $N_A^- = 1 - N_B^-$ and after rearranging, we get

$$\tilde{J}_A^K = \frac{\sqrt{t}}{2t}\left[\lambda^K C_A^K - N_A^+ \int\limits_{\lambda^{-\infty}}^{\lambda^K} \frac{Y_N}{V_m} d\lambda + N_A^- \int\limits_{\lambda^K}^{\lambda^{+\infty}} \frac{(1-Y_N)}{V_m} d\lambda\right] \tag{S–22b}$$

From Boltzmann parameter in Equation (S–4), we know that $\lambda^K = \frac{x^K}{\sqrt{t}}$ or $x^K = \lambda^K\sqrt{t}$.

Using it, the velocity of the Kirkendall marker plane can be expressed as

$v^K = \frac{dx^K}{dt} = \frac{d(\lambda^K\sqrt{t})}{dt} = \lambda^K \frac{d(\sqrt{t})}{dt} = \frac{\lambda^K}{2\sqrt{t}} = \frac{\lambda^K\sqrt{t}}{2t}$

Also, differentiating Boltzmann parameter with respect to $x$, from Equation (S–11) we know that $\sqrt{t}\,d\lambda = dx$.

Putting $\frac{\lambda^K\sqrt{t}}{2t} = v^K$ and $\sqrt{t}\,d\lambda = dx$ in Equations (S–22), we get

$$\tilde{J}_B^K = v^K C_B^K - \frac{1}{2t}\left[N_B^+ \int\limits_{x^{-\infty}}^{x^K} \frac{Y_N}{V_m} dx - N_B^- \int\limits_{x^K}^{x^{+\infty}} \frac{(1-Y_N)}{V_m} dx\right] \tag{S–23a}$$

$$\tilde{J}_A^K = v^K C_A^K - \frac{1}{2t}\left[N_A^+ \int\limits_{x^{-\infty}}^{x^K} \frac{Y_N}{V_m} dx - N_A^- \int\limits_{x^K}^{x^{+\infty}} \frac{(1-Y_N)}{V_m} dx\right] \tag{S–23b}$$

Following Darken's Analysis [23], we know that $\tilde{J}_B^K = J_B + v^K C_B^K$ and $\tilde{J}_A^K = J_A + v^K C_A^K$. Therefore, we can get an expression for intrinsic flux of component B and A, *i.e.*, $J_B$ and $J_A$, respectively, as follows:

$$J_B = \tilde{J}_B^K - v^K C_B^K \Rightarrow J_B = -\frac{1}{2t}\left[N_B^+ \int_{x-\infty}^{x^K} \frac{Y_N}{V_m} dx - N_B^- \int_{x^K}^{x+\infty} \frac{(1-Y_N)}{V_m} dx\right] \tag{S–24a}$$

$$J_A = \tilde{J}_A^K - v^K C_A^K \Rightarrow J_A = -\frac{1}{2t}\left[N_A^+ \int_{x-\infty}^{x^K} \frac{Y_N}{V_m} dx - N_A^- \int_{x^K}^{x+\infty} \frac{(1-Y_N)}{V_m} dx\right] \tag{S–24b}$$

Using Fick's first law [21], we can write $D_B = \frac{-J_B}{\left(\frac{\partial C_B}{\partial x}\right)_{x^K}}$ and $D_A = \frac{-J_A}{\left(\frac{\partial C_A}{\partial x}\right)_{x^K}}$. Therefore, we can write an expression for intrinsic diffusion coefficient of component B and A, *i.e.*, $D_B$ and $D_A$, respectively, as follows:

$$D_B = \frac{1}{2t}\left(\frac{\partial x}{\partial C_B}\right)_K \left[N_B^+ \int_{x-\infty}^{x^K} \frac{Y_N}{V_m} dx - N_B^- \int_{x^K}^{x+\infty} \frac{(1-Y_N)}{V_m} dx\right] \tag{S–25a}$$

$$D_A = \frac{1}{2t}\left(\frac{\partial x}{\partial C_A}\right)_K \left[N_A^+ \int_{x-\infty}^{x^K} \frac{Y_N}{V_m} dx - N_A^- \int_{x^K}^{x+\infty} \frac{(1-Y_N)}{V_m} dx\right] \tag{S–25b}$$

Using $\bar{V}_A dC_A + \bar{V}_B dC_B = 0$, we get $\frac{\partial C_B}{\partial x} = -\frac{\bar{V}_A}{\bar{V}_B}\frac{\partial C_A}{\partial x} \Rightarrow \frac{\partial x}{\partial C_B} = -\frac{\bar{V}_B}{\bar{V}_A}\frac{\partial x}{\partial C_A}$.

Utilizing $\left(\frac{\partial x}{\partial C_B}\right)_K = -\frac{\bar{V}_B^K}{\bar{V}_A^K}\left(\frac{\partial x}{\partial C_A}\right)_K$ in Equations (S–25), the ratio of intrinsic diffusivities can be written as

$$\frac{D_B}{D_A} = \frac{\bar{V}_B^K}{\bar{V}_A^K}\left[\frac{N_B^+ \int_{x-\infty}^{x^K} \frac{Y_N}{V_m} dx - N_B^- \int_{x^K}^{x+\infty} \frac{(1-Y_N)}{V_m} dx}{-N_A^+ \int_{x-\infty}^{x^K} \frac{Y_N}{V_m} dx + N_A^- \int_{x^K}^{x+\infty} \frac{(1-Y_N)}{V_m} dx}\right] \tag{S–26}$$

Paul [6] and van Loo [29] differently derived this relation for intrinsic diffusivities. Following Darken–Manning Analysis [23, 24], the intrinsic ($D_i$) and tracer ($D_i^*$) diffusion coefficients are related as follows

$D_A = \frac{V_m}{\bar{V}_B} D_A^* \Phi (1 + W_A)$ and $D_B = \frac{V_m}{\bar{V}_A} D_B^* \Phi (1 - W_B)$, where the terms $W_i = \frac{2N_i(D_A^* - D_B^*)}{M_0(N_A D_A^* + N_B D_B^*)}$ arise from the vacancy–wind effect, a constant $M_0$ depends on the crystal structure. $\Phi$ is the thermodynamic factor which (according to the Gibbs–Duhem relation) is same for both the components A and B in a binary system. Therefore, the ratio of intrinsic diffusion coefficients in terms of tracer diffusion coefficients can be expressed by

$$\frac{D_B}{D_A} = \frac{\bar{V}_B^K}{\bar{V}_A^K} \frac{D_B^* (1 - W_B)}{D_A^* (1 + W_A)} \tag{S–27}$$

Equating the $(D_B/D_A)$ ratios in both the above Equations (S–26) and (S–27), we get

$$\frac{\bar{V}_B^K}{\bar{V}_A^K} \frac{D_B^* (1 - W_B)}{D_A^* (1 + W_A)} = \frac{\bar{V}_B^K}{\bar{V}_A^K} \left[ \frac{N_B^+ \int_{x^{-\infty}}^{x^K} \frac{Y_N}{V_m} dx - N_B^- \int_{x^K}^{x^{+\infty}} \frac{(1-Y_N)}{V_m} dx}{-N_A^+ \int_{x^{-\infty}}^{x^K} \frac{Y_N}{V_m} dx + N_A^- \int_{x^K}^{x^{+\infty}} \frac{(1-Y_N)}{V_m} dx} \right] \tag{S–28}$$

Therefore, the ratio of tracer diffusion coefficients (which is indirectly measured by the diffusion couple experiment) can be expressed by

$$\frac{D_B^*}{D_A^*} = \left[ \frac{N_B^+ \int_{x^{-\infty}}^{x^K} \frac{Y_N}{V_m} dx - N_B^- \int_{x^K}^{x^{+\infty}} \frac{(1-Y_N)}{V_m} dx}{-N_A^+ \int_{x^{-\infty}}^{x^K} \frac{Y_N}{V_m} dx + N_A^- \int_{x^K}^{x^{+\infty}} \frac{(1-Y_N)}{V_m} dx} \right] \frac{(1 + W_A)}{(1 - W_B)} \tag{S–29a}$$

Very often for the case of line compounds or intermetallic phases with complex crystal structure, there arises a need to neglect the role of the vacancy–wind effect [24], *i.e.*, to consider $W_i \approx 0$, since it cannot be determined because of the unknown structure factor $M_0$.

Considering $W_i \approx 0$, we can rewrite the ratio of tracer diffusion coefficients as

$$\frac{D_B^*}{D_A^*} = \left[ \frac{N_B^+ \int_{x^{-\infty}}^{x^K} \frac{Y_N}{V_m} dx - N_B^- \int_{x^K}^{x^{+\infty}} \frac{(1-Y_N)}{V_m} dx}{-N_A^+ \int_{x^{-\infty}}^{x^K} \frac{Y_N}{V_m} dx + N_A^- \int_{x^K}^{x^{+\infty}} \frac{(1-Y_N)}{V_m} dx} \right] \tag{S–29b}$$

# Chapter 4

# Solid–state diffusion–controlled growth of the intermediate phases from room temperature to an elevated temperature in the Cu–Sn and the Ni–Sn systems

Investigation of the temperature dependent growth of phases in the Cu–Sn and the Ni–Sn systems is reported in this chapter, over a wide temperature range, for the first time, from room temperature (RT) to 215 °C, which is the maximum possible temperature at which the solid–state diffusion couple experiments could be successfully conducted.

## 4.1    Introduction and Statement of the Problem

As we already discussed in Chapter 1, Cu and Ni are the two most important under bump metallization (UBM) layers widely used in flip–chip bonding for making electro–mechanical contact by soldering with a Sn–based alloy. Cu is used for good bonding which is achieved by the formation of Cu–Sn based intermetallic phases during soldering. Afterwards, during storage at RT and service at elevated temperature, these compounds continue to grow further by solid–state diffusion–controlled process, although the growth process and phase evolutions might be different in different temperature ranges. With the thrust for miniaturization, the whole solder might be consumed by the growth of these brittle compounds introducing reliability concern for the UBM/solder joints. Ni is believed to slow down the growth rate due to its inherently slower reaction kinetics during soldering with Sn–based solders and therefore act as a barrier layer.

---

This chapter is written based on the article:
[1] V.A. Baheti, S. Kashyap, P. Kumar, K. Chattopadhyay, A. Paul: Solid–state diffusion–controlled growth of the intermediate phases from room temperature to an elevated temperature in the Cu–Sn and the Ni–Sn systems, Journal of Alloys and Compounds 727 (2017) 832-840.

Numerous articles are published every year reporting the growth of intermetallic compounds in the Cu–Solder (Pb–free, Sn–based) and Cu–Sn systems. A review of the scientific issues and outcomes can be found in References [2, 17] and Chapter 5 of this thesis. Number of studies are also concentrated on the growth of intermetallic compounds in the Ni–Sn system [2]. However, as rightly pointed out by Yuan et al. [39], there are not many studies available comparing the temperature dependent growth of the intermediate phases covering a low to high temperature range. It should be noted that Yuan et al. [39] studied the growth of the phases in the Cu–Sn system in the temperature range of 130–200 °C only. Hence, a dedicated study is still required covering from RT to an elevated temperature for comparing the growth of the phases. Moreover, there is no study available comparing the growth of the product phases in these two systems (*i.e.*, Cu–Sn and Ni–Sn) based on quantitative analysis. Therefore, in this segment of the work, for the first time, we conduct the experiments from RT to an elevated temperature (up to which the solid–state diffusion couple experiments can be conducted) and compare the diffusion–controlled growth of the phases in these two systems. Additionally, the difference in growth behaviour of the phases is compared for bulk and electroplated Sn on bulk Cu and Ni substrates.

## 4.2    Results and Discussion

Because of the difference in nature of the growth of phases, our analysis is first presented for the Cu–Sn system, and following, these are compared with the analysis for the Ni–Sn system. Also, most of the studies reported are concentrated mainly on either bulk or electroplated (EP) diffusion couple without the comparison of growth process between these two conditions. It should be noted here that making a good contact at the interface of two dissimilar materials at an elevated temperature is relatively easy; however, at low temperatures, it is almost impossible to prepare a bulk diffusion couple with an efficient bonding by interdiffusion in the solid–state. Therefore, all the studies at RT or other low temperatures are conducted with EP Sn, since a bonding is created at the very initial stage of the electroplating itself. Moreover, in certain cases, because of industrial relevance, Sn is deposited following acid plating method only as used in this study [11]. Therefore, at the elevated temperatures (*i.e.*, 215–125 °C), we first compare the growth process for bulk and EP diffusion couples. Following, we shall discuss the growth process at lower temperatures (*i.e.*, 100 °C – RT) using EP diffusion couples.

### 4.2.1 Growth of the phases in the Cu–Sn system

We consider the binary Cu–Sn phase diagram published by Saunders and Miodownik [40]. The maximum temperature of the diffusion couple experiments (*i.e.*, 215 °C) is marked with a dashed line on the phase diagram shown in Figure 4.1.

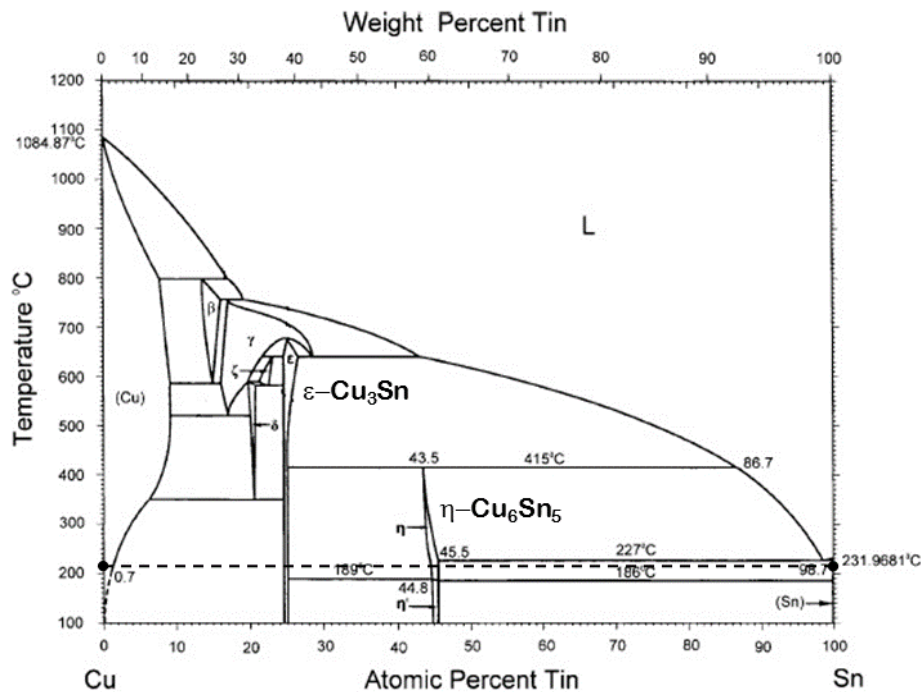

Figure 4.1: The Cu–Sn phase diagram adapted from Saunders and Miodownik [40].

Figure 4.2a and b show the microstructure of Cu/Sn bulk and Cu/Sn EP diffusion couples annealed at 215 °C for 100 hrs. Both the phases, Cu₃Sn and Cu₆Sn₅,

which are present in the Cu–Sn phase diagram [40] at this temperature, are found to grow in the interdiffusion zone. The thickness of both the phases are more or less similar in both types of diffusion couples. However, an additional feature is found in the Cu/Sn EP diffusion couples. Here, a phase mixture of $Cu_6Sn_5$ and Sn is found in a region between single phase layer of $Cu_6Sn_5$ and unaffected Sn end–member, which can be clearly seen in Figure 4.3a. To gain further insights, the phase mixture region was milled with FIB. The single phase region of $Cu_6Sn_5$ along with isolated phase is shown in Figure 4.3b, which is clearer in Figure 4.3c. It is to be noted here that the images are captured where the phase mixture is found prominently, which is not grown evenly throughout the diffusion couple.

At this point, it should be noted here that a phase mixture is not allowed to grow for a thermodynamically controlled interdiffusion process in a binary system [18]. This can be understood in two different ways:

(i)     According to the Gibb's phase rule, $F = C - P + 2$, where $F$ is the degrees of freedom, $C$ is the number of components and $P$ is the number of phases. Since diffusion couple experiments are conducted at constant temperature and pressure, the phase rule can be rewritten as $F = C - P$. In a binary system $C = 2$ and since one degree of freedom is already fixed for composition variation, a two phase mixture ($P = 2$) is not allowed to grow.

(ii)    It can also be understood from the view–point of the chemical potential [18]. By any chance, if a phase mixture could grow in an interdiffusion zone of a binary system, this part would stay in equilibrium because of the same values of chemical potential of a particular component in this region. Therefore, the components would not be able to interdiffuse. This is not allowed in a diffusion couple of two dissimilar materials, till the system as a whole, reaches to the thermodynamically equilibrium state.

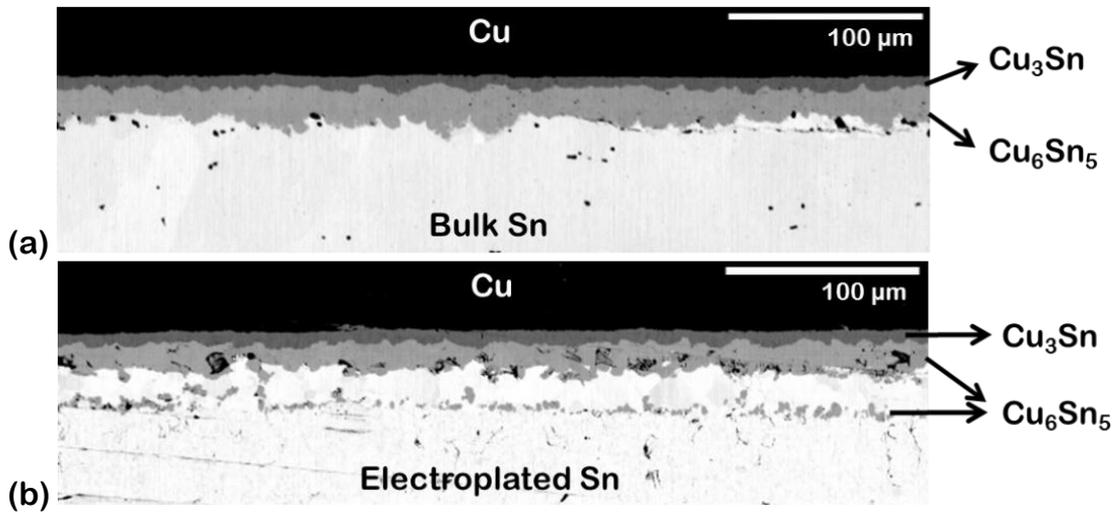

Figure 4.2: BSE image showing the interdiffusion zone of Cu/Sn diffusion couple annealed at 215 °C for 100 hrs: (a) bulk couple and (b) electroplated couple.

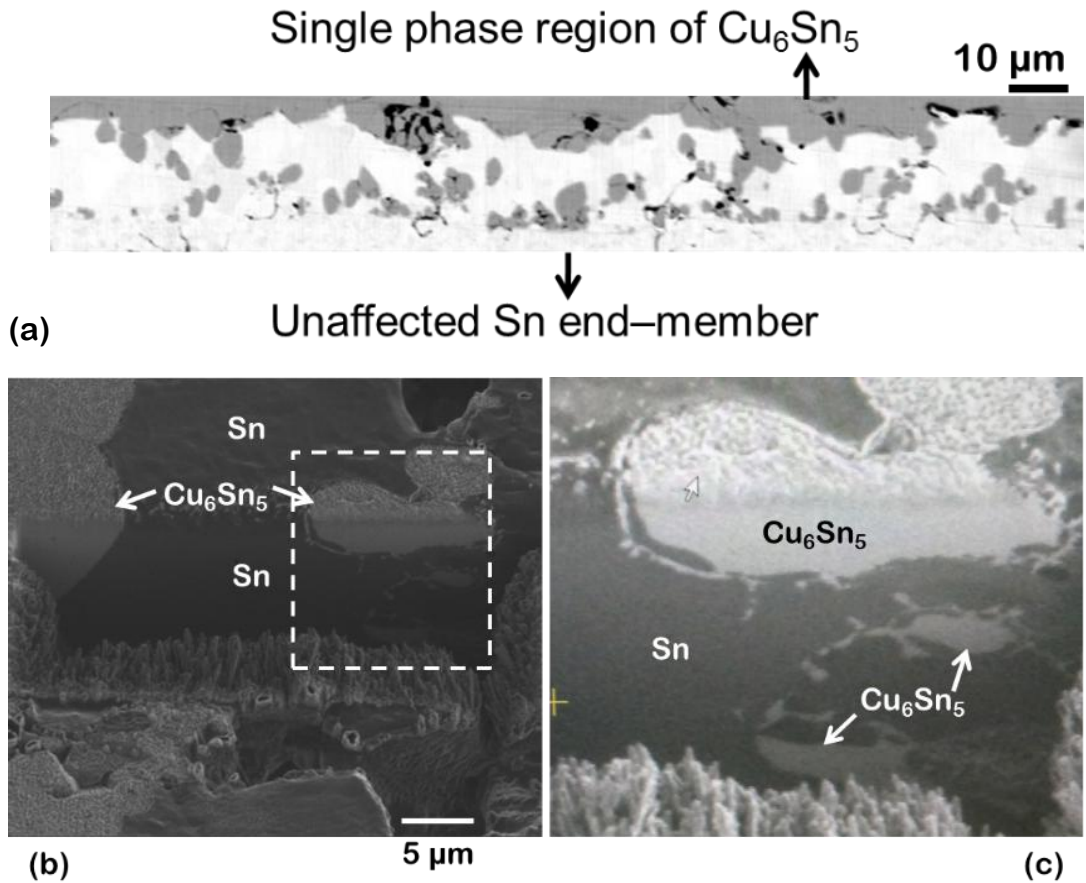

Figure 4.3: Micrographs of the Cu/Sn electroplated diffusion couple annealed at 215 °C for 100 hrs: (a) BSE image showing the phase mixture region between $Cu_6Sn_5$ and Sn; FIB image showing (b) phase mixture region and (c) a focused region, as indicated by dotted square in (b), after polishing in FIB.

Sometimes a phase is precipitated during cooling of the diffusion couple from an elevated temperature of annealing [41]. Precipitates of $(Ni,Pt)_3Al$ phase was found in the $Ni(Al,Pt)$ solid solution near the $Ni(Al,Pt)/(Ni,Pt)_3Al$ interface because of decrease in solubility limit of Al in $Ni(Al,Pt)$ solid solution at lower temperature. However, the size of precipitate was very fine because of the short time that is usually available during cooling of the sample and due to the decrease in diffusion rates of components (with decrease in temperature) for the growth of the precipitates. Therefore, it is unlikely that such big precipitates are formed during cooling of Cu/Sn EP diffusion couple. If precipitates could grow because of this reason, then these would be found in both bulk and EP diffusion couples of Cu/Sn. On the other hand, a phase mixture can be found in a ternary system, which can be understood very easily based on the Gibb's phase rule (as stated above). With the number of components $C = 3$ and composition as a degree of freedom, a phase mixture ($P = 2$) can grow in a diffusion couple experiment at constant temperature and pressure. Moreover, a phase mixture in a ternary or higher component system can stay thermodynamically in equilibrium state to facilitate the diffusion of components [18]. Therefore, it is evident that Cu/Sn bulk couple is indeed a binary system, whereas, the Cu/Sn EP couple must have evolved as a ternary or multicomponent system. It is already known that different types of impurities are added during electroplating. However, because of a very low concentration of different types of impurities present, it is difficult to know about the component which is responsible for the growth of the phase mixture. The change in growth process because of a very low concentration of other component is discussed by van Loo [29]. The same behaviour is found even at 200 °C. At lower temperatures, this is not found prominently which might be because of low growth kinetics of this region. As discussed in the next section, a similar behaviour is found even in the Ni–Sn system.

Diffusion–controlled growth of the phases in Cu/Sn bulk diffusion couples have been studied by several authors through time dependent experiments, mainly at higher temperatures [2, 39, 42]. Recently, Yuan et al. [39] conducted experiments in a very wide temperature range of 130–200 °C and they found the parabolic nature of growth for both the phases, $Cu_3Sn$ and $Cu_6Sn_5$. In this segment of the study, we conducted these experiments for Cu/Sn EP diffusion couples at 150 °C (which falls in the range considered by Yuan et al. [39]) and for the first time at 100 °C, which is lower than the temperature they considered. Because of very slow growth of the phases, it is very difficult to conduct these experiments at a temperature lower than 100 °C.

The growth of the phases at 150 °C after annealing for 25 hrs for both Cu/Sn bulk and Cu/Sn EP diffusion couples is shown in Figure 4.4. It can be seen that there is no difference in the growth rate of both the phases in these diffusion couples.

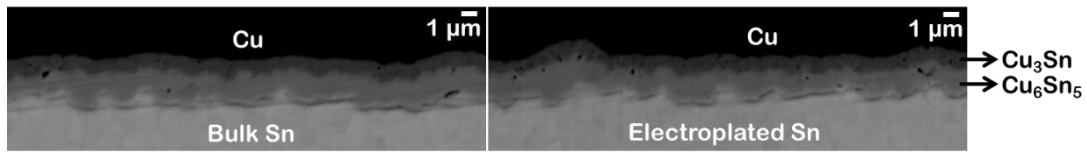

Figure 4.4: BSE image of both Cu/Sn bulk and Cu/Sn electroplated diffusion couple annealed at 150 °C for 25 hrs.

Time dependent growth of both the phases are shown in Figure 4.5. This is plotted with respect to [18]:

$$(\Delta x)^2 - (\Delta x_i)^2 = 2k_P(t - t_o) \qquad (4.1)$$

where $\Delta x$ is the thickness of a particular phase layer after the annealing for time $t$ at the temperature of interest, $\Delta x_i$ is the intercept at the thickness axis for $t = 0$ and it is sometimes positive because of the various reasons as explained in Reference [18], and $t_o$ is the incubation time, when the phase layer does not grow immediately after the start of annealing. The parabolic growth of both the phases is evident from the linear fit in a plot of $(\Delta x)^2$ vs. $t$, as shown in Figure 4.5.

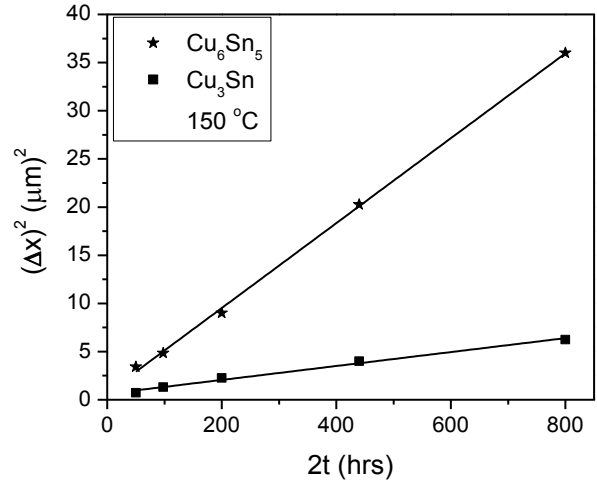

Figure 4.5: Time dependent growth of both the phases $Cu_3Sn$ and $Cu_6Sn_5$ in Cu/Sn diffusion couples annealed at 150 °C.

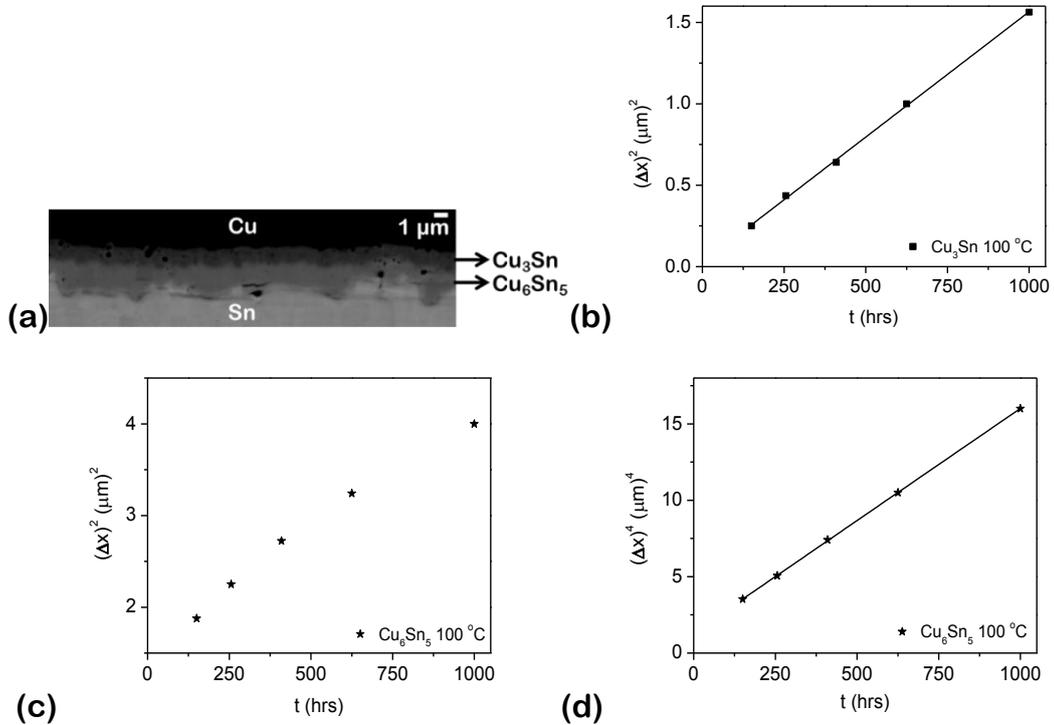

Figure 4.6: Cu/Sn electroplated diffusion couple annealed at 100 °C (a) BSE image showing the presence of both phases $Cu_3Sn$ and $Cu_6Sn_5$ in the interdiffusion zone after 1000 hrs annealing. Time dependent growth of the phase: (b) $Cu_3Sn$, $(\Delta x)^2$ vs. $t$; (c) $Cu_6Sn_5$, $(\Delta x)^2$ vs. $t$ and (d) $Cu_6Sn_5$, $(\Delta x)^4$ vs. $t$.

A similar time dependent experiment was then conducted at 100 °C, which has not been reported at this temperature till date. The growth of the phases after annealing for 1000 hrs is shown in Figure 4.6a. Figure 4.6b and c show the time dependent growth of the phases, Cu₃Sn and Cu₆Sn₅, with respect to $(\Delta x)^2$ vs. $t$. The growth of the Cu₃Sn phase is parabolic in nature, which is evident from linear fit of the data (Figure 4.6b), while the data could not be fit linearly for the Cu₆Sn₅ phase (Figure 4.6c). However, when the same is plotted with respect to $(\Delta x)^4$ vs. $t$, it could be fit linearly, as shown in Figure 4.6d, indicating a difference in the diffusion–controlled growth mechanism of this phase.

Diffusion in a polycrystalline material is categorized in 3 different regimes depending on the contribution from lattice $(D_l)$ and grain boundary $(D_g)$ diffusion [18, 43]. When the diffusion depth is comparable for both the lattice and the grain boundary and $\sqrt{D_l t}$ is much higher than the average grain size ($d$), it is categorized as Type A regime. It is generally found in higher temperature range in which the concentration of point defects assisting the lattice diffusion is high. In this regime, parabolic growth of the phase ($\Delta x \propto t^{0.5}$) is witnessed, as it is found at 150 °C in this study and in the temperature range of 130–200 °C by Yuan et al. [39]. With the decrease in temperature, when the diffusion depth via lattice is much lower compared to the grain boundaries and it falls in the range of $100\delta < \sqrt{D_l t} < d/20$, where $\delta$ is the grain boundary thickness, it falls in the Type B regime. In this regime, time dependent growth of the phase layer follows $\Delta x \propto t^{0.25}$, as it is found for the Cu₆Sn₅ phase at 100 °C in this study. With the decrease in temperature, the growth kinetics may shift to Type C regime, when lattice diffusion is negligible such that $20\sqrt{D_l t} < \delta$ and the phase layer grows because of the grain boundary diffusion. In this regime, again the parabolic growth kinetics is found such that $\Delta x \propto t^{0.5}$. Since time dependent growth of the phases cannot be determined with a certain degree of confidence in this regime because of very low growth kinetics in RT–75 °C range, it is not known if the growth of the phase falls in Type C regime at temperature lower than 100 °C in the Cu–Sn system. Since the growth kinetics of $Cu_6Sn_5$ at 100 °C follow $(\Delta x)^4$ vs. $t$ as shown in Figure 4.6d, it is apparent that the growth of this phase at this temperature falls in the Type B regime.

Following, experiments were conducted at 75 and 50 °C for 1000 hrs. Only the $Cu_6Sn_5$ phase is found in the interdiffusion zone with the thickness of 1.5 μm at 75 °C and 1 μm at 50 °C after annealing for 1000 hrs. One of the representative micrographs of a Cu/Sn diffusion couple annealed at 75 °C for 1000 hrs is shown in Figure 4.7, which shows the absence of $Cu_3Sn$ phase. Therefore, this study indicates that the onset temperature for the growth of $Cu_3Sn$ phase is in the temperature range of 75–100 °C. Similar results are reported in few other manuscripts also, as discussed next. For example, Bandyopadhyay and Sen [44] conducted the experiments at 60 and 80 °C, and at both temperatures, they found only the $Cu_6Sn_5$ phase in the Cu/Sn diffusion couple. Halimi et al. [45] found the presence of $Cu_3Sn$ at 90 °C and the same was found by Hwang et al. [46] at 85 °C; however, only after the consumption of Sn, *i.e.*, in a condition similar to an incremental diffusion couple of Cu/$Cu_6Sn_5$, in which only the $Cu_3Sn$ phase can grow. A similar behaviour was noticed earlier in a Cu(Ni)/Sn diffusion couple at higher temperature [8]. With the addition of 5 at.% Ni in Cu, only the $(Cu,Ni)_6Sn_5$ phase grows in the interdiffusion zone of Cu(5Ni)/Sn diffusion couple at 200 °C; however, the $(Cu,Ni)_3Sn$ phase could grow as a very thin layer in an incremental diffusion couple of Cu(5Ni)/$(Cu,Ni)_6Sn_5$ after the removal of Sn [8]. These studies indicate that it is because of the kinetics reason, and not the nucleation issues, due to which the $Cu_3Sn$ is not found at a temperature below 100 °C in a Cu/Sn diffusion couple; since there is no change in the Cu/Cu$_6$Sn$_5$ interface at which the Cu$_3$Sn phase grows after consumption or removal of Sn, as discussed above. In a multiphase growth, a phase with much lower growth kinetics might not be found because of consumption by the other neighbouring phase with much higher growth kinetics [47].

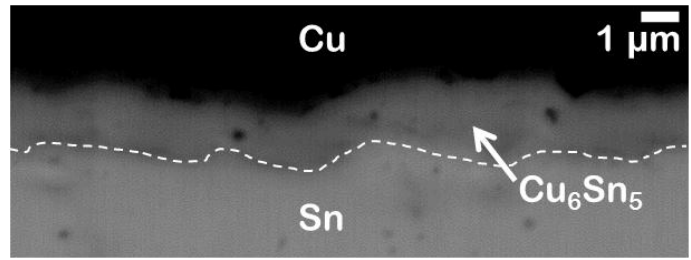

Figure 4.7: BSE image of the interdiffusion zone of Cu/Sn electroplated diffusion couple, showing the presence of only Cu$_6$Sn$_5$ phase (and absence of Cu$_3$Sn phase) after annealing at 75 °C for 1000 hrs.

A similar behaviour of growth, *i.e.*, the presence of only Cu$_6$Sn$_5$, is found after storage at room temperature (RT) in a Cu/Sn EP diffusion couple. Initially, it grows at only few places, which is evident from the SEM micrograph after 17 days, as shown in Figure 4.8a. This non uniform growth continues almost for 1 year (a similar behaviour is found even in Ni/Sn EP couple, which is discussed in the next section).

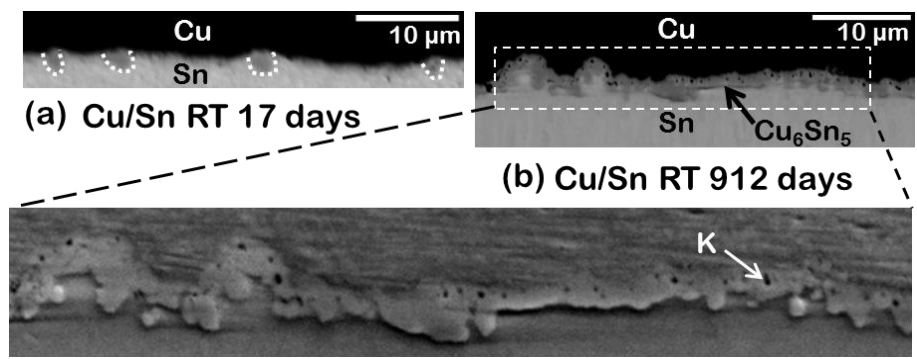

Figure 4.8: BSE micrograph of the Cu/Sn electroplated diffusion couple after storage at room temperature for (a) 17 days, dotted curves denote a few locations across the Cu/Sn interface indicating the presence of Cu$_6$Sn$_5$ phase; and (b) 912 days, *i.e.*, 2.5 years, K denotes the Kirkendall pores, which appear clearer in focused SE image.

A continuous layer of the product phase is found after 2.5 years (912 days), as shown in Figure 4.8b. With such a behaviour of growth and because of a wavy and very thin layer of the phase, it is not possible to examine the growth mechanism based on time dependent experiments. However, the location of the Kirkendall marker plane can indicate the relative rate of the diffusion of components. Line of pores or the duplex morphology, *i.e.*, different microstructure on two sides of the Kirkendall marker plane (because of the growth of a phase from different interfaces) efficiently indicates the location of this plane [12, 18]. A line of pores can be seen clearly (in focused SE image shown in Figure 4.8b) throughout the phase layer indicating the location of the Kirkendall marker plane. Tu and Thompson [48] reported a similar finding following an analysis based on the Rutherford backscattering spectroscopy. Even a similar location is found to demarcate a duplex morphology with sublayers of fine and relatively bigger grains by He and Ivey [49]. Although this observation [49] was not related with the Kirkendall effect by them; however, it efficiently indicates the location of the Kirkendall plane. Since it is located near the $Cu/Cu_6Sn_5$ interface, the diffusion of Cu must have a bigger role on the growth of $Cu_6Sn_5$ phase at RT. On the other hand, at higher temperatures, Sn is found to have a higher diffusion rate compared to Cu [17]. This indicates that there must be a change in the growth mechanism of this phase with the change in temperature. Interestingly, this phase goes through a polymorphic transformation from hexagonal to monoclinic [40]. As a result, there might be a difference in the diffusion–controlled growth process.

### 4.2.2 Growth of the phases in the Ni–Sn system

We consider the latest binary Ni–Sn phase diagram published by Schmetterer et al. [50]. An elevated temperature range of 125–215 °C for the diffusion couple experiments is marked with dash–dot–dot lines on the phase diagram shown in Figure 4.9. Figure 4.10a and b shows the interdiffusion zone of both Ni/Sn bulk and Ni/Sn EP diffusion couples annealed at 200 °C for 100 hrs. Following the binary Ni–Sn phase diagram [50], all the thermodynamically stable phases, *i.e.*, Ni$_3$Sn, Ni$_3$Sn$_2$ and Ni$_3$Sn$_4$ should have grown. However, only the Ni$_3$Sn$_4$ phase is found in the interdiffusion zone which might be because of sluggish growth kinetics of other phases. This statement finds support since the missing phases were found to grow in incremental diffusion couples at higher temperatures, *i.e.*, 580–800 °C [51].

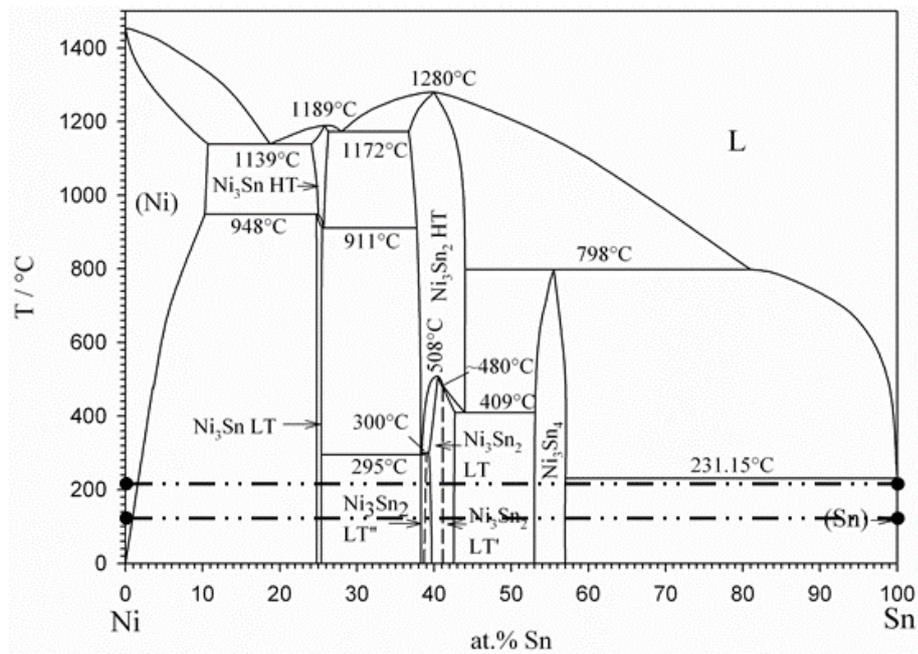

Figure 4.9: Binary Ni–Sn phase diagram adapted from Schmetterer et al. [50].

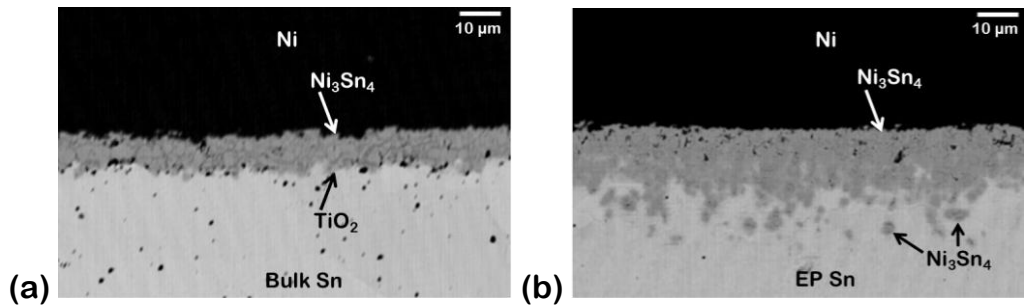

Figure 4.10: BSE image of the diffusion couple annealed at 200 °C for 100 hrs: (a) Ni/Sn bulk couple, where the location of TiO$_2$ inert particle is found close to Ni$_3$Sn$_4$/Sn interface; and (b) Ni/Sn electroplated couple.

Unlike the Cu/Sn diffusion couples (discussed in the previous section), a difference in the growth rate and therefore the thickness is found in these two types of Ni/Sn couples, *i.e.*, bulk and EP. However, there is a similarity in EP couples (of Cu/Sn and Ni/Sn) such that a phase mixture of Ni$_3$Sn$_4$ and Sn is found between a single phase layer Ni$_3$Sn$_4$ and end–member Sn. FIB milling was done in a phase mixture region, as indicated in Figure 4.11a. Figure 4.11b clearly indicates the presence of isolated Ni$_3$Sn$_4$ phase inside Sn. In the Ni/Sn bulk diffusion couple, TiO$_2$ particles were used as the inert markers, which were found very close to the Ni$_3$Sn$_4$/Sn interface. This indicates that the product phase grows mainly because of diffusion of Sn and the diffusion rate of Ni is negligible. Ambiguity exists in the literature over the composition range of the Ni$_3$Sn$_4$ phase, which is reported as 53–57 at.% Sn in a latest article [50]. Based on our EPMA analysis, we find it as 55–60 at.% Sn, which is similar to the range indicated from analysis by Mita et al. [52]. The results are found to be similar up to 125 °C in Ni/Sn EP couples and up to 175 °C in Ni/Sn bulk couples.

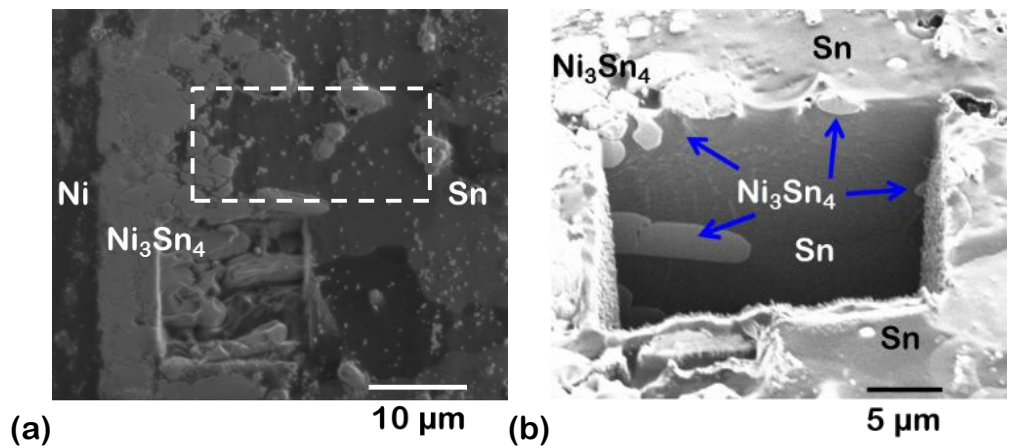

Figure 4.11: Micrographs of the Ni/Sn electroplated diffusion couple annealed at 200 °C for 100 hrs: (a) FIB image showing region selected for milling and (b) top region shown in (a) after polishing in FIB.

With decrease in annealing temperature to 150 °C or below, we noticed difficulties for the growth of Ni$_3$Sn$_4$ phase in bulk diffusion couples with high incubation time. Similar observation was reported by Mita et al. [52] in Ni/Sn bulk couple annealed at 160 °C. Therefore, we continued our further studies with EP Sn, since Sn makes a bond with Ni during electroplating itself and therefore, the product phase starts growing almost immediately at the start of the annealing process.

However, in the interdiffusion zone developed at 150 °C (as shown in Figure 4.12) after 400 hrs of annealing, isolated Ni is found at a few places inside the product phase. It is to be noted here that the image is especially captured where unreacted Ni is found and it is not necessarily present everywhere. Since the product phase grows by diffusion of Sn, it is well possible that Sn diffuses with higher rate through grain boundaries before consuming the whole lattice by relatively slow diffusion rate through lattice. Also, the lattice diffusion decreases at faster rate compared to the grain boundary diffusion with the decrease in temperature due to higher activation energy required for diffusion via lattice.

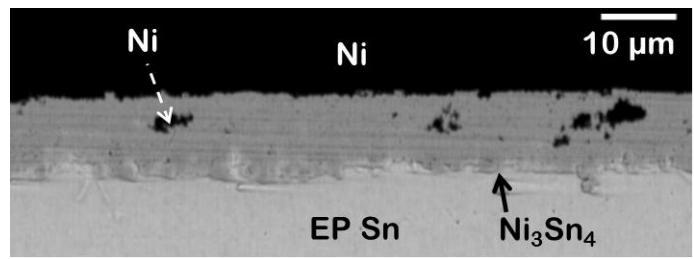

Figure 4.12: BSE image of Ni/Sn electroplated diffusion couple annealed at 150 °C for 400 hrs, showing the presence of pure Ni at a few places inside the product phase as confirmed by EPMA analysis.

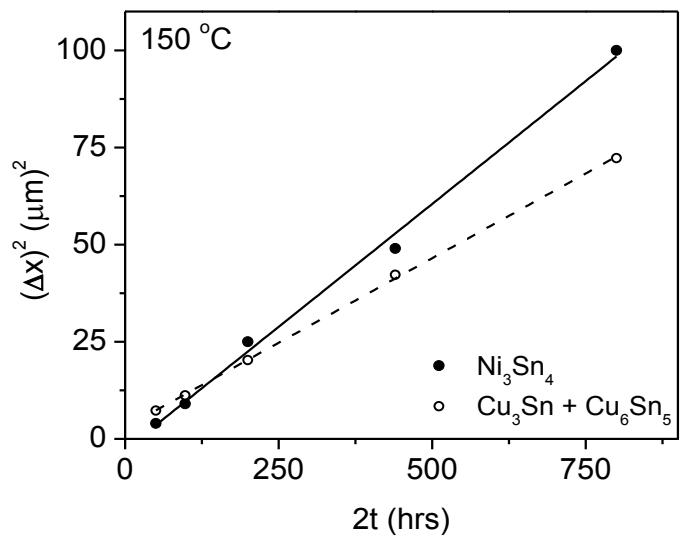

Figure 4.13: Time dependent growth of the $Ni_3Sn_4$ phase in Ni/Sn electroplated diffusion couple annealed at 150 °C, along with the total phase layer thickness of Cu/Sn couple for comparison.

Time dependent growth of the product phase is conducted at 150 °C for Ni/Sn EP couples and, as shown in Figure 4.13 (in which total of thicknesses of the phase layers in the Cu–Sn system is also shown for discussion, as presented in the next section), a linear fit is found when plotted with respect to $(\Delta x)^2$ vs. $t$, indicating diffusion–controlled growth of the $Ni_3Sn_4$ phase.

There is a general consensus about the existence and crystal structure (monoclinic) of the $Ni_3Sn_4$ phase at higher temperatures. This is confirmed from the X–ray diffraction (XRD) analysis by Tang et al. [53]. However, there exists an ambiguity about the evolution of phase(s) at lower temperatures, as pointed out by Belyakov [54], and he found the presence of metastabe phase $NiSn_4$ (which is of our interest) during solidfictaion of Sn–Ni alloys, in Ni/Sn soldered joints and also in EP–Ni/EP–Sn diffusion couples. Therefore, we extend our investigation of structural characterization at lower temperatures (*i.e.*, 100 °C – RT) in the Ni/Sn EP couples. It should be noted here that our EPMA point analysis of the interdiffusion zone, *i.e.*, product layer grown in the Ni/Sn EP couples annealed in the temperature range of 50–100 °C for 1000 hrs, shows different compositions at different places, with overall composition in the range of 60–80 at.% Sn. Since the composition range of $Ni_3Sn_4$ is found to be 55–60 at.% Sn, it indicates a difference in the phase evolutions at these temperatures when compared to temperatures $\geq$ 125 °C. To gain further insights on the same, TEM analysis is conducted and an example of the same at 50 °C is shown in Figure 4.14. In this study, all the electron diffraction patterns (DPs) are either indexed to metastable $NiSn_4$ (*oC*20) [55] or equilibrium $Ni_3Sn_4$ (*mC*14) [56] phase. TEM analysis indicates the presence of many small grains (for example, as can be seen in the bright–field images) and in particular, coexistence of the metastable $NiSn_4$ phase along with the equilibrium $Ni_3Sn_4$ phase. This is established based on DPs acquired from many grains in the interdiffusion zone. In the example shown at 50 °C, DP is indexed with zone axis [201] of the orthorhombic NiSn$_4$ phase and with zone axis [$\bar{1}$01] of the monoclinic Ni$_3$Sn$_4$ phase. Since the grains of both the phases are small and they co–exist together, the random composition analysis in EPMA indicates such a wide range of compositions at different places in the interdiffusion zone of Ni/Sn EP couples at 50–100 °C, as discussed above.

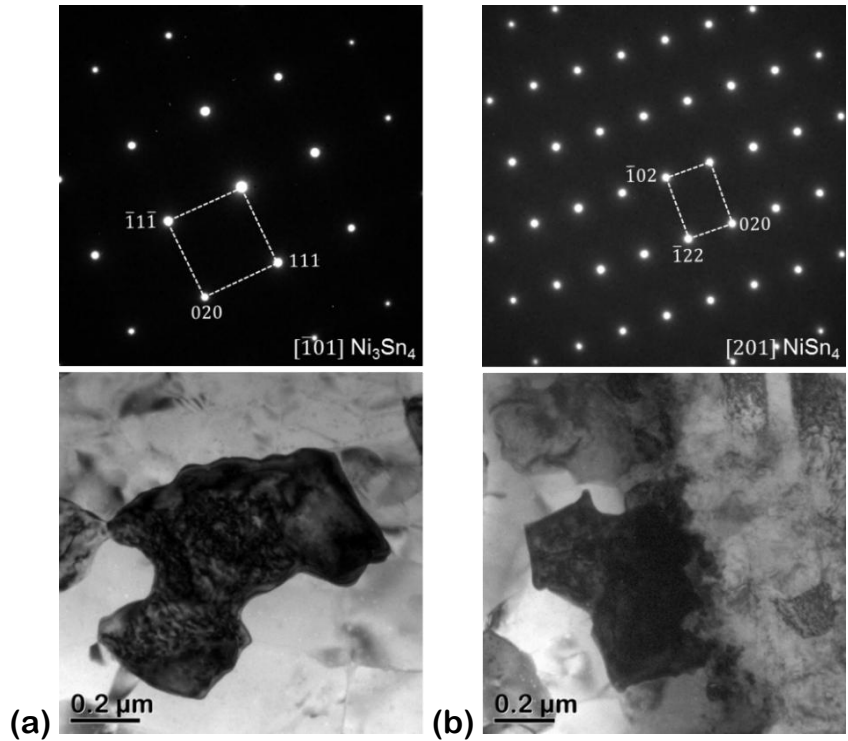

Figure 4.14: TEM analysis of Ni/Sn electroplated diffusion couple annealed at 50 °C. Diffraction pattern (top) along with the corresponding bright–field image (bottom) of the (a) Ni$_3$Sn$_4$ and (b) NiSn$_4$ phases.

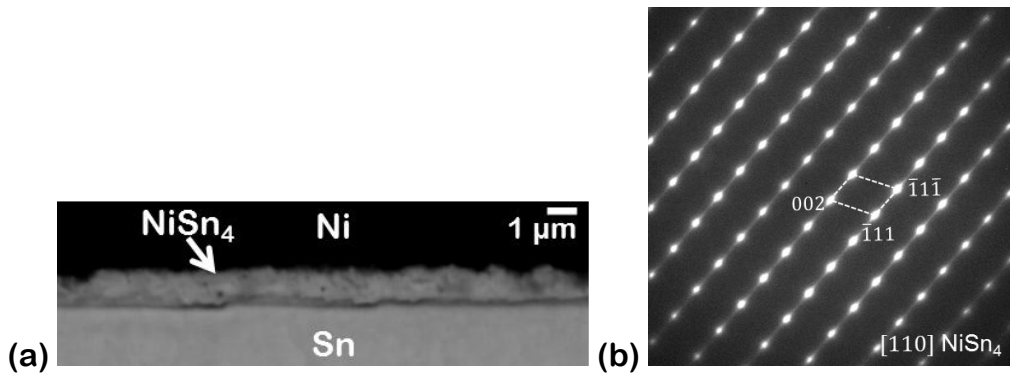

Figure 4.15: Micrograph of the Ni/Sn electroplated diffusion couple after storage at room temperature for 912 days, *i.e.*, 2.5 years: (a) BSE image and (b) diffraction pattern of the NiSn$_4$ phase.

Following, we conducted our experiments of storage at RT. Figure 4.15a shows the SEM micrograph of the interdiffusion zone of Ni/Sn EP couple after storage for 2.5 years (912 days). We found that the average thickness of the phase layer is comparable to the thickness of $Cu_6Sn_5$ in the Cu–Sn system (Figure 4.8) after storage for the same time. EPMA analysis equipped with field emission gun (FEG) indicates that the average composition is close to 80 at.% Sn, which indicates that phase layer might have mainly $NiSn_4$. Further investigation is continued in TEM to get insights into the crystal structure of the product phase. DP acquired from one of the grains is indexed with zone axis [110] of the orthorhombic $NiSn_4$ phase, as shown in Figure 4.15b. Belyakov [54] examined the interdiffusion zone till 6,200 hrs after storage and indicated that this metastable $NiSn_4$ phase might be transformed to the stable $Ni_3Sn_4$ phase after longer ageing times. However, our study for 21,888 hrs (*i.e.*, 912 days or 2.5 years) still indicates the presence of this metastable phase.

### 4.2.3 Estimation and comparison of the diffusion coefficients

For the purpose of comparison of the growth rate of the product phases between the Cu–Sn and the Ni–Sn systems, we plot them with respect to $(\Delta x^{Cu_3Sn} + \Delta x^{Cu_6Sn_5})^2$ vs. $t$ and $(\Delta x^{Ni_3Sn_4})^2$ vs. $t$. It can be seen in Figure 4.13 that the growth rate in the Ni/Sn EP couple is comparable to that in the Cu/Sn EP couple. We can estimate the diffusion parameters in the temperature range of 125–215 °C, since the product phases grow parabolically with time in this range. The integrated interdiffusion coefficients, $\widetilde{D}_{int}$ (in m²/s), are estimated following the relation developed by Wagner [26], Equation (3.33), considering the molar volume of the phases as $V_m^{Cu_3Sn} = 8.59 \times 10^{-6}$ and $V_m^{Cu_6Sn_5} = 10.59 \times 10^{-6}$ m³/mol [17]. After estimating the $\widetilde{D}_{int}$ of the phases at different temperatures, these are plotted, as shown in Figure 4.16, following the Arrhenius relation, *i.e.*, plot of $\left(log_{10}\ \widetilde{D}_{int}\right)$ with respect to $(1/T)$, where $T$ is the temperature in Kelvin.

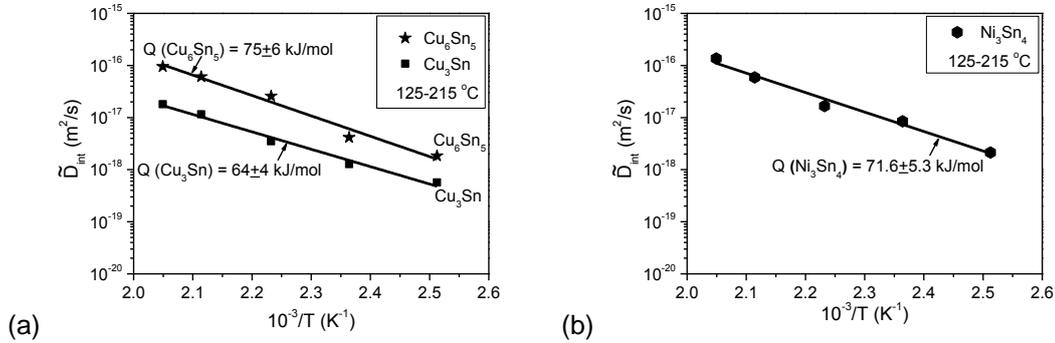

Figure 4.16: Arrhenius plot of the integrated interdiffusion coefficients in the phase (a) $Cu_3Sn$ and $Cu_6Sn_5$, and (b) $NiSn_4$.

In the temperature range of our interest in this study, *i.e.*, 125–215 °C, the activation energies are estimated as 64±4 kJ/mol for $Cu_3Sn$ and 75±6 kJ/mol for $Cu_6Sn_5$ (Figure 4.16a). The data are estimated from the bulk diffusion couples and the growth rates are found to be similar in both bulk and electroplated diffusion couples. These activation energy values are found to be in reasonable agreement with those determined by Paul et al. [17] from the solid–state incremental diffusion couples (*i.e.*, to grow $Cu_3Sn$, $Cu/Cu_6Sn_5$ were coupled and to grow $Cu_6Sn_5$, $Cu_3Sn/Sn$ were coupled) as 74 kJ/mol for $Cu_3Sn$ in the temperature range of 225–350 °C and 81 kJ/mol for $Cu_6Sn_5$ in the temperature range of 150–200 °C. A similar value of the activation energy, *i.e.*, 71.6±5.3 kJ/mol is estimated for the $Ni_3Sn_4$ phase in the same temperature range (Figure 4.16b). The data for the parent phase with continuous layer is estimated in the electroplated couples, since the interdiffusion zone did not grow properly in the bulk diffusion couples in the whole temperature range considered for these estimations. The relatively low values of the activation energies in both the systems indicate a prominent role of the grain boundary diffusion for the growth of phases [18].

## 4.3 Conclusions

Experiments are conducted in the Cu–Sn and the Ni–Sn systems covering a wide temperature range to study the temperature dependent growth of phases (for the first time) from room temperature to the maximum temperature possible at which the solid–state diffusion couples could be prepared, *i.e.*, 215 °C. A few key observations of this segment of the study can be stated as follows:

(i)     Both the intermediate phases, $Cu_3Sn$ and $Cu_6Sn_5$ available in the Cu–Sn system, are found to grow at and above 100 °C. In Cu/EP–Sn diffusion couple, a phase mixture of $Cu_6Sn_5$ and Sn grows at higher temperatures, which does not grow in the Cu/Sn bulk diffusion couple. At lower temperatures (*i.e.*, ≤ 75 °C), up to the room temperature only the $Cu_6Sn_5$ phase grows in the interdiffusion zone.

(ii)    Out of 3 stable intermediate phases, only the $Ni_3Sn_4$ phase grows in the Ni–Sn system at an elevated temperature. Similar to the Cu–Sn system, a phase mixture of $Ni_3Sn_4$ and Sn grows in the Ni/EP–Sn diffusion couple, which does not grow in the Ni/Sn bulk diffusion couple. At 50–100 °C, a metastable phase $NiSn_4$ is found to grow along with the stable $Ni_3Sn_4$ phase, as indicated by EPMA and TEM analysis. At room temperature, only the $NiSn_4$ phase is found.

(iii)   Ni is used as a barrier layer to retard the growth rate of the product phases in the Cu–Sn system, which seems to be true based on the comparison of growth behaviour of the phases in Cu/Sn and Ni/Sn bulk diffusion couples; however, when we study the growth of the phases individually at all the temperatures in the solid–state condition, the growth rate of the product phase(s) in the Ni/Sn EP couples are found to be more or less similar to that in the Cu/Sn couples.

The growth rate of the phases even at the room temperature is found to be significant and therefore it will be similar during storage and service life of the microelectronic device.

*To summarize, the temperature dependent growth of the product phases in the Cu–Sn and the Ni–Sn systems is studied in this chapter and found to be as follows:*

*Cu–Sn system:*

      *RT–75 °C*     $\longrightarrow$     *Only $Cu_6Sn_5$*

      *100–215 °C*     $\longrightarrow$     *Both $Cu_3Sn$ and $Cu_6Sn_5$*

*Ni–Sn system:*

      *RT*     $\longrightarrow$     *Only metastable phase $NiSn_4$*

      *50–100 °C*     $\longrightarrow$     *Coexistence of $NiSn_4$ and $Ni_3Sn_4$*

      *125–215 °C*     $\longrightarrow$     *Only equilibrium phase $Ni_3Sn_4$*

# Chapter 5

# Bifurcation of the Kirkendall marker plane and the effect of impurities on the growth of the Kirkendall voids in the Cu–Sn system

The technologically important Cu–Sn system is studied extensively over many decades. However, one of the important predictions by Paul et al. [17] on the presence of the bifurcation of the Kirkendall marker plane is yet to be validated experimentally. This would change the viewpoint of diffusion–controlled growth mechanism of the phases and the formation of the Kirkendall voids introducing the electro–mechanical failure of the flip–chip bonds. These two issues are discussed in detail in this chapter. First (in Section 5.2), an experimental proof is reported elucidating the bifurcation of the Kirkendall marker plane in the Cu–Sn system. Next (in Section 5.3), we discuss the role of impurities on the formation of the Kirkendall voids in the $Cu_3Sn$ phase.

## 5.1   Introduction and Statement of the Problem

In the last decade, the discovery of splitting of the Kirkendall markers into more than one plane (bifurcation and trifurcation) led to the development of many new theories in the solid–state diffusion [18]. One of such theories, a physico–chemical approach [47] (developed by one of this thesis advisors, Paul) relates the microstructural evolution with the rates of diffusing components and helps to predict the location of the Kirkendall marker planes in a particular system. This was applied to the technologically important Cu–Sn system [17] since the growth of the brittle $Cu_3Sn$ and $Cu_6Sn_5$ intermetallic phases along with the Kirkendall voids (by Frenkel effect) in the interdiffusion zone of Cu under bump metallization (UBM) and

---

This chapter is written based on the article:
[1] V.A. Baheti, S. Kashyap, P. Kumar, K. Chattopadhyay, A. Paul: Bifurcation of the Kirkendall marker plane and the role of Ni and other impurities on the growth of Kirkendall voids in the Cu–Sn system, Acta Materialia, 131 (2017) 260-270.

Sn–based solder is a major concern for the electronics industry. Analysis of the microstructural evolution with respect to the location of the Kirkendall marker plane is an important aspect for understanding the physico–mechanical properties of these brittle phases and thereby the reliability of an electronic component. Numerous articles are published every year in the Cu–Sn or Cu–Solder (Sn–based) systems studying these aspects, and the details of these can be found in References [17, 42, 57-60] and references therein. However, a clear understanding of the growth mechanism of these brittle phases is yet to be developed. Interestingly, Paul et al. [17] reported a mismatch between the predicted (based on physico–chemical approach [47]) and the experimentally found locations of the Kirkendall marker plane. Identifying the locations of these planes is important for understanding the finer details of the diffusion–controlled growth process of both the phases. A single Kirkendall marker plane is found (inside the $Cu_6Sn_5$ phase [17, 42, 57]) in these previously studied Cu/Sn diffusion couples. However, based on the estimated diffusion coefficients from incremental diffusion couples, Paul et al. [17] predicted a bifurcation of the Kirkendall marker plane, *i.e.*, the presence of the Kirkendall marker plane in both the phases, namely $Cu_3Sn$ and $Cu_6Sn_5$. Latter the same was validated following a different theoretical analysis by Svoboda et al. [61].

At the same time, numerous articles are published on the growth of the Kirkendall voids, as reported in References [58-60, 62-66] and references therein. Due to drive for miniaturization of microelectronic systems and devices, the growth of voids in the $Cu_3Sn$ phase is one of the main reasons of electro–mechanical failure in microelectronic components. Immediately after the discovery of the Kirkendall effect [25], researchers could correlate the growth of voids in an interdiffusion zone with this effect [62-66]. A flux of vacancies is created because of a difference in the diffusion rates of components, which are supersaturated and nucleated heterogeneously if not absorbed by the sinks (such as dislocations, grain boundaries and interfaces) [63, 65, 66]. The presence of impurities is known to play an adverse role on the growth of these voids [58, 63]. Because of industrial relevance, most of the studies are conducted with electroplated Cu in which impurities can be included from the electroplating bath leading to the very high growth rate of these voids in the interdiffusion zone of Cu UBM and Sn–based solder [58-60]. However, there is an ambiguity about the exact role of impurities, since these studies are not compared extensively for the known and different concentration of impurities in Cu.

Therefore, the aim of the present chapter is two–fold. The first aim is to examine if a bifurcation of the Kirkendall marker plane is indeed present (as predicted [17]) in the Cu–Sn system. If it is present, then what is the reason for not detecting it in the previous experiments [17, 42, 57]? Following, based on the estimation of the diffusion coefficients at the Kirkendall marker plane(s), the second aim is to understand the role of impurities on the growth of the Kirkendall voids by considering Cu with different concentration of impurities. This is analysed based on the effect of concentration of impurities on void size distribution.

## 5.2 Bifurcation of the Kirkendall marker plane in the Cu–Sn system

It should be noted here that the presence of more than one Kirkendall marker plane is a special phenomenon and found only in very few systems. It allows developing a finer understanding of the phenomenological diffusion process [18]. There are different ways to detect the location of this plane in an interdiffusion zone. The conventional method is of course by the use of inert particles [18]. On the other hand, as established based on the physico–chemical approach [47], the microstructural features efficiently indicate these locations without using any inert particles. Since the experiments using the inert particles failed to show the presence of bifurcation of the Kirkendall marker plane in the Cu–Sn system [17, 42, 57] (as predicted [17]), the microstructural evolution of both the phases, $Cu_3Sn$ and $Cu_6Sn_5$, is examined for our analysis.

Before going for further explanation, it is necessary to understand the growth mechanism of the phases following the physico–chemical approach [47] and what kind of microstructural evolution is expected depending on the location of the Kirkendall marker plane. This can be explained with the help of a schematic diagram as shown in Figure 5.1.

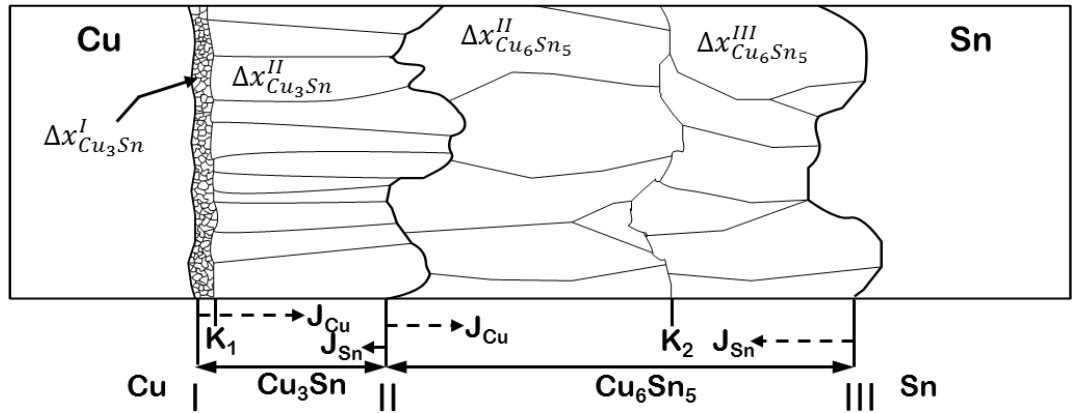

Figure 5.1: Schematic representation of the growth mechanism of the phases and expected microstructural evolution depending on the location of the Kirkendall marker planes, $K_1$ and $K_2$ in $Cu_3Sn$ and $Cu_6Sn_5$, respectively.

We consider the presence of the Kirkendall marker plane in both the phases, *i.e.*, $K_1$ and $K_2$ for $Cu_3Sn$ and $Cu_6Sn_5$, respectively. Interface I, II and III refers to Cu/$Cu_3Sn$, $Cu_3Sn$/$Cu_6Sn_5$ and $Cu_6Sn_5$/Sn interfaces, respectively. $\Delta x^I_{Cu_3Sn}$ and $\Delta x^{II}_{Cu_3Sn}$ are the thickness of the sublayers in the $Cu_3Sn$ phase, while $\Delta x^{II}_{Cu_6Sn_5}$ and $\Delta x^{III}_{Cu_6Sn_5}$ are the thickness of sublayers in the $Cu_6Sn_5$ phase such that the total phase layer thickness is $\Delta x_{Cu_3Sn} = \Delta x^I_{Cu_3Sn} + \Delta x^{II}_{Cu_3Sn}$ and $\Delta x_{Cu_6Sn_5} = \Delta x^{II}_{Cu_6Sn_5} + \Delta x^{III}_{Cu_6Sn_5}$. $J_i$ denotes the intrinsic fluxes of diffusing components, *i.e.*, Cu and Sn, in both the $Cu_3Sn$ and $Cu_6Sn_5$ phases.

The reaction–dissociation of the components at different interfaces and therefore the morphological evolution is described below qualitatively. The quantitative mathematical analysis of the diffusion coefficient dependent on the microstructural evolution can be learnt from Reference [47].

*The growth of the $Cu_3Sn$ phase:*

(i)      Cu is released from Cu end–member at the interface I (Cu/$Cu_3Sn$). It diffuses through $Cu_3Sn$ and then reacts with $Cu_6Sn_5$ at the interface II ($Cu_3Sn/Cu_6Sn_5$) for the growth of $Cu_3Sn$ from the same interface.

(ii)      Sn dissociates from $Cu_6Sn_5$ at the interface II to produce $Cu_3Sn$ and Sn. The same Sn then diffuses through $Cu_3Sn$ and reacts with Cu at the interface I to produce $Cu_3Sn$.

*The growth of the $Cu_6Sn_5$ phase:*

(iii)      $Cu_3Sn$ dissociates at the interface II to produce $Cu_6Sn_5$ and Cu. The same Cu then diffuses through $Cu_6Sn_5$ and reacts with Sn at the interface III ($Cu_6Sn_5$/Sn) to produce $Cu_6Sn_5$.

(iv)      Sn is released from the Sn end–member at the interface III. It diffuses through $Cu_6Sn_5$ and then reacts with $Cu_3Sn$ at the interface II for the growth of $Cu_6Sn_5$.

Note that the reaction–dissociation process controls the microstructural evolution; however, the growth of phases depends on the diffusion rates of components. From the discussion above, it must be evident that (depending on the diffusion rates of Cu and Sn) the $Cu_3Sn$ phase grows from interfaces I and II, while the $Cu_6Sn_5$ phase grows from interfaces II and III. $Cu_3Sn$ at the interface I (with sublayer thickness of $\Delta x_{Cu_3Sn}^{I}$) and $Cu_6Sn_5$ at the interface III (with sublayer thickness of $\Delta x_{Cu_6Sn_5}^{III}$) grow without getting consumed by the neighbouring phases. However, the growth process is complex at the interface II as both the phases try to grow at the cost of the other phase. The presence of a single or bifurcation of the Kirkendall marker plane depends on the growth and consumption rates of the phases at the interface II. If both the phases have their growth rate higher than the consumption rate, then $\Delta x_{Cu_3Sn}^{II}$ and $\Delta x_{Cu_6Sn_5}^{II}$ will have positive values. In such a situation, both the phases will grow with two sublayers. Therefore, the bifurcation of the Kirkendall marker ($K_1$ in $Cu_3Sn$ and $K_2$ in $Cu_6Sn_5$) and hence splitting of inert marker particles should be found in this Cu–Sn system [47]. Since both the phase layers grow differently from two different interfaces, a duplex morphology should be found demarcated by $K_1$ and $K_2$, as shown schematically in Figure 5.1. On the other hand, if the thickness of any of the sublayers $\Delta x_{Cu_3Sn}^{II}$ or $\Delta x_{Cu_6Sn_5}^{II}$ is negative (*i.e.*, consumption rate of a phase at the interface II is higher than the growth rate), a single Kirkendall marker plane should be found. For example, if $\Delta x_{Cu_6Sn_5}^{II}$ is positive and $\Delta x_{Cu_3Sn}^{II}$ is negative, then the marker plane will be present only in the $Cu_6Sn_5$ phase. A negative value of $\Delta x_{Cu_3Sn}^{II}$ means that the sublayer $\Delta x_{Cu_3Sn}^{II}$ along with some part of $Cu_3Sn$ which is grown from interface I (*i.e.*, $\Delta x_{Cu_3Sn}^{I}$) is consumed because of the growth of $Cu_6Sn_5$ at the interface II. In such a situation, a duplex morphology in the $Cu_6Sn_5$ phase (demarcated by $K_2$) and one type of morphology (because of growth only from interface I) is expected to be found in the $Cu_3Sn$ phase. This will lead to the presence of a single Kirkendall marker plane in the Cu–Sn system.

As already mentioned, the splitting of the Kirkendall marker plane is a special phenomenon and found only in very few systems. Most of the systems, in general, grow with a single Kirkendall marker plane. Therefore, as a general trend, efforts are not made to locate more than one marker plane. On the other hand, if the diffusion parameters in different phases are known, one can calculate the location of the Kirkendall marker planes in a particular system [47]. Paul et al. [17] estimated these parameters in the $Cu_3Sn$ and $Cu_6Sn_5$ phases by incremental diffusion couple experiments in which couples were prepared such that a single phase grows in the interdiffusion zone. For example, Cu and $Cu_6Sn_5$ were coupled for the growth of $Cu_3Sn$, whereas $Cu_3Sn$ and Sn were coupled for the growth of $Cu_6Sn_5$ [17]. Subsequently, these diffusion parameters were used to calculate the thickness of the sublayers in both the phases, $Cu_3Sn$ and $Cu_6Sn_5$, in a Cu/Sn diffusion couple. Surprisingly, positive values of all the sublayers were calculated indicating the presence of a bifurcation of the Kirkendall marker plane in the Cu–Sn system [17]. However, the inert markers could locate this plane only in the $Cu_6Sn_5$ phase showing the presence of a single Kirkendall marker plane [17, 42, 57]. To examine this disparity, as an alternate method, microstructure evolution is examined in this segment of the study to locate the Kirkendall marker plane.

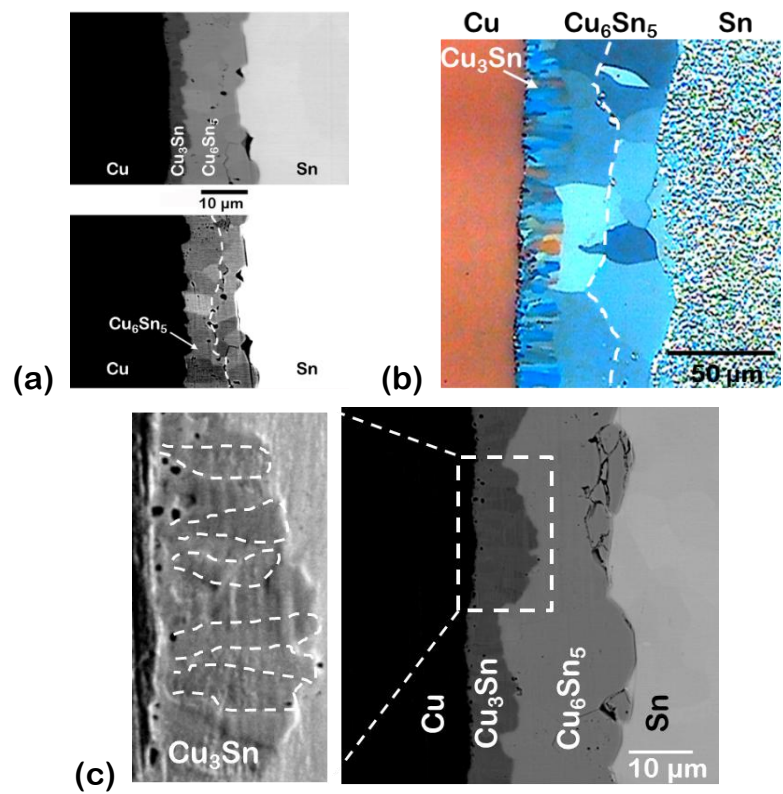

Figure 5.2: The growth of $Cu_3Sn$ and $Cu_6Sn_5$ phases in the interdiffusion zone of Cu/Sn diffusion couple annealed at: (a) 200 °C for 81 hrs, BSE micrograph; (b) 215 °C for 1600 hrs, polarized light optical micrograph [Courtesy of Dr. A.A. Kodentsov, Eindhoven University of Technology, The Netherlands]; (c) 200 °C for 400 hrs, BSE image (right) and focused SE image (left) highlighting elongated $Cu_3Sn$ grains. The location of the Kirkendall marker plane is indicated by duplex morphology inside the $Cu_6Sn_5$ phase as denoted by a dashed line in (a) and (b).

The micrographs of Cu/Sn diffusion couple are shown in Figure 5.2, revealing the microstructure in the $Cu_3Sn$ and $Cu_6Sn_5$ phases. Figure 5.2a is the BSE micrograph of the interdiffusion zone of Cu/Sn diffusion couple, which shows the presence of both the phases $Cu_3Sn$ and $Cu_6Sn_5$ after annealing at 200 °C for 81 hrs. The grains of the $Cu_6Sn_5$ phase could be resolved by increasing the contrast–brightness of the same image (in which the $Cu_3Sn$ phase is not visible). A duplex morphology is clearly visible inside the $Cu_6Sn_5$ phase demarcated by the Kirkendall marker plane, as shown by the dashed line. Previously, a similar location of the marker plane was detected by the use of inert markers [6, 17]. A polarized light optical micrograph of a Cu/Sn diffusion couple annealed at 215 °C for 1600 hrs, as shown in Figure 5.2b (kindly provided by Dr. A.A. Kodentsov, Eindhoven University of Technology, The Netherlands), also shows a similar nature of the grain morphology inside the $Cu_6Sn_5$ phase in which the Kirkendall markers (*i.e.*, inert particles) were also found along the dashed line [6]. This indicates that $\Delta x_{Cu_6Sn_5}^{II}$ could grow despite being consumed by $Cu_3Sn$ at the interface II (Figure 5.1). Since $Cu_6Sn_5$ grains cover from the interfaces (II and III) to the Kirkendall marker plane ($K_2$), it is evident that once grains nucleated, they grew continuously without further nucleation. This is common in majority of the systems with intermetallic compounds, most probably because of the high activation energy barrier for nucleation [18]. The grain morphology of the $Cu_3Sn$ phase could not be resolved in the image shown in Figure 5.2a. However, the polarized light micrograph, as shown in Figure 5.2b, indicates that the elongated grains might be present covering almost the whole $Cu_3Sn$ phase layer. The grains in the same phase could be faintly detected in a BSE image of Cu/Sn diffusion couple annealed at 200 °C for 400 hrs, as shown in Figure 5.2c (grain boundaries are marked by dashed lines in a focused SE image). It is difficult to resolve the grain morphology at regions very near to the Cu/Cu₃Sn interface (interface I); however, rest of the Cu₃Sn phase thickness is covered by long grains covering almost the whole phase layer. The Kirkendall plane (denoted by $K_1$ in schematic Figure 5.1) should not be found in the Cu₃Sn phase if the elongated grains indeed cover the whole Cu₃Sn phase layer. Since inert particles used to detect the location of the Kirkendall marker plane were not found in this phase, it was accepted that only one Kirkendall marker plane (inside the Cu₆Sn₅ phase) is present in the Cu–Sn system [17, 42, 57]. However, an important question remains unanswered with this consideration. Diffusion coefficients measured in an incremental diffusion couple of Cu/Cu₆Sn₅ in which only the Cu₃Sn phase grows at the interface indicates that Cu has much higher (almost ~30 times) diffusion rate compared to Sn in the Cu₃Sn phase [17]. It means that the growth rate of Cu₃Sn phase from the Cu₃Sn/Cu₆Sn₅ (interface II) must be much higher compared to the small growth rate of this phase from the Cu/Cu₃Sn (interface I). The Kirkendall marker plane ($K_1$ in Figure 5.1) could be absent in this phase only if the consumption rate of Cu₃Sn at the interface II is very high (*i.e.*, $\Delta x_{Cu_3Sn}^{II}$ has a negative value). It would mean that Cu₆Sn₅ consumes the whole amount of this sublayer of Cu₃Sn that is grown from the interface II ($\Delta x_{Cu_3Sn}^{II}$) along with some part of the phase which grows from the interface I ($\Delta x_{Cu_3Sn}^{I}$). In such a situation, the overall thickness of the Cu₃Sn phase (*i.e.* $\Delta x_{Cu_3Sn} = \Delta x_{Cu_3Sn}^{I} + \Delta x_{Cu_3Sn}^{II}$) should be much less than what is found in the Cu–Sn system. Moreover, the calculation of the thickness of the sublayers utilizing the diffusion coefficients estimated from the incremental diffusion couples indicates that all the 4 sublayer values ($\Delta x_{Cu_3Sn}^{I}$, $\Delta x_{Cu_3Sn}^{II}$, $\Delta x_{Cu_6Sn_5}^{II}$, $\Delta x_{Cu_6Sn_5}^{III}$) are positive [17]. That means, we should find a bifurcation of the Kirkendall marker plane, one each in the Cu₃Sn and Cu₆Sn₅ phases. Since the growth rate of the Cu₃Sn phase is expected to be small from the interface I ($\Delta x^I_{Cu_3Sn}$) and the $Cu_3Sn$ grains could not be resolved near the $Cu/Cu_3Sn$ interface, as shown in Figure 5.2 captured using SEM; we extended our investigation to TEM (higher resolution than SEM) with the expectation that the location of the Kirkendall marker plane might be detected in the $Cu_3Sn$ phase based on the presence of a duplex morphology.

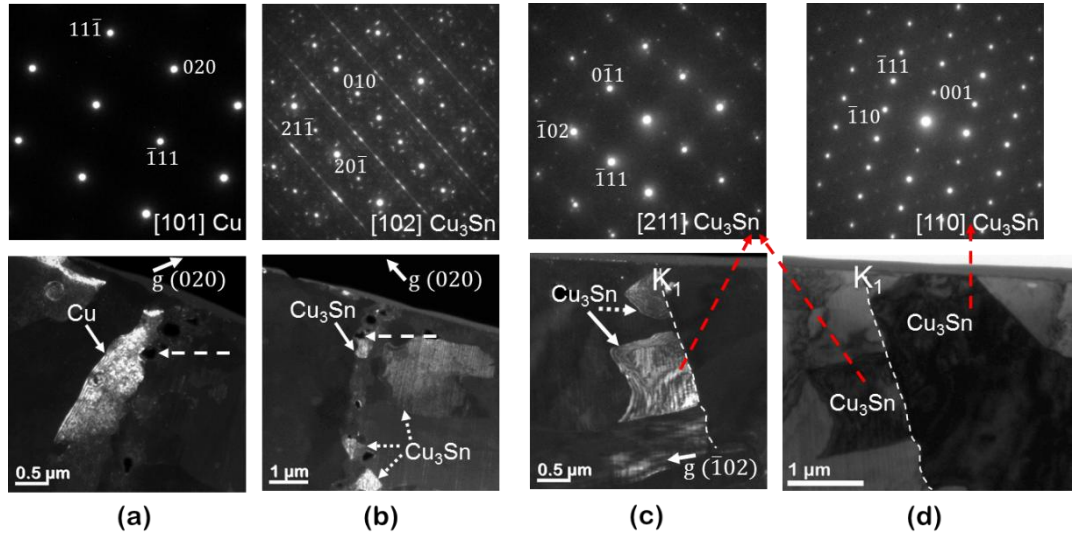

Figure 5.3: Diffraction patterns (top) along with the corresponding TEM micrographs (bottom) of (a) Cu and (b, c, d) at different locations of the $Cu_3Sn$ phase. Micrographs shown in (a, b, c) are dark–field images and (d) is bright–field image.

Figure 5.3 shows microstructures at different locations in the $Cu_3Sn$ phase. Respective dark–field (DF) or bright–field (BF) TEM micrographs along with their indexed DP (recognizing the phases) are shown. The $Cu_3Sn$ phase is an orthorhombic long period superstructure with 80 atoms in a unit cell (oC80, *Cmcm*). It can be seen as combination of 10 units with a prototype of orthorhombic $Cu_3Ti$ (oP8, *Pmnm*) ordered lattice. All the diffraction pattern (DP) of the $Cu_3Sn$ phase is indexed based on oP8 crystal structure [67]. Weak superlattice reflections can be seen in DP of the $Cu_3Sn$ phase due to ordering. Figure 5.3a shows the DP acquired from one of the Cu grain (near the $Cu/Cu_3Sn$ interface) along with the corresponding DF image. DP can be indexed with zone axis [101] of Cu and the DF image is acquired using (020) reflection. Dashed arrow indicates the same hole (very close to the $Cu/Cu_3$Sn interface) in both Figure 5.3a and b. On the right side of the interface, many small grains of $Cu_3Sn$ are found, as shown in Figure 5.3b. DP acquired from grain just below the dashed arrow is indexed with zone axis [102] of the $Cu_3Sn$ phase. Apart from weak superlattice reflections, extra double diffraction spots can be also seen in this DP (Figure 5.3b). The other grains of the $Cu_3Sn$ phase (indicated by dotted arrows), which might be oriented along the same zone axis [102], can be seen in the DF image (at lower magnification) acquired using (020) reflection. Therefore, there are many small grains of the $Cu_3Sn$ phase close to the $Cu/Cu_3Sn$ interface and following relatively bigger grains are found. Figure 5.3c shows the DP acquired from the grain (indicated by the arrow) along with the corresponding DF image. DP is indexed with zone axis [211] of the $Cu_3Sn$ phase. Other grain of the $Cu_3Sn$ phase (indicated by a dotted arrow) might also be oriented along the same zone axis [211] as can be seen in the DF image acquired using ($\bar{1}02$) reflection. These grains (Figure 5.3c) are relatively bigger than the grains observed close to $Cu/Cu_3Sn$ interface (Figure 5.3b). However, overall, all these grains are much smaller than the grains detected in the SEM micrograph in Figure 5.2c. Therefore, if the Kirkendall marker plane ($K_1$) in the $Cu_3Sn$ phase exists, it could be demarcated by two sublayers, one with smaller grains and another with much bigger grains. This is indeed found, as shown in the BF micrograph in Figure 5.3d. The DF and BF images, as shown in Figure 5.3c and d respectively, are acquired from the same region. DP acquired from the very big grain (indicated by a red dotted arrow in Figure 5.3d) is indexed with zone axis [110] of the $Cu_3Sn$ phase. This means that the Kirkendall marker plane is indeed present in this phase, which demarcates two different sublayers with different grain morphologies. The sublayer between interface I ($Cu/Cu_3Sn$) and $K_1$ has many small grains indicating the repeated nucleation of grains along with growth of this sublayer because of diffusion of Sn originated from the interface II ($Cu_3Sn/Cu_6Sn_5$). Another sublayer between $K_1$ and interface II has much bigger grains covering the thickness of whole sublayer, which is grown because of reaction (of Cu with $Cu_6Sn_5$) and dissociation (of Sn from $Cu_6Sn_5$) at the interface II. Instead of nucleating repeatedly, the product phase continuously joins with the existing $Cu_3Sn$ grains in this sublayer. This indicates that the nucleation of $Cu_3Sn$ might be easier at the interface I (on Cu) as compared to the interface II (on $Cu_6Sn_5$). Most of the intermetallic compounds in various systems grow with large grains covering the thickness from one interface to the Kirkendall marker plane or one interface to another (if the Kirkendall marker plane is not present or it is present very close to one of the interfaces [68]). However, there are few systems [69, 70], in which grains in a particular sublayer of intermetallic phase grow with many smaller grains indicating the ease of nucleation similar to what we have found between interface I and $K_1$. In a previous study based on the TEM analysis, Tian et al. [71] found that the Kirkendall marker plane (detected by inert marker) in the $Cu_2(In,Sn)$ phase indeed demarcates two sublayers with different types of grain morphologies, which supports the concept of the physico–chemical approach [47].

Now the question is why the bifurcation of the Kirkendall marker plane in the Cu–Sn system could not be detected when inert particles were used as the Kirkendall markers in the previous studies [17, 42, 57]. It should be noted here that a system might fulfill the conditions for bifurcation of the marker plane; however, the markers (inert particles) will be able to split into two different phases only if both the phases start growing together at the very initial stage of diffusion annealing. Bulk diffusion couples, in general, show the simultaneous growth; however, the sequential growth of the phases in the Cu–Sn system is reported [72], which was indeed suspected based on the thermodynamic viewpoint [17]. The Cu$_3$Sn phase grows only after the growth of Cu$_6$Sn$_5$ phase. Therefore, once the markers (inert particles) are trapped inside the Cu$_6$Sn$_5$ phase, they cannot move into the Cu$_3$Sn phase. On the other hand, the microstructural analysis has evolved as a reliable technique for the detection of the Kirkendall marker plane [18]. This is now commonly practiced in the systems in which markers cannot be used, such as material in applications or thin films [18]. This is also recently followed in many refractory metal–silicon systems [68]; this is because of the fact that diffusion couples could not be bonded successfully when inert marker particles were used at the interface between these hard materials, often creating a gap between them.

Now, we estimate the diffusion parameters, to facilitate our discussion on the growth of the Kirkendall voids. The ratio of the tracer diffusivities, $D_i^*$ of the phases with narrow homogeneity range in this system, are estimated following van Loo's method [29], Equation (3.30c). We have neglected the role of vacancy–wind effect [18, 24], since the structure factor required for the estimation of this effect is not known. Considering different types (with different annealing times) and different locations of the diffusion couples, we estimated the values at K$_1$ (in Cu$_3$Sn) and K$_2$ (in Cu$_6$Sn$_5$) as:

$$\left[\frac{D_{Sn}^*}{D_{Cu}^*}\right]_{K_1(Cu_3Sn)} = 0.033 \quad \Longrightarrow \quad \left[\frac{D_{Cu}^*}{D_{Sn}^*}\right]_{K_1(Cu_3Sn)} = 30 \pm 10$$

$$\left[\frac{D_{Sn}^*}{D_{Cu}^*}\right]_{K_2(Cu_6Sn_5)} = 2.3 \pm 1$$

It should be noted here that the calculation of tracer (or intrinsic) diffusion coefficients by diffusion couple technique introduces high error when the ratio is outside the range of 0.1–1 [18]. Moreover, the error in calculation is high especially in this system because of the waviness of phase layers. When a single phase layer grows in an incremental diffusion couple, $Cu_3Sn$ grows with more or less flat interfaces, whereas both the interfaces of $Cu_6Sn_5$ are found to be highly wavy [17]. Therefore, $Cu_3Sn$ is flat at the interface I, however, very wavy at the interface II, since the sublayer that is grown from interface II is dependent on the growth of $Cu_6Sn_5$, as discussed before. Thermodynamics constrain the system for the growth with flat interfaces [29]; however, orientation–dependent anisotropic growth [29] of $Cu_6Sn_5$ makes it wavy, which must be clear from grains revealed in this phase, as shown in Figure 5.2. Nevertheless, the experimental evidence of finding the bifurcation of the Kirkendall marker plane in the Cu–Sn system will bring finer understanding on the growth of the phases based on the theoretical analysis on vacancy creation and annihilation in this technologically important system, which draws special attention of many groups [73-76].

## 5.3 Effect of impurity content on the growth of the Kirkendall voids in $Cu_3Sn$

From the estimated ratio of tracer diffusion coefficients $\left[\frac{D_{Cu}^*}{D_{Sn}^*}\right]$, it is clear that Cu has much higher diffusion rate compared to Sn in the $Cu_3Sn$ phase, whereas, this difference is not that high in the $Cu_6Sn_5$ phase. Considering a constant molar volume of a phase, the flux of vacancies ($J_V$) can be related to the intrinsic fluxes ($J_i$) of components by $J_V = -(J_{Cu} + J_{Sn})$. Note here that $J_{Cu}$ and $J_{Sn}$ have opposite signs (Figure 5.1). Therefore, the flux of vacancies is significantly higher in the $Cu_3Sn$ phase. In fact, even 1% relative excess non–equilibrium vacancy concentration can be enough to create voids if not absorbed by the sinks [63]. It is also apparent that the vacancies are not absorbed completely in the $Cu_3Sn$ phase since the Kirkendall voids (could be recognized as dark spots) are found in this phase, as shown in Figure 5.4. Since the main focus in this section is to discuss the growth of voids in the $Cu_3Sn$ phase, henceforth the micrograph of this phase is only shown, and the $Cu_6Sn_5$ phase is shown only when it is required for any other discussion.

The presence of impurities plays a very important role in the growth of the Kirkendall voids. For example, it is known that the presence of impurities can increase the concentration of vacancies by decreasing the enthalpy of vacancy formation [77]. Because of relevance to the manufacturing process of making contacts between Cu as one of the layers of UBM and Sn–based solder, most of the studies [58-60] on the growth of the Kirkendall voids in the $Cu_3Sn$ phase are mainly focused on electroplated (EP) Cu. Generally, for a smooth and bright layer, different additional constituents (both organic and inorganic) are used in Cu electroplating bath, which are known to promote the inclusion of fairly high concentration of impurities in Cu, such as S, Cl, C, O, N and H. Out of these, S is found to play an adverse role in the growth of the Kirkendall voids [58, 78]. Interestingly, in a few studies, it is reported that the Kirkendall voids are not found or found with very small numbers when Cu with higher purities are coupled with Sn or Sn–based solder [79, 80]. It is indeed expected to find the low growth of the voids because of the smaller concentration of impurities in Cu. However, according to the certificate provided by Alfa Aesar (USA), the S content is around 0.2 ppm in 99.9999 wt.% Cu used in this segment of the study. Therefore, the voids are not expected to be completely absent even when the very high purity of Cu is used. In this segment of the study, we have considered different purities of Cu starting from 99.9999 wt.% to commercially pure (98–99 wt.% [14, 15]) for the comparison of the growth of voids with EP Cu.

As already explained in the previous section, the $Cu_3Sn$ phase grows by reaction–dissociation of components from the $Cu/Cu_3Sn$ and $Cu_3Sn/Cu_6Sn_5$ interfaces, whereas the $Cu_6Sn_5$ phase grows from the $Cu_3Sn/Cu_6Sn_5$ and $Cu_6Sn_5/Sn$ interfaces (Figure 5.1). Therefore, it is expected that the condition of Sn should not affect the growth of the Kirkendall voids in the $Cu_3Sn$ phase directly or significantly. To cross–check this statement, one diffusion couple is prepared between bulk 99.9 wt.% Cu and EP Sn (Figure 5.4b) for comparison of the growth of the $Cu_3Sn$ phase with a diffusion couple of bulk 99.9 wt.% Cu and bulk 99.99 wt.% Sn (Figure 5.4a). After comparison of many voids at different locations in the interdiffusion zone, we did not find any significant difference. Therefore, now onwards, the results for the different purity of Cu and bulk 99.99 wt.% Sn are shown. Following the practice in electronics industry, the commercially available plating bath provided by Growel (India) [13] is used for electroplating Cu in this segment of the study. Since the impurity concentration in EP Cu is found to be much higher than 99.9 wt.%, commercially pure Cu is considered for the purpose of comparison of bulk Cu and EP Cu.

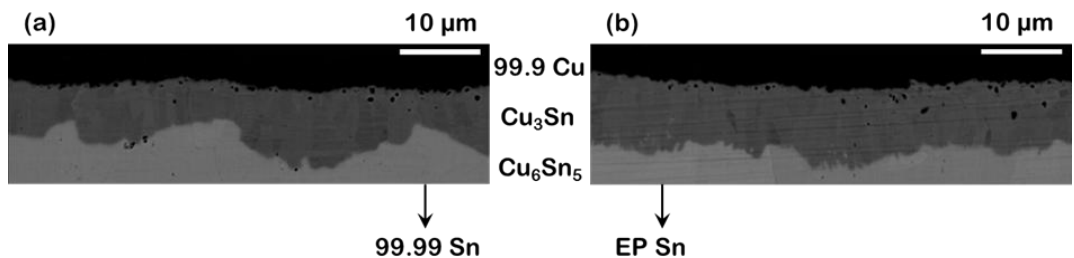

Figure 5.4: BSE micrographs showing comparison of the growth of the Kirkendall voids in diffusion couples of: (a) 99.9 wt.% bulk Cu and 99.99 wt.% bulk Sn, and (b) 99.9 wt.% bulk Cu and electroplated (EP) Sn annealed at 200 °C for 400 hrs.

When the different purities of Cu such as 99.9999, 99.999 and 99.9 wt.% are coupled with 99.99 wt.% Sn, the thicknesses of the phase layers are found to be more or less the same for a particular annealing time, indicating insignificant difference in the growth rate of both the phases. However, there is a difference in the growth of the Kirkendall voids in the $Cu_3Sn$ phase. Figure 5.5 shows the voids for different purities of Cu after annealing for 144 hrs at 200 °C. The $Cu_3Sn$ phase is more or less free from voids for 99.9999 wt.% Cu. A small number of voids are found when the diffusion couple is prepared with 99.999 wt.% Cu. However, they are not distributed evenly. Voids are found only at a few places, which indicates that impurities might not be distributed evenly. With the increase in impurity of Cu to 99.9 wt.%, the number of voids are found to be much higher and also distributed more or less evenly. Hence, to facilitate our systematic analysis with higher annealing time experiments, we have chosen only extremes cases of Cu purities, *i.e.*, 99.9999 and 99.9 wt.%. After increasing the annealing time to 400 hrs (at the same temperature), as shown in Figure 5.6a, the Kirkendall voids are found for both the purities of Cu, although the number of voids is significantly higher for 99.9 wt.% Cu.

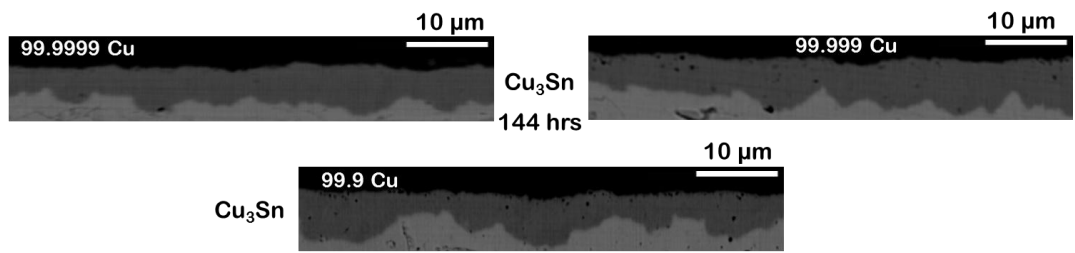

Figure 5.5: BSE micrographs of the Cu₃Sn phase in the interdiffusion zone of Cu/Sn bulk diffusion couples of 99.9999, 99.999 and 99.9 wt.% Cu with 99.99 wt.% Sn annealed at 200 °C for 144 hrs.

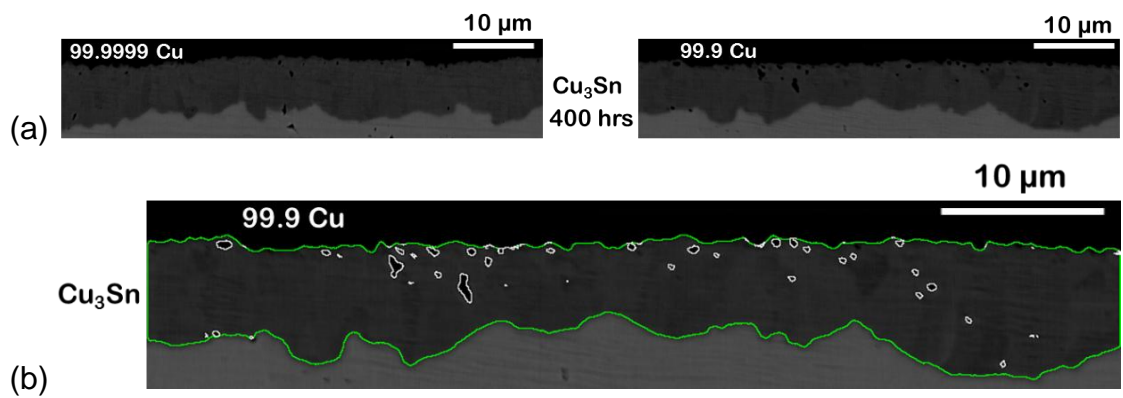

Figure 5.6: BSE micrographs of the Cu₃Sn phase in the interdiffusion zone of Cu/Sn bulk diffusion couples annealed at 200 °C for 400 hrs: (a) Diffusion couples of 99.9999 and 99.9 wt.% Cu with 99.99 wt.% Sn. (b) An example of analysis of the Kirkendall void size distribution by MIPAR on the micrograph of 99.9 wt.% Cu shown in figure (a).

To get further insights on the growth of the Kirkendall voids, the void statistics, *i.e.*, distribution of number of voids and their sizes viz. equivalent diameter (in terms of average size with $\pm$ 0.05 µm range) are estimated by using MIPAR (Materials Image Processing and Automated Reconstruction) [16], a powerful image analysis software [81]. An example of detection of these voids by the software (for 99.9 wt.% Cu diffusion couple annealed for 400 hrs at 200 °C, *i.e.*, Figure 5.6a) is shown in Figure 5.6b. First, the $Cu_3Sn$ phase boundary can be identified in the MIPAR software, as shown by green colour line, and then the voids inside the phase can be detected, as shown by white colour lines. To avoid error, we have not considered the void sizes less than 0.15 µm. Moreover, all the figures are checked carefully for the voids to be selected correctly, by making use of some of the inbuilt, simple and unique features in the MIPAR software. After the analysis, if any wrong consideration of other black spots as voids is identified, the software allows to remove them manually from the list of equivalent diameters estimated. In fact, while using this software, it is also possible to view two different images together at the same time, viz. the original image (Figure 5.6a) and the image with detected voids (Figure 5.6b), which allows to ascertain if the voids are detected correctly. Measurements are done using many images so that at least 200 voids are considered for each condition, and then, these are normalized for 1000 µm² area of the $Cu_3Sn$ phase.

To understand the effect of annealing times (for the same Cu purity) on the growth of the Kirkendall voids, the void size and number distributions are compared for 99.9 wt.% Cu after annealing for 144 and 400 hrs (at the same temperature), as shown in Figure 5.7. The total number of voids are found to be 136 and 150 per 1000 µm² after annealing for 144 and 400 hrs, respectively. The maximum size of a void is found to be around 0.9 µm after 144 hrs and 1.1 µm after 400 hrs for 99.9 wt.% Cu. Moreover, the number of voids of bigger sizes increase with the increase in annealing time. Overall number and size of voids will depend on the (i) relaxation of vacancies by sinks (K sinks) [82], (ii) heterogeneous nucleation of new voids because of supersaturation of vacancies and (iii) relaxation of vacancies in existing voids (F sinks) [82]. Relaxation of vacancies by sinks will not create a void. Nucleation of new voids will increase the number of voids, whereas the addition of vacancies in the existing void will increase the overall size of a void. Therefore, we can conclude that the excess vacancies, which are not absorbed by sinks, can generate new voids as well as increase the size of existing voids with the increase in annealing time.

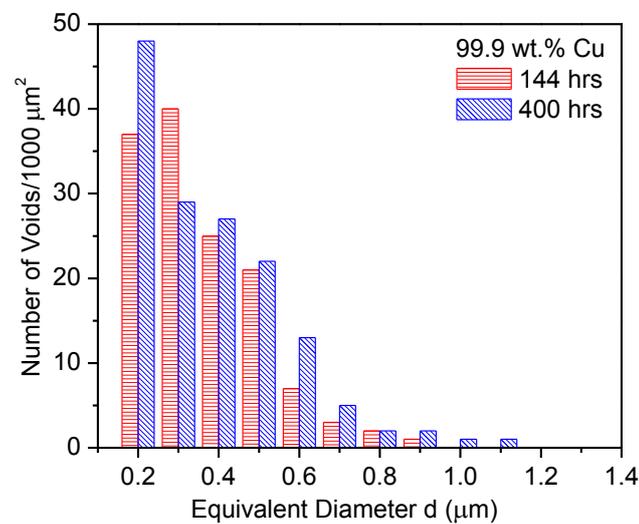

Figure 5.7: Void size distribution (at 200 °C) comparing two annealing times, 144 and 400 hrs for 99.9 wt.% Cu.

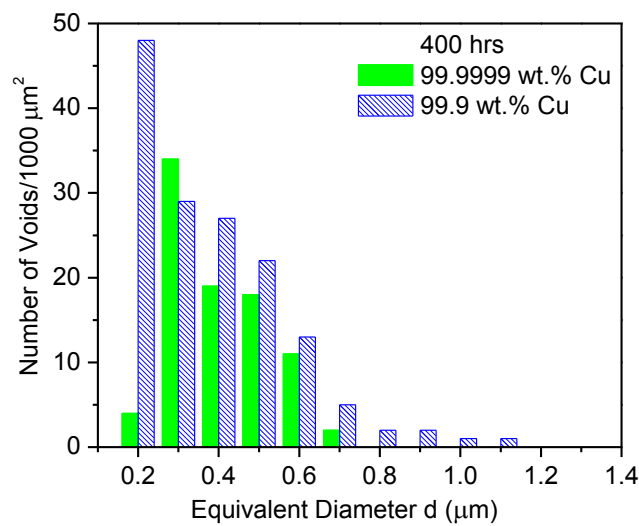

Figure 5.8: Void size distribution (at 200 °C) comparing two different purities, 99.9999 and 99.9 wt.% Cu for 400 hrs.

To understand the effect of impurity in Cu (after the same annealing time) on the growth of the Kirkendall voids, voids distributions are plotted in Figure 5.8 for 99.9999 and 99.9 wt.% Cu. Both the diffusion couples are annealed for the same time of 400 hrs. The clear differences in void size distribution can be seen in the micrographs of the $Cu_3Sn$ phase, as shown in Figure 5.6a. Further insights on the same (*i.e.*, growth of voids) are given by the distribution plot. The total number of voids for 99.9999 wt.% Cu is found to be 88 per 1000 $\mu m^2$ as compared to 150 voids per 1000 $\mu m^2$ for 99.9 wt.% Cu after annealing for 400 hrs. The estimated number of voids of the smallest size range (*i.e.*, the average equivalent diameter of 0.2 $\mu m$) for 99.9999 wt.% Cu is found to be very less, indicating the lower rate of nucleation of voids, which might be due to lack of availability of impurities for nucleation. Therefore, excess vacancies might prefer to join the existing voids. Void size and number distribution for 99.9 wt.% Cu indicates that voids could nucleate with a higher rate in this case and even the flux of excess vacancies must be much higher so that numbers of almost all sizes of voids are higher for 99.9 wt.% Cu when compared to the voids for 99.9999 wt.% Cu.

Next, we consider the growth of the Kirkendall voids for electroplated (EP) Cu, as shown in Figure 5.10a. Since the growth rate of the Kirkendall voids is much higher in this case, the annealing time is restricted to 100 hrs at 200 °C. As shown in Figure 5.9a, void size distribution is compared with the highest impurity of bulk Cu considered till now, *i.e.*, 99.9 wt.% Cu, which was annealed for 144 hrs (at the same temperature). Although the diffusion couple prepared with EP Cu is annealed for a smaller time (Figure 5.10a) compared to 99.9 wt.% Cu (Figure 5.6a), the void numbers and sizes are clearly much higher for EP Cu which can also be understood from the void distribution plot. The total number of voids is estimated as 435 per 1000 $\mu m^2$ for EP Cu after 100 hrs of annealing when compared to 136 voids per 1000 $\mu m^2$ for 99.9 wt.% Cu after 144 hrs of annealing. The highest average void size (equivalent diameter) is found to be 1.3 $\mu m$ for EP Cu. For a reasonable comparison, considering different annealing times of these diffusion couples in which the phases grow parabolically with time [39, 42], the void numbers are plotted with respect to the equivalent diameter (*d*) normalized by the square root of the annealing time *i.e.* $d/t^{1/2}$ ($\mu m/hr^{1/2}$) as shown in Figure 5.9b. This brings even higher difference for these two different impurities of Cu. Therefore, we can safely state that the impurity concentration in EP Cu must be much higher than 99.9 wt.% Cu. Shimizu et al. [83] stated based on their analysis that the total concentration of impurities in EP Cu could be close to 1 wt.%, which means the purity of EP Cu could be much higher than 99.0 wt.% Cu.

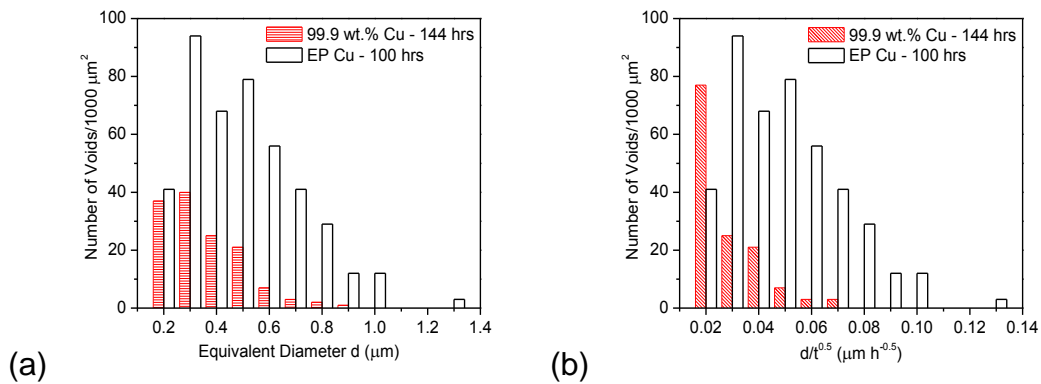

Figure 5.9: Void size distribution (at 200 °C) comparing two different purities, *i.e.*, 99.9 wt.% Cu and EP Cu (a) for two different annealing times, and (b) with respect to the normalized time.

Considering some of the embedded impurities in EP Cu could be volatile in nature (which may be released during annealing under vacuum), the effect of vacuum pre–heat treatment of the Cu on the growth of the Kirkendall voids in the $Cu_3Sn$ phase is studied previously [59, 84]. This could be beneficial if the temperature of heat treatment is restricted to a limit such that an electronic component can withstand it. Yin and Borgesen [59] heat treated EP Cu at 650 °C and found the negligible growth of the Kirkendall voids after that. Similar experimental observations are reported at 400–600 °C by Kim and Yu [84]. This indicates that Cu is indeed purified with a pre–heat treatment step. However, the temperature of the annealing is very high in both these studies [59, 84]. In the present study, to examine if this beneficial step is useful even at lower pre–heat treatment temperature, we first heat treated EP Cu foil in high vacuum ($\sim 10^{-4}$ Pa) at the same temperature and time as that of diffusion annealing, *i.e.*, at 200 °C for 100 hrs. This is designated as EP (HT) in this chapter. Following, this is electroplated with Sn to prepare a diffusion couple (similar to a couple of EP Cu and EP Sn shown in Figure 5.10a). This is further annealed at 200 °C 100 hrs for studying a difference in the growth of the Kirkendall voids in $Cu_3Sn$ phase layers for both these cases, *i.e.*, EP Cu and EP (HT) Cu. It can be seen in Figure 5.10b that the voids are mostly accumulated near the Cu/$Cu_3Sn$ interface for EP (HT) Cu. This is much clearer in the SE image, as shown in Figure 5.10c. A previous study by Singh et al. [85] indicates that impurities are transported to the Cu surface (leaving Cu beneath the surface as pure) following heat treatment of Cu alone and this could be the reason to find all the voids at the Cu/$Cu_3Sn$ interface. Although it is impossible to estimate voids distribution for EP (HT) Cu; however, looking at the microstructure (Figure 5.10b and c), it seems that the number of voids could be less for EP (HT) Cu when compared to EP Cu as plated (Figure 5.10a). This indicates the decrease in impurity concentration because of vacuum pre–heat treatment step. This further indicates the significant role of impurities on the growth of the Kirkendall voids, since in the case of EP Cu (Figure 5.10a) voids are spread over the whole $Cu_3Sn$ phase layer while in the case of EP (HT) Cu (Figure 5.10b and c) voids are negligible inside the $Cu_3Sn$ phase (and found only at Cu/Cu$_3$Sn interface). The accumulation of voids for EP (HT) Cu makes the interface very weak, which is not good for the electro–mechanical reliability of an electronic component during service.

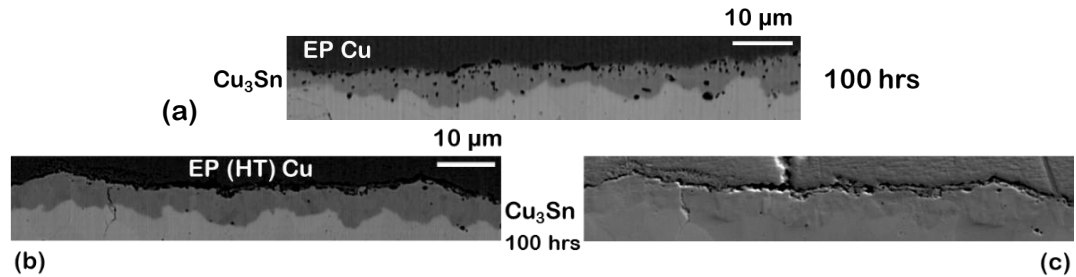

Figure 5.10: Micrographs of the interdiffusion zone of Cu/Sn diffusion couple annealed at 200 °C for 100 hrs: (a) EP Cu with EP Sn, BSE image, (b) EP (HT) Cu with EP Sn, BSE image and (c) SE image of the same. HT refers to vacuum pre–heat treatment step.

To compare these results with bulk Cu, since the void concentration for EP Cu is found to be higher than 99.9 wt.% bulk Cu, commercial pure (CP) Cu which is known to have the impurity content of around 1–2 wt.% [14, 15] is coupled with bulk Sn and annealed at 200 °C for 400 hrs, as shown in Figure 5.11a and b.

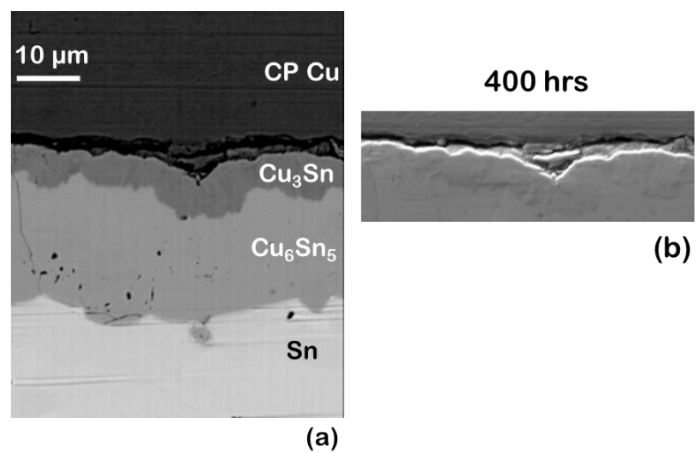

Figure 5.11: Micrographs of a diffusion couple of commercial pure (CP) Cu with 99.99 wt.% bulk Sn annealed at 200 °C for 400 hrs: (a) BSE image showing both the Cu$_3$Sn and Cu$_6$Sn$_5$ phases and (b) corresponding SE image, at the same magnification, showing only the Cu$_3$Sn phase.

Looking at the back–scattered electron (BSE) image (Figure 5.11a), it feels as if there is a gap at the Cu/Cu$_3$Sn interface. However, the secondary electron (SE) image, as shown in Figure 5.11b, clearly shows that there is actually a thin crack or fine separation present at the Cu/Cu$_3$Sn interface. The Kirkendall voids are not found inside the Cu$_3$Sn phase, which indicates that all the voids are accumulated at the interface for CP Cu creating bigger holes and making the interface very weak. This might cause the separation of Cu/Cu$_3$Sn interface during cross–sectioning of the diffusion couple and metallographic preparation. Considering the thickness of the Cu$_3$Sn phase after including the black contrast in the BSE image (Figure 5.11a) for CP Cu (since SE image shown in Figure 5.11b indicates the presence of the Cu$_3$Sn phase almost up to the Cu/Cu$_3$Sn interface), we find the thicknesses of both the phase layers as more or less the same for CP Cu when compared to 99.9 wt.% Cu (Figure 5.2c) after annealing for 400 hrs (at the same temperature). It indicates that the presence of the Kirkendall voids at the interface did not affect the supply of Cu from the end–member. It is well possible that the decrease in cross–sectional area is compensated by the surface diffusion of Cu through the voids.

## 5.4 Conclusions

For the very first time, a systematic study is conducted on the two very important aspects related to the Kirkendall effect in a technologically important Cu–Sn system: (a) to detect bifurcation of the Kirkendall marker plane based on the microstructural evolution and (b) to study the role of impurities on the growth rate of the Kirkendall voids in the Cu$_3$Sn phase. Much finer understanding on the growth process of the phases based on microstructural evolution and the Kirkendall voids are developed based on this study. Few keys findings can be stated as follows:

(i)  The experimental evidence of the presence of bifurcation of the Kirkendall marker plane is shown in the Cu–Sn system. The splitting of the markers was predicted before by Paul et al. [17]; however, the conventional marker experiment utilizing inert particles failed to detect this [17, 42, 57]. In this segment of the study, the locations of these planes in both $Cu_3Sn$ and $Cu_6Sn_5$ are detected by the microstructural evolution. The reason for not detecting this by the conventional marker experiment is the sequential growth of the phases in the Cu–Sn system, due to which the inert particles, entrapped in the initially growing $Cu_6Sn_5$ phase, cannot move into the $Cu_3Sn$ phase which grows later.

(ii)    This segment of the study strengthens the concept of the physico–chemical approach [47], which explains that one can efficiently detect (even in the case of sequential phase growth) the locations of the Kirkendall marker planes based on the analysis of microstructural evolution. There is no need to use inert particles at the interface. This is important since one cannot easily introduce inert particles in a system prepared with thin films or the materials in applications. The growth mechanism of the phase based on the relative mobilities of components can still be understood by locating this plane based on the microstructural evolution, which was not possible earlier.

(iii)   Quantitative diffusion analysis indicates that there is not much difference in the diffusion rates of components in the $Cu_6Sn_5$ phase. On the other hand, Cu has almost 30 times higher diffusion rate compared to Sn in the $Cu_3Sn$ phase. Since the Kirkendall voids grow in the $Cu_3Sn$ phase, it is evident that the flux of vacancies created due to the difference in diffusion rates of components are not absorbed completely by the sinks.

(iv)    The sublayer between the $Cu/Cu_3Sn$ interface and $K_1$ in the $Cu_3Sn$ phase grows by reaction of diffusing Sn with Cu at this interface (interface I). Sn is generated by dissociation of $Cu_6Sn_5$ at the $Cu_3Sn/Cu_6Sn_5$ interface leading to the growth of $Cu_3Sn$ in another sublayer between $K_1$ and this interface (interface II). This sublayer is also grown because of diffusion of Cu from the interface I at much higher rate and then by reaction with $Cu_6Sn_5$ at the interface II. As a result, the thickness of the sublayer that is grown from Cu at the interface I is much smaller compared to the thickness of another sublayer that is grown from $Cu_6Sn_5$ at the interface II.

(v)     Consequently, as found in the present study, the growth of the voids in the $Cu_3Sn$ sublayer that is grown from Cu is higher because of the presence of impurities in Cu and high flux of excess vacancies near the interface I, which are not absorbed by the sinks. Since the voids could grow even in the other $Cu_3Sn$ sublayer that is grown from $Cu_6Sn_5$ at the interface II, it is evident that impurities could transport inside the product phase from the Cu end–member and the concentration of excess vacancies must be enough in this sublayer also, which is required for the formation and growth of the voids. This is important for the theoretical studies, since the growth of voids in the sublayer that is grown from $Cu_6Sn_5$ is not considered or explained in the theoretical studies published earlier by others.

(vi)    The presence of impurities plays an important role in the growth of voids. The growth rate increases significantly when the impurity concentration is equal to or more than 0.1 wt.% in Cu. The presence of a very high number of voids for EP Cu indicates the incorporation of fairly high concentration of impurities during electroplating. Voids size distribution indicates that the nucleation of new voids can happen along with the growth of existing voids.

(vii)   A theoretical study by Svoboda and Fischer [73] indicates that an incoherent interface acts as non–ideal source and sink for vacancies, leading to most effective site for nucleation of voids. These are detached from the interface during interface migration and stay dispersed inside the product phase. Surface diffusion because of the presence of pores is also possible such that voids are detached from the interface after growth of fresh $Cu_3Sn$ at the $Cu/Cu_3Sn$ interface, which is recently shown by Gusak et al. [86].

(viii)  The role of surface diffusion of both the components is evident from the fact that the growth rates of the phases are not much different even when many voids are accumulated mainly near to the $Cu/Cu_3Sn$ interface in the case of CP Cu leading to decrease in the interfacial area through which the components diffuse. However, at this point it is not clear to us that why the voids spread over the whole phase layer for 99.9999–99.9 wt.% Cu and EP Cu and these are accumulated near the interface for EP (HT) Cu and CP Cu.

This system draws a special attention for theoretical analysis [73, 82, 86, 87] because of technological importance. The finding of the bifurcation of the Kirkendall marker plane and the systematic analysis of the growth of the Kirkendall voids with increasing concentration of impurity will help to establish the exact underlying mechanism based on simulations, which is otherwise difficult to establish based on just experimental studies.

*To summarize, we present experimental evidence of the bifurcation of the Kirkendall marker plane based on the microstructural evolution of both $Cu_3Sn$ and $Cu_6Sn_5$, and unambiguously show an accelerating effect of impurities in Cu on the formation of the Kirkendall voids in the Cu–Sn system.*

# Chapter 6

# Investigation of the growth of phases in the Cu(Ni)–Sn system

In previous two chapters, we have discussed the growth (kinetics and mechanism) of phases in the binary Cu–Sn and Ni–Sn systems. It is a well–known fact that the presence of a third element might strongly change the growth kinetics of the phases in the Cu–Sn binary system. Therefore, the aim of this chapter is to examine the effect of Ni content in Cu on the growth of phases and the formation of the Kirkendall voids in the Cu–Sn system.

## 6.1    Introduction and Statement of the Problem

As explained in previous chapters, Cu and Ni are two important elements in under bump metallization (UBM) layers used for making electro–mechanical bonding in microelectronic components. The intermetallic compounds, $Cu_3Sn$ and $Cu_6Sn_5$, grow at the interface of UBM and Sn–based solder alloy during soldering. Subsequently, these phases grow further by the solid–state reaction–diffusion process during service. Therefore, it may influence the physico–mechanical properties of the flip–chip bonding interface significantly.

At present, there are numerous studies available discussing the Cu–Sn system; however, limited studies are conducted in the Cu–Ni–Sn system [4-6]. The presence of Ni in Cu is found to have a strong influence on the growth rate of these product phases. However, none of the previous studies have rationalized the growth mechanism of the phases by estimating the diffusion parameters. Therefore, the aim of this segment of the study is to conduct experiments in Cu(Ni)–Sn system in the solid–state for estimation of the important diffusion parameters and for understanding the growth mechanism of the phases. In addition, the detailed analysis of observations pertaining to microstructural changes (e.g., grain morphology) and the calculated thermodynamic driving forces (using experimental data obtained in this study) are used for understanding the diffusion–controlled growth mechanism of the product phases. Following, the atomic mechanism of diffusion in ternary phases is discussed. Additionally, the effect of Ni addition on the crystal structure of the product phases is analysed based on the selected–area diffraction patterns acquired using a transmission electron microscope (TEM). Further, the effect of Ni content on the growth of the Kirkendall voids is discussed (*i.e.*, an extension of the discussion in Chapter 5 on the role of impurities in Cu in the formation of the Kirkendall voids), which is one of the major concerns of the electronics industry.

## 6.2 Effect of Ni addition in Cu on the growth of phases

As shown in Figure 6.1, two intermetallic compounds $Cu_3Sn$ and $Cu_6Sn_5$ grow at 200 °C in the interdiffusion zone of a binary Cu/Sn diffusion couple. The growth of both the phases is affected because of the addition of Ni in Cu. This can be understood from the average thicknesses ($\Delta x$) of phases, as mentioned in Figure 6.1. For the purpose of comparison, the diffusion couples with 0–5 at.% Ni are shown with the same magnification. Since the thickness of the phase layer in 8 at.% Ni is much higher, it is shown with a lower magnification. The concentrations of Ni are chosen such that we can study the change in growth rate of the phases in two different situations related to the presence and the absence of $(Cu,Ni)_3Sn$ in the interdiffusion zone. It can be seen that with the addition of 0.5 at.% Ni, the thickness of $Cu_6Sn_5$ increases (when compared to binary Cu/Sn); however, there is a slight decrease in the thickness of Cu₃Sn. Overall, there is an increase in the thickness of the whole interdiffusion zone. This trend continues till the addition of 2.5 at.% Ni. However, with the addition of 3 at.% Ni and beyond (Figure 6.1), the interdiffusion zone consists of only (Cu,Ni)₆Sn₅ phase and there is significant increase in its growth rate.

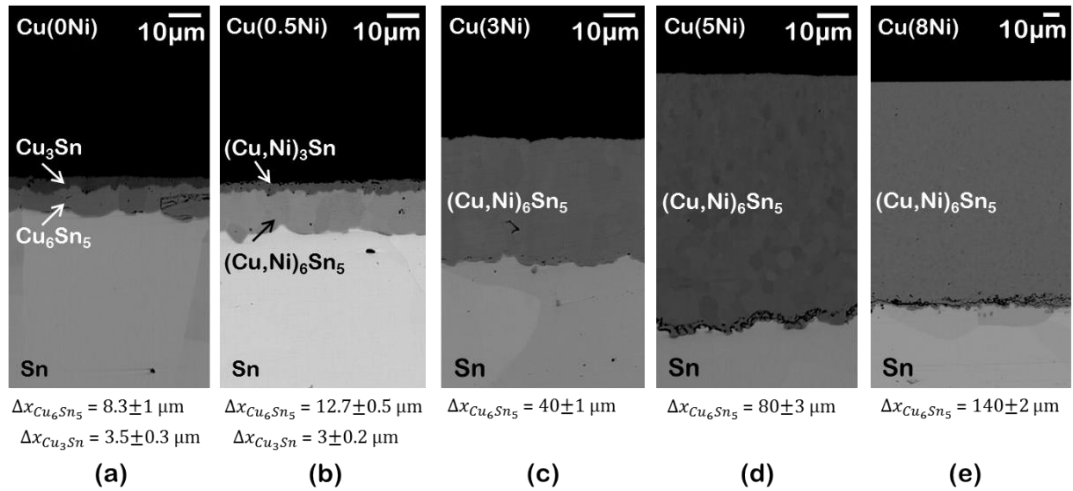

Figure 6.1: BSE images of the interdiffusion zone along with the average thicknesses (Δx) of different phases in various Cu(Ni)/Sn diffusion couples annealed at 200 °C for 81 hrs: (a) Cu/Sn, (b) Cu(0.5Ni)/Sn, (c) Cu(3Ni)/Sn, (d) Cu(5Ni)/Sn and (e) Cu(8Ni)/Sn. Ni content is mentioned in atomic percentage (at.%).

To confirm the absence of (Cu,Ni)₃Sn upon addition of 3 at.% Ni in Cu, the interface at Cu(3Ni)/(Cu,Ni)₆Sn₅ is examined in TEM and validated with the diffraction pattern (DP) analysis, as shown in Figure 6.2. The red arrow indicates the same region near the Cu(3Ni)/(Cu,Ni)₆Sn₅ interface. DP acquired from one of the Cu(3Ni) grain is shown along with the corresponding dark–field (DF) and bright–field (BF) images. DP is indexed with zone axis [001] of Cu(3Ni) and the DF image is acquired using (200) reflection. On the right side of the interface (denoted by red arrow), nano–beam electron diffraction pattern (NBDP) acquired from one of the (Cu,Ni)₆Sn₅ grain is shown along with the corresponding BF image, where Cu(3Ni)/(Cu,Ni)₆Sn₅ phase boundary is marked by a dotted line. DP is indexed with zone axis [10$\bar{1}$1] of the (Cu,Ni)₆Sn₅ phase. The comparison of these TEM micrographs clearly show the presence of $(Cu,Ni)_6Sn_5$ phase adjacent to Cu(3Ni) grain.

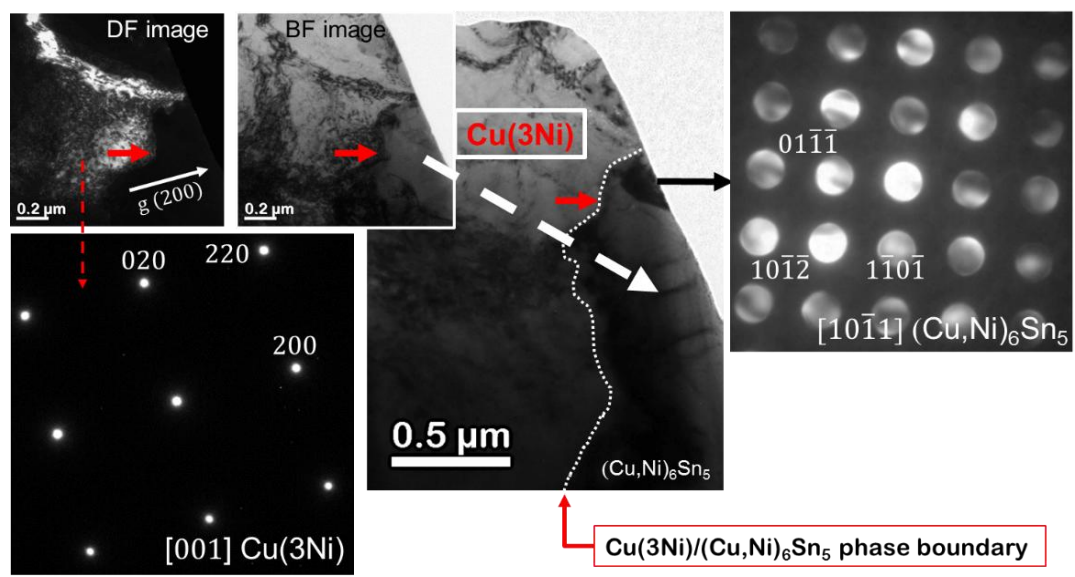

Figure 6.2: TEM analysis revealing the absence of the $(Cu,Ni)_3Sn$ phase in the Cu(3Ni)/Sn diffusion couple annealed at 200 °C. Diffraction patterns along with the corresponding TEM micrographs for Cu(3Ni) alloy and $(Cu,Ni)_6Sn_5$ phase are shown.

Before further explanation of diffusion–controlled growth kinetics of the phases, it is important to check if there is any change in the crystal structure of these product phases because of Ni addition, that might play an important role in the diffusion mechanism of components at the atomic level.

## 6.3    Structural Characterization using TEM

In this study, the DP of the Cu(Ni) alloys is indexed to pure Cu having FCC crystal structure, indicating similar *d*–spacing. At the temperature of interest (200 °C), it is known that $Cu_3Sn$ has an orthorhombic and $Cu_6Sn_5$ has a hexagonal crystal structure in the Cu–Sn system.

As shown in Figure 6.3, the $Cu_3Sn$ phase is an orthorhombic (oC80, *Cmcm*) long period superstructure, which is a combination of 10 units with a prototype of orthorhombic $Cu_3Ti$ (oP8, *Pmnm*) ordered lattice. All the DP of the $Cu_3Sn$ phase (irrespective of the Ni content) is indexed based on oP8 crystal structure [67]. The DPs acquired from the $(Cu,Ni)_3Sn$ phase in the binary Cu/Sn and ternary Cu(0.5Ni)/Sn diffusion couple are shown in Figure 6.3, which are indexed with zone axis [102] of the phase. Apart from weak superlattice reflections (due to ordering in this phase), extra double diffraction spots can be also seen in the DP.

The $Cu_6Sn_5$ phase is hexagonal (hP4, $P6_3/mmc$) structure with the prototype of $NiAs–Ni_2In$. As shown in Figure 6.4, a partially filled $B8_2$ $Ni_2In$ crystal structure fulfills the atomic arrangement of $Cu_6Sn_5$ [88]. But it has the basic unit as $B8_1$ NiAs (hP4) structure [89]. To illustrate it further, Cu occupies the 2a Wyckoff sublattices (2 atoms per unit cell) and Sn occupies the 2c site (2 atoms per unit cell). Following, a partial filling of 2d sites is required (0.4 atom fraction of Cu) for meeting the stoichiometric composition and measured density [88]. All the DP of the $Cu_6Sn_5$ phase (irrespective of the Ni content) is indexed based on the h$P4$ crystal structure [89]. DP acquired from the $(Cu,Ni)_6Sn_5$ phase in Cu(0Ni)/Sn, Cu(0.5Ni)/Sn and Cu(3Ni)/Sn diffusion couples are shown in Figure 6.4. All the DP are indexed with zone axis $[10\bar{1}1]$ of the $Cu_6Sn_5$ phase.

This confirms that we have found the equilibrium crystal structure for both the phases in binary Cu/Sn couple. Moreover, from the comparison of the same zone axis DP (acquired at the same camera length) for each phase, it is evident that:

(i)     There is no change in the crystal structure because of Ni addition in both the phases. At the temperature of our interest, the same crystal structure (as found in this study) is reported previously by Nogita [90] for the $(Cu,Ni)_6Sn_5$ phase in solidified Sn–0.7Cu–0.05Ni (wt.%) solder alloys.

(ii)    Since the $R$–spacing of all the reflections (diffraction spots) are similar in DP, it indicates similar $d$–spacing of the respective planes and therefore no significant change is observed in the lattice parameters for both the phases, when compared to the binary Cu–Sn phases. This is stated following the standard relation: $Rd = \lambda L$, where $R$ is the distance between the direct and diffracted beams, $d$ is the distance between atomic planes corresponding to specific reflection, $\lambda$ is wavelength and $L$ is the camera length such that $\lambda L$ is camera constant.

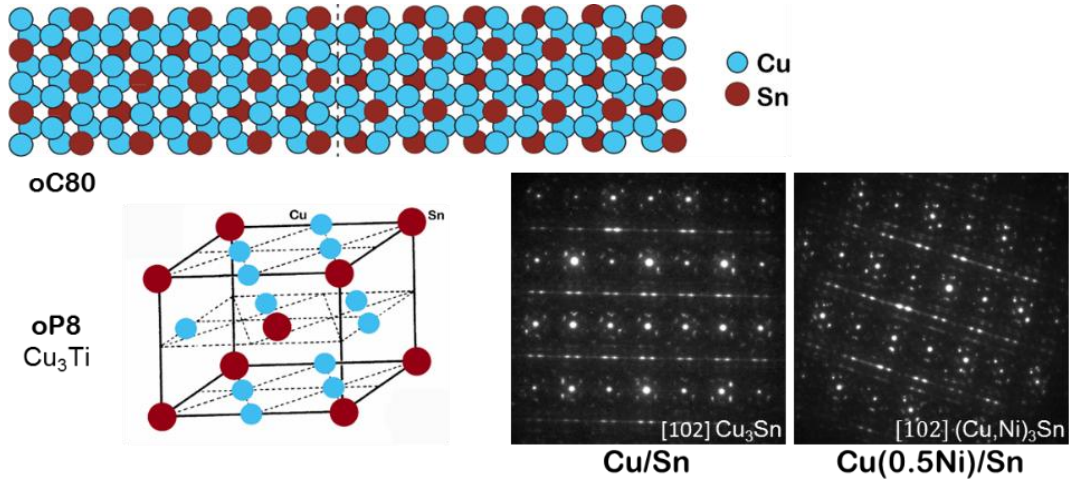

Figure 6.3: Crystal structure of the $Cu_3Sn$ phase along with the diffraction pattern acquired from the phase grown in the interdiffusion zone of Cu/Sn and Cu(0.5Ni)/Sn diffusion couples annealed at 200 °C.

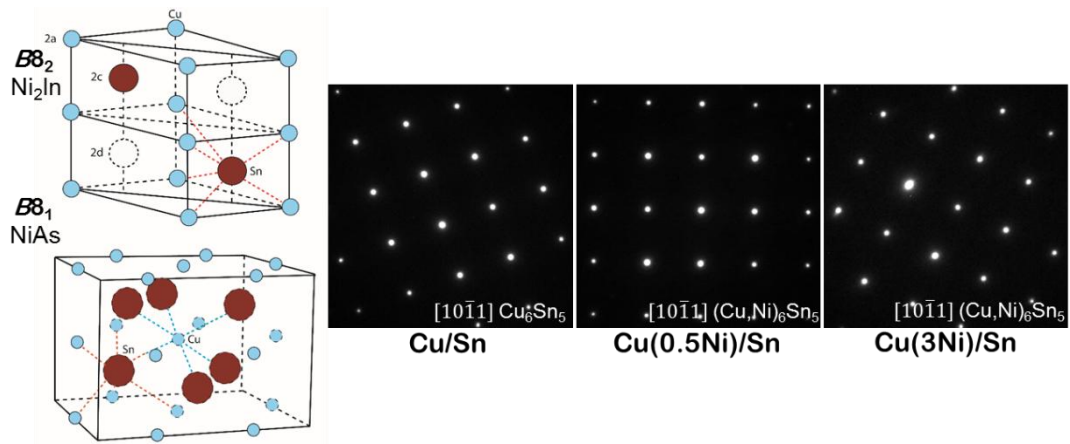

Figure 6.4: Crystal structure of the $Cu_6Sn_5$ phase along with the diffraction pattern acquired from the phase grown in the interdiffusion zone of Cu/Sn, Cu(0.5Ni)/Sn and Cu(3Ni)/Sn diffusion couples annealed at 200 °C.

Interestingly, few studies [91, 92] in the binary Cu–Sn system have found both these phases with similar composition to have different crystal structure than the well–known structure of the $Cu_3Sn$ and $Cu_6Sn_5$ phases present in the equilibrium phase diagram. However, the samples in these studies were prepared by melting route, which may have produced non–equilibrium crystal structure. On the other hand, in general, bulk diffusion couples are known to produce thermodynamically equilibrium phases (both composition and structure) [12].

## 6.4 Effect of Ni addition in Cu on the Kirkendall voids

Next, we focus our attention to the Kirkendall voids visible as dark spots in the $Cu_3Sn$ phase, as shown in Figure 6.1. The growth of the Kirkendall voids in the $Cu_3Sn$ phase because of supersaturation of vacancies is a major concern in an electronics industry, which significantly degrades the electro–mechanical bonding of the contacts. Since these voids are found in the $Cu_3Sn$ phase in which Cu has almost 30 times higher diffusion rate than Sn (*i.e.*, $[D_{Cu}^*/D_{Sn}^*]$ as estimated at the end of Section 5.2), it is evident that the vacancies created because of the difference in flux of diffusing components (in opposite directions) are not completely consumed by sinks [82]. Previously, in Chapter 5, we have shown that the growth rate of the voids are strongly dependent on the concentration of impurities present in Cu. In this segment of the study, we compare the growth of these voids because of the addition of 0.5 at.% Ni in 99.999 wt.% Cu.

For comparison, first 99.999 wt.% Cu is coupled with Sn for 81 hrs at 200 °C, as shown in Figure 6.5a. Only the $Cu_3Sn$ phase is shown in which these Kirkendall voids are found to grow. A close examination of the interdiffusion zone of the whole diffusion couple indicates that there are relatively void free regions at many places and a few voids are concentrated at few places. It is possible that the impurities are not distributed uniformly, which act as the site for heterogeneous nucleation of voids because of supersaturated vacancies.

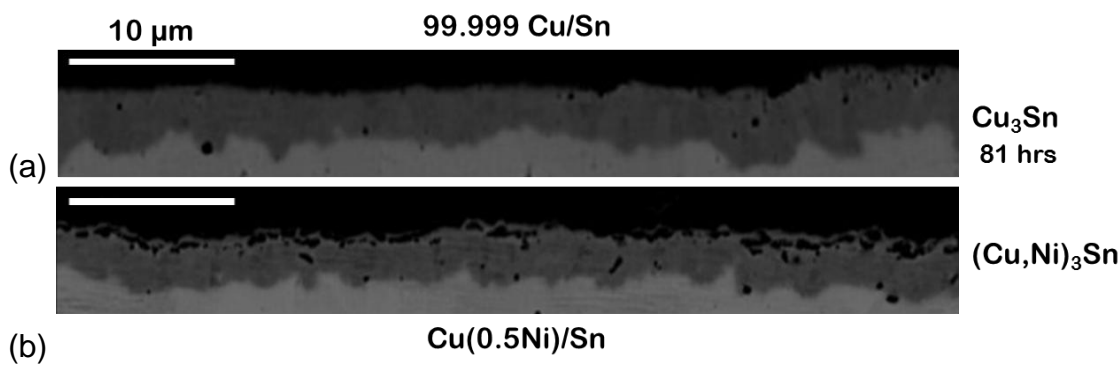

Figure 6.5: BSE micrographs for the comparison of growth of the Kirkendall voids in diffusion couples of (a) Cu/Sn and (b) Cu(0.5Ni)/Sn, annealed at 200 °C for 81 hrs. Purity is the same for both cases, *i.e.*, 99.999 wt.% Cu and 99.99 wt.% Sn.

The same purity of Cu is used for the preparation of Cu(0.5Ni) alloy. As already mentioned earlier, the thickness of (Cu,Ni)$_3$Sn decreases slightly; however, a significant increase in the void concentration is found, as shown in Figure 6.5b, indicating an adverse effect of Ni addition on the growth rate of the Kirkendall voids in this phase. It is generally believed that S and other (organic and inorganic) impurities, which are added during electroplating of Cu, play an important role for the significant increase in growth rate of the Kirkendall voids [58]. It is also speculated that voids created in such a situation are not grown because of the Kirkendall effect but they grow because of the presence of organic impurities (or its complex) in the electroplated Cu [59]. In this study, we find that the addition of 0.5 at.% Ni in bulk Cu, which is equivalent to the concentration of impurities incorporated in an electroplated Cu (as discussed earlier in Chapter 5), produces the voids with comparable rates. Therefore, the growth of these voids must be influenced by the presence of other inorganic impurities also along with the organic impurities. This further indicates that the different impurities and alloying addition can play an adverse role on the growth of the Kirkendall voids.

## 6.5    Microstructural Evolution

Since there is no change in the crystal structure, as discussed in Section 6.3, our next aim is to examine the effect of Ni addition on the microstructure, *i.e.*, grain morphologies, which may strongly influence the diffusion rates of components. During interdiffusion, components can diffuse via both lattice and grain boundaries. We cannot experimentally measure the change of concentration of lattice defects (vacancies and antisites) due to Ni addition. Added complexity comes from the fact that the concentrations of these defects are very small and different sublattices can have different concentration of defects, which may get affected differently because of the alloying addition of a third element. However, we can study the change in grain size (and hence grain boundary area) which is important for grain boundary diffusion. Additionally, a preliminary study conducted much earlier by one of this thesis advisors (Paul) [6] indicates the refinement of grains in $Cu_6Sn_5$ with the addition of Ni content, which might play an important role. Therefore, we concentrate our microstructural analysis only in this phase. We first analyse the microstructural evolution in the Cu/Sn and Cu(0.5Ni)/Sn diffusion couples in which the $(Cu,Ni)_6Sn_5$ phase grows along with the $(Cu,Ni)_3Sn$ phase.

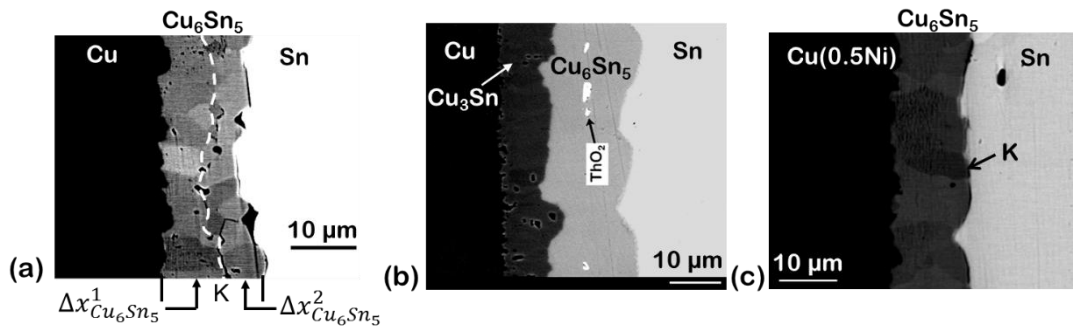

Figure 6.6: Micrographs showing the grains of the $Cu_6Sn_5$ phase across the interdiffusion zone in the diffusion couple of (a) Cu/Sn annealed at 200 °C for 81 hrs, where K denotes the Kirkendall marker plane indicated by white dotted line (b) Cu/Sn annealed at 215 °C for 225 hrs, where K is detected by inert $ThO_2$ particles [6]. (c) BSE image of Cu(0.5Ni)/Sn diffusion couple annealed at 200 °C for 81 hrs.

Based on the physico–chemical approach [47], it is already established that the microstructural evolution indicates the location of the Kirkendall marker plane efficiently. This is evident from the presence of duplex morphology in the $Cu_6Sn_5$ phase for Cu/Sn diffusion couple shown in Figure 6.6a. This is the same couple, as shown in Figure 6.1a. The contrast–brightness in the interdiffusion zone is adjusted to reveal the microstructure in the $Cu_6Sn_5$ phase and therefore the $Cu_3Sn$ phase is not visible. The dotted line shows the location of the Kirkendall marker plane, which indicates the growth of the phase differently from two different interfaces, viz. $Cu_3Sn/Cu_6Sn_5$ and $Cu_6Sn_5/Sn$. A similar location of this plane in Cu/Sn couple was detected by the use of inert markers ($ThO_2$ particles) previously by one of this thesis advisors (Paul), as shown in Figure 6.6b [6].

Similarly, as shown in Figure 6.6c, the contrast–brightness of the Cu(0.5Ni)/Sn diffusion couple is adjusted to reveal the microstructure of $(Cu,Ni)_6Sn_5$. The presence of the other phase $(Cu,Ni)_3Sn$ can be seen in Figure 6.1b, which is not visible in this figure. The location of the Kirkendall marker plane (indicated by K in Figure 6.6c) detected by the presence of $TiO_2$ particles was found very close to the $Cu_6Sn_5/Sn$ interface. Therefore, long grains covering almost the whole $Cu_6Sn_5$ phase layer is found in a micrograph captured by SEM. A further inspection revealed the presence of a line of pores at few places very close to the $(Cu,Ni)_6Sn_5/Sn$ interface, which also indicates the location of the marker plane [18]. Therefore, there is an ambiguity whether (i) the marker plane is located at the interface, or (ii) it is inside the product phase and very close to the interface. Since the microstructural evolution indicates the location of this plane efficiently (as discussed in Chapter 5) and because of the lack of resolution in SEM image near this interface, the sample is further analysed in TEM.

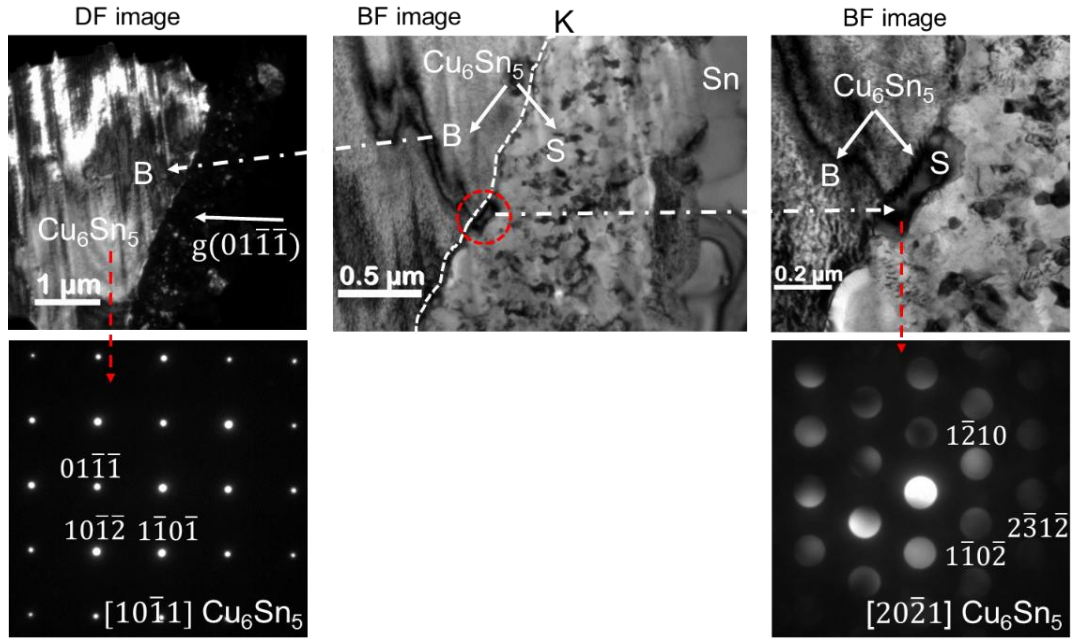

Figure 6.7: Grains of the $Cu_6Sn_5$ phase across the interdiffusion zone in Cu(0.5Ni)/Sn diffusion couple annealed at 200 °C for 81 hrs. TEM analysis with DP and respective DF – BF image, where K is found very close to the $(Cu,Ni)_6Sn_5$/Sn interface.

Figure 6.7 shows the respective BF or DF TEM micrographs along with their indexed DP (recognizing the phase). The bigger grain similar to that visible in the BSE micrograph (Figure 6.6c) is denoted by "B". The central BF image is of the region very close to the Sn/(Cu,Ni)$_6Sn_5$ interface. The DF image is acquired using $(01\bar{1}\bar{1})$ reflection and DP acquired from "B" grain is indexed with zone axis $[10\bar{1}1]$ of the $(Cu,Ni)_6Sn_5$ phase. On the right side of this grain, smaller grains of the $(Cu,Ni)_6Sn_5$ phase, as denoted by "S" are present. These grains (which might be oriented along the same zone axis $[10\bar{1}1]$ of the $(Cu,Ni)_6Sn_5$ phase) can be revealed after close examination of the DF image at the same magnification. To gain further insights, NBDP (nano–beam electron diffraction pattern) were acquired from the grains in this region and one such example is shown in Figure 6.7, which is indexed with zone axis $[20\bar{2}1]$ of the $(Cu,Ni)_6Sn_5$ phase. This DP is acquired from one of the grains adjacent to "B" grain (as indicated by an arrow in figure). Therefore, the Kirkendall marker plane (as denoted by K) demarcates two sublayers with big and small grains. This could not be detected by the conventional method because of the relatively bigger size of the inert $TiO_2$ particles used ($\sim 1$ μm) and the lack of resolution in SEM.

This means that in a binary Cu–Sn system, the Kirkendall marker plane (K) is found in the middle of the $Cu_6Sn_5$ phase (Figure 6.6a and b), while K is found near the $(Cu,Ni)_6Sn_5/Sn$ interface in the Cu(0.5Ni)/Sn diffusion couple (Figure 6.6c and d). Following the physico–chemical approach [47], we know that the sublayer $\left(\Delta x^1_{Cu_6Sn_5}\right)$ between $Cu_3Sn/Cu_6Sn_5$ interface and K grows because of diffusion of Sn towards $Cu_3Sn$ and the sublayer $\left(\Delta x^2_{Cu_6Sn_5}\right)$ between K and $Cu_6Sn_5/Sn$ interface grows because of diffusion of Cu (and Ni when it is added) towards Sn. Since both the sublayers have comparable thickness in a binary system (*i.e.*, in the absence of Ni), it is apparent that the diffusion of both the components has contributed to the growth of this phase layer. However, in the Cu(0.5Ni)/Sn diffusion couple (*i.e.*, when Ni is added), $\Delta x^1_{(Cu,Ni)_6Sn_5} \gg \Delta x^2_{(Cu,Ni)_6Sn_5}$, which indicates that the relative diffusion rate of Sn is much higher compared to Cu and Ni. Therefore, the $(Cu,Ni)_6Sn_5$ phase grows mainly because of diffusion of Sn.

Next, we consider the diffusion couples of Cu(3Ni)/Sn, Cu(5Ni)/Sn and Cu(8Ni)/Sn in which $(Cu,Ni)_3Sn$ does not grow in the interdiffusion zone, as shown in Figure 6.1. Grain morphologies of these couples are shown in Figure 6.8. Because of very high thickness and small grains, only the interdiffusion zone is shown for (Cu8Ni)/Sn diffusion couple. The Kirkendall marker plane again, like Cu(0.5Ni)/Sn diffusion couple, is found very near to the $(Cu,Ni)_6Sn_5/Sn$ interface in all these diffusion couples indicating the growth of the product phase mainly because of diffusion of Sn. Since the thickness of the product phase increases rapidly, it is apparent that the diffusion rate of Sn increases significantly with the increase in Ni content. Additionally, the following important fact should be noted here: In these 3 couples, the product phase consists of many small grains instead of elongated grains from one interface to the Kirkendall marker plane. In general, in most of the systems, grains of the intermetallic compounds do not nucleate repeatedly, possibly because of the high activation energy barrier [18]. This is found when $(Cu,Ni)_6Sn_5$ grows in the presence of $(Cu,Ni)_3Sn$ in Cu/Sn and Cu(0.5Ni)/Sn diffusion couples (*see* Figure 6.6). However, when $(Cu,Ni)_3Sn$ is missing in the interdiffusion zone and, therefore, $(Cu,Ni)_6Sn_5$ grows directly over Cu(Ni), it is able to nucleate repeatedly. Additionally, the average grain size decreases with the increase of Ni content in the Cu(Ni) alloy. A similar repeated nucleation of grains and therefore the growth of a sublayer with many small grains is found even in $Cu_3Sn$ which grows from Cu in Cu/Sn diffusion couple, as shown in Figure 6.9 and earlier in Figure 5.3. A dashed arrow in Figure 6.9 indicates the same hole (very close to the $Cu/Cu_3Sn$ interface) around which grains of Cu and $Cu_3Sn$ are shown in both DF images. The identification of the phases is evaluated based on analysis of DP shown in Figure 6.9.

On the other hand, as shown in Figure 5.2, grains of the $Cu_3Sn$ phase which are grown from $Cu_6Sn_5$ at the $Cu_3Sn/Cu_6Sn_5$ interface are elongated covering the whole $Cu_3Sn$ sublayer between this interface and the Kirkendall marker plane in the $Cu_3Sn$ phase. Therefore, considering the microstructure in different diffusion couples, we can conclude that both the phases nucleate repeatedly and have smaller grains when they grow from Cu. On the other hand, the same phases have elongated grains covering the whole sublayer (*i.e.*, between the interface and the Kirkendall plane) when these are grown from another intermetallic compound. Therefore, the nucleation barrier for the growth or the interfacial energy at the $Cu/Cu_3Sn$ or $Cu/Cu_6Sn_5$ must be low. Since the average sizes of grains in the $Cu_6Sn_5$ phase decrease, it indicates that there must be a decrease in the nucleation barrier with the increase in Ni content.

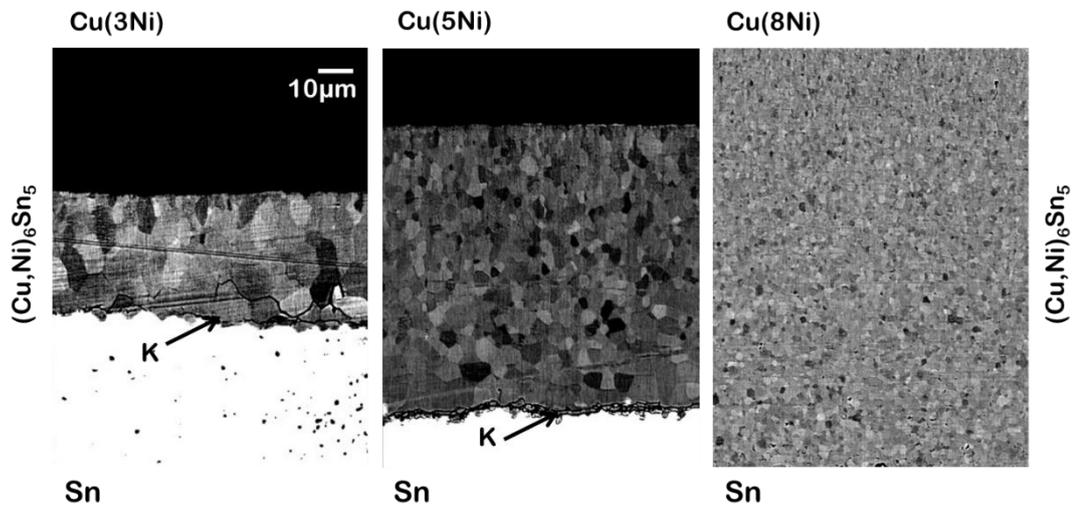

Figure 6.8: BSE micrographs showing the grains of the $(Cu,Ni)_6Sn_5$ phase across the interdiffusion zone in the diffusion couple of (a) Cu(3Ni)/Sn (b) Cu(5Ni)/Sn and (c) Cu(8Ni)/Sn, annealed at 200 °C for 81 hrs. K denotes the location of the Kirkendall marker plane.

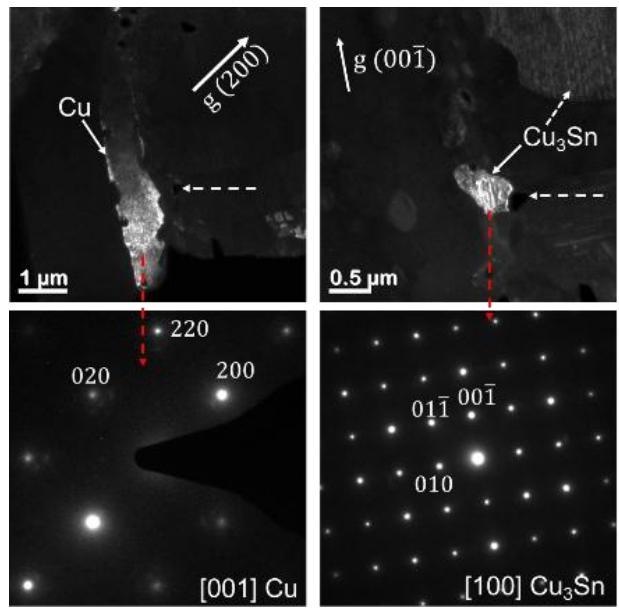

Figure 6.9: TEM micrographs showing Cu (left) and $Cu_3Sn$ (right) grains at the Cu/$Cu_3Sn$ interface in binary Cu/Sn diffusion couple annealed at 200 °C.

## 6.6   Parabolic Growth

Before estimating the diffusion parameters, it is necessary to confirm the parabolic growth of the phases by examining the time dependent growth of the phases. The parabolic growth of the phases in the temperature range of our interest is already confirmed in the Cu–Sn system, as reported in Chapter 4. In the Cu(Ni)–Sn system, time dependent growth at 200 °C in Cu(2.5Ni)/Sn and Cu(5Ni)/Sn diffusion couples was conducted as shown in Figure 6.10. Thickness ($\Delta x$) and time ($t$) with respect to $(\Delta x)^2$ vs. $2t$ are plotted only for the (Cu,Ni)$_6$Sn$_5$ phase since the thickness of the (Cu,Ni)$_3$Sn phase was very small. Linear fit of the data indicates the parabolic growth nature, *i.e.*, diffusion–controlled growth of the (Cu,Ni)$_6$Sn$_5$ phase. The slope of the graph is the parabolic growth constant, $k_p = (\Delta x)^2/2t$. However, the parabolic growth constant is not a material constant, as it depends on end–member compositions [18]. Therefore, the discussion on the diffusion–controlled growth mechanism of the phases should be done based on the estimation of the diffusion parameters, which are material constants.

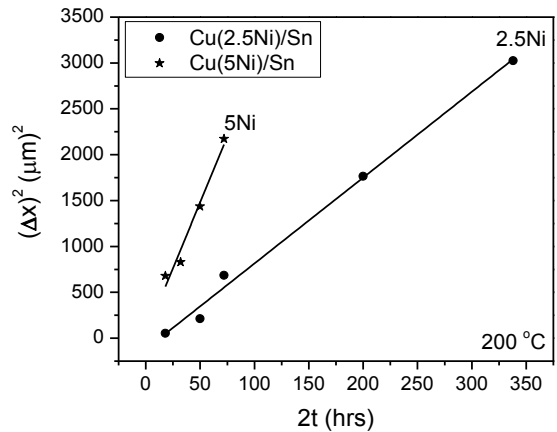

Figure 6.10: Parabolic growth law followed by the (Cu,Ni)$_6$Sn$_5$ phase in Cu(2.5Ni)/Sn and Cu(5Ni)/Sn diffusion couples at 200 °C.

## 6.7 Effect of Ni addition in Cu on the diffusion parameters

As the parabolic growth of the phases is confirmed in Section 6.6, we can now straightforwardly estimate the diffusion coefficients. The integrated interdiffusion diffusion coefficients are estimated in $\beta$ phase with narrow homogeneity range ($\widetilde{D}_{int}^{\beta}$), since the actual concentration gradient cannot be determined from the measured composition profile. This is defined as the interdiffusion coefficient ($\widetilde{D}$) integrated over the unknown narrow composition range $\Delta N_i^\beta = \left( N_i^{\beta_2} - N_i^{\beta_1} \right)$ of the product phase $\beta$ [26], as follows:

$$\widetilde{D}_{int}^{\beta,i} = \int_{N_i^{\beta_1}}^{N_i^{\beta_2}} \widetilde{D}\, dN_i = \widetilde{D}\Delta N_i^\beta \tag{6.1}$$

We can assume that there is not significant variation of the interdiffusion coefficient in a small composition range of the phase. $\widetilde{D}_{int}^{\beta,i}$ can be estimated directly from the composition profile using the following equation [18]:

$$\widetilde{D}_{int}^{\beta,i} = \frac{\left( N_i^\beta - N_i^- \right)\left( N_i^+ - N_i^\beta \right)}{\left( N_i^+ - N_i^- \right)} \frac{\left( \Delta x^\beta \right)^2}{2t}$$

$$+ \frac{\Delta x^\beta}{2t} \left[ \frac{N_i^+ - N_i^\beta}{N_i^+ - N_i^-} \int_{x^{-\infty}}^{x^{\beta_1}} \frac{V_m^\beta}{V_m} (N_i - N_i^-)\, dx + \frac{N_i^\beta - N_i^-}{N_i^+ - N_i^-} \int_{x^{\beta_2}}^{x^{+\infty}} \frac{V_m^\beta}{V_m} (N_i^+ - N_i)\, dx \right] \tag{6.2}$$

This is expressed with respect to the composition profile of component $i$. $N_i^-$ and $N_i^+$ are the mole fractions of un–affected left– and right–hand side of the end–members, $N_i^\beta$ is the stoichiometric composition of the phase of interest, $V_m$ is the molar volume, $\Delta x$ is the thickness of the phase and $t$ is the isothermal annealing time. Since we did not find any significant change in the lattice parameters of the product phases as well as Cu(Ni) alloy because of Ni addition (Section 6.3), we have considered the molar volumes of the phases equal to that estimated in a binary Cu–Sn system, irrespective of the Ni content, *i.e.*, $V_m^{Cu_3Sn} = 8.59 \times 10^{-6}$ and $V_m^{Cu_6Sn_5} = 10.59 \times 10^{-6}$ m³/mol [17]. Note here that in a binary system, $\widetilde{D}_{int}^{\beta,i}$ will be the same when it is estimated using the composition profiles of any of the components, viz. Cu or Sn. On the other hand, in a ternary system, 3 different components will have their own $\widetilde{D}_{int}^{\beta,i}$ in a particular phase. However, in the Cu(Ni)–Sn system, Ni occupies the same sublattice as Cu in the product phases. As shown in Figure 6.11, it can be easily seen from the composition profile (measured by EPMA) of Cu(8Ni)/Sn diffusion couple that the product phase $(Cu,Ni)_6Sn_5$ maintains the stoichiometry of $(Cu+Ni){:}Sn \equiv 6{:}5$. Therefore, we can compare the diffusion coefficients based on the data calculated using Sn composition profile, which does not change because of the addition of Ni. The steps of the estimation procedure of the integrated interdiffusion coefficient in a multiphase growth can be learnt from Reference [18]. Using Equation (6.2), integrated interdiffusion diffusion with respect to Sn are calculated in the $(Cu,Ni)_6Sn_5$ phase since it grows with enough thickness over the temperature range considered in this study. Calculated values are shown in Figure 6.12 with respect to Arrhenius plot:

$$\widetilde{D}_{int}^{Cu_6Sn_5,Sn} = \widetilde{D}_{int,o}^{Cu_6Sn_5,Sn} \times exp\left[\left(-\frac{Q}{R}\right)\frac{1}{T}\right] \tag{6.3}$$

where $\widetilde{D}_{int,o}^{Cu_6Sn_5,Sn}$ (m$^2$/s) is the pre–exponential factor, $Q$ (J/mol) is the activation energy, $R$ (8.314 J/mol.K) is the gas constant and $T$ is the temperature in K.

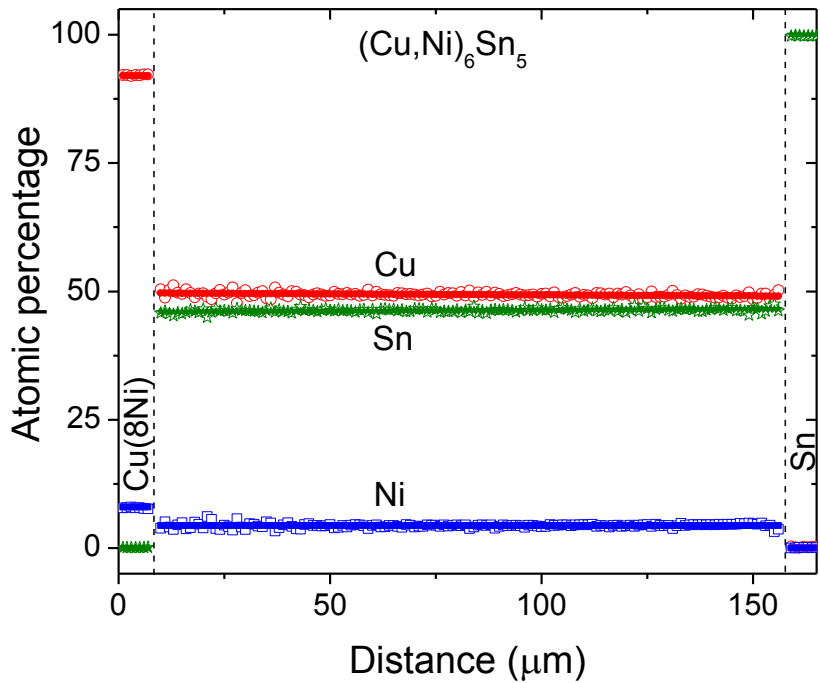

Figure 6.11: Composition profile developed across the interdiffusion zone of Cu(8Ni)/Sn diffusion couple annealed at 200 °C for 81 hrs. Sn profile is used for the analysis of important diffusion parameters.

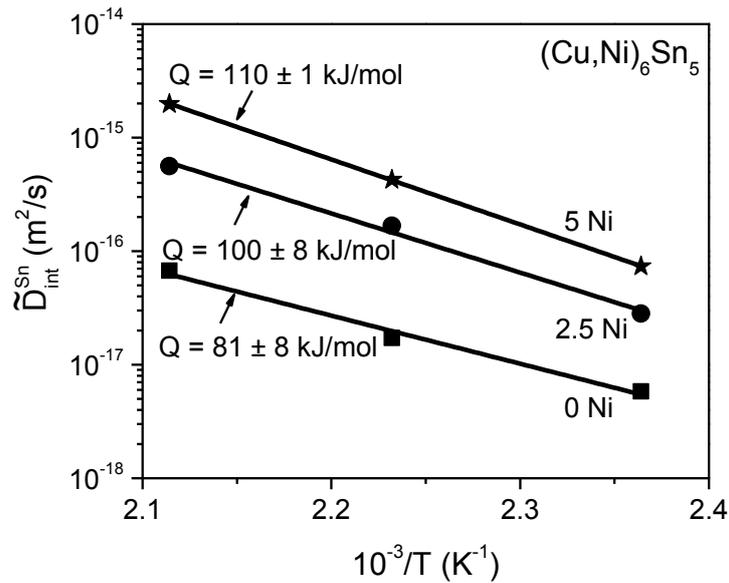

Figure 6.12: Arrhenius plots of the integrated interdiffusion coefficients of Sn in the $(Cu,Ni)_6Sn_5$ phase along with activation energies estimated from temperature dependent experiments. For comparison, the data of the Cu–Sn system is also shown.

It can be seen that the integrated interdiffusion coefficient increases with the Ni content explaining the higher growth rate of the $(Cu,Ni)_6Sn_5$ phase. Additionally, there is an increase in the activation energy (*see* Section 6.9 for more discussion on it). To visualize it further, the same data are plotted with respect to Ni content, as shown in Figure 6.13. Four additional sets of diffusion coefficients (calculated using Equation (6.2)) in Cu(0.5Ni)/Sn, Cu(3Ni)/Sn, Cu(7.5Ni)/Sn and Cu(8Ni)/Sn diffusion couples at 200 °C are also shown. It can be seen that at all the temperatures, diffusion coefficient increases with the increase in Ni content. Data used to plot Figure 6.13 are also listed in Table 6.1, which are used later for comparing results with the $Cu_3Sn$ phase.

It was already mentioned at the end of Section 6.6 that the measurement of thickness of a phase layer does not give information about the actual growth mechanism of a phase. Therefore, to understand the reason for a decrease in thickness of $(Cu,Ni)_3Sn$ when compared with the thickness of $(Cu,Ni)_6Sn_5$, it is necessary to estimate the appropriate diffusion data. It should be noted that diffusion data is calculated for the $(Cu,Ni)_3Sn$ phase only at 200 °C because at lower temperatures, the thickness of this phase was too small, and hence difficult to estimate data. Even at this temperature, thickness is very small; however, it is still worth calculating in order to get an idea about the growth mechanism of this phase. The diffusion data, calculated using Equation (6.2), are plotted with respect to Ni content in Figure 6.14.

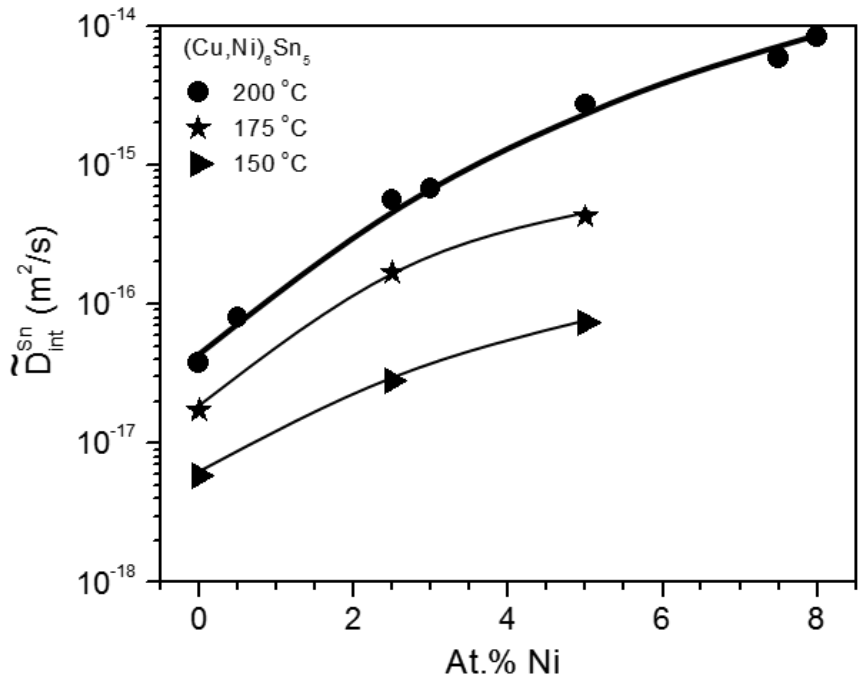

Figure 6.13: Variation (at different temperatures) of integrated interdiffusion coefficients of Sn in the $(Cu,Ni)_6Sn_5$ phase as a function of Ni content.

| Diffusion Couple | $Cu_6Sn_5$ | | | $Cu_3Sn$ |
|---|---|---|---|---|
| | 150 °C | 175 °C | 200 °C | 200 °C |
| Cu/Sn [17] | $5.8 \times 10^{-18}$ | $1.7 \times 10^{-17}$ | $6.7 \times 10^{-17}$ | $1.81 \times 10^{-17}$ |
| Cu/Sn | | | $3.8 \times 10^{-17}$ | $9.45 \times 10^{-18}$ |
| Cu(0.5Ni)/Sn | --- | --- | $7.9 \times 10^{-17}$ | $1.14 \times 10^{-17}$ |
| Cu(2.5Ni)/Sn | $2.8 \times 10^{-17}$ | $1.6 \times 10^{-16}$ | $5.6 \times 10^{-16}$ | $1.96 \times 10^{-17}$ |
| Cu(3.0Ni)/Sn | --- | --- | $6.8 \times 10^{-16}$ | --- |
| Cu(5.0Ni)/Sn | $7.3 \times 10^{-17}$ | $4.2 \times 10^{-16}$ | $2.7 \times 10^{-15}$ | --- |
| Cu(7.5Ni)/Sn | --- | --- | $5.8 \times 10^{-15}$ | --- |
| Cu(8.0Ni)/Sn | --- | --- | $8.3 \times 10^{-15}$ | --- |

Table 6.1: Integrated interdiffusion coefficient of Sn ($m^2/s$) in the $Cu_6Sn_5$ phase at different temperatures and the same for Sn in the $Cu_3Sn$ phase at 200 °C.

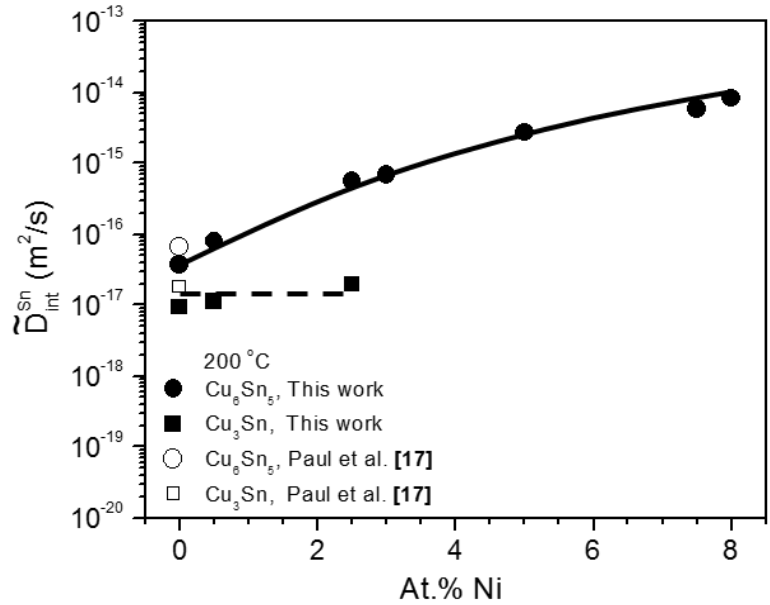

Figure 6.14: Variation (at 200 °C) of integrated interdiffusion coefficients of Sn in both Cu3Sn and Cu6Sn5 is shown for comparison, as a function of Ni content.

For the purpose of comparison, data (shown before in Figure 6.13) for (Cu,Ni)6Sn5 at 200 °C are again plotted in Figure 6.14. A good match of the binary Cu/Sn couple data estimated in this study with a previous study on the binary Cu–Sn system [17] is evident. It can be seen that although the thickness of (Cu,Ni)3Sn decreases compared to (Cu,Ni)6Sn5 (Figure 6.1), the diffusion coefficient remains more or less constant up to 2.5 at.% Ni. On the contrary, we can see in Figure 6.14 that the diffusion coefficient of the (Cu,Ni)6Sn5 phase increases. This is the reason for the observed increase in the thickness of (Cu,Ni)6Sn5 when compared to (Cu,Ni)3Sn. This can be further understood based on Equation (6.2). The relation for the integrated interdiffusion coefficient has two parts. The first part considers the layer thickness of the phase of interest (Cu3Sn), whereas the other part inside the square bracket considers the layer thickness of the neighbouring phases present in an interdiffusion zone before and after the phase of interest. Therefore, when the diffusion coefficient for a particular phase does not change while the thickness of the neighbouring phases increases, then the thickness of the phase of interest should decrease. In other words, although the diffusion coefficient of Sn for $Cu_3Sn$ does not change up to 2.5 at.% Ni, but the thickness of the neighbouring phase ($Cu_6Sn_5$) is found to increase with Ni addition. Hence, we expect the layer thickness of $Cu_3Sn$ to decrease which is also observed in this segment of the study such that the $Cu_3Sn$ phase is not present at 3 at.% Ni and beyond. This can also be used to understand the growth mechanism of the phases by reaction–diffusion process, *i.e.*, physico–chemical approach [47], as discussed earlier qualitatively in Section 5.2. At the $(Cu,Ni)_3Sn/(Cu,Ni)_6Sn_5$ interface, both the phases grow by consuming the other phase. Therefore, even if the integrated interdiffusion coefficient does not change for a particular phase of interest and it increases in the neighbouring phase, then the thickness should decrease for the phase of interest. Higher the growth rate of a phase, here $(Cu,Ni)_6Sn_5$, means higher the consumption rate of the other phase, here $(Cu,Ni)_3Sn$. This explains the decrease in thickness of $(Cu,Ni)_3Sn$ as compared to $(Cu,Ni)_6Sn_5$ (Figure 6.1), even though the diffusion coefficient for $(Cu,Ni)_3Sn$ does not change up to 2.5 at.% Ni.

The above discussion can be further summarized by comparing the diffusion data listed in Table 6.1 for $(Cu,Ni)_3Sn$ and $(Cu,Ni)_6Sn_5$. At 200 °C, there is not much difference between the integrated interdiffusion coefficients for $Cu_3Sn$ and $Cu_6Sn_5$ leading to comparable thickness in the binary Cu–Sn system. When 0.5 at.% Ni is added in Cu, there is not much difference in the data of integrated interdiffusion coefficient for $(Cu,Ni)_3Sn$ in comparison to that in pure Cu. However, the integrated interdiffusion coefficient increases markedly for $(Cu,Ni)_6Sn_5$, thus leading to a higher difference in the layer thickness between the two product phases. Due to the same reason, the difference is further increased when more Ni (2.5 at.% Ni) is added in Cu.

Few important points to be noted are repeated for summarizing here. The integrated interdiffusion coefficient of $(Cu,Ni)_3Sn$ remains more or less the same with the increase in Ni content up to 2.5 at.%. However, the thickness of this phase decreases because of significant increase in the growth rate and integrated interdiffusion coefficient of the $(Cu,Ni)_6Sn_5$ phase. Following the physico–chemical approach [47], once both the phases are present in the interdiffusion zone, they try to grow by consuming the other phase at the $(Cu,Ni)_3Sn/(Cu,Ni)_6Sn_5$ interface. Although the diffusion rate of Sn does not change in the $(Cu,Ni)_3Sn$ phase, this phase is consumed at a higher rate because of the increase in diffusion rate of Sn in the $(Cu,Ni)_6Sn_5$ phase. Therefore, we concentrate our analysis on $(Cu,Ni)_6Sn_5$ for the increase in diffusion rates of components because of Ni addition.

Ideally, in a ternary system, intrinsic diffusion coefficients can be estimated if two diffusion couples intersect at the composition of the Kirkendall marker plane on Gibb's triangle [18]. It is extremely difficult to design experiments fulfilling this even in a phase with wide homogeneity range. A very narrow homogeneity range of the phases in this system makes it impossible to achieve such a situation. However, we can still estimate the intrinsic diffusion coefficient after simplifying the equations particularly because of the certain diffusion characteristics in this system, as explained next.

For comparison of the data, let us first start with the binary system in which we can estimate the intrinsic diffusion coefficients of the components rather straightforwardly. Interdiffusion flux ($\tilde{J}_i$) is related to the intrinsic fluxes ($J_i$) in a *n* component system by [18]:

$$\tilde{J}_i = J_i - N_i \sum_{k=1}^{n} J_k \qquad (6.4)$$

When interdiffusion flux is estimated using a composition profile of a particular component, let say component $B$, in a binary $A$–$B$ system such that $(N_A + N_B) = 1$, it leads to

$$\tilde{J}_B = N_A J_B - N_B J_A \qquad (6.5)$$

From Fick's first law, we can write

$$-\tilde{D}\frac{dC_B}{dx} = -N_A D_B \frac{dC_B}{dx} + N_B D_A \frac{dC_A}{dx} \qquad (6.6)$$

We consider a constant molar volume $V_m^\beta$ (since the lattice parameter variation with composition is not known in a phase with narrow homogeneity range) estimated at the stoichiometric composition, such that partial molar volume $\bar{V}_i^\beta = V_m^\beta$. Further, we assume a linear variation of the composition gradient inside the phase (as assumed by Wagner [26] since it cannot be measured in a phase with narrow homogeneity range) such that $\frac{dC_B}{dx} = \frac{1}{V_m^\beta}\frac{dN_B^\beta}{dx} = \frac{1}{V_m^\beta}\frac{\Delta N_B^\beta}{\Delta x^\beta}$. Since $dN_A + dN_B = 0$ in a binary system, we can rewrite the Equation (6.6) as:

$$-\tilde{D}\Delta N_B^\beta = -N_A D_B \Delta N_B^\beta - N_B D_A \Delta N_B^\beta \qquad (6.7)$$

Replacing right–hand side of Equation (6.1) by Equation (6.7), we get

$$\tilde{D}_{int}^{\beta,B} = (N_A D_B + N_B D_A)\Delta N_B^\beta \qquad (6.8a)$$

Therefore, the integrated interdiffusion coefficient estimated in the Cu₆Sn₅ phase with respect to the composition profile of Sn can be related as:

$$\tilde{D}_{int}^{\beta,Sn} = \left(N_{Cu}^{Cu_6Sn_5} D_{Sn} + N_{Sn}^{Cu_6Sn_5} D_{Cu}\right)\Delta N_{Sn}^\beta \qquad (6.8b)$$

Since the integrated interdiffusion coefficients are a kind of average of the intrinsic diffusion coefficients over a composition range of the phase, it does not indicate the diffusion coefficients of individual components and therefore the atomic mechanism of diffusion [12]. We can gain further insights into the same, only if we can estimate the intrinsic diffusion coefficients. Following the van Loo's method [29], the ratio of the intrinsic fluxes or the intrinsic diffusion coefficients at the Kirkendall marker plane ($x_K$) considering a constant molar volume of a particular phase can be directly estimated as [18]:

$$\frac{J_B}{J_A} = \frac{D_B}{D_A} = \left[ \frac{N_B^+ \int_{x^{-\infty}}^{x_K} \frac{Y_B}{V_m} dx - N_B^- \int_{x_K}^{x^{+\infty}} \frac{(1-Y_B)}{V_m} dx}{-N_A^+ \int_{x^{-\infty}}^{x_K} \frac{Y_B}{V_m} dx + N_A^- \int_{x_K}^{x^{+\infty}} \frac{(1-Y_B)}{V_m} dx} \right] \tag{6.9a}$$

where $Y_B = \frac{N_B^\beta - N_B^-}{N_B^+ - N_B^-}$. We have neglected the role of the vacancy–wind effect [24], which cannot be determined because of the unknown structure factor [18]. The intrinsic fluxes and intrinsic diffusion coefficients of components are related individually by [18]:

$$J_A = -D_A \left( \frac{dC_A}{dx} \right)_K = -\frac{1}{2t} \left[ N_A^+ \int_{x^{-\infty}}^{x_K} \frac{Y_B}{V_m} dx - N_A^- \int_{x_K}^{x^{+\infty}} \frac{(1-Y_B)}{V_m} dx \right] \tag{6.9b}$$

$$J_B = -D_B \left( \frac{dC_B}{dx} \right)_K = -\frac{1}{2t} \left[ N_B^+ \int_{x^{-\infty}}^{x_K} \frac{Y_B}{V_m} dx - N_B^- \int_{x_K}^{x^{+\infty}} \frac{(1-Y_B)}{V_m} dx \right] \tag{6.9c}$$

where concentration $C_i = \frac{N_i}{V_m}$. In the binary Cu–Sn system, the ratio of the intrinsic diffusion coefficients at the Kirkendall marker plane inside the Cu$_6$Sn$_5$ phase (as shown by dotted line in Figure 6.6a) is estimated from Equation (6.9a) as $\frac{D_{Sn}}{D_{Cu}} = 2.3 \pm 1$. We cannot estimate the absolute values of the intrinsic diffusion coefficients directly because of unknown concentration gradient inside a phase with narrow homogeneity range. However, we can still estimate the average values of these parameters considering a linear composition gradient. From the experimentally measured composition (using EPMA) in this study and another work [42], we consider the composition range of this phase as $\Delta N_B^{Cu_6Sn_5} \approx 0.01$, *i.e.*, 1 at.%. Using these details (*i.e.*, composition range and the ratio of intrinsic diffusivities), the absolute values of the intrinsic diffusion coefficients in (Cu,Ni)$_6$Sn$_5$ from Equation (6.8b) are estimated as $D_{Sn} = 5.1 \times 10^{-15}$ and $D_{Cu} = 2.2 \times 10^{-15}$ m²/s in the binary system.

In a ternary system $(N_A + N_B + N_C = 1)$, the interdiffusion flux estimated with respect to the component $B$ can be related to the intrinsic fluxes of components following Equation (6.4) as [18]:

$$\tilde{J}_B = (1 - N_B)J_B - N_B(J_A + J_C) \tag{6.10a}$$

With respect to the composition profile and interdiffusion flux of Sn in the $(Cu,Ni)_6Sn_5$ phase, we can rewrite this equation as:

$$\tilde{J}_{Sn} = (1 - N_{Sn})J_{Sn} - N_{Sn}(J_{Cu} + J_{Ni}) \tag{6.10b}$$

This equation can be simplified because of the diffusion behaviour of components in this system. As already mentioned, when Ni is added, the Kirkendall marker plane moves very close to the $(Cu,Ni)_6Sn_5$/Sn interface indicating the growth of this phase mainly by the diffusion of Sn. Moreover, the Ni content in the product phase retains the ratio of atomic fractions of Cu and Ni used in the Cu(Ni) alloy. For example, the average compositions of Ni and Cu in the $(Cu,Ni)_6Sn_5$ phase which is grown in the interdiffusion zone of a Cu(8Ni)/Sn diffusion couple, as shown in Figure 6.11, are measured as 4.35 and 49.37 at.%, respectively. Therefore, we have $\frac{Ni}{Cu+Ni} \approx 8$ at.% in the product phase, which is similar to the Ni content in the Cu(8Ni) end–member. Therefore, it is evident that the product phase grows mainly by diffusion of Sn and other components, viz. Cu and Ni, are added because of the reaction of Cu(Ni) alloy with the diffusing component (*i.e.*, Sn) at the Cu(Ni)/(Cu,Ni)$_6$Sn$_5$ interface keeping the similar ratio as that of the alloy. Since the diffusion of Cu and Ni is negligible, in comparison of $J_{Sn}$, we can assume $(J_{Cu} + J_{Ni}) \approx 0$. Therefore, we can safely consider the cross diffusion coefficients, which account the flux of a component because of concentration gradient of another component [18], as zero. Considering these facts and simplifying Equation (6.10b) following similar steps as explained for the binary system, the integrated and intrinsic diffusion coefficients in the ternary Cu(Ni)/Sn diffusion couples can be related as:

$$\widetilde{D}_{int}^{Cu_6Sn_5,Sn} \approx (1 - N_{Sn}^{Cu_6Sn_5})D_{Sn}\Delta N_{Sn}^{Cu_6Sn_5} \qquad (6.11)$$

It should be noted here that although Equation (6.11) is written for obtaining approximated value, there will be negligible difference between this approximated value with the actual value (if we could estimate) simply because $J_{Sn} \gg (J_{Cu} + J_{Ni})$.

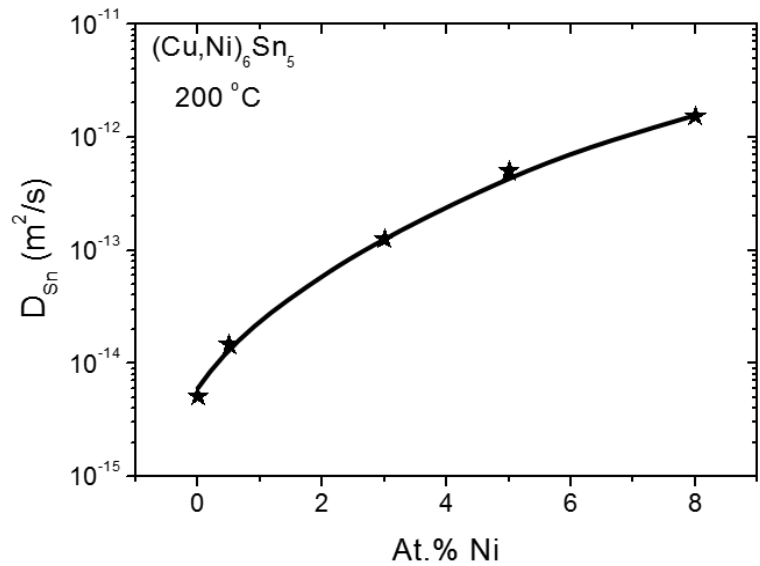

Figure 6.15: Variation (at 200 °C) of intrinsic flux of Sn
in the (Cu,Ni)$_6$Sn$_5$ phase as a function of Ni content.

The estimated $D_{Sn}$ for different Ni content in Cu(Ni) end–members are shown in Figure 6.15. Therefore, when we compare $D_{Sn}$ with the diffusion coefficients measured in a binary Cu–Sn system, it is apparent that the diffusion coefficient of Cu decreases drastically, whereas the diffusion coefficient of Sn increases significantly with the increase of Ni content in the Cu(Ni) alloy. The estimated intrinsic diffusion coefficients of Sn ($D_{Sn}$) and intrinsic flux of Sn ($J_{Sn}$) in the Cu$_6$Sn$_5$ phase at 200 °C along with the integrated interdiffusion data (for comparison at the same temperature) are listed in Table 6.2.

| Diffusion Couple | $\widetilde{D}_{int}^{(Cu,Ni)_6Sn_5,Sn}$ (m²/s) | $D_{Sn}$ (m²/s) in (Cu,Ni)₆Sn₅ | $J_{Sn}$ (mol/m²) |
|---|---|---|---|
| Cu/Sn | 3.8×10⁻¹⁷ | 5.1×10⁻¹⁵ | −0.17 |
| Cu(0.5Ni)/Sn | 7.9×10⁻¹⁷ | 1.4×10⁻¹⁴ | −0.30 |
| Cu(3.0Ni)/Sn | 6.8×10⁻¹⁶ | 1.2×10⁻¹³ | −0.84 |
| Cu(5.0Ni)/Sn | 2.7×10⁻¹⁵ | 5.0×10⁻¹³ | −1.70 |
| Cu(8.0Ni)/Sn | 8.3×10⁻¹⁵ | 1.5×10⁻¹² | −2.99 |

Table 6.2: Integrated diffusion coefficients estimated utilizing the composition profile of Sn, intrinsic diffusion coefficients and intrinsic flux of Sn in Cu₆Sn₅ at 200 °C.

Relatively low range of activation energies estimated in this study (shown in Figure 6.12) indicates that the grain boundary diffusion might be playing an important role in the growth of the phases. Moreover, the thermodynamic driving force for diffusion of different species should also be influenced by alloying. Therefore, in order to better understand the role of Ni content on growth of the product phases, calculation of thermodynamics parameters (Section 6.8) and analysis of the grain morphology (Section 6.5) were carried out.

## 6.8   Thermodynamic–kinetic analysis *

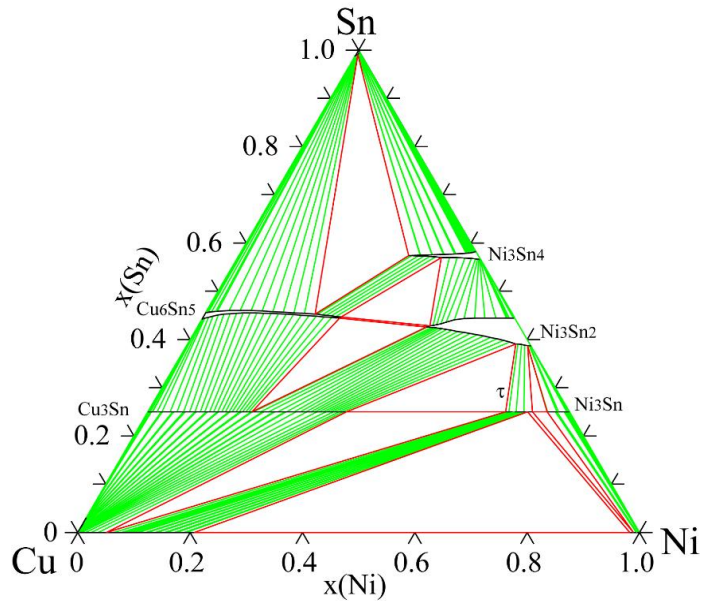

Figure 6.16: Calculated Cu–Ni–Sn phase diagram at 200 °C using the assessed data–set from [93] and the additional experimental data from this study.

Detailed description of the thermodynamic assessment of the Cu–Ni–Sn system can be found from Reference [93], so it will not be repeated here. In this, the stability region of $(Cu,Ni)_3Sn$ was modified based on the experimental data reported in this work. The isothermal section at 200 °C calculated by using the set of data is shown in Figure 6.16. It can be immediately noted that based on the ternary phase diagram, in the Cu(7.5Ni)/Sn diffusion couple, the $(Cu,Ni)_3Sn$ phase is not thermodynamically stable anymore and instead, the ternary phase should be next to the Cu(7.5Ni) alloy. However, as the formation of the ternary phase ($\tau$) is expected to require relatively high temperatures and long annealing times, its formation in the diffusion couples produced in this study is not expected. Nevertheless, owing to these uncertainties, the data from the couples having more than 7.5 at.% of Ni were disregarded when the growth mechanism of phases was analysed based on thermodynamic calculations. It is also to be noted that based on the angle of the tie–lines in the two–phase region Cu(Ni)–$(Cu,Ni)_3Sn$ (Figure 6.16), there must be a higher amount of Ni in the $(Cu,Ni)_3Sn$ phase than what is present originally in the Cu(Ni) alloy. Further, for the mass balance to be obeyed, the amount of Ni must decrease towards the $(Cu,Ni)_3Sn/(Cu,Ni)_6Sn_5$ interface. Evidence of this is not very clearly seen from the experimental composition profiles measured for Cu(2.5Ni)/Sn couple because of the very small thickness of the $(Cu,Ni)_3Sn$ phase. In fact, only three points could be measured inside the phase. However, the thermodynamically expected trend could still be seen. The effect is clearer in the Cu(0.5Ni)/Sn couple, as shown in Figure 6.17, since in this case, the phase layer thickness of $(Cu,Ni)_3Sn$ was significant and it was possible to measure composition variation inside the phase much more quantitatively. Figure 6.17a guides the phase boundaries based on Sn composition profile, as evident from the jumps across the phase boundaries. Figure 6.17b shows the Ni variation inside different phases. Thus, the experimentally observed variation in the Ni content inside (Cu,Ni)₃Sn discussed above is well explained by considering local equilibrium requirements together with mass–balance considerations.

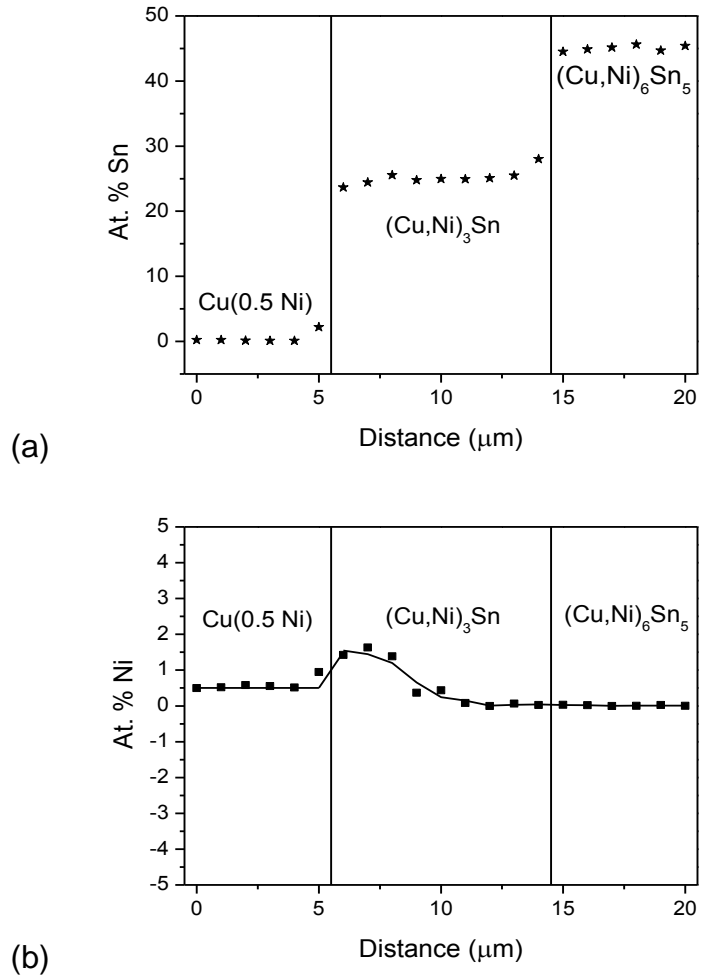

Figure 6.17: The composition profile developed across the interdiffusion zone of Cu(0.5Ni)/Sn diffusion couple annealed at 200 °C for 144 hrs: (a) showing phase boundaries by dotted lines, which is determined based on the jump in Sn composition across the interface and (b) showing evidence of Ni uphill diffusion inside the (Cu,Ni)₃Sn phase, which is consistent with the observed diffusion path trends based on the thermodynamic calculations.

By utilizing the assessed thermodynamic data (and assuming that local equilibrium is established in the system [94]), we have calculated the chemical potentials for Sn and Cu at different interfaces in several diffusion couples. The data shown for 3 and 8 at.% Ni are after extracting from the rest. The results are listed in

Table 6.3. As can be seen, the driving forces for diffusion of Sn and Cu through Cu$_3$Sn and Cu$_6$Sn$_5$ change as a function of Ni content in the system. This will in turn have an effect on the diffusion fluxes and the growth behaviour of the phases in the system.

| Diffusion Couple | $\Delta\mu_{Sn}^{(Cu,Ni)_6Sn_5}$ (J/mol) | $\Delta\mu_{Cu}^{(Cu,Ni)_6Sn_5}$ (J/mol) | $\Delta\mu_{Sn}^{(Cu,Ni)_3Sn}$ (J/mol) | $\Delta\mu_{Cu}^{(Cu,Ni)_3Sn}$ (J/mol) |
|---|---|---|---|---|
| Cu/Sn | −3441.9 | −2801.7 | −29156.6 | −9718.8 |
| Cu(0.5Ni)/Sn | −3451.1 | −2804.3 | −29466.6 | −9730.6 |
| Cu(2.5Ni)/Sn | −3614.3 | −2902.3 | −31883.1 | −9775.8 |
| Cu(3.0Ni)/Sn | −3621.8 | −2922.5 | | |
| Cu(5.0Ni)/Sn | −3651.9 | −3003.5 | −34064.4 | −9808.3 |
| Cu(7.5Ni)/Sn | −3829.2 | −3012.2 | | |
| Cu(8.0Ni)/Sn | −3864.7 | −3013.9 | | |

Table 6.3: Driving forces (*i.e.*, the difference of chemical potentials at the interfaces of phases at 200 °C) for diffusion of Cu and Sn through ε–Cu$_3$Sn and η–Cu$_6$Sn$_5$.

The common tangent construction can be used to better visualize the effects of the changes in the driving forces for diffusion of Cu through ε–Cu$_3$Sn and Sn through η–Cu$_6$Sn$_5$. As the stability of different phases changes differently with Ni addition and Ni content is not homogeneous over the whole diffusion couple, the Gibb's free energy diagram in Figure 6.18 is qualitatively drawn. The exact values for the changes in the driving forces at 200 °C can be found from Table 6.3. In the figure, $\Delta\mu_{Cu}^{Cu_3Sn}$ and $\Delta\mu_{Sn}^{Cu_3Sn}$ are the chemical potential difference of Cu and Sn between interfaces Cu$_3$Sn/Cu and Cu$_6$Sn$_5$/Cu$_3$Sn, which drives the diffusion of Cu through Cu$_3$Sn, while $\Delta\mu_{Cu}^{Cu_6Sn_5}$ and $\Delta\mu_{Sn}^{Cu_6Sn_5}$ are the chemical potential difference of Cu and Sn between the interfaces Sn/Cu$_6$Sn$_5$ and Cu$_6$Sn$_5$/Cu$_3$Sn, which drives the diffusion of Sn through the Cu$_6$Sn$_5$ phase layer. Experimental studies in the Cu–Sn system indicates that Cu has a much higher diffusion rate as compared to Sn (∼30 times) in the Cu$_3$Sn phase, whereas, Sn has ∼2.3 higher diffusion rate as compared to Cu in the Cu$_6$Sn$_5$ phase. Therefore, any change in diffusion rate of Cu through Cu$_3$Sn will induce differences in the growth rate. On the other hand, in the Cu₆Sn₅ phase, changes in diffusion rate of both the components must be considered.

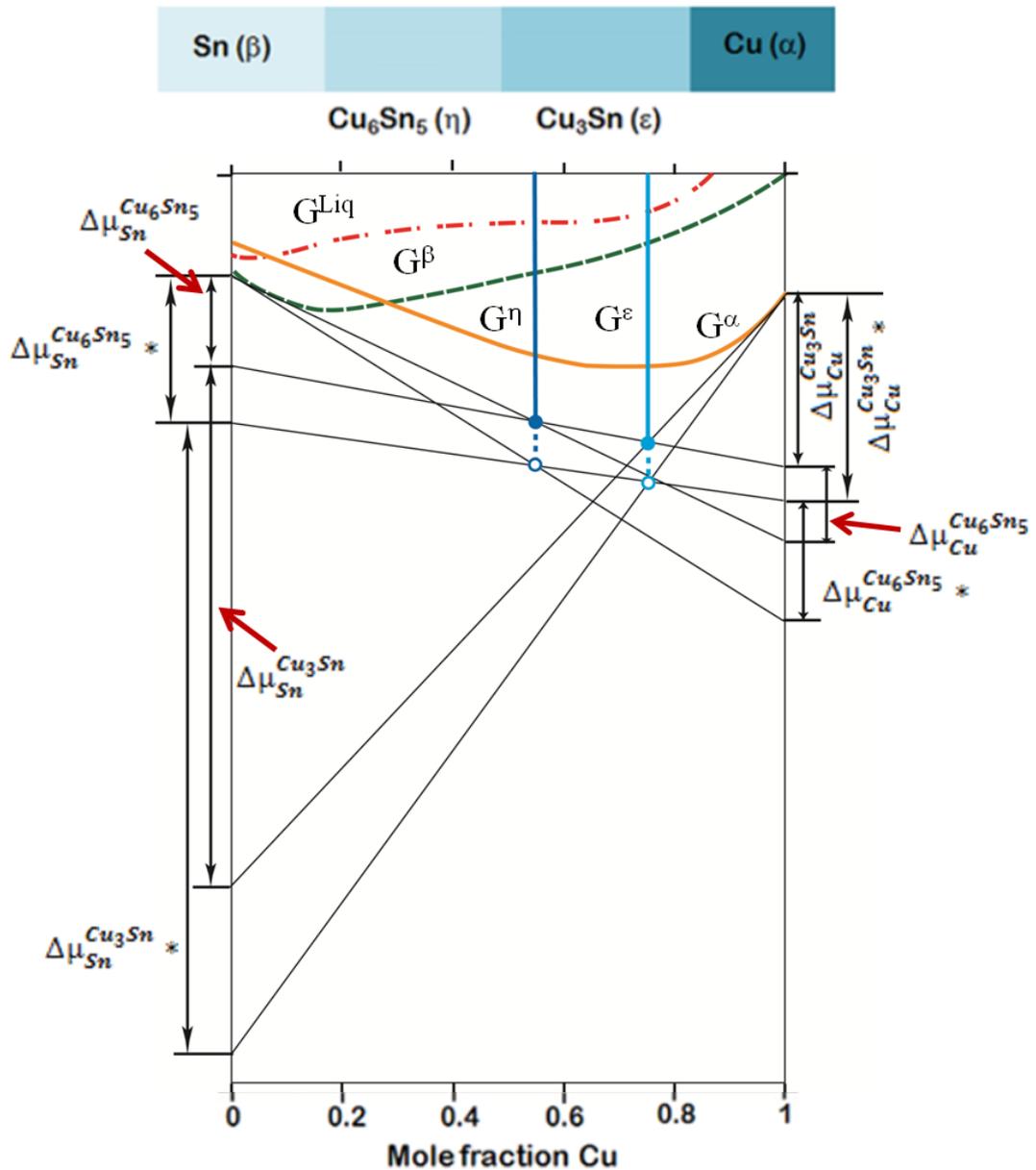

Figure 6.18: Superimposed Gibb's free energy diagrams from Cu–Sn and Cu–Ni–Sn systems showing the changes in the driving forces for diffusion of Cu and Sn through ε–Cu₃Sn and η–Cu₆Sn₅ phases at 200 °C, as Ni is added to the binary system.

It is easy to realize that changes in the stabilities of the η–Cu₆Sn₅ and ε–Cu₃Sn phases (in their Gibbs free energy) will change the values of $\Delta\mu_{Sn}^{Cu_6Sn_5}$ and $\Delta\mu_{Cu}^{Cu_3Sn}$, and hence change the driving forces for the diffusion of components in the system when Ni is added in Cu. This situation is superimposed in Figure 6.18. At this point, it is to be noted that the Gibbs energy diagram in Figure 6.18 is not a binary one, but a vertical section from a ternary three–dimensional Cu–Ni–Sn Gibbs energy diagram. In ternary systems, the tie–lines are not usually in the plane of the vertical sections [95]. The following simplifications are made in the analysis:

(i) Because of Ni addition, the stability of all phases will change. However, it is very difficult to visualize the changes in driving forces by considering the changes in all phases, especially while considering the non–homogeneous composition variation in the phases. Therefore, we have only drawn the changes in the stability of the compounds $\eta$–$Cu_6Sn_5$ and $\varepsilon$–$Cu_3Sn$.

(ii) Our analysis indicates that Ni addition stabilizes the phases differently. The free energy changes more rapidly in $\eta$–$Cu_6Sn_5$ compared to $\varepsilon$–$Cu_3Sn$. This is taken into account in the analysis.

(iii) The discussion that follows is qualitative and does not reflect the exact changes in stabilities of the phases or driving forces.

The aim of this analysis is to explain the trend of change in driving forces. The exact numerical values for the present case can be found from Table 6.3. From Figure 6.18, we can see that when Ni is added, the stability of $(Cu,Ni)_6Sn_5$ increases at higher rate than that of $(Cu,Ni)_3Sn$. This will have an effect on the chemical potentials (and thereby activities) in such a way that the driving force of both the components through both the phases increases from, for example, $\Delta\mu_{Sn}$ to $\Delta\mu_{Sn}*$ and $\Delta\mu_{Cu}$ to $\Delta\mu_{Cu}*$. As the diffusion flux is linearly proportional to the driving force, the material flux can be expected to grow in both the phases to give a higher growth rate and integrated interdiffusion coefficient. However, based on our analysis in this work, we have seen that the integrated interdiffusion coefficient does not change for $(Cu,Ni)_3Sn$ up to 2.5 at.% Ni. In the $Cu_6Sn_5$ phase, it increases; however, the rate of increase is much higher than what we expect based on the changes in the thermodynamic driving force.

It should be noted here that there are mainly two factors, namely the thermodynamic driving force and the defect concentrations, such as point defects (important for lattice diffusion) and grain boundary area (important for grain boundary diffusion), because of which diffusion rate can change with the addition of the third element. As already discussed, it might change the thermodynamic driving forces of components leading to a change in the diffusion coefficient. There are systems in which this factor plays a major role. For example, the change in driving force because of variation of Sn content explains the exceptionally increased growth rate of $Nb_3Sn$ in the Cu(Sn)–Nb system [96]. There are many other systems in which thermodynamics can explain the trend of change in diffusion coefficient with respect to composition [5, 97-100]. On the other hand, it is also not rare to find that the variation of driving force cannot explain the trend in variation of the diffusion coefficient. For example, in β–NiAl phase, thermodynamic factor accounting for the driving force decreases; however, diffusion coefficient increases because of deviation on either side of the stoichiometric composition. In that case, the increase in the number of point defects assisting lattice diffusion of components played a bigger role [101]. Similarly, the change in defect concentration overrules the minor change in driving force on diffusion rates of Ni and Al because of Pt addition in the β–(Ni,Pt)Al phase [102, 103]. Also, the addition of third element Mo was found to change the defect concentration in $Nb_5Si_3$ [104] affecting the growth rate and diffusion coefficient [105]. Therefore, it is necessary to examine the change in defect concentration because of the addition of Ni in the Cu–Sn system, which is already done based on the microstructural analysis, *i.e.*, grain morphologies in Section 6.5.

**6.9    Discussion on diffusion–controlled growth mechanism of (Cu,Ni)₆Sn₅ phase**

To understand the diffusion mechanism, we need to consider a few important estimated parameters:

(i)     As shown in Figure 6.15, the intrinsic diffusion coefficient of Sn in (Cu,Ni)$_6$Sn$_5$ changes by more than two orders of magnitude because of the addition of Ni by 8 at.%. Although the flux changes with time continuously during the diffusion annealing, the total amount of flux that is transferred during the whole annealing time is relevant to our discussion. The estimated intrinsic flux of Sn ($J_{Sn}$) after 81 hrs of annealing, as listed in Table 6.2, increases by ~17 times for the same annealing time.

(ii)    The differences of chemical potential of Sn at two different interfaces of (Cu,Ni)$_6$Sn$_5$, $\Delta\mu_{Sn}^{(Cu,Ni)_6Sn_5}$ (as listed in Table 6.3) changes by around 1.12 times.

(iii)   As measured from Figures 6.6 and 6.8, the average grain size of (Cu,Ni)$_6$Sn$_5$ up to the addition of 3 at.% Ni is ~8–9 μm. This reduces to ~4.5 μm for 5 at.% Ni and ~2 μm for 8 at.% Ni.

(iv)    The activation energy for integrated interdiffusion coefficient for Cu$_6$Sn$_5$ phase in a binary system is 81 kJ/mol. This increases to 100 kJ/mol for 2.5 at.% Ni and 110 kJ/mol for 5 at.% Ni addition in Cu, as shown in Figure 6.12.

Intrinsic flux of a component "$i$" can be expressed as $J_i = -\frac{D_i^*}{RT}C_i\frac{d\mu_i}{dx}$, where $D_i^*$ is the tracer diffusion coefficient and $\mu_i$ is the chemical potential. Therefore, the flux depends on the tracer diffusion coefficient (which is controlled by defects present in the growing product phase) and the thermodynamic driving force. There are systems in which one of them play a major role, as already discussed at the end of Section 6.8.

In the $(Cu,Ni)_6Sn_5$ phase, it is not possible to estimate the exact chemical potential gradient in a phase with narrow homogeneity range and also because of its continuous change locally along with growth of the product phase. However, one can still get an idea about its role by comparing the estimated flux and the difference in chemical potential at phase boundaries. Since $J_{Sn}$ is increased by $\sim 17$ times and $\Delta\mu_{Sn}^{(Cu,Ni)_6Sn_5}$ is increased by only $\sim 1.12$ times because of addition of 8 at.% Ni, the change in defect concentration must have a more important role in increasing the diffusion rate and therefore the growth kinetics of the product phase.

During interdiffusion process, the components diffuse via both lattice and grain boundaries in a bulk material and we actually measure an apparent value expressed as $J_{app} = (1-\delta)J_l + \delta J_g \approx J_l + \delta J_g$, where $J_l$ and $J_g$ are the fluxes through lattice and grain boundaries, respectively. $\delta$ is the volume fraction of grain boundaries, which is much lower than the volume fraction of lattice. Diffusion process in a particular system might be controlled by one type of diffusion when it dominates over the other. The activation energy for lattice diffusion is higher than grain boundary diffusion since it counts the activation energy for both the formation of point defects (such as vacancies and antisites) and the migration of components. On the other hand, only the migration energy is required for the grain boundary diffusion. As mentioned above, a relatively low value of the activation energy for the integrated interdiffusion coefficient of $Cu_6Sn_5$ in the binary Cu–Sn indicates a major role of the grain boundary diffusion. Therefore, the increased diffusion rate of Sn because of Ni addition would indicate the increased contribution from grain boundary diffusion (note that higher grain boundary area corresponds to higher flux because of the grain boundary diffusion) by grain refinement (as shown for $(Cu,Ni)_6Sn_5$ phase in Figures 6.6 and 6.8). This was indeed found when Ti was added in the Cu(Ga)/V system for the growth of $V_3Ga$ intermetallic superconductor [106]. However, no grain size difference is found up to 3 at.% Ni addition in Cu. It starts decreasing only for the addition of 5 and 8 at.% Ni. However, a significant increase in the flux of Sn (Figure 6.15) and therefore the diffusion coefficient is found even for the Ni addition up to 3 at.% (*see* Figure 6.14). Moreover, just the grain boundary diffusion cannot explain the increase in the activation energy for diffusion with the addition of Ni. This might be an indication that the concentration of point defects and therefore the diffusion via lattice also increases with the increase in Ni content leading to increase in the activation energy (Figure 6.12) estimated from the apparent diffusion coefficients. It should be noted here that the activation energy for integrated interdiffusion coefficient would be the same as the activation energy for intrinsic diffusion coefficient of Sn, *see* Equation (6.11). Considering the melting point of $Cu_6Sn_5$, $T_m = 415$ °C (688 K) [40], the temperature range of the diffusion couple experiments (*i.e.*, 150–200 °C) for the estimation of activation energy with respect to the homologous temperature ($T/T_m$) is in the range of 0.6–0.7. Therefore, the available concentration of point defects must be sufficient for the diffusion of components via lattice.

The contribution from lattice diffusion can be further realized when we consider the diffusion of Cu. The intrinsic fluxes crossing the Kirkendall marker plane in the binary couple are estimated (following Equations 6.9b and 6.9c) as $J_{Sn} = -0.17$ and $J_{Cu} = -0.07$ mol/m$^2$ after 81 hrs of annealing. Similarly, for 0.5 at.% Ni addition, where Ni content is very small compared to Cu content in the $(Cu,Ni)_6Sn_5$ phase, we estimate the fluxes as $J_{Sn} = -0.30$ and $J_{Cu} + J_{Ni} \approx J_{Cu} = -0.02$ mol/m$^2$. Therefore, the flux of Cu decreases significantly because of small addition of Ni and this trend follows even for increased content of the same. If Cu mainly diffuses via grain boundaries, then there is no reason why its diffusion rate should decrease after the addition of Ni, especially when the grain boundary area and the driving force (*see* Table 6.3) increase with the increase in Ni content. On the other hand, if it diffuses via lattice then the diffusion rate may decrease because of the change in concentration of point defects. It should be again pointed out here that there is no experimental technique present at this point of time, which can measure the change in such a low concentration of point defects (vacancies and antisites) on different sublattices in an intermetallic compound other than one attempt in $MoSi_2$ [107]. Additional complications come from the fact that different sublattices can have different concentration of vacancies and antisites and in general affected differently by the addition of a third component. Theoretical calculation of these defects is conducted only in few systems, which are validated by the estimation of the diffusion coefficients [18]. On the other hand, in different works of our group, it has been demonstrated that a change in diffusion coefficient can efficiently indicate a change in the defect concentration on different sublattices qualitatively [68, 108]. This is important since many other physical and mechanical properties are affected by the change of concentration of these defects.

To illustrate this in the $Cu_6Sn_5$ phase, let us consider the atomic arrangements in the crystal of this phase, which is important for the discussion of diffusion of components by the sublattice diffusion mechanism [18]. It can be seen in Figure 6.4 that every Cu atom is surrounded by 6 Sn atoms and every Sn atom is surrounded by 6 Cu atoms (*i.e.*, there are no Sn–Sn and Cu–Cu bonds in the case of $B8_1$ NiAs structure). Therefore, both the components cannot diffuse via lattice unless antisites are present along with vacancies. Otherwise, a particular atom will jump to a wrong sublattice designated for another component, which is not allowed unless an antisite is present. The presence of extra Cu atoms at the 2d sites (in the case of partially filled $B8_2$ $Ni_2In$ structure) creates few Cu–Cu bonds, which may facilitate some extra diffusion of Cu. Moreover, two different components will have different concentration of vacancies and antisites. When an alloying component is added, it affects the defect concentration of different components differently [18]. Since the flux of Cu decreases drastically because of Ni addition, it is apparent that Cu diffuses via lattice but the concentration of defects on its sublattice decreases. On another hand, there might be an increase in the concentration of defects of Sn since the intrinsic flux of Sn increases because of Ni addition along with the flux through grain boundaries.

## 6.10 Conclusions

In this chapter, the role of Ni on the growth of the product phases in the Cu(Ni)–Sn system is studied based on the solid–state diffusion couple experiments, theoretical calculations (using present experimental data) of thermodynamic driving forces for diffusion, and microstructural characterization using SEM and TEM. The important outcomes of this segment of the study can be stated as follows:

- Time dependent experiments indicate diffusion–controlled growth of the product phase in the Cu(Ni)–Sn system.

- Growth kinetics of the product phases $(Cu,Ni)_3Sn$ and $(Cu,Ni)_6Sn_5$ are strongly affected by the addition of Ni. It decreases for $(Cu,Ni)_3Sn$ phase up to the addition of 2.5 at.% Ni and the phase does not grow for the Ni content $\geq$ 3 at.% in Cu. On the other hand, the growth kinetics increases significantly for the $(Cu,Ni)_6Sn_5$ phase with the addition of Ni. One of this thesis advisors (Paul) [6] found similar results, with respect to the evolution of the phases and hence the growth rate of the phases, in Cu/Sn and Cu(5Ni)/Sn solid–state diffusion couples annealed at 215 °C.

- Growth kinetics of a phase depends on the growth rate of other phases. The estimated integrated interdiffusion coefficients for the $Cu_3Sn$ phase is not affected up to 2.5 at.% Ni addition in Cu. However, the growth rate of this phase decreases because of the increase in the integrated interdiffusion coefficient and the growth kinetics of $Cu_6Sn_5$ (both of them increases continuously with the increase in Ni content). The same is explained based on the physico–chemical approach.

- The location of the Kirkendall marker plane indicates that the $Cu_6Sn_5$ phase grows by diffusion of both Cu and Sn in the binary Cu–Sn system. However, when Ni is added, the phase grows mainly by diffusion of Sn.

- Thermodynamic calculations of driving forces explain that the diffusion rates of components increase in the product phases with the increase in Ni content. However, driving forces alone cannot explain the observed much higher change in the growth rate of the product phases.

- In the Cu–Sn binary system, quantitative diffusion analysis indicates that the intrinsic diffusion coefficient of Sn is only 2.3 times higher than Cu in the $Cu_6Sn_5$ phase. However, a very small addition of Ni changes the diffusion rates of components in the same phase significantly. The diffusion rate of Cu and Ni become much smaller and the product phase grows mainly because of diffusion of Sn. This trend continues with the addition of Ni with higher amounts.

- There is a strong influence of Ni addition on the microstructure of $(Cu,Ni)_6Sn_5$. In the absence of Ni, $Cu_6Sn_5$ grows with a duplex morphology in which both the sublayers (with equivalent thickness) consist of the grains covering from one interface to the Kirkendall marker plane. With the addition of 0.5 at.% Ni, grains cover almost the whole phase layer. For the addition of 3 at.% Ni and above grains nucleate repeatedly and become smaller when these are grown directly from Cu in the absence of $(Cu,Ni)_3Sn$.

- The possible atomic mechanism of diffusion is discussed based on the different estimated parameters (*i.e.*, integrated interdiffusion coefficients, intrinsic fluxes and intrinsic diffusion coefficients of the components), thermodynamic driving forces and the average grain sizes.

- At 200 °C, the growth of phases (by diffusion process) with equilibrium crystal structure is confirmed in the binary couple (which refutes the previously studied reports in the Cu–Sn system on materials prepared by melting route) and the same crystal structure is found in the ternary couples also. Even no significant change is found in the lattice parameters of the ternary phases, based on the TEM analysis.

- The growth rate of the Kirkendall voids in the $(Cu,Ni)_3Sn$ phase increases significantly because of 0.5 at.% addition of Ni in Cu. Beyond 3 at.% Ni voids are not present because of the absence of this phase layer in the interdiffusion zone. It is an advantage since the presence of voids degrades the electro–mechanical reliability of the contact.

- It is very common to find cracks in the $Cu_6Sn_5$ phase in the absence of Ni, which indicates that this phase must be very brittle. On the other hand, cracks are not found when Ni is added, which indicates that there might be an improvement of the mechanical properties. This is indeed found for the elastic modulus and hardness in the study conducted by Mu et al. [109]. With the thrust for miniaturization, the whole solder ball is expected to be converted to the intermetallic product phases. Therefore, with the addition of more than 3 at.% Ni, it could be considered as beneficial because of the absence of the Kirkendall voids (which is due to the absence of the $(Cu,Ni)_3Sn$ phase) and enhanced mechanical properties of the (Cu,Ni)$_6$Sn$_5$ phase despite the increase in growth rate of this phase. This is important for the production of a stronger electro–mechanical bond.

*To summarize, we investigate the influence of Ni content in Cu on the diffusion–controlled growth of the (Cu,Ni)$_3$Sn and (Cu,Ni)$_6$Sn$_5$ phases. An analysis of the growth kinetics, crystal structure, microstructural evolution, thermodynamic driving forces and the diffusion rates of components in the Cu(Ni)–Sn system is presented. This sheds light on the exceptional affect of the presence of Ni in Cu on the growth kinetics of phases in the Cu–Sn system.*

# Chapter 7

# Solid–state diffusion–controlled growth of the phases

# in Au–Sn system

Investigation of the growth kinetics of phases in the Au–Sn system is reported in this chapter, over a wide temperature range, for the first time, from room temperature to the maximum possible high temperature at which the solid–state diffusion couple experiments could be successfully conducted. The temperature dependent growth of the phases is studied using bulk and electroplated couples.

## 7.1 Introduction and Statement of the Problem

As mentioned in Chapter 1, Cu, Ni and Au are used as under bump metallization (UBM) layers in soldering assembly of microelectronics packaging. Au is used for shelf–life protection of Cu (used for good bonding) and Ni (used as barrier layer) from oxidation and corrosion. Often, Sn–based alloy is used as solder in current microelectronic devices. During soldering, a good metallurgical bonding is achieved by the formation of UBM–Sn based intermetallic phases, which affect the performance of an electronic component. Afterwards, these compounds continue to grow further by solid–state diffusion–controlled process during storage at room temperature and service at an elevated temperature. Additionally, the Au–Sn system is also very important for flux–less soldering and high temperature interconnect technology [110, 111].

Numerous studies in the Au–Sn system are available at room temperature (RT) because of high growth rate of the phases; for example, References [112-117] and references therein. An excellent overview of RT studies can be found in

---

This chapter is written based on the article:

[1] V.A. Baheti, S. Kashyap, P. Kumar, K. Chattopadhyay, A. Paul: Solid–state diffusion–controlled growth of the phases in the Au–Sn system, Philosophical Magazine 98(1) (2018) 20-36.

Reference [2]. Comparatively, to our knowledge, very limited number of studies are conducted at higher temperatures (HT), and therein also mostly at a single temperature, for example in References [6, 118, 119]. One study is conducted in a temperature range of 120–200 °C [120]. However, the analysis is conducted based on the activation energy of parabolic growth constants, which are not material constants in a multiphase growth. The understanding of the diffusion mechanism should be developed based on the estimated diffusion coefficients and activation energies. RT experiments are mostly conducted with thin–film diffusion couples, whereas HT experiments are conducted with bulk diffusion couples. The characteristics of phase evolution and growth of the phases at RT and HT could be different, as these are already reported in Chapter 4 for the Cu–Sn and the Ni–Sn systems. Additionally, the condition of materials, *i.e.*, bulk or electroplated also may affect the growth process differently. Therefore, experiments in the Au–Sn system should also be conducted at various other temperatures for understanding the growth mechanism of phase(s) and the diffusion mechanism of components based on diffusion coefficients, relative mobilities of the components and estimated activation energies. Therefore, our aim of this chapter is to study the temperature dependent growth of the product phases in both bulk and electroplated diffusion couples, for the first time, covering the wide temperature range of RT to HT and to develop a better understanding of the diffusion–controlled growth mechanism of the phases in the Au–Sn system.

## 7.2    Results and Discussion

Before explaining results on the evolution of phases, it is important to refer the Au–Sn phase diagram, as shown in Figure 7.1 (kindly provided by Dong et al. [121]). There are 5 intermetallic compounds, $Au_{10}Sn$, $Au_5Sn$ with two different crystal structures ($\zeta$ and $\zeta'$), $AuSn$, $AuSn_2$, and $AuSn_4$. With the aim of conducting experiments in the solid–state condition, the maximum temperature is selected as 200 °C such that a small overshoot of temperature during heating cycle for a small time does not cross the eutectic temperature of 211 °C at ~95 at.% Sn.

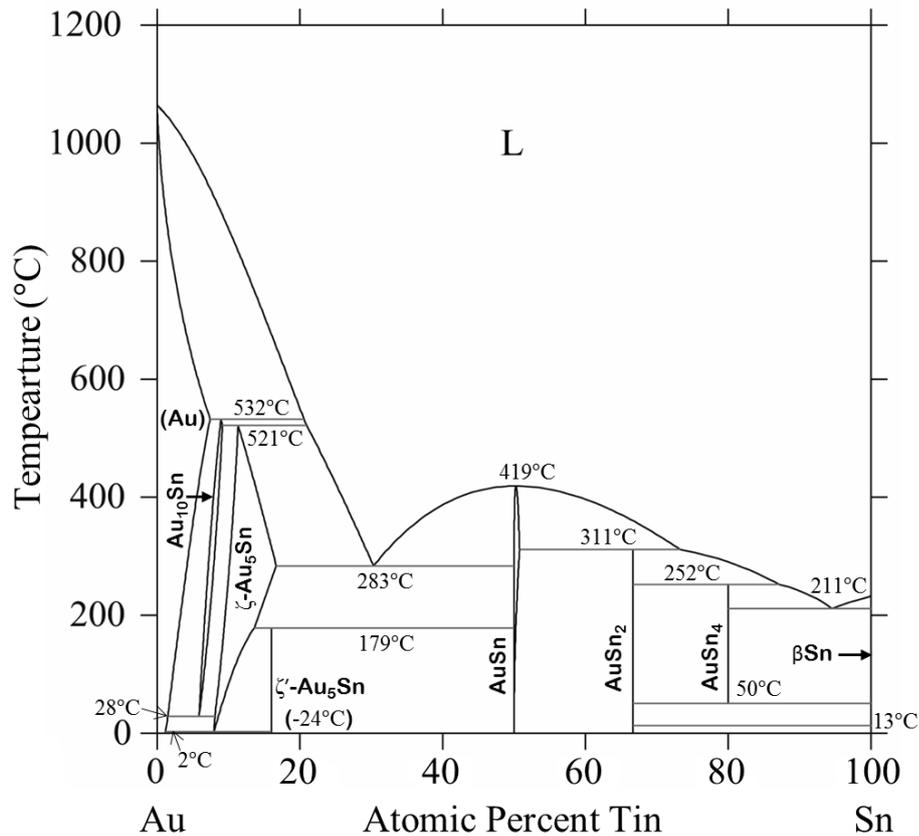

Figure 7.1: Binary Au–Sn phase diagram adapted from Dong et al. [121].

### 7.2.1 Phase evolution and structural characterization

Bulk diffusion (BD) couples are prepared in the higher temperature range 200–125 °C because of difficulty in achieving an efficient bonding by interdiffusion in the solid–state at and below 100 °C. On the other hand, Au/EP–Sn diffusion (EPD) couples are studied from high temperature to room temperature, since a bond between two dissimilar materials is created during the electroplating stage itself. Our study indicates that there is no difference with respect to the evolution of phases in these two types of diffusion couples, *i.e.*, BD and EPD, although there is a difference in the growth kinetics of the product phases. To show the growth of various phases, an interdiffusion zone of BD couple annealed at 200 °C for 4 hrs is shown in Figure 7.2a. Except $Au_{10}Sn$ phase, all other thermodynamically stable phases are found. Very frequently in a multiphase interdiffusion zone, one or more phases are not found to grow mainly because of sluggish growth kinetics. Similar to the Au–Sn system, in general, the missing phases are rich in high melting point component, *i.e.*, these missing phases are closer to the high melting point component (Au in the present case) in the phase diagram [68]. Because of a very small contrast difference of the Au–rich phases, $Au_5Sn$ could not be detected easily in the BSE micrograph shown in Figure 7.2a. The presence of the $Au_5Sn$ phase can be realized from the composition profile (measured by EPMA), as shown in Figure 7.2b, of a diffusion couple annealed at the same temperature (*i.e.*, 200 °C) but for a longer time of 49 hrs. Solid circles in the graph show the compositions of the phases reported in the phase diagram, whereas the open squares represent the actual compositions measured in EPMA. The composition range of $Au_5Sn$ indicates that ζ is grown in the interdiffusion zone, which has a narrow homogeneity range. The composition range of 10–14 at.% Sn in ζ at 200 °C is found to agree well with the phase diagram. On the other hand, the composition of all other detected line compounds, namely AuSn, $AuSn_2$ and $AuSn_4$, are measured (at all temperatures in both type of diffusion couples) as around 1 at.% Sn higher than the stoichiometric composition of these phases. Furthermore, the results with respect to the evolution of phases in BD couple are same as above at 175 °C also. We have not shown a comparison between BD and EPD couples because of failure of producing reliable EPD couples in this temperature range, *i.e.*, above 150 °C. The interdiffusion zone gets separated during cross–sectioning and further metallographic preparation due to the presence of excessive voids near the $AuSn_4$/Sn interface.

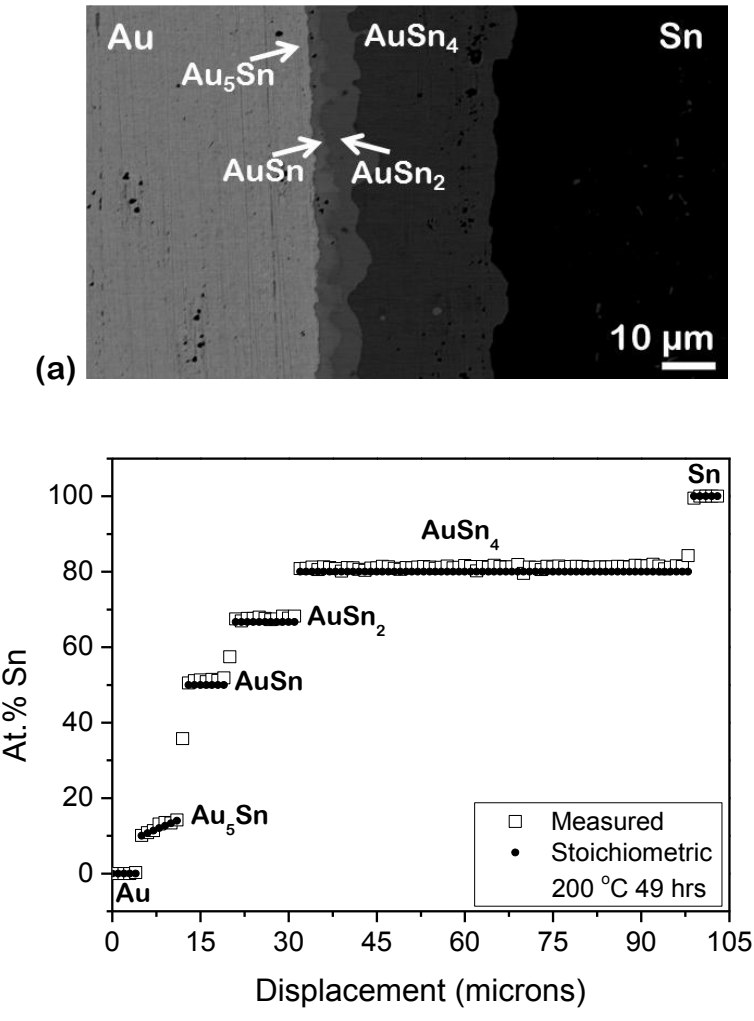

(a)

(b)

Figure 7.2: Au/Sn bulk diffusion couple annealed at 200 °C: (a) for 4 hrs, BSE image and (b) for 400 hrs, composition profile of the interdiffusion zone.

In the temperature range of 150–100 °C, along with $Au_{10}Sn$, the $Au_5Sn$ phase also could not be detected using SEM in the interdiffusion zone of Au/Sn couples. The presence of the other 3 phases, namely $AuSn$, $AuSn_2$, and $AuSn_4$, is clearly revealed by Figure 7.3. For the purpose of comparison, both BD and EPD couples annealed at 150 °C for 4 hrs are shown. The higher growth rate of the whole interdiffusion zone in EPD couple compared to BD couple is evident. A line of pores indicating the location of the Kirkendall marker plane [12, 18] is found in the $AuSn_4$ phase layer. A similar location of the marker plane was detected earlier inside $AuSn_4$ phase based on the presence of the inert markers as well as the duplex morphology [6, 118].

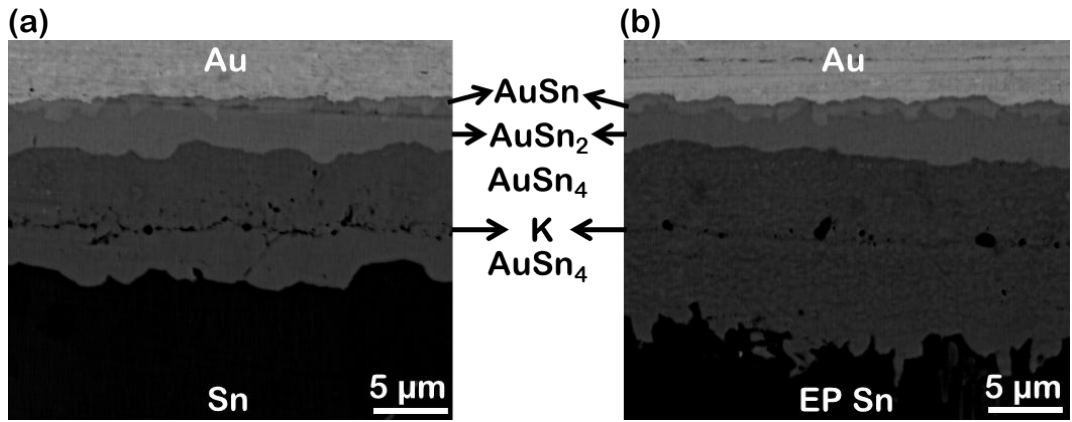

Figure 7.3: BSE micrographs of the diffusion couple annealed at 150 °C for 4 hrs: (a) Au/Sn bulk couple and (b) Au/Sn electroplated couple. K denotes the location of the Kirkendall pores in the $AuSn_4$ phase in both these couples.

With the further decrease in annealing temperature, in the range of 75 °C to RT, another phase AuSn could also not be detected in SEM. Therefore, the interdiffusion zone is mainly found to consist of $AuSn_2$ and $AuSn_4$. The SEM micrographs of the EPD couple stored at RT are shown in Figure 7.4. The product phases grow with the reasonably high rate even at this temperature. Both the phases $AuSn_2$ and $AuSn_4$ could be detected easily using SEM even after the storage for just 30 days, as shown in Figure 7.4a. $AuSn_2$ is grown with an uneven layer, whereas $AuSn_4$ is found to have irregular growth along with the presence of spikes. With the increase in storage time to 912 days (2.5 years), there is an increase in thickness of both the phases, as shown in Figure 7.4b. Additionally, the $AuSn_2$ phase is found as an even layer; however, $AuSn_4$ still grows irregularly and with long spikes. In a recent study based on the TEM analysis, Tang et al. [115] found the growth of AuSn phase at RT after the consumption of Sn end–member, like previous studies [2]. The consumption of one of the end–members makes it an incremental diffusion couple (as discussed earlier in Section 2.5) and in such a situation the phases with slow growth kinetics are found to grow in an interdiffusion zone. In fact, incremental diffusion couple experiments are commonly conducted to study the diffusion–controlled growth of a particular phase as a single product phase in an interdiffusion zone or to study the growth of the phases which do not grow in a diffusion couple when prepared with pure components as end–member [68].

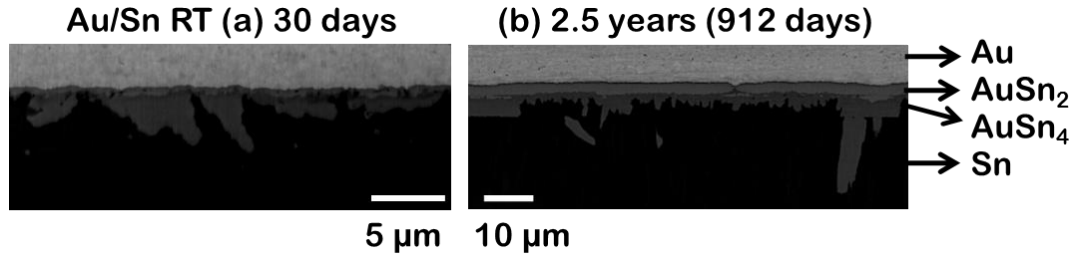

Figure 7.4: BSE images of the Au/Sn electroplated diffusion couple after storage at room temperature for (a) 30 days and (b) 912 days, *i.e.*, 2.5 years.

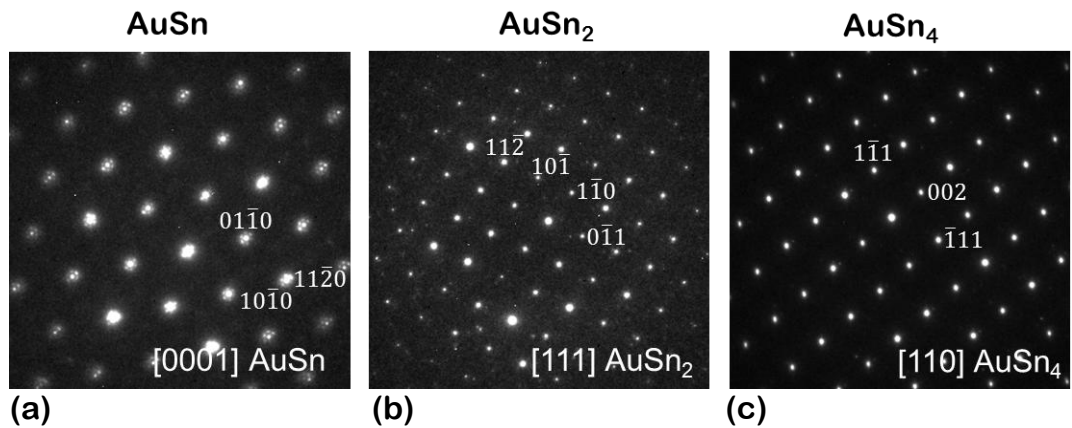

Figure 7.5: Diffraction pattern acquired from the interdiffusion zone for the phase (a) AuSn, (b) AuSn$_2$ and (c) AuSn$_4$, grown in the Au/Sn couple annealed at 150 °C.

TEM analysis for structural characterization is conducted, following the approach similar to that reported in Chapters 5 and 6, in the interdiffusion zone of Au/Sn diffusion couples annealed at 150 °C (Figure 7.3). At this temperature in the Au–Sn system, it is known that AuSn has a hexagonal, while AuSn$_2$ and AuSn$_4$ have an orthorhombic crystal structures. Our analysis indicates that there is no difference in the structure of these phases in both BD and EPD couples. In this study, all the diffraction patterns (DP) are indexed based on the crystal structures as follows: hP4 for AuSn, oP24 for AuSn$_2$ and oC20 for AuSn$_4$, such that the lattice parameters of these phases are also consistent with that reported in Reference [122]. A few representative DPs acquired from the AuSn, AuSn$_2$ and AuSn$_4$ phases are shown in

Figure 7.5. DP is indexed with zone axis: [0001] of AuSn, [111] of AuSn$_2$ and [110] of AuSn$_4$. The above analysis confirms that we have found the equilibrium crystal structure for all the phases in binary Au/Sn couple, irrespective of the condition of Sn in this study.

### 7.2.2 Diffusion parameters

Time dependent experiments are conducted with BD couples at 200 °C for 4 – 81 hrs, as shown in Figure 7.6. Thickness ($\Delta x$) and time ($t$) with respect to $(\Delta x)^2$ vs. $2t$ are plotted only for AuSn$_4$, AuSn$_2$ and AuSn phases, since the thickness of the Au$_5$Sn phase is very small and it is difficult to produce any meaningful plot for the same without very high error. A linear fit of data indicates the parabolic nature, *i.e.*, diffusion–controlled growth of the phases.

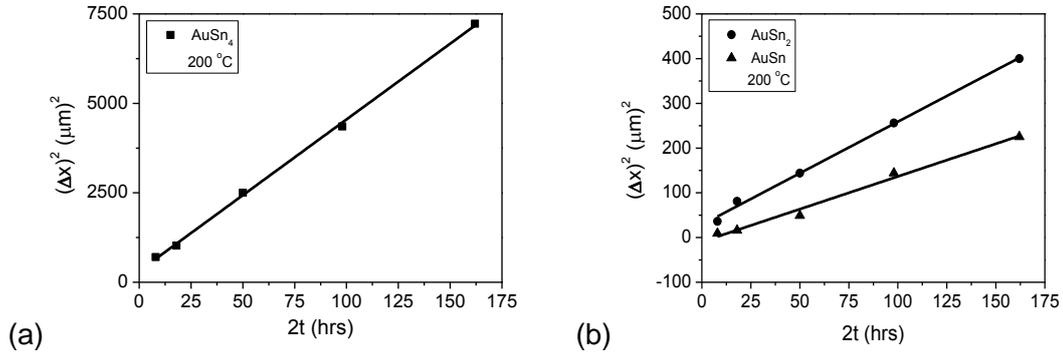

Figure 7.6: Parabolic growth law followed at 200 °C by the phase
(a) AuSn$_4$, and (b) AuSn$_2$ and AuSn.

Since AuSn, AuSn$_2$ and AuSn$_4$ grow as line compounds, we can estimate the integrated interdiffusion coefficients ($\widetilde{D}_{int}$), which are estimated following the relation developed by Wagner [26], Equation (3.33). The molar volume of the phases are estimated using the lattice parameter data available in literature [122] as $V_m^{AuSn} = 13.44$, $V_m^{AuSn_2} = 14.38$ and $V_m^{AuSn_4} = 14.99$ ($\times 10^{-6}$ m$^3$/mol). The estimated $\widetilde{D}_{int}$ of the phases at various temperatures, as listed in Table 7.1a, are plotted in Figure 7.7 with respect to $(1/T)$. Note that these are estimated for the AuSn phase in BD couples only, except at one temperature (150 °C) in EPD couple, as this phase does not grow with sufficient thickness for calculation in experiments conducted at lower temperatures ($\leq 125$ °C) for the both types of couples (*i.e.*, BD and EPD). The activation energy can be estimated from the slope of the Arrhenius plot.

Table 7.1 (a): Integrated interdiffusion coefficients in $AuSn_4$, $AuSn_2$ and $AuSn$

| Diffusion Couple | Au/Sn EPD couple | | | Au/Sn BD couple | | |
|---|---|---|---|---|---|---|
| Temperature (°C) | $\widetilde{D}_{int}^{AuSn_4}$ (m²/s) | $\widetilde{D}_{int}^{AuSn_2}$ (m²/s) | $\widetilde{D}_{int}^{AuSn}$ (m²/s) | $\widetilde{D}_{int}^{AuSn_4}$ (m²/s) | $\widetilde{D}_{int}^{AuSn_2}$ (m²/s) | $\widetilde{D}_{int}^{AuSn}$ (m²/s) |
| 50 | $1.2\times10^{-18}$ | $4.4\times10^{-19}$ | | | | |
| 75 | $1.1\times10^{-17}$ | $6.0\times10^{-18}$ | | | | |
| 100 | $6.5\times10^{-17}$ | $2.1\times10^{-17}$ | | | | |
| 125 | $2.6\times10^{-16}$ | $8.4\times10^{-17}$ | | $1.3\times10^{-16}$ | $6.6\times10^{-17}$ | |
| 150 | $1.2\times10^{-15}$ | $2.6\times10^{-16}$ | $6.5\times10^{-17}$ | $6.2\times10^{-16}$ | $2.1\times10^{-16}$ | $5.3\times10^{-17}$ |
| 175 | | | | $1.9\times10^{-15}$ | $6.1\times10^{-16}$ | $1.9\times10^{-16}$ |
| 200 | | | | $5.0\times10^{-15}$ | $1.1\times10^{-15}$ | $4.2\times10^{-16}$ |

Table 7.1 (b): Tracer diffusion coefficients in the $AuSn_4$ phase

| Diffusion Couple | TP [+] (J/mol) | $AuSn_4$ in Electroplated | | | $AuSn_4$ in Bulk | | |
|---|---|---|---|---|---|---|---|
| Temperature (°C) | | $\dfrac{D_{Sn}^*}{D_{Au}^*}$ | $D_{Au}^*$ (m²/s) | $D_{Sn}^*$ (m²/s) | $\dfrac{D_{Sn}^*}{D_{Au}^*}$ | $D_{Au}^*$ (m²/s) | $D_{Sn}^*$ (m²/s) |
| 100 | −242.49 | 3.73 | $5.3\times10^{-16}$ | $2.0\times10^{-15}$ | | | |
| 125 | −364.33 | 3.74 | $1.5\times10^{-15}$ | $5.7\times10^{-15}$ | 9.24 | $4.4\times10^{-16}$ | $4.0\times10^{-15}$ |
| 150 | −486.17 | 3.80 | $5.2\times10^{-15}$ | $2.0\times10^{-14}$ | 9.26 | $1.6\times10^{-15}$ | $1.5\times10^{-14}$ |
| 175 | −608.01 | | | | 9.38 | $4.4\times10^{-15}$ | $4.2\times10^{-14}$ |
| 200 | −729.84 | | | | 18.5 | $5.9\times10^{-15}$ | $1.1\times10^{-13}$ |

[+] Thermodynamic Parameter (TP) in $AuSn_4$, *i.e.* $N_{Sn}\Delta_d G_{Sn} = N_{Au}\Delta_d G_{Au}$ (in J/mol).

Table 7.1: Diffusion parameters are listed at various temperatures in various phases for both bulk and electroplated couples: (a) integrated interdiffusion coefficients and (b) tracer diffusion coefficients.

By comparing the estimated data for different phases in both BD and EPD couples, it can be stated that $\widetilde{D}_{int}^{AuSn_4} > \widetilde{D}_{int}^{AuSn_2} > \widetilde{D}_{int}^{AuSn}$ without much difference in their activation energies. These data indicate the relative growth rate of the phases (for example, *see* Figure 7.3). Further, the diffusion coefficients are higher when measured from diffusion profiles in EPD couples as compared to BD couples because of the higher growth rate of phases in EPD couple. The difference is the most prominent in the case of $AuSn_4$ phase.

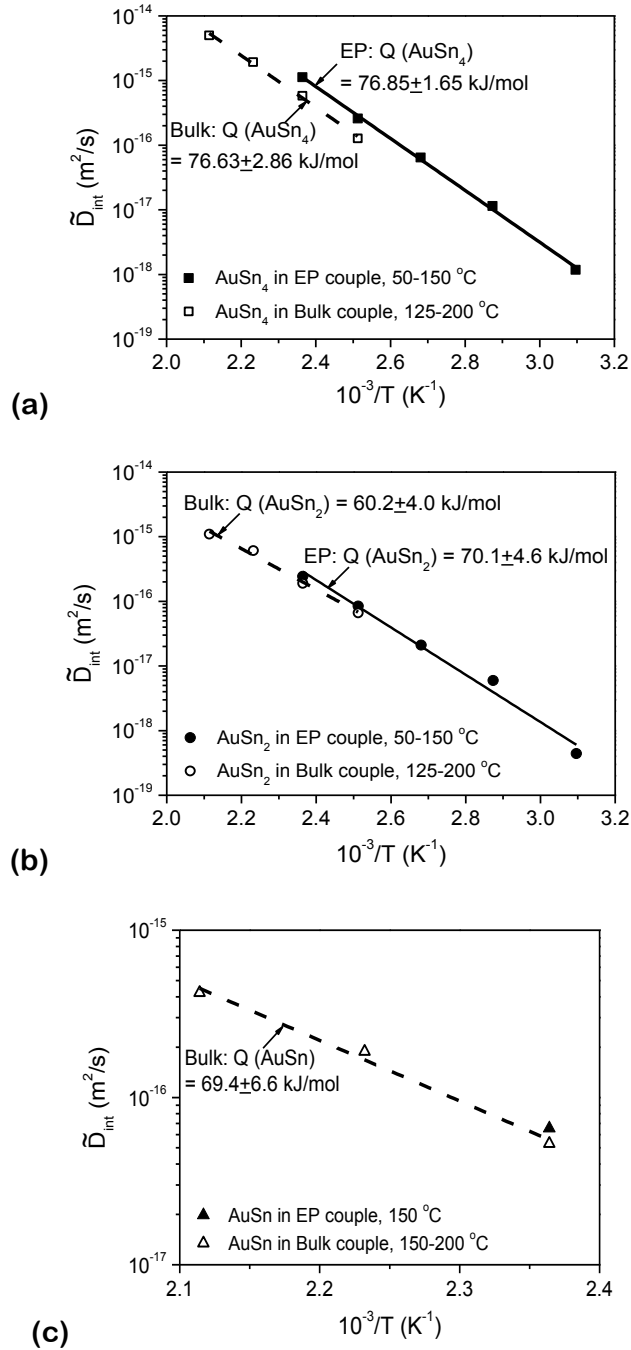

Figure 7.7: Arrhenius plot of the integrated interdiffusion coefficients in the phase (a) $AuSn_4$, (b) $AuSn_2$ and (c) AuSn, for both bulk and electroplated couples.

The ratio of the intrinsic fluxes $(J_i)$ of components is equal to the ratio of the intrinsic diffusion coefficients $(D_i)$ considering a constant molar volume of a line compound, since the composition variation of molar volume in such a small composition range is not known. Note that we cannot estimate the absolute values of $D_i$ because of the unknown concentration gradient in a line compound. Nevertheless, neglecting the unknown vacancy–wind effect, the ratio of the intrinsic diffusion coefficients (for a constant molar volume of $\beta$ phase, such that partial molar volume $\bar{V}_i^\beta = V_m^\beta$) can be written as equal to the ratio of the tracer diffusion coefficients [18]. Following van Loo's method [29], Equation (3.30c), these are estimated in AuSn4 because of the presence of the Kirkendall marker plane in this phase. These are listed in Table 7.1b, which are required for further analysis. It can be seen that the ratio of diffusivities $\frac{D_{Sn}^*}{D_{Au}^*} = \frac{D_{Sn}}{D_{Au}}$ is less in EPD couples as compared to the BD couples.

The diffusion couple technique is an indirect method for the estimation of the tracer diffusion coefficients, as already demonstrated in many systems [18]. $\widetilde{D}_{int}$ and $D_i^*$ can be related by the below equation [18] as:

$$\widetilde{D}_{int} = -(N_{Au}D_{Sn}^* + N_{Sn}D_{Au}^*)\frac{N_{Au}\Delta_d G_{Au}}{RT} = -(N_{Au}D_{Sn}^* + N_{Sn}D_{Au}^*)\frac{N_{Sn}\Delta_d G_{Sn}}{RT} \quad (7.1)$$

where $N_i$ is the composition of components in the phase of interest, and $\Delta_d G_{Au}$ and $\Delta_d G_{Sn}$ are the driving forces for diffusion of Au and Sn such that $N_{Au}\Delta_d G_{Au} = N_{Sn}\Delta_d G_{Sn}$. It should be noted here that $\widetilde{D}_{int}$ and $D_i^*$ are material constants, and hence should give the same values within the limits of experimental error when calculated from a diffusion couple with different end–members. Therefore, as explained in References [12, 18], the driving forces (for diffusion of components through AuSn4) at various temperatures are estimated from the free energy versus composition diagram utilizing the free energy values of AuSn2, AuSn4 and pure Sn at various temperatures as available in the supplementary data file of Reference [121]. It means that for the estimation of the driving force, we can assume as if AuSn$_4$ is grown between AuSn$_2$ and Sn surrounding it in the Au–Sn phase diagram (Figure 7.1), which is explained schematically in Figure 7.8. Note here that although the driving forces for diffusion of components through a phase are different, they are related as $N_A \Delta_d G_A = N_B \Delta_d G_B$ in a binary $A$–$B$ system [18]. The values of this thermodynamic parameter at various temperatures are listed in Table 7.1b along with the absolute values of tracer diffusion coefficients, which are estimated utilizing the values of $\widetilde{D}_{int}$ (listed in Table 7.1a) and $\frac{D_{Sn}^*}{D_{Au}^*}$ (in Table 7.1b) in Equation (7.1).

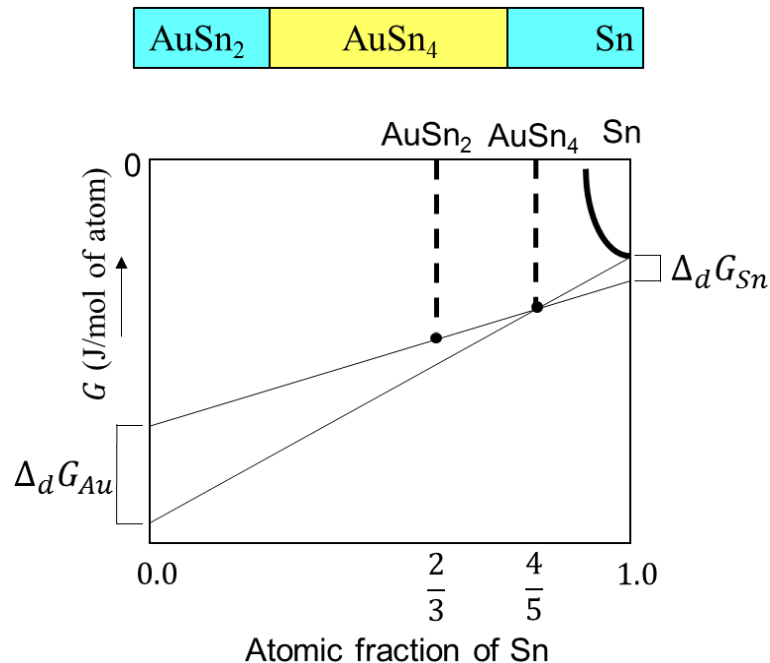

Figure 7.8: Schematic illustration of free energy diagram explaining the calculation of driving forces for diffusion of components through the AuSn$_4$ phase, considering its growth between the AuSn$_2$ phase and Sn.

### 7.2.3 Diffusion mechanism

We need to consider two important points from the estimated data for understanding the diffusion–controlled growth mechanism of the AuSn$_4$ phase, in which we could measure the important diffusion parameters:

(i)     $\widetilde{D}_{int}$ measured in EPD couples are higher than the values measured in BD couples (Figure 7.7, Table 7.1a).

(ii)    On the other hand, the ratio of diffusivities $\frac{D_{Sn}^*}{D_{Au}^*}$ in AuSn4 is less in the case of EPD couples as compared to the values measured in BD couples (Table 7.1b).

In a diffusion couple, components diffuse via both lattice and grain boundaries simultaneously and, therefore, we actually measure an apparent diffusion coefficient from the measured apparent flux $J_{app} = (1 - \delta)J_l + \delta J_g \approx J_l + \delta J_g$, where $J_l$ and $J_g$ are the fluxes through the lattice and the grain boundaries, respectively, and $\delta$ is the volume fraction of grain boundaries [18]. In many systems, depending on the annealing temperature for a particular phase, the diffusion process is controlled mainly by the transport of components through lattice or grain boundaries. However, in certain systems, both lattice and grain boundary diffusion might play an important role, as it is found in the case of Cu(Ni)–Sn system for which the results are reported in Chapter 6.

The estimated activation energy of diffusion coefficient indicates the dominant diffusion mechanism. The activation energy for grain boundary diffusion is much less as compared to the lattice diffusion. It is the summation of the formation energy of point defects and the migration energy of components for the lattice diffusion, while, on the other hand, the activation energy for grain boundary diffusion is equal to only the energy for migration of components. Moreover, the migration energy for grain boundary diffusion is smaller than that for lattice diffusion because of the relatively open structure of grain boundaries [18]. The relatively low values of the activation energies estimated, as shown in Figure 7.7, indicates the dominant role of the grain boundary diffusion in the Au–Sn system.

Therefore, to understand the reason for higher values of the $\widetilde{D}_{int}$ in the case of EPD couples when compared to BD couples, the grain size distribution plot is constructed based on the grains of the AuSn$_4$ phase examined in TEM as shown in Figure 7.9, since the difference of estimated $\widetilde{D}_{int}$ is prominent in this phase as compared to the other phases. Note that the grains of the phase are identified after analysing the acquired DPs (recognizing the phase). One example of bright–field images showing the grain of the AuSn$_4$ phase is given in the inset for both EPD and BD couples. It is evident from the distribution plot that the grains of the AuSn$_4$ phase are found to be relatively smaller in the case of EPD couple when compared to BD couple. Smaller grain size, and therefore higher grain boundary area, leads to the higher amount of diffusing flux, explaining the higher values of $\widetilde{D}_{int}$ in the case of EPD couples when compared to BD couples.

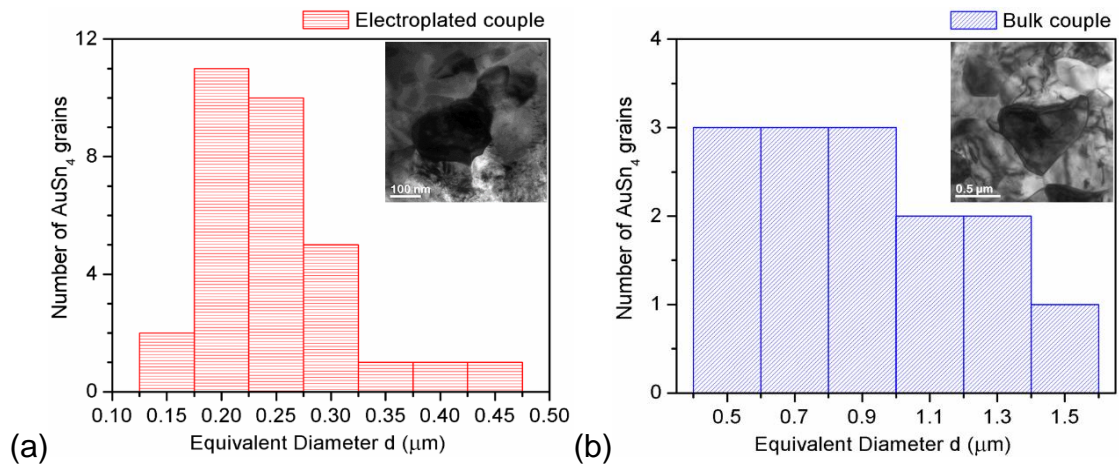

Figure 7.9: Grain size distribution of AuSn$_4$ grains from the interdiffusion zone of diffusion couple annealed at 150 °C: (a) Au/Sn electroplated couple and (b) Au/Sn bulk couple, along with the corresponding bright–field image shown in inset.

The tracer diffusivities of components in the AuSn$_4$ phase (as listed in Table 7.1b) provide further insights into the diffusion mechanism since the tracer diffusion coefficients (which are not dependent on the thermodynamic driving force) indicate the contribution of defects to the diffusion process. With the decrease in grain size of AuSn$_4$ in EPD couples, the diffusion rate of Au increases with a faster rate than Sn, leading to low ratios of diffusivities $\frac{D_{Sn}^*}{D_{Au}^*}$ in EPD couples compared to the BD couples. Although the exact contribution of different modes of diffusion (*i.e.*, via grain boundaries and lattice) cannot be estimated, we can make an assessment relating the concept of the sublattice diffusion mechanism in intermetallic compounds [18] with the estimated tracer diffusion data, the homologous temperature of the experiments and crystal structure of AuSn$_4$.

The activation energy indicates a dominant role of the grain boundary diffusion; however, the role of the lattice diffusion cannot be completely ignored since the experiments are conducted at high homologous temperatures ($T/T_m$, where $T$ is the annealing temperature and $T_m$ is the melting point in Kelvin of the phase of interest). The AuSn$_4$ phase, for which we are discussing the diffusion mechanism, has a melting point of 252 °C. Therefore, the temperature range of experiments (100–200 °C) in which the tracer diffusion coefficients of components are estimated falls in the range of homologous temperature of 0.7–0.9, which is much high than 0.5 $T_m$. Therefore, point defects must be present with enough concentration to assist the lattice diffusion.

Now, we need to consider the crystal structure to understand the complications in diffusion of components following the sublattice diffusion mechanism. It should be noted here that the possibility of the presence of different types of point defects (vacancies and antisites) with different concentrations on different sublattices make the discussion very complex. These cannot be estimated experimentally and even calculated theoretically unless bond energies are known [18]. However, one can make a fair prediction of the type of defects present based on the estimated diffusion coefficients, as done successfully in other systems [68, 108].

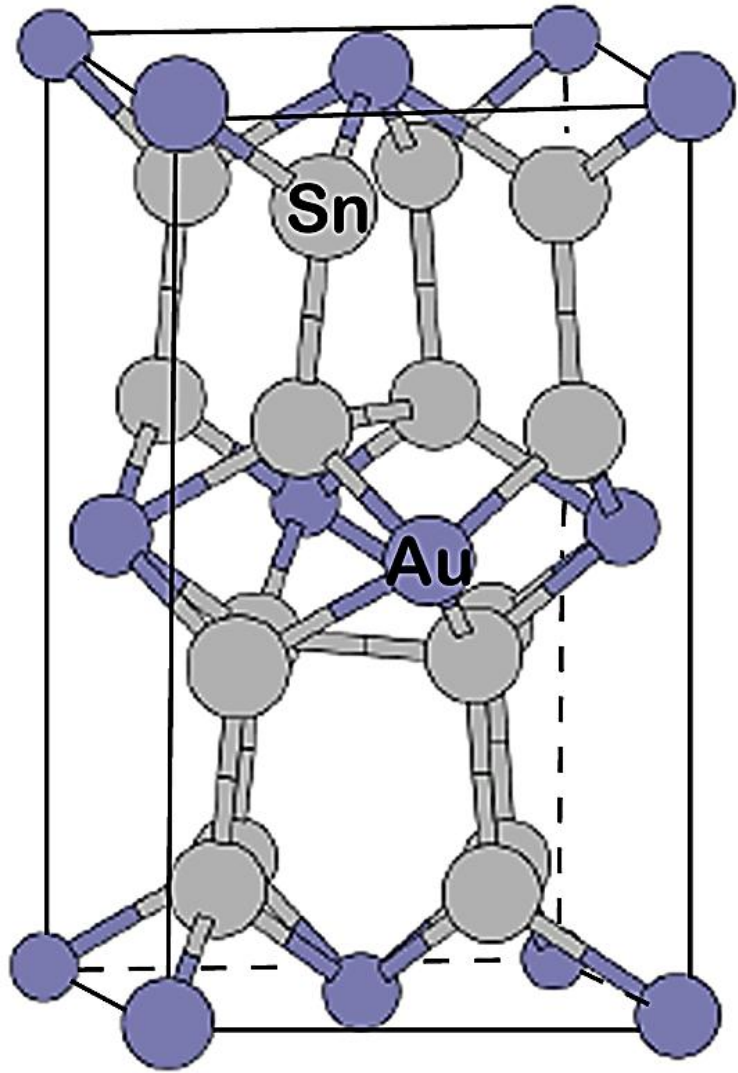

Figure 7.10: Crystal structure of the AuSn₄ phase.

The oC20 crystal structure of AuSn₄ is shown in Figure 7.10. It can be seen that many Sn–Sn bonds are available; however, Au–Au bonds are not present. Therefore, according to the sublattice diffusion mechanism in intermetallic compounds [18], Sn can diffuse easily via its own sublattice by exchanging position with the vacancies on the same sublattice. On the other hand, Au cannot diffuse straightforwardly via its sublattice due to the absence of Au–Au bonds. By exchanging the position with vacancies in the next sublattice position, Au will have to move to the sublattice of Sn, which is not allowed. This is possible only if Au–antisites (Au$_{Sn}$) are present, which can occupy the sublattice of Sn. The deviation of the composition of a phase from stoichiometry is mostly compensated by the presence of antisites, which can increase the diffusion rate of component [18]. In the presence of Au structural antisites, the composition of $AuSn_4$ phase would deviate positively towards Au–rich side in the phase diagram. As already mentioned and shown in Figure 7.1, this phase is reported as a line compound (*i.e.*, with negligible homogeneity range) at the stoichiometric composition [121]. However, our EPMA analysis in all the diffusion couples indicates that the composition of this phase deviates positively from stoichiometry towards the Sn–rich side by ∼1 at.% relative to the reported value. This must be compensated by the presence of Sn–antisites ($Sn_{Au}$). These defects (occupying the sublattice of Au) will increase the lattice diffusion rate of Sn in addition to the diffusion of the same component via Sn–vacancies present on its own sublattice. Therefore, we can conclude that the diffusion rate of Au via lattice must be negligible compared to its high diffusion rate via grain boundaries (as discussed before). Although the activation energy indicates the dominant role of grain boundaries, the lattice diffusion of Sn cannot be ruled out because of availability of Sn–Sn bonds, high homologous temperature at which vacancies must be present on its sublattice to assist the diffusion of Sn and the presence of Sn antisites because of the positive deviation of composition of Sn from the stoichiometry. With grain refinement in the $AuSn_4$ phase in EPD couple, only the diffusion of components through grain boundaries will be affected. Therefore, the diffusion rate of Au (which diffuses mainly via grain boundaries) increases more than Sn in the case of EPD couple compared to the BD couple. This lowers the value of $\frac{D^*_{Sn}}{D^*_{Au}}$ in the EPD couple. If both the components would diffuse only via grain boundaries and since the diffusion rate of Sn is higher than Au, the diffusion rate of Sn would increase at least equally (if not more) as compared to the diffusion rate of Au due to grain refinement.

## 7.3    Conclusions

For the first time, the solid–state diffusion–controlled growth of the phases in the Au–Sn system is studied over a very wide temperature range, *i.e.*, from 200 °C to RT (room temperature), to examine the temperature dependent growth of phases. Few key findings of this segment of the study can be stated as follows:

(i)     At 200 °C, except $Au_{10}Sn$, all other thermodynamically stable intermetallic compounds, *i.e.*, $Au_5Sn$, $AuSn$, $AuSn_2$, and $AuSn_4$, are found to grow in the interdiffusion zone. With the decrease in annealing temperature, *i.e.*, in the temperature range of 150–100 °C, the $Au_5Sn$ phase also does not grow along with $Au_{10}Sn$. With a further decrease in the annealing temperature, *i.e.*, 75 °C to RT, only $AuSn_2$ and $AuSn_4$ are found grow in the interdiffusion zone. The growth rate of the phases is significant even at the RT.

(ii)    Although there is no difference in the evolution of phases in the case of EPD and BD couples, the growth kinetics is found to be higher in EPD couples when compared to BD couples. The difference in growth rate of the $AuSn_4$ phase is significant for these couples.

(iii)   The estimated activation energies for various phases indicate a dominant role of the grain boundary diffusion.

(iv)    The selected–area diffraction pattern analysis of these product phases indicates that they grow with the equilibrium crystal structure in both EPD and BD couples. Grain size distribution plotted based on TEM analysis indicates that the grains of $AuSn_4$, *i.e.*, the phase in which the difference in growth rate is maximum, are smaller in EPD couple as compared to the BD couple. This explains the higher growth rate of $AuSn_4$ (in EPD couple as compared to BD) because of a higher amount of diffusing flux through grain boundaries in EPD couple.

(v)     Following, the comparison of the estimated tracer diffusion coefficients in these two types of couples (*i.e.*, EPD and BD) and the consideration of sublattice diffusion mechanism with respect to the crystal structure of AuSn$_4$ and the high homologous temperature (0.7–0.9) of diffusion annealing experiments indicate that Au diffuses mainly via the grain boundaries, whereas Sn diffuses via both the lattice and the grain boundaries.

*To summarize, the temperature dependent growth of phases in the Au–Sn system is studied in this chapter and found to be as follows:*

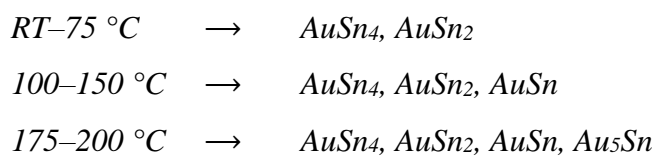

*RT–75 °C     $\longrightarrow$     AuSn$_4$, AuSn$_2$*

*100–150 °C     $\longrightarrow$     AuSn$_4$, AuSn$_2$, AuSn*

*175–200 °C     $\longrightarrow$     AuSn$_4$, AuSn$_2$, AuSn, Au$_5$Sn*

*The difference in growth rate of phases is observed in bulk and electroplated couples. It is most prominent for the AuSn$_4$ phase, which is explained based on the analysis which relates the estimated tracer diffusion coefficients with the homologous temperature, grain size, crystal structure and the concept of the sublattice diffusion mechanism in the intermetallic compounds.*

# Chapter 8

# Growth of phases in the solid–state from room temperature to an elevated temperature in the Pd–Sn and the Pt–Sn systems

The temperature dependent growth kinetics of phases in the Pd–Sn and the Pt–Sn systems is reported in this chapter, over a wide temperature range in the solid–state using bulk and electroplated diffusion couples, and the results are compared with the Au–Sn system.

## 8.1    Introduction and Statement of the Problem

As discussed in Chapter 1, Cu, Ni and Au are commonly used as under bump metallization (UBM) in the flip–chip bonding. Cu is added for good bonding, whereas Ni acts as barrier layer during soldering. Au is primarily used for the shelf–life protection of Cu and Ni against corrosion and oxidation during storage. Being noble metals, like Au, Pd and Pt are currently being considered as potential candidates for replacing Au. During soldering with a Sn–based solder alloy, intermetallic compounds often form, which is important for a good metallurgical bonding at UBM/solder joints. However, subsequent growth of these brittle product phases in the solid–state, during storage at room temperature (RT) and service at an elevated temperature, degrades the electro–mechanical performance of the electronic components. Moreover, due to the thrust for miniaturization, the whole solder might be consumed by the growth of these phases, introducing major reliability concern in the microelectronic packages.

---

This chapter is written based on the article:

Numerous interdiffusion studies on the growth of phases and their growth mechanism are available in the Cu–Sn, Ni–Sn and Au–Sn systems. An overview of the scientific issues in these systems and related outcomes can be found in References [2, 17], and Chapters 4, 5 and 7 of this thesis. Comparatively, very limited number of growth and interdiffusion studies have been conducted in the Pd–Sn and the Pt–Sn systems [123-126]. Many of them have reported the studies conducted following different methods in different temperature ranges and using different condition of materials. For example, Tu and Rosenberg [123] conducted RT experiments with various thin–film diffusion couples using X–ray diffraction (XRD). Chromik and Cotts [127] studied the growth of product phases in the Pd–Sn multilayer thin films by differential scanning calorimetry (DSC) and XRD analysis. Sharma et al. [128] studied the growth of phases between Pd and reflowed Sn–based solders in the temperature range of 156–210 °C. Ravi and Paul [124] conducted bulk diffusion couple experiments in the range of 125–200 °C for the Pd–Sn system.

On the other hand, with respect to Pt, Meagher et al. [129] studied the solid–state diffusion–controlled growth of the phases in the Pt–Sn(Pb) solder; however, the samples with cylindrical geometry were used. This sample geometry is not suitable for the quantitative diffusion analysis since the relations for estimating various diffusion parameters are developed for planar interface. Kim and Kim [125] found the growth of product phase to be typical diffusion–controlled at Pt/liquid solder interfaces studied by them, where the product phase $PtSn_4$ was identified based on TEM analysis of Pt/$PtSn_4$ interface and XRD analysis. However, Wang and Liu [126] reported a reaction–controlled growth of the same phase, as identified using XRD analysis, in the Pt/liquid Sn couple. Both these studies [125, 126] have reported different growth mechanism of $PtSn_4$ in the liquid–state and, in addition, studies are not available in the solid–state for the Pt–Sn system, expect the one at lower temperature by Tu and Rosenberg [123].

Based on the above discussion, it is evident that previous studies have followed different materials and experimental conditions, viz. bulk or thin film, planar or cylindrical samples, solid or liquid state of Sn or solder. The previous studies in the other systems, as reported in Chapters 4 and 7, have shown the difference in growth of the product phases by changing the condition of materials (bulk and electroplated). Additionally, different studies followed completely different temperature range for their studies, leading to ambiguity about the temperature dependent growth rate of the phases, and hence it is important to cover the whole temperature range in the solid–state. Therefore, our aim in this segment of the study is to conduct experiments in the solid–state from RT to an elevated temperature, for both bulk and electroplated diffusion couples, to understand the growth mechanism of the product phases in the Pd–Sn and the Pt–Sn systems.

## 8.2    Results and Discussion

Because of the difference in the nature of growth of phases, our results and analysis are presented separately for the Pd–Sn and the Pt–Sn systems.

### 8.2.1    The Pd–Sn system

Before discussing results on the evolution of phases, it is important to refer the Pd–Sn phase diagram [130] shown in Figure 8.1. There are 11 phases in the temperature range (*i.e.*, 25 (RT) – 215 °C) of experiments, as marked with dash–dot lines on the phase diagram. In the Pd–Sn system, electroplated diffusion (EPD) couples are studied from 215 °C to RT, since a bonding between 2 dissimilar materials is expected to be created during the very initial stage of electroplating itself.

On the other hand, bulk diffusion (BD) couples are studied in the higher temperature range 215–100 °C because of the joining issues at and below 75 °C.

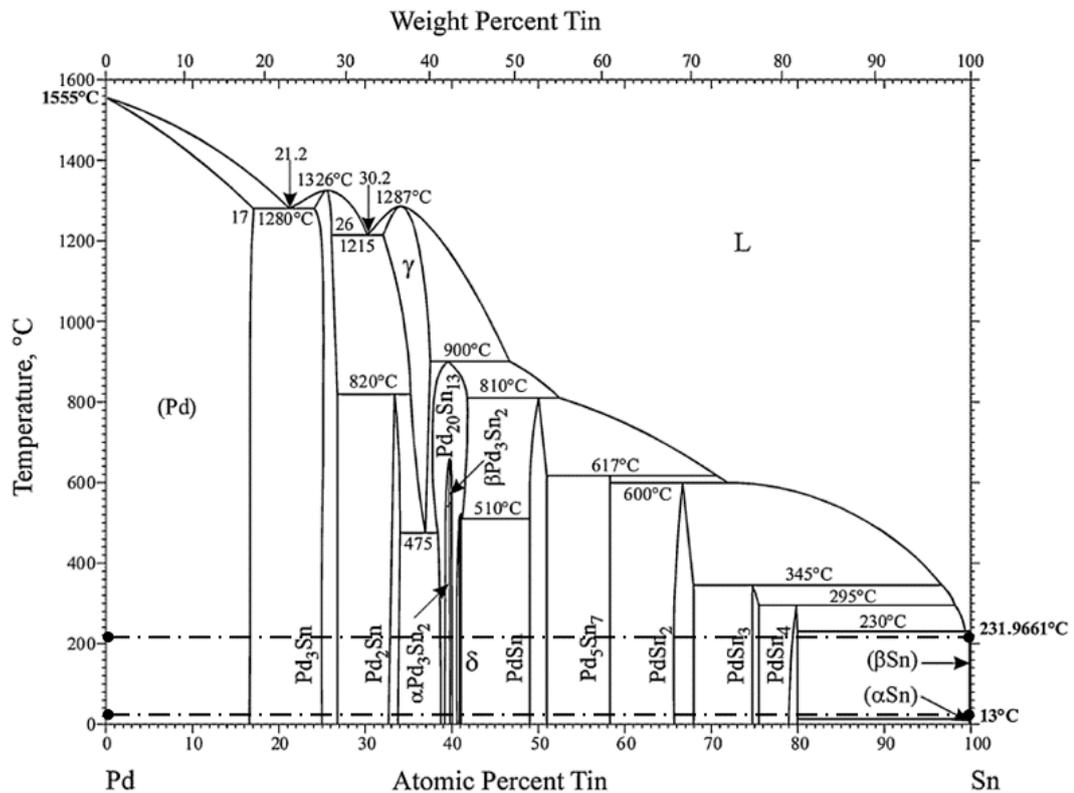

Figure 8.1: Binary Pd–Sn phase diagram adapted from H. Okamoto [130].

### 8.2.1.1 *Phase Evolution*

To discuss the growth of various phases, an interdiffusion zone (IDZ) of a Pd/Sn EPD couple annealed at 215 °C for 4 hrs is shown in Figure 8.2a, along with the focused region at Pd/IDZ in Figure 8.2b. The corresponding composition profile (measured by EPMA) developed across the IDZ is shown in Figure 8.2c. It can be seen that PdSn$_4$ grows with much higher thickness than PdSn$_3$, whereas PdSn$_2$ grows with negligible thickness. The measured compositions of Sn (in at.%) in PdSn$_4$ and PdSn$_3$ are found to be slightly higher than the stoichiometric compositions of these phases. Since the presence of all 3 phases could not be resolved in Figure 8.2a, a focused area at Pd/IDZ is shown in Figure 8.2b. A dotted line demarcating the PdSn$_3$/PdSn$_4$ interface is shown on the micrograph in Figure 8.2a. Therefore, only 3 out of 11 thermodynamically stable phases (Figure 8.1) are found in the IDZ. In a multiphase growth, a few phases are frequently not found to grow in an IDZ because of their sluggish growth kinetics [18]. In general, these missing phases are closer to the high melting point component (Pd in this case) in the phase diagram. In the Pd–Sn system, we did not find any difference in the growth rate of phases and the microstructural features of IDZ in the BD and the EPD couples in temperature range of 215–100 °C.

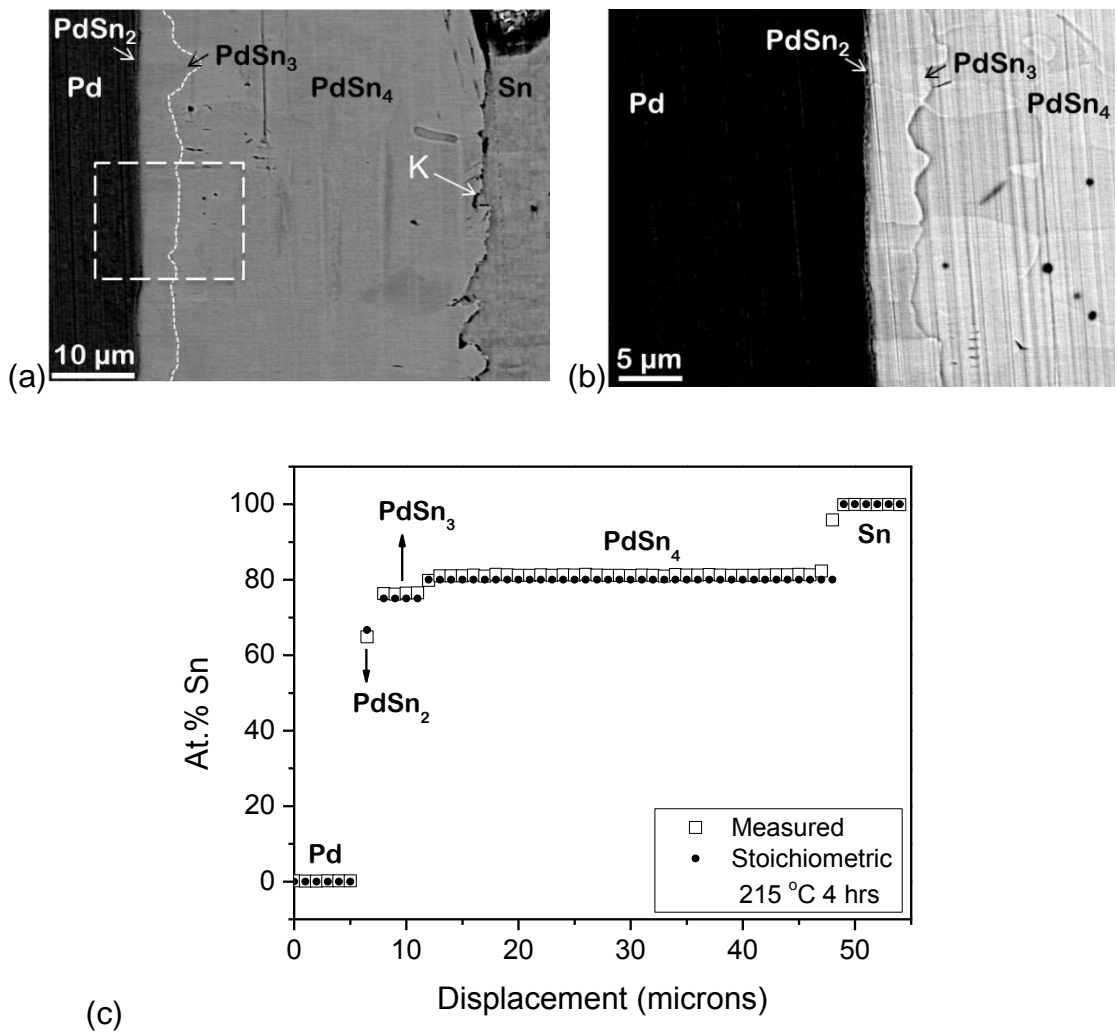

Figure 8.2: Pd/Sn electroplated diffusion couple – annealed at 215 °C for 4 hrs: (a) BSE image, where a dotted line indicates the phase boundary at PdSn₃/PdSn₄ interface. The profile shown in (c) along with the help of slight contrast visible on the computer screen is used to draw a dotted line. K denotes location of the Kirkendall marker plane; (b) adjoining BSE image of the region at Pd/IDZ interface showing PdSn₂ and PdSn₃ phases and (c) composition profile of the interdiffusion zone.

With the decrease in annealing temperature, *i.e.*, at 75 °C and below, the thickness of the product phases decreases rapidly and one such example, which is annealed at 50 °C, is shown in Figure 8.3a. With the aim of growing the phases with higher thickness, this couple is annealed for a longer period, *i.e.*, 400 hrs. In this couple also, the same 3 phases (*i.e.*, PdSn$_4$, PdSn$_3$ and PdSn$_2$) are found to grow; however, it is difficult to resolve the PdSn$_3$/PdSn$_4$ interface in the micrograph. The same trend follows even at RT, as shown in Figure 8.3b, which was stored for 2.5 years (912 days). At this temperature, no phase layer was found to grow after 30 days indicating an incubation period before the start of the growth of phases. To cross–check this observation, one experiment was conducted at 50 °C for a shorter annealing time of 36 hrs and a similar observation (*i.e.*, no growth of phase layer) was noticed in the Pd/Sn EPD couple, which indicates the presence of an incubation period for the growth of phases at lower temperatures in the Pd–Sn system.

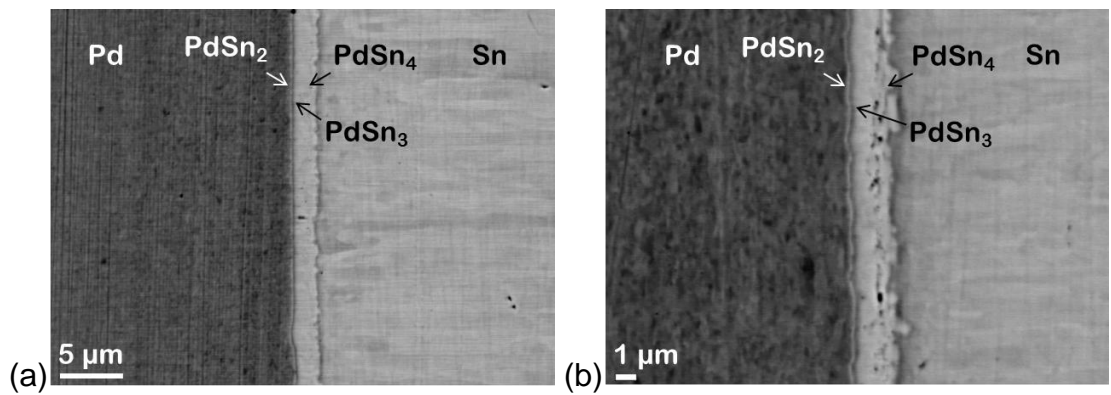

Figure 8.3: BSE images of the interdiffusion zone in Pd/Sn electroplated diffusion couple after (a) annealing at 50 °C for 400 hrs and (b) storage for 2.5 years.

### 8.2.1.2 *Parabolic Growth and Diffusion Parameters*

For ascertaining the parabolic growth of the product phases, time dependent growth kinetics experiments are conducted for 4–49 hrs at 200 °C in Pd/Sn couples, as shown in Figure 8.4a. The data are plotted following Equation (4.1) and only for the PdSn$_4$ phase since the thickness of other two phases is very small and it is difficult to make any meaningful plot without very high error. A linear fit of the data indicates the parabolic nature of growth, *i.e.*, diffusion–controlled growth of the PdSn4 phase. As mentioned earlier, we have found similar growth kinetics of the product phases in the Pd/Sn BD and EPD couples in this study. This means that the growth of phases in the Pd–Sn system is not affected by the condition of Sn being bulk or electroplated.

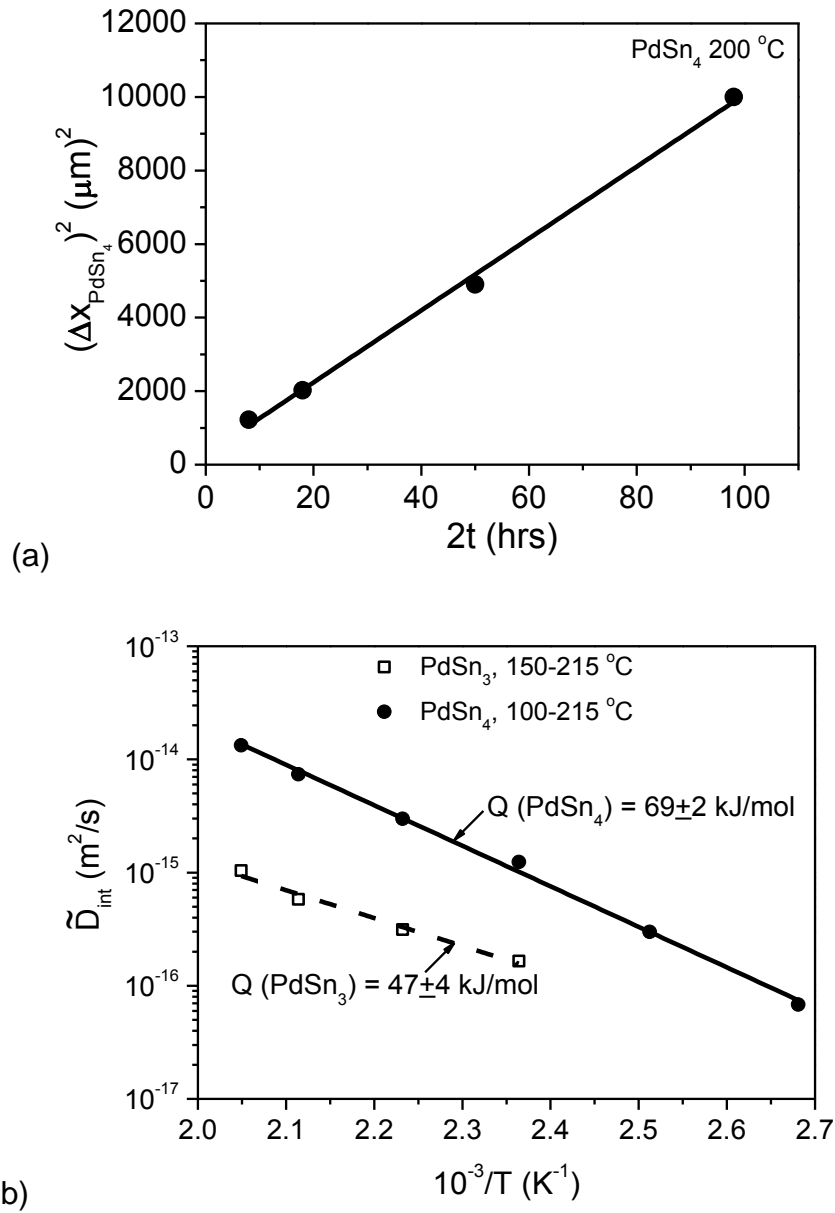

(a)

(b)

Figure 8.4: Pd/Sn electroplated diffusion couples: (a) Parabolic growth law followed by the PdSn4 phase at 200 °C, and (b) Arrhenius plot for integrated interdiffusion coefficients in PdSn4 and PdSn3.

Since parabolic growth constant $k_p$ depends on the end–member compositions and is not a material constant (unlike diffusion parameters), the discussion on diffusion–controlled growth should be done based on the estimation of diffusion parameters. As the phases in this system are known to have a very narrow homogeneity range, the integrated interdiffusion coefficients of a component B in a phase can be estimated following the relation developed by Wagner [26], *i.e.*, Equation (3.33). The molar volume of the phases are $V_m^{PdSn_4} = 14.18 \times 10^{-6}$ m³/mol and $V_m^{PdSn_3} = 13.52 \times 10^{-6}$ m³/mol [124]. Since the PdSn₂ phase layer thickness is negligible in all Pd/Sn couples, it is almost impossible to estimate $\widetilde{D}_{int}^{PdSn_2}$. The estimated $\widetilde{D}_{int}^{PdSn_4}$ and $\widetilde{D}_{int}^{PdSn_3}$ are plotted in Figure 8.4b with respect to the Arrhenius relation. It can be seen that $\widetilde{D}_{int}^{PdSn_4}$ is higher than $\widetilde{D}_{int}^{PdSn_3}$, which is directly related to the higher growth rate of PdSn₄ phase as compared to that of PdSn₃ phase. The $Q$ values are estimated as 47±4 and 69±2 kJ/mol for the PdSn₃ (150–215 °C) and PdSn₄ (100–215 °C) phases, respectively. Note that these are estimated in EPD couples. $\widetilde{D}_{int}^{PdSn_3}$ is not estimated below 150 °C, whereas, $\widetilde{D}_{int}^{PdSn_4}$ is not estimated below 100 °C to avoid error in the estimation of data because of small thickness of the product phases and also due the presence of incubation period.

### *8.2.1.3        Growth mechanism of the PdSn₄ phase*

To gain further insights on the growth mechanism, inert Y₂O₃ particles were used as the Kirkendall markers prior to annealing at 200 °C in Pd/Sn BD couples. They are found almost at the PdSn₄/Sn interface. A closer examination revealed the presence of a line of pores very close to the PdSn₄/Sn interface in various Pd/Sn EPD couples, which also indicates the location of the Kirkendall marker plane [18]. These findings indicate that the PdSn$_4$ phase grows mainly due to diffusion of Sn, and negligible diffusion of Pd, through this phase in both EPD and BD couples.

During the process of interdiffusion, components could diffuse via both the lattice and the grain boundaries, and therefore, we actually measure an apparent diffusion coefficient. The relatively low value of the activation energy calculated from $\tilde{D}_{int}$ indicates a dominant role of the grain boundary diffusion. However, the contribution of the lattice diffusion cannot be ignored since the experiments are conducted at high homologous temperature, $T/T_m$, where $T_m$ is the melting point of the phase of interest. For example, the melting point of PdSn$_4$ is 295 °C (568 K) [130]. Therefore, the value of the activation energy for diffusion in this phase is estimated in the homologous temperature range of 0.66−0.86. In this range, the concentration of point defects must be sufficient to facilitate significant diffusion through lattice [18].

To explain the above inference further, PdSn$_4$ has a crystal structure of oC20, similar to the AuSn$_4$ phase (Figure 7.10), in which it has several Sn–Sn bonds and no Pd–Pd bonds. Therefore, following the sublattice diffusion mechanism, Sn can easily diffuse via its own sublattice by exchanging position with the available vacancies. However, Pd cannot diffuse simply by exchanging position with a vacancy on the sublattice of Sn (since all Pd atoms are surrounded by Sn only), because it will then move to the sublattice of Sn, which is not allowed. This is possible only if the Pd antisites are present, which could be occupied on the sublattice of Sn. The negligible diffusion rate of Pd as compared to Sn indicates the absence of these antisite defects. In fact, measured composition profile in this phase, as shown in Figure 8.2c, indicates a positive deviation from stoichiometry towards Sn–rich side, indicating the presence Sn–antisites rather than Pd–antisites to compensate the off–stoichiometry composition [18]. This will increase the diffusion rate of Sn. Additionally, it is evident from the location of the Kirkendall marker plane that the diffusion rate of Pd via grain boundaries is negligible, unlike the diffusion of Au in the $AuSn_4$ phase (as reported in Chapter 7), which has the same crystal structure (*i.e.*, oC20). For the Au–Sn system, as discussed earlier in Chapter 7, quantitative diffusion analysis in $AuSn_4$ indicates that Sn could diffuse via both lattice and grain boundaries, whereas Au could diffuse with significant rate via only grain boundaries.

### 8.2.2   The Pt–Sn system

In the Pt–Sn system, BD couples are studied in the higher temperature range 215–175 °C because of the joining issues at 150 °C and below, while EPD couples are studied from 215 °C to RT. Before discussing results, it is important to refer the Pt–Sn phase diagram [131] shown in Figure 8.5. There are 5 phases, $Pt_3Sn$, $PtSn$, $Pt_2Sn_3$, $PtSn_2$ and $PtSn_4$, in the temperature range (*i.e.*, 125–215 °C) of successful experiments, as marked with dashed lines on the phase diagram.

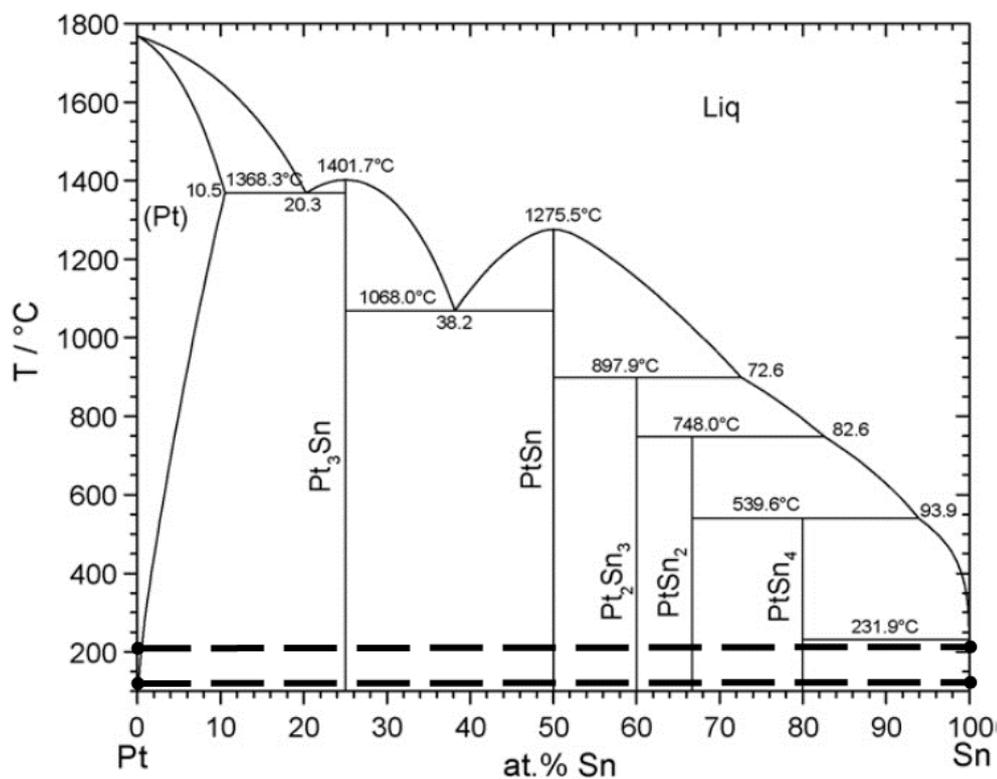

Figure 8.5: Binary phase diagram of the Pt–Sn system
adapted from V. Grolier and R. Schmid–Fetzer [131].

#### 8.2.2.1 *Phase Evolution*

To discuss the difference in growth behaviour of the product phase, the BSE images of Pt/Sn couples, both BD and EPD couples, annealed at 200 °C for 100 hrs are shown in Figure 8.6a and b, respectively. Out of 5 intermetallic compounds present in the phase diagram (Figure 8.5), only PtSn$_4$ grows in the IDZ. A uniform layer of PtSn$_4$ is grown in the BD couple. On the other hand, in the EPD couple, IDZ has two parts; a single phase layer of PtSn$_4$ and a zone of two phase mixture of PtSn$_4$ and Sn. There is insignificant difference in the thickness of the single and the uniform phase layers of PtSn$_4$ in EPD and BD couples, respectively.

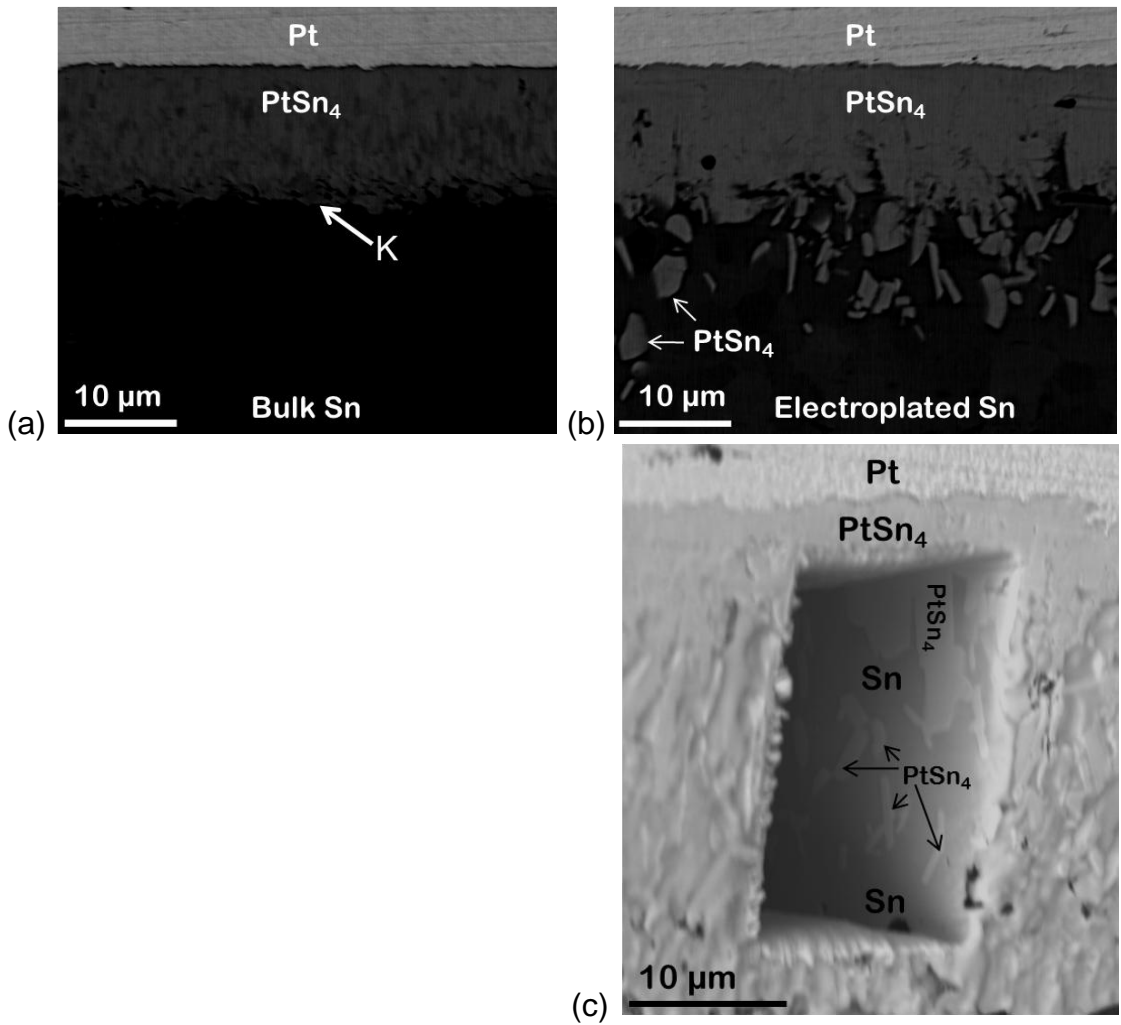

Figure 8.6: BSE images of the Pt/Sn diffusion couple annealed at 200 °C for 100 hrs: (a) bulk couple, where K denotes the location of marker plane, (b) electroplated couple and (c) a phase mixture region from electroplated couple polished in FIB showing growth of isolated PtSn$_4$ phase in Sn.

To gain further insights into the growth behaviour, this phase mixture region was milled in FIB along with some part of a single phase layer. The presence of isolated $PtSn_4$ phase in Sn end–member is evident from the micrograph shown in Figure 8.6c. At lower temperatures, these additional microstructural features are not found. A very similar phenomena is also reported in Chapter 4 for the Cu–Sn and the Ni–Sn systems. It is already explained in Chapter 4 that, based on thermodynamics point of view, a phase mixture is not allowed to grow in a binary system; however, it is allowed to grow in a ternary or multicomponent system [18]. Since this is not found in BD couple and found in EPD couple, it is evident that the addition of impurities during electroplating plays a role behind the growth of this phase mixture. Previously, the presence of very small concentration of impurities influencing the growth of the product phases is shown in another system by van Loo [29]. A phase mixture evolves, when supersaturated impurities are left after the diffusion of Sn towards Pt/IDZ interface and if these are not completely dissolved in the product phase. This microstructural feature is not found in the Pd–Sn system, which indicates that the impurities are dissolved in the product phase in this system. Since several different organic and inorganic impurities are added during electroplating with very low concentration, it is impossible to detect the impurities which are responsible for this.

With the decrease in annealing temperature, the growth rate of the product phase decreases drastically. As shown in Figure 8.7, the thickness of the product phase is negligible when annealed at 100 °C even after annealing for much longer time of 1000 hrs. Therefore, as expected, at RT no growth is found even after the maximum ageing time of 2.5 years.

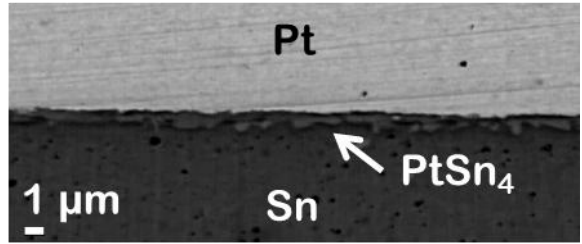

Figure 8.7: BSE image of the Pt/Sn electroplated diffusion couple annealed at 100 °C for 1000 hrs.

#### 8.2.2.2 *Linear Growth of the PtSn₄ phase and its growth mechanism*

Time dependent experiments are conducted at 200 °C for 25–625 hrs in Pt/Sn BD couples. Thickness ($\Delta x$) and time ($t$) are plotted in Figure 8.8. A linear fit of the data indicates the reaction–controlled growth of the PtSn₄ phase. Similar reaction–controlled growth nature has been reported earlier by Wang and Liu [126] for reaction of Pt with liquid Sn.

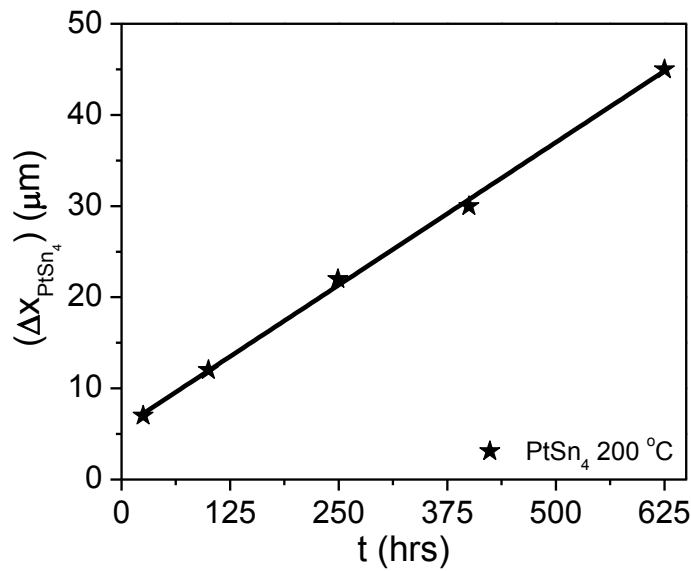

Figure 8.8: Linear growth law followed by the PtSn₄ phase at 200 °C.

It indicates that, unlike the Pd–Sn and the Au–Sn systems, the formation of the product phase is the rate–controlling process in the Pt–Sn system instead of the diffusion rates of components, which is rather common in most of the systems. In such a situation, we cannot estimate the diffusion parameters; however, the marker experiment can still shed light on the relative diffusion rates of components. Hence, to gain further insights on the growth mechanism, inert $TiO_2$ particles were used as markers at the contact interface of Pt/Sn prior to annealing at 200 °C in BD couples. The location of marker plane after the annealing treatment is found almost at the $PtSn_4$/Sn interface. This indicates that the growth of $PtSn_4$ phase would be mainly by the diffusion of Sn atoms (faster diffusing species), and diffusion of Pt atoms through the $PtSn_4$ phase layer could be considered as negligible. Therefore, similar to the $PdSn_4$ phase with the same crystal structure, it is evident from the location of marker plane that Pt diffusion rate (via grain boundaries) is very negligible, which is stated following the same arguments as discussed earlier for the Pd–Sn system.

### 8.2.3 Comparison of Au–Sn, Pd–Sn and Pt–Sn systems

For comparison, the time dependent growth of total IDZ at 200 °C in Au–Sn (from Chapter 7), Pd–Sn and Pt–Sn systems are plotted together in Figure 8.9. As shown in Figure 7.2a, 4 phases $AuSn_4$, $AuSn_2$, AuSn and $Au_5Sn$ grow in the Au–Sn system at this temperature. Similarly, as discussed in this chapter, 3 phases $PdSn_4$, $PdSn_3$ and $PdSn_2$ grow in the Pd–Sn system and only one phase $PtSn_4$ grows in the Pt–Sn system. It can be seen in Figure 8.9 that there is not much difference in the growth rate of total IDZ in the Pd–Sn and the Au–Sn systems, with slightly higher growth rate in the former system. On the other hand, the growth rate is much slower in the Pt–Sn system. The shape of the curves is different since the product phases in the Pd–Sn and the Au–Sn systems grow parabolically with time (*i.e.*, diffusion–controlled), whereas the product phase in the Pt–Sn system grows linearly with time (*i.e.*, reaction–controlled).

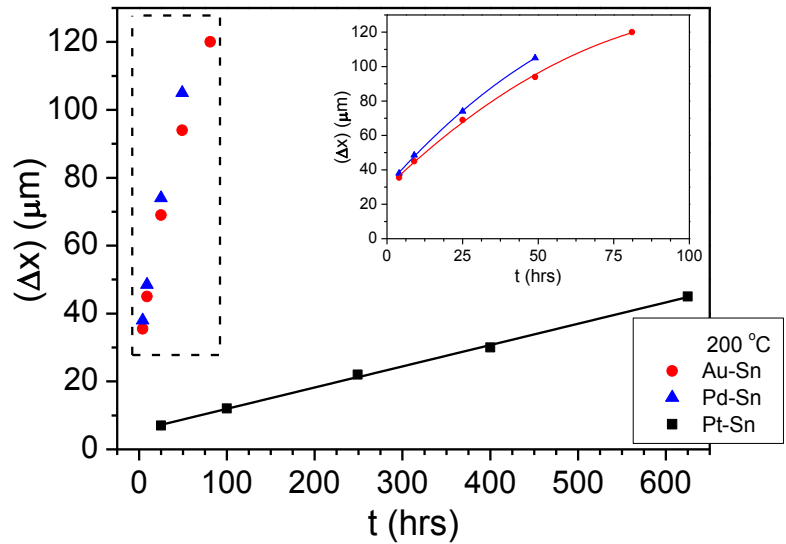

Figure 8.9: Comparison of time dependent growth of the total interdiffusion zone in the Au–Sn, Pd–Sn and Pt–Sn systems by considering the bulk diffusion couples annealed at 200 °C.

Since the growth rate is reasonably high in the Au–Sn and the Pd–Sn systems even during RT storage and the growth of brittle intermetallic compounds is unwanted because of reliability concern, Pt could suitably replace Au in flip–chip bonding. However, the above statement is true only if these noble metals are not dissolved in Cu. The study reported in Chapter 9 indicates that the presence of Pd or Pt in Cu increases the growth of the Kirkendall voids in the $Cu_3Sn$ phase, which are deleterious for reliability of flip–chip bonding. On the other hand, we found in Chapter 9 that the presence of Au in Cu does not have significant effect on the growth of the Kirkendall voids. This study should be complemented with detailed analyses of electrical and mechanical properties for a right selection of a material used as oxidation and corrosion protection during shelf–life.

## 8.3 Conclusions

The growth of the phases is studied in the Pd–Sn and the Pt–Sn systems, which are being considered as possible replacement of the Au–Sn system in the microelectronic packages. For the very first time, experiments covering the whole temperature range in the solid–state (*i.e.*, RT to 215 °C) are conducted in these two systems, to observe the temperature dependent growth rate of phases. A few key observations of this segment of the study can be stated as follows:

(i)     In the Pd–Sn system, 3 phases, namely $PdSn_4$, $PdSn_3$ and $PdSn_2$, are found to grow in EPD and BD couples. Mainly $PdSn_4$ covers the whole IDZ. The growth of this product phase is diffusion–controlled.

(ii)    In the Pt–Sn system, $PtSn_4$ is the only phase found to grow in EPD and BD couples. The growth of this product phase is reaction–controlled.

(iii)   The analysis in $PdSn_4$ and $PtSn_4$ phases indicate that the diffusion rates of Pd and Pt are negligible and these product phases grow mainly by the diffusion of Sn.

(iv)    The product phases in the Pd–Sn system grow at very fast rate (which is similar to the Au–Sn system) as compared to the Pt–Sn system. Additionally, the growth rate of product phase, namely $PtSn_4$, in the Pt–Sn system is negligible at or below 100 °C and it does not grow during storage at RT.

*To summarize, the temperature dependent growth rates of IDZ (interdiffusion zone) in the Pd–Sn and the Pt–Sn systems is studied in this chapter and compared with the Au–Sn system. The growth rate of IDZ is linear (which is uncommon) in the Pt–Sn system, while it is found to be parabolic (which is most common) in the Pd–Sn and the Au–Sn systems. At all temperatures, Pt/Sn IDZ has slower growth than Pd/Sn IDZ and Au/Sn IDZ.*

# Chapter 9

# Effect of Au, Pd and Pt addition in Cu on the growth of intermetallic compounds and the Kirkendall voids in the Cu–Sn system

In previous two chapters, we have discussed the growth of phases in the binary Au–Sn, Pd–Sn and Pt–Sn systems. In this chapter, the effect of the addition of Au, Pd and Pt in Cu on the growth of the Kirkendall voids and the product phases in the Cu–Sn system is studied and it is compared to that of the effect of addition of Ni in Cu on the same aspects.

## 9.1    Introduction and Statement of the Problem

As we already mentioned in Chapter 1, Cu, Ni and Au are integral parts of under bump metallization (UBM) in the flip–chip bonding. Cu produces good bonding by forming $Cu_3Sn$ and $Cu_6Sn_5$ compounds during soldering with a Sn–based alloy. Because of high growth rate of these phases, Ni is added as a barrier layer. A very thin layer of Au is used for the protection of other layers of UBM from corrosion and oxidation during storage. Pd and Pt are considered as replacement of Au.

It is a well–known fact that the presence of other components might affect the growth of the Kirkendall voids and the growth kinetics of phases in the Cu–Sn system significantly. Many studies related to these aspects are conducted with the different alloying additions in Sn or Sn–based solders [7]; however, limited studies are done to examine the effect of the addition of third element in Cu on the formation of the Kirkendall voids and the growth of the product phases. During soldering and service at an elevated temperature, alloys of Cu are developed because of interdiffusion

---

This chapter is written based on the article:
[1] V.A. Baheti, P. Kumar, A. Paul: Effect of Au, Pd and Pt addition in Cu on the growth of intermetallic compounds and the Kirkendall voids in the Cu–Sn system, Journal of Materials Science: Materials in Electronics 28(22) (2017) 17014-17019.

between the UBM layers. Previously, we have reported very strong effects of impurities in Cu on the growth of the Kirkendall voids in $Cu_3Sn$ (in Chapter 5), and addition of Ni in Cu on the growth of voids as well as the growth kinetics of phases (in Chapter 6). Therefore, the aim of this segment of the study is to examine the role of Au, Pd and Pt addition in Cu on the growth kinetics of phases and the formation of the Kirkendall voids, which are very important for reliability of flip–chip bonding. To the best of our knowledge, such a study has never been reported in the open literature.

## 9.2    Results and Discussion

Here, the effects of addition of a few important metallic components used in UBM (such as Au, Pd and Pt) on the growth of the product phases and the formation of the Kirkendall voids in the Cu–Sn system is reported. For the comparison of these two aspects with the effect of addition of Ni in Cu, the results in the Cu(Ni)–Sn system are considered. Following, the interdiffusion coefficients are estimated for a better comparison of the diffusion–controlled growth rates of the phases.

### 9.2.1    Growth of phases in the interdiffusion zone (IDZ)

#### *9.2.1.1          Addition of Au in Cu*

Figure 9.1 shows a comparison of the growth of phases for different contents of Au in Cu in the Cu(Au)/Sn diffusion couples annealed at 200 °C for 81 hrs. As shown in Figure 9.1a, two phases $Cu_3Sn$ and $Cu_6Sn_5$ grow in IDZ of binary Cu/Sn couple. It can be seen in Figure 9.1b and c that both the phases grow for Au content up to 5 at.%. There is a slight decrease in the thickness of $(Cu,Au)_3Sn$ and increase in thickness of $(Cu,Au)_6Sn_5$ upon addition of Au up to 5 at.% such that there is an increase in thickness of the total IDZ. When the Au content is increased to 8 at.%, as shown in Figure 9.1d, a thin layer of ternary phase (T1 with ~20 at.% Sn) is grown in IDZ next to Cu(8Au) end–member. This is clearer in the secondary electron (SE) image shown in Figure 9.1e. Therefore, the diffusion path [18] consists of Cu(8Au)→T1→(Cu,Au)$_3$Sn→(Cu,Au)$_6$Sn$_5$→Sn, which can be correlated with the Au–Cu–Sn ternary phase diagram shown in Figure 9.2 [132]. The other two phases, (Cu,Au)$_3$Sn and (Cu,Au)$_6$Sn$_5$, have the similar range of Sn content as in the binary phases.

### 9.2.1.2        *Addition of Pd in Cu*

Figure 9.3 shows the effect of Pd addition in Cu on the growth kinetics of Cu$_3$Sn and Cu$_6$Sn$_5$ phases and the growth of the Kirkendall voids in the Cu(Pd)/Sn diffusion couples annealed at 200 °C for 400 hrs. A higher annealing time was used in this system as compared to Cu(Au)/Sn, since otherwise, we had difficulties to resolve or measure compositions of few phases in EPMA. Note here that the difference in the thicknesses of Cu$_3$Sn and Cu$_6$Sn$_5$ phases in the Cu/Sn diffusion couple, shown in Figures 9.1a and 9.3a, is because of the difference in annealing times. There is not much difference in the growth kinetics of phases because of the addition of 1 at.% Pd compared to the binary Cu–Sn system. However, the thickness of (Cu,Pd)$_6$Sn$_5$ increases and (Cu,Pd)$_3$Sn decreases when 5 at.% Pd is added in Cu; though there is not much change in the thickness of the total IDZ. The (Cu,Pd)$_3$Sn phase is not present when 8 at.% Pd is added in Cu and the IDZ consists of a phase mixture of (Cu,Pd)$_6$Sn$_5$ and discrete islands of (Pd,Cu)Sn$_4$. To confirm this trend, one experiment was conducted with 15 at.% Pd in Cu and a similar IDZ, as of 8 at.% Pd in Cu, with respect to the presence of phases is found. However, the volume fraction of (Pd,Cu)Sn$_4$ and the thickness of the overall IDZ is found to increase drastically.

### 9.2.1.3        *Addition of Pt in Cu*

In the Cu(Pt)–Sn system, because of failure of the experiments, we could produce diffusion couple only with 1 at.% Pt in Cu, which is shown in comparison with binary Cu/Sn diffusion couple in Figure 9.4. It can be seen that there is not much difference in the thickness of $(Cu,Pt)_3Sn$; however, there is almost 50% increase in the thickness of $(Cu,Pt)_6Sn_5$ as compared to the binary system.

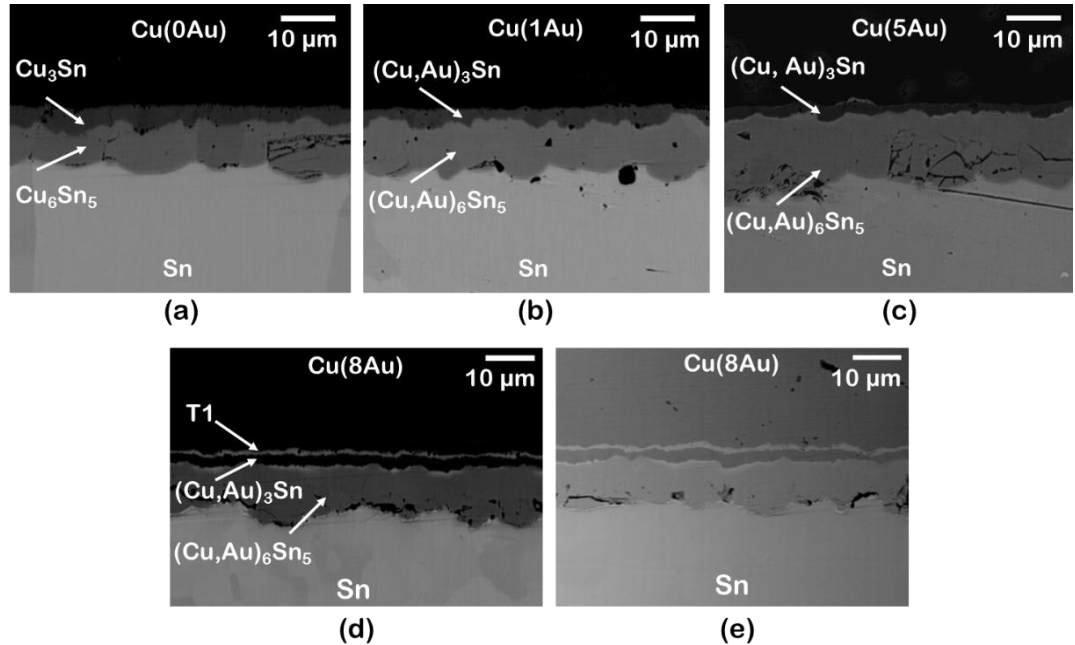

Figure 9.1: Interdiffusion zone of the diffusion couple after annealing at 200 °C for 81 hrs: (a) Cu(0Au)/Sn, (b) Cu(1Au)/Sn, (c) Cu(5Au)/Sn, (d) BSE and (e) SE image of Cu(8Au)/Sn.

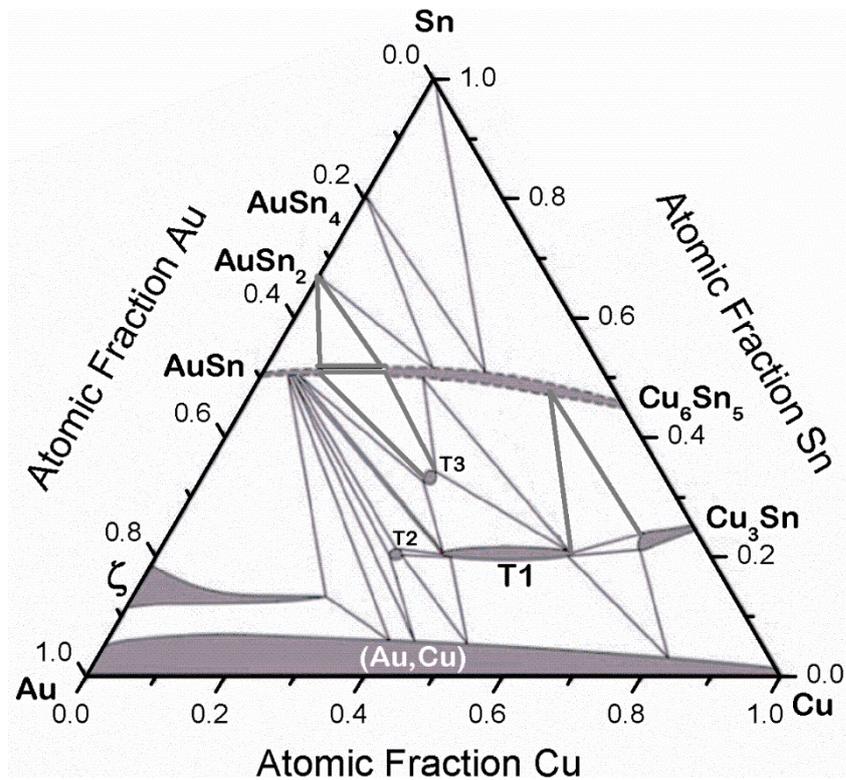

Figure 9.2: The Cu–Au–Sn phase diagram at 200 °C [132].

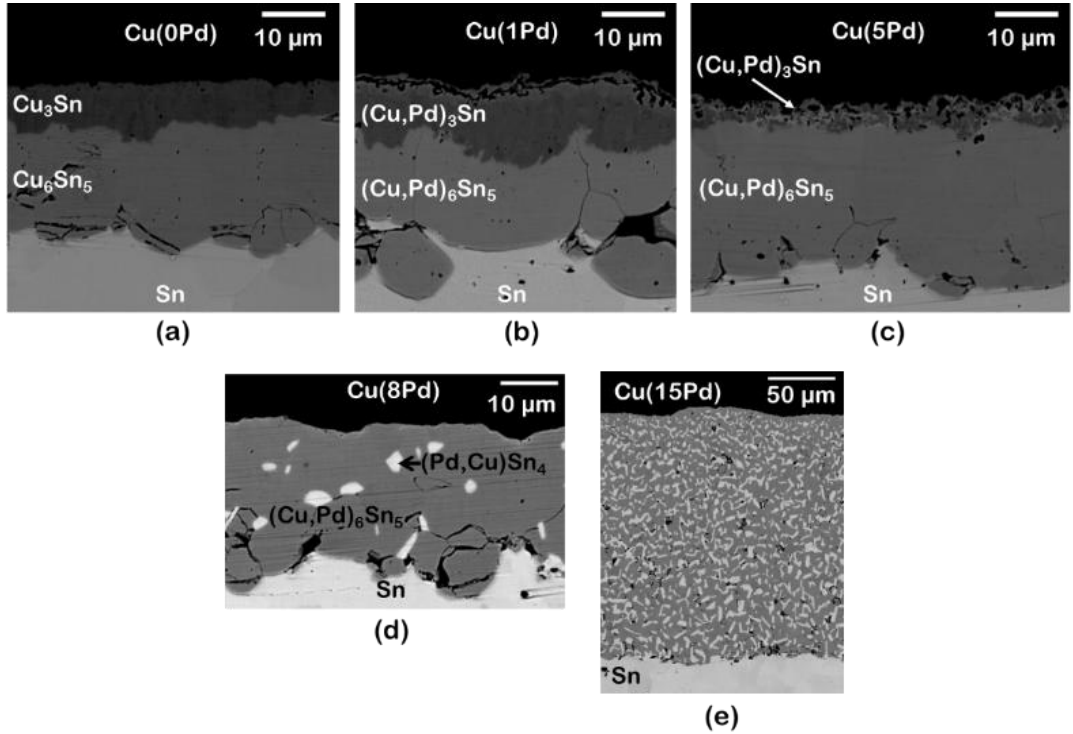

Figure 9.3: Interdiffusion zone of the diffusion couple after annealing at 200 °C for 400 hrs: (a) Cu(0Pd)/Sn, (b) Cu(1Pd)/Sn, (c) Cu(5Pd)/Sn, (d) Cu(8Pd)/Sn and (e) Cu(15Pd)/Sn. The bright phase is (Pd,Cu)Sn$_4$ and the light phase is (Cu,Pd)$_6$Sn$_5$, in the micrographs (d) and (e).

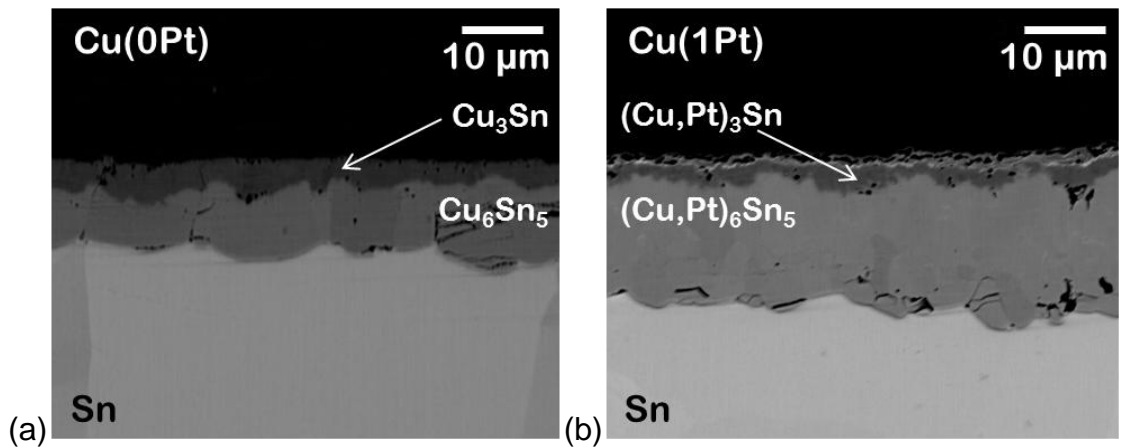

Figure 9.4: Interdiffusion zone of the diffusion couple after annealing at 200 °C for 81 hrs: (a) Cu(0Pt)/Sn and (b) Cu(1Pt)/Sn.

We have considered the Cu(Ni)–Sn system for further analysis and comparison of the growth of phases and the formation of voids. The growth of phases is shown in Chapter 6. We found that there is decrease in the growth rate of (Cu,Ni)$_3$Sn and an exceptional increase in the growth rate of (Cu,Ni)$_6$Sn$_5$ with the addition of Ni in Cu. Beyond 2.5 at.% Ni, only the (Cu,Ni)$_6$Sn$_5$ phase grows in IDZ.

### 9.2.2    Growth of the Kirkendall voids in the $Cu_3Sn$ phase

One of the problems of the Cu–Sn system is recognized as the formation of the Kirkendall voids, which is a major concern for electro–mechanical contact in the flip–chip bonding. These voids are grown because of a very high difference in the diffusion rates of components (in opposite directions) leading to the generation of net flux of vacancies, which are not consumed by sinks [18]. The voids are found in the $Cu_3Sn$ phase, in which (as estimated in Chapter 5) Cu has $30\pm10$ times higher diffusion rate than Sn, in the Cu–Sn binary system. The role of different factors on the growth of these voids is described extensively in Chapter 5. In the analysis reported in Chapter 5, we found that the purity of Cu has a strong influence on the growth of these voids. When 99.999 wt.% Cu is coupled with Sn, the growth of voids is not much and the voids are not distributed evenly. As shown in Figure 9.5a, the image is captured particularly in an area in which voids are found for this case. At other places, the $Cu_3Sn$ phase layer is relatively free of voids. Since the growth of voids is strongly related to the presence of impurities, it is well possible that the impurities in 99.999 wt.% Cu (used in this study) are not distributed uniformly. Only the $Cu_3Sn$ phase is shown in Figure 9.5, since our focus is to compare the growth of the Kirkendall voids in this phase.

As shown in Figure 9.5b, with the addition of 1 at.% Au in Cu, the voids are again found to be distributed unevenly similar to 99.999 wt.% Cu and we did not find a significant difference in the growth of the Kirkendall voids. Note that the same purity of Cu is used to produce all the alloys. However, as shown in Figure 9.5c and d, the growth of voids increases significantly upon addition of 1 at.% Pd in Cu and 1 at.% Pt in Cu, respectively. Since the annealing time for Cu(Pd)/Sn experiments reported in Figure 9.3 are different than the other systems considered for the analysis of the product phases in IDZ, one experiment using Cu(1Pd)/Sn is conducted at 200 °C for 81 hrs for comparison of voids with other systems (Figure 9.5c). As expected, by comparing Figure 9.3b (400 hrs) and Figure 9.5c (81 hrs), it is clear that the growth of voids decreases with the decrease in annealing time. For comparison, the growth of voids in Cu(0.5Ni)/Sn couple from Section 6.4 is shown in Figure 9.5e. Although the impurity content (in wt.%) is different in Cu, the growth rate of voids after 81 hrs in the case of 1 at.% Pd in Cu is comparable to the 0.5 at.% Ni in Cu for the same annealing time. This indicates that different impurities (depending on their content in Cu) can play a different role on the growth rate of the Kirkendall voids in the $Cu_3Sn$ phase.

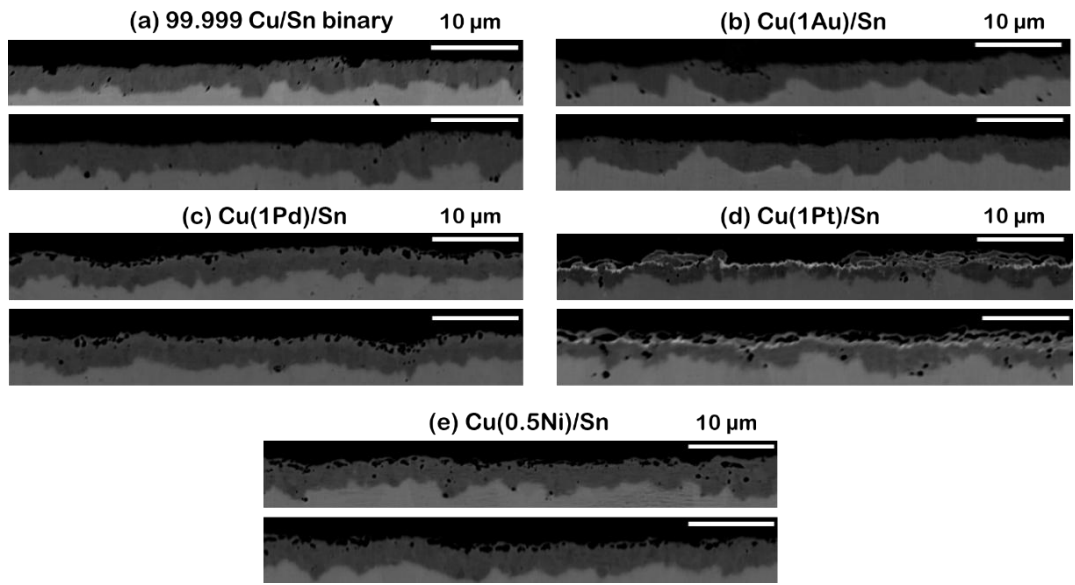

Figure 9.5: BSE micrographs of $(Cu,M)_3Sn$ for comparison of the growth of the Kirkendall voids (visible as dark spots) in diffusion couples annealed at 200 °C for 81 hrs: (a) Cu/Sn, (b) Cu(1Au)/Sn, (c) Cu(1Pd)/Sn, (d) Cu(1Pt)/Sn and (e) Cu(0.5Ni)/Sn. Purity is same for all the cases, *i.e.*, 99.999 wt.% Cu, 99.99 wt.% Sn, and 99.95 wt.% M = Au, Pd, Pt and Ni. The whole interdiffusion zones are shown in previous figures.

It is speculated for a very long time that the growth of these voids are very strongly related to the incorporation of S and other organic compounds when Cu is deposited by electroplating [58]. Occasionally, it is commented that these are not the Kirkendall voids but the voids created because of the presence of organic components in electroplated Cu [59]. We have shown before in Chapter 5 that the growth of voids increases drastically for commercially pure (CP) Cu in which impurities are present in high concentration (1–2 wt.%) and almost equivalent to the impurity present in electroplated (EP) Cu. In this study, we have shown that even the addition of the third inorganic metallic component in Cu, such as Pd, Pt and Ni, increases the growth of voids significantly. This means that the growth of these voids must be influenced by the presence both types of impurities, *i.e.*, inorganic and organic.

### 9.2.3 Estimation of the interdiffusion coefficients

Because of the difference in growth of phases in different Cu(M)–Sn systems, these should be compared based on the estimation of the interdiffusion coefficients. Since the phases grow with narrow homogeneity range and therefore it is difficult to estimate the concentration gradient from the measured composition profiles, we can estimate the integrated interdiffusion coefficient, $\widetilde{D}_{int}$. Our EPMA point analysis indicates that both $Cu_3Sn$ and $Cu_6Sn_5$ maintain the stoichiometric composition irrespective of the metallic alloying additions (M = Au, Pd, Pt, Ni) such that (Cu+M):Sn ≡ 3:1 in the $(Cu,M)_3Sn$ phase and (Cu+M):Sn ≡ 6:5 in the $(Cu,M)_6Sn_5$ phase. It means that the alloying components replace Cu in the product phase. In a ternary system, three values of $\widetilde{D}_{int}$ can be determined, each for one component in a particular phase [18]. However, since the concentration of Cu and M varies depending on the composition of the Cu(M) alloy, one can logically compare the diffusion data estimated using the composition profile of Sn, in a similar manner as it is done for the $(Cu,Ni)_6Sn_5$ phase following Equation (6.2). We estimate these values in the $(Cu,M)_6Sn_5$ phase only, since it grows with reasonable thickness in all the couples. These are estimated in diffusion couples up to 5 at.% Au and Pd, since only (Cu,M)$_3$Sn and (Cu,M)$_6$Sn$_5$ grow in IDZ. For the purpose of comparison, we include the previously estimated data in the Cu(Ni)–Sn system from Chapter 6 and a single data estimated for 1 at.% Pt in the Cu(Pt)–Sn system. The molar volume of the phases are (considered same as binary phases) $V_m^{Cu_3Sn} = 8.59{\times}10^{-6}$ m$^3$/mol and $V_m^{Cu_6Sn_5} = 10.59{\times}10^{-6}$ m$^3$/mol [17]. The estimated data are shown in Figure 9.6. As expected from the change in growth rate of phases (and total IDZ), we have $\widetilde{D}_{int}^{(Cu,Ni)_6Sn_5,Sn} > \widetilde{D}_{int}^{(Cu,Pt)_6Sn_5,Sn} > \widetilde{D}_{int}^{(Cu,Au)_6Sn_5,Sn} > \widetilde{D}_{int}^{(Cu,Pd)_6Sn_5,Sn}$. In addition, $\widetilde{D}_{int}$ increases monotonically with the concentration of M in Cu.

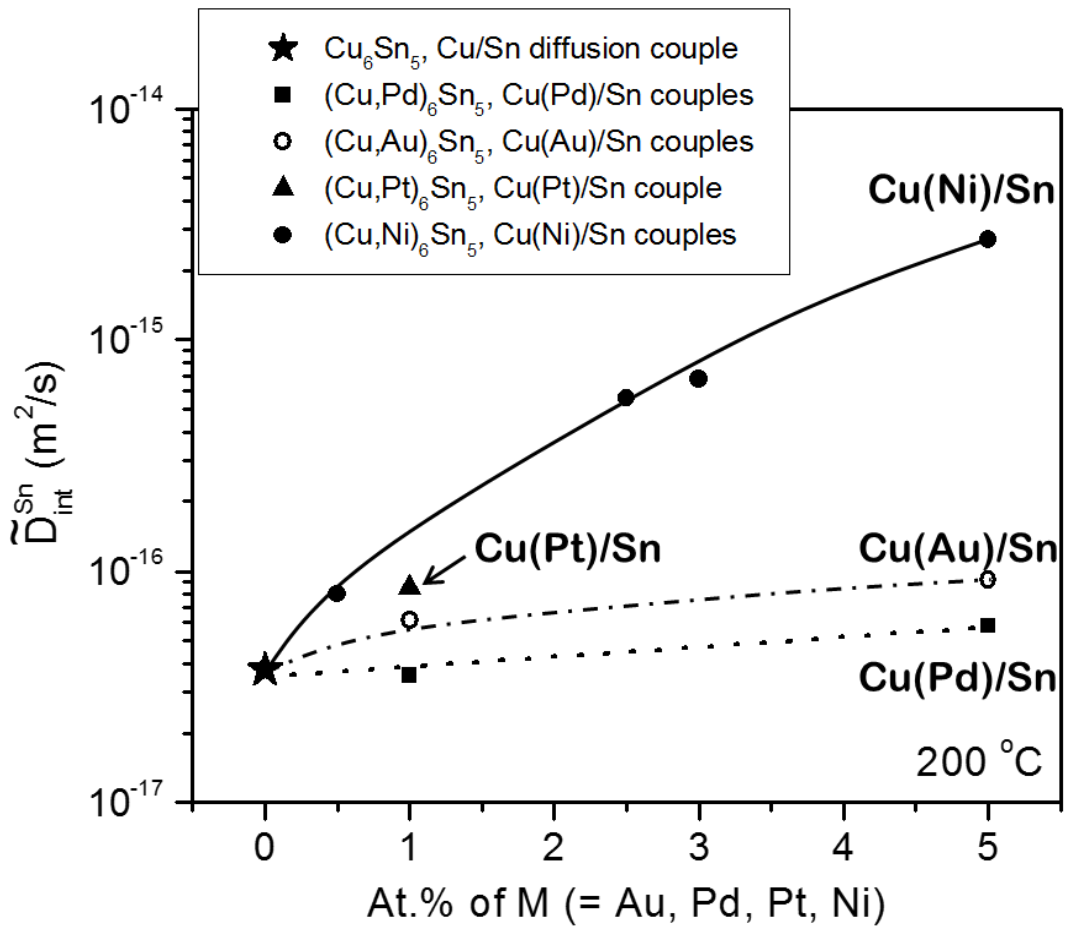

Figure 9.6: Variation of integrated interdiffusion coefficients of Sn as a function of M (= Au, Pd, Pt, Ni) content for the product phase Cu$_6$Sn$_5$ at 200 °C.

A strong influence of Ni content in terms of the increase in integrated interdiffusion coefficient of Sn with the increase in Ni content can be understood based on the higher growth rate of the $(Cu,Ni)_6Sn_5$ phase, as reported in Chapter 6. For example, at 5 at.% Ni, the $(Cu,Ni)_6Sn_5$ phase grows with the average phase layer thickness of $80\pm3$ μm after annealing for 81 hrs, as shown in Figure 6.1. On the other hand, as shown in Figure 9.1 for the same annealing time, at 5 at.% Au, the $(Cu,Au)_6Sn_5$ phase grows with the average phase layer thickness of only $14\pm0.5$ μm. This is reflected as the much drastic increase in the interdiffusion coefficient for $Cu_6Sn_5$ with Ni addition in Cu when compared to Au or Pd addition in Cu. We have found in this study that the thicknesses of $(Cu,Au)_6Sn_5$ and $(Cu,Pd)_6Sn_5$ increase when Au and Pd are added in Cu, respectively. This trend is reflected (depending on the overall growth rate of phases) as an increase in the interdiffusion coefficients for the $(Cu,Au)_6Sn_5$ and $(Cu,Pd)_6Sn_5$ phases with the increase in Au and Pd content, respectively. We also found in this study that in Cu(Au)/Sn couples, there is an increase in thickness of both the $(Cu,Au)_6Sn_5$ phase and the total IDZ. On the other hand, we found that in Cu(Pd)/Sn couples, although there is an increase in thickness of the $(Cu,Pd)_6Sn_5$ phase, there is not much change in the thickness of the total IDZ. This is the reason for a little bit higher interdiffusion coefficient for the $(Cu,Au)_6Sn_5$ when compared to the $(Cu,Pd)_6Sn_5$ phase. Moreover, we found that only $(Cu,Pt)_6Sn_5$ has almost 50% increase in thickness when compared to binary $Cu_6Sn_5$ phase and $(Cu,Au)_6Sn_5$ or $(Cu,Pd)_6Sn_5$. Therefore, based on the similar arguments, as discussed above, it can be understood that $\widetilde{D}_{int}^{(Cu,Ni)_6Sn_5,Sn} > \widetilde{D}_{int}^{(Cu,Pt)_6Sn_5,Sn} > \widetilde{D}_{int}^{(Cu,Au)_6Sn_5,Sn} > \widetilde{D}_{int}^{(Cu,Pd)_6Sn_5,Sn}$.

## 9.3 Conclusions

For the first time, the effect of Au, Pd and Pt addition in Cu is studied on the growth of phases and the Kirkendall voids in the Cu–Sn system. Based on this study, we can make 3 very important conclusions:

(1) The addition of Au in Cu does not significantly affect the formation of the Kirkendall voids, as compared to the Cu/Sn binary couple.

(2) The addition of Pd as well Pt in Cu has a strong effect on the growth the Kirkendall voids. Since voids degrade the electro–mechanical contact in the flip–chip bonding, these components may not be suitable to replace Au because of the reliability concern in an electronic component.

(3) Previous studies were mainly concentrated on the role of organic impurities in Cu on the growth of the Kirkendall voids. In this study, we have discussed that even mixing of inorganic metallic components (viz., Pd, Pt, Ni) in Cu can lead to the growth of voids, equivalent to their growth because of organic impurities in Cu.

*To summarize, we demonstrate that both inorganic and organic impurities can play a significant role on the formation and growth of the Kirkendall voids, which is found to increase drastically with Pd, Pt and Ni addition in Cu, and similar behaviour is also found for the case of electroplated Cu and commercial pure Cu, in this study.*

# Chapter 10

# Conclusions

In this PhD research work, for the very first time, a systematic study is conducted in the M–Sn systems (M = Cu, Ni, Au, Pd, Pt) covering the whole possible temperature range in the solid–state, *i.e.*, room temperature (RT) to 215 °C, so as to study the temperature dependent growth of various product phases in the diffusion couples, which were either aged up to a maximum time of 2.5 years at RT or annealed for 4–1000 hrs. For the purpose of comparison, the M/Sn bulk couples are studied along with the M/EP–Sn couples. In electroplated diffusion (EPD) couples, the whole temperature range is examined as the bonding (between the two end–members) is established during electroplating itself; however, in the bulk diffusion (BD) couples it is generally 125–215 °C due to difficulties in joining or proper bonding at lower temperatures. Electroplatings of Cu (for studying growth of voids) and Sn (for making EPD couples) are done by commercial acidic solution used in industries, due to its lower power consumption and compatibility with photoresists. We have considered solid–state diffusion annealing in this study as it is known to mimic the growth of phases in the real systems, after soldering.

In the Cu/Sn diffusion couples, both $Cu_6Sn_5$ and $Cu_3Sn$ grow at temperatures $\geq 100$ °C and only the $Cu_6Sn_5$ phase is found at lower temperatures. A transition from Type B (at lower temperature) to Type A (at higher temperature) regime of grain boundary diffusion is observed for the growth of $Cu_6Sn_5$; such a growth behaviour in the Cu–Sn system has never been studied in the past. The integrated interdiffusion coefficients are estimated in the temperature range of 125–215 °C (Type A regime) because of the observation of parabolic growth. The activation energies are estimated to be 64±4 and 75±6 kJ/mol for $Cu_3Sn$ and $Cu_6Sn_5$, respectively.

The presence of bifurcation of the Kirkendall marker plane, a very special phenomenon discovered recently, is found in a technologically important Cu–Sn system. It was predicted based on estimated diffusion coefficients; however, it could not be detected in the conventional inert marker experiments. As reported in this study, we could detect the locations of these planes based on the microstructural features examined in SEM and TEM. This strengthens the concept of the physico–chemical approach that relates microstructural evolution with the diffusion rates of components and imparts finer understanding of the growth mechanism of phases. The estimated diffusion coefficients at the Kirkendall marker planes indicates that the reason for the growth of the Kirkendall voids is the non–consumption of excess vacancies (by the sinks) which are generated due to unequal diffusion rates of components.

Systematic experiments using different purity of Cu indicates the importance of the presence of impurities on the growth of voids, which increases drastically for $\geq 0.1$ wt.% impurity. The growth of voids increases drastically for electroplated (EP) Cu and commercially pure (CP) Cu. Void size and number distribution analysis indicates the nucleation of new voids along with the growth of existing voids with the increase in annealing time. The newly found location of the Kirkendall marker plane in the $Cu_3Sn$ phase indicates that voids grow on both the sides of this plane which was not considered earlier for developing theoretical models.

In the Ni/Sn diffusion couples, only $Ni_3Sn_4$ grows at higher temperatures. At 50–100 °C, metastable $NiSn_4$ phase is found along with an equilibrium $Ni_3Sn_4$ phase. At RT, only the metastable $NiSn_4$ phase is found. The activation energy for the integrated interdiffusion coefficient of $Ni_3Sn_4$ is estimated as $71.6\pm5.3$ kJ/mol, which is similar to the values estimated in the Cu–Sn system. The growth rate of the product phases, for EPD couples, is found to be a little higher in the Ni–Sn system as compared to the Cu–Sn system. During storage at RT and service at elevated temperatures, the phases in the Ni–Sn system have comparable growth rate when compared to the same in Cu–Sn system.

In the Au/Sn diffusion couples, the number of product phases in the interdiffusion zone decreases with the decrease in annealing temperature. These phases grow with significantly high rates even at RT. The growth rate of the $AuSn_4$ phase is found to be higher in the case of EPD couple because of the relatively small grains and hence high contribution of the grain boundary diffusion when compared to the BD couple. The diffraction pattern analysis indicates that we have the same equilibrium crystal structure of the phases in EPD and BD couples. The analysis in the $AuSn_4$ phase relating the estimated tracer diffusion coefficients with grain size, crystal structure, the homologous temperature of experiments and the concept of the sublattice diffusion mechanism in the intermetallic compounds indicate that Au diffuses mainly via grain boundaries, whereas Sn diffuses via both grain boundaries and lattice.

In the Pd/Sn couples, the $PdSn_4$ phase covers almost whole interdiffusion zone, while $PtSn_4$ is the only phase found in the Pt/Sn couples. The growth rate of the product phase in the Pt–Sn system is found to be much lower compared to the Pd–Sn system and also the Au–Sn system, which is currently used in the microelectronics industry. The time dependent experiments in the Pd–Sn system indicate that the growth rate is parabolic in nature, *i.e.*, it is controlled by the diffusion rates of components through the product phases. However, this is linear and reaction–controlled in the Pt–Sn system, which indicates that the formation of the compound is the rate–limiting step compared to the diffusion rates of components. The marker experiments indicate that both the phases, $PdSn_4$ and $PtSn_4$, grow mainly by the diffusion of Sn along with the negligible diffusion of Pd and Pt, respectively. Furthermore, the analysis considering the same crystal structure (oC20) of these phases along with the concept of sublattice diffusion mechanism indicates that the diffusion rates of both Pd and Pt are negligible via both the lattice and the grain boundaries, when compared to the reasonably high diffusion rate of Au in the $AuSn_4$ phase mainly via grain boundaries.

In this research work, a systematic study is conducted, mainly at 200 °C, in the ternary Cu(M)–Sn systems (M = Ni, Au, Pd, Pt) using bulk diffusion couples, where M represents the addition of an alloying element in Cu.

A strong influence of Ni content in Cu on the diffusion–controlled growth of the $(Cu,Ni)_3Sn$ and $(Cu,Ni)_6Sn_5$ phases by coupling different Cu(Ni) alloys with Sn in the solid–state is reported. The growth of the Kirkendall voids increases drastically with the addition of 0.5 at.% Ni in Cu, which is found to be similar to that of EP Cu and CP Cu, indicating the adverse role of both inorganic and organic impurities. The diffraction pattern analysis indicates that the presence of Ni does not change the crystal structure of $(Cu,Ni)_3Sn$ and $(Cu,Ni)_6Sn_5$. However, it strongly affects the microstructural evolution and diffusion rates of components. The growth rate of $(Cu,Ni)_3Sn$ decreases without changing the diffusion coefficient because of the increase in growth rate of $(Cu,Ni)_6Sn_5$, which grows by consuming $(Cu,Ni)_3Sn$ at the $(Cu,Ni)_3Sn/(Cu,Ni)_6Sn_5$ interface. With the increase in Ni content, driving forces for the diffusion of components through both the $Cu_3Sn$ and $Cu_6Sn_5$ phases increase for both the components (*i.e.*, Sn and Cu), but at different rates. However, the magnitude of these changes alone is not large enough to explain the high difference in the observed growth rate of the product phases due to Ni addition. For 3 at.% or higher Ni addition in Cu, only the $(Cu,Ni)_6Sn_5$ phase grows in the interdiffusion zone. The elongated grains of $(Cu,Ni)_6Sn_5$ are found when it is grown from $(Cu,Ni)_3Sn$. This indicates that the newly formed intermetallic compound joins with the existing grains of the phase. On the other hand, smaller grains are found when this phase grows directly from Cu in the absence of $(Cu,Ni)_3Sn$, indicating the ease of repeated nucleation. Grain size of $(Cu,Ni)_6Sn_5$ decreases with further increase in Ni content, which indicates a further reduction of activation barrier for nucleation. The relations for the estimation of relevant diffusion parameters are established, considering the diffusion mechanism in the Cu(Ni)–Sn system, which is otherwise impossible in the phases with narrow homogeneity range in a ternary system. The flux of Sn increases, whereas the flux of Cu decreases drastically with the addition of very small amount of Ni, such as 0.5 at.% Ni, in Cu. Analysis of the atomic mechanism of diffusion indicates the contribution from both lattice and grain boundary for the growth of $(Cu,Ni)_6Sn_5$.

The growth rates of the product phases $Cu_3Sn$ and $Cu_6Sn_5$ in the Cu–Sn system are affected differently by the addition of Au, Pd and Pt in Cu. The addition of 8 at.% of Au and Pd in Cu produce different (and additional) phases in the interdiffusion zone of ternary Cu(Au)/Sn and Cu(Pd)/Sn diffusion couples, respectively, when compared to the binary Cu/Sn diffusion couple. The addition of Au has very little effect; however, the addition of Pd and Pt significantly increases (both size and number) the formation of the Kirkendall voids. This study clearly indicates that addition of these components (*i.e.*, inorganic impurities) have similar detrimental role like organic impurities in Cu on the growth of the Kirkendall voids, which leads to the degradation of electro–mechanical contact in an electronic component.

There had been intense research on developing a relation to estimate composition dependent diffusion parameters, which does not involve determining the initial contact plane; for example, the relations developed by den Broeder (graphical approach) and Wagner (analytical approach). The den Broeder relation for interdiffusion coefficient is derived analytically, starting from Fick's laws, using a similar line of treatment which is used to derive the Wagner relation. Following this, the relations of integrated interdiffusion coefficient and intrinsic diffusion coefficients of components (which were not available before) utilizing the concentration normalized variable $Y_C$ are derived. For comparing these relations with Wagner's approach, the derivation of relations utilizing the composition normalized variable $Y_N$ are also given. It is shown that the relations derived following both the approaches (*i.e.*, using $Y_C$ and $Y_N$) are theoretically similar, which can be understood considering the constant molar volume in a diffusion couple.

*To conclude, this thesis:*

(i)     it develops a basic understanding of relations used for the estimation of important diffusion parameters.

(ii)    the scientific contributions will help to understand, in particular evolution of phases and their growth kinetics in various UBM–Sn (or Sn–based solder) systems which are relevant to microelectronics industry.

(iii)   in the experiments conducted for this research work, the growth of the Kirkendall voids in the $Cu_3Sn$ phase is found to increase drastically for the case of EP Cu, CP Cu and with Ni, Pd, Pt addition in Cu, which clearly indicates that both inorganic and organic impurities present in Cu play a significant role of on the growth of the Kirkendall voids.

# List of Publications